\documentclass[a4paper,11pt]{book}
\usepackage[latin1]{inputenc}

\usepackage[english]{babel}
\usepackage{bbold}
\usepackage{amsmath} 
\usepackage{wasysym} 
\usepackage{graphicx}
\usepackage{vmargin}
\usepackage{amsthm}
\usepackage{amssymb}
\usepackage{mathrsfs}
\usepackage{amsfonts}
\usepackage{subfig}
\usepackage{enumerate}
\usepackage{slashed}
\usepackage{empheq}
\usepackage{makeidx}
\usepackage{pbsi}
\usepackage{aurical}
\usepackage{booktabs}
\makeindex

\usepackage[titletoc]{appendix}

\def\bea#1\eea{\begin{align}#1\end{align}}
\def\bean#1\eean{\begin{align*}#1\end{align*}}

\def\sffrac#1#2{\leavevmode\kern.1em
\raise.3ex\hbox{\the\scriptscriptfont0 #1}\kern-.1em/\kern-.15em
\lower.2ex\hbox{\the\scriptscriptfont0 #2}}
\newtheoremstyle{myobsstyle} % name
    {\topsep}                    % Space above
    {\topsep}                    % Space below
    {\sffamily\small}                  % Body font
    {}                           % Indent amount
    {\bfseries}                  % Theorem head font
    {.}                          % Punctuation after theorem head
    {.5em}                       % Space after theorem head
    {}  % Theorem head spec (can be left empty, meaning normal)

\newtheoremstyle{mystyle} % name
    {\topsep}                    % Space above
    {\topsep}                    % Space below
    {\sffamily}                  % Body font
    {}                           % Indent amount
    {\bfseries}                  % Theorem head font
    {.}                          % Punctuation after theorem head
    {.5em}                       % Space after theorem head
    {}  % Theorem head spec (can be left empty, meaning normal)

\theoremstyle{mystyle}

\theoremstyle{mystyle}

\theoremstyle{mystyle}

\theoremstyle{myobsstyle}

\theoremstyle{mystyle}

\setmarginsrb{25mm}{25mm}{25mm}{25mm}%
             {10mm}{10mm}{10mm}{10mm}

\renewcommand{\chaptermark}[1]{%
  \ifnum\value{chapter}>0
    \markboth{Chapter \thechapter{}: #1}{}%
  \else
    \markboth{#1}{}%
  \fi}

\usepackage{fancyhdr}
\renewcommand{\chaptermark}[1]{\markboth{\sffamily#1}{}}

\fancypagestyle{plain}{\fancyhead{}}

\allowdisplaybreaks

\usepackage{simplewick}

\usepackage{tikz}
\usetikzlibrary{arrows,shapes}
\usetikzlibrary{trees}
\usetikzlibrary{matrix,arrows} 				% For commutative diagram
\usetikzlibrary{positioning}				% For "above of=" commands
\usetikzlibrary{calc,through}				% For coordinates
\usetikzlibrary{decorations.pathreplacing}  % For curly braces
\usepackage{pgffor}							% For repeating patterns

\usetikzlibrary{decorations.pathmorphing}	% For Feynman Diagrams
\usetikzlibrary{decorations.markings}
\tikzset{
    vector/.style={decorate, decoration={snake}, draw},
	provector/.style={decorate, decoration={snake,amplitude=2.5pt}, draw},
	antivector/.style={decorate, decoration={snake,amplitude=-2.5pt}, draw},
    fermion/.style={draw=black, postaction={decorate},
        decoration={markings,mark=at position .55 with {\arrow[draw=black]{>}}}},
    fermionbar/.style={draw=black, postaction={decorate},
        decoration={markings,mark=at position .55 with {\arrow[draw=black]{<}}}},
    fermionnoarrow/.style={draw=black},
    gluon/.style={decorate, draw=black,
        decoration={coil,amplitude=4pt, segment length=5pt}},
    scalar/.style={dashed,draw=black, postaction={decorate},
        decoration={markings,mark=at position .55 with {\arrow[draw=black]{>}}}},
    scalarbar/.style={dashed,draw=black, postaction={decorate},
        decoration={markings,mark=at position .55 with {\arrow[draw=black]{<}}}},
    scalarnoarrow/.style={dashed,draw=black},
    electron/.style={draw=black, postaction={decorate},
        decoration={markings,mark=at position .55 with {\arrow[draw=black]{>}}}},
	bigvector/.style={decorate, decoration={snake,amplitude=4pt}, draw},
}

\tikzstyle{block} = [draw, rectangle, 
    minimum height=3em, minimum width=6em]

	\usepackage[Bjornstrup]{fncychap}
	\usepackage{titlesec}
\usepackage[hidelinks,bookmarksnumbered]{hyperref}
\hypersetup{colorlinks=false}

\newcommand{ \Tr }{\text{Tr}}

\newcommand{ \de }{\partial}
\renewcommand{\d}{d}

\newcommand{ \as }{\alpha_{s}}

\usepackage{framed, color}
\definecolor{shadecolor}{rgb}{0.8,0.8,0.8}

\begin{document}

\thispagestyle{empty}
% \frontmatter
\pagenumbering{roman}

\begin{titlepage}

\begin{center}
\begin{figure}[htbp]
\includegraphics[width=16cm]{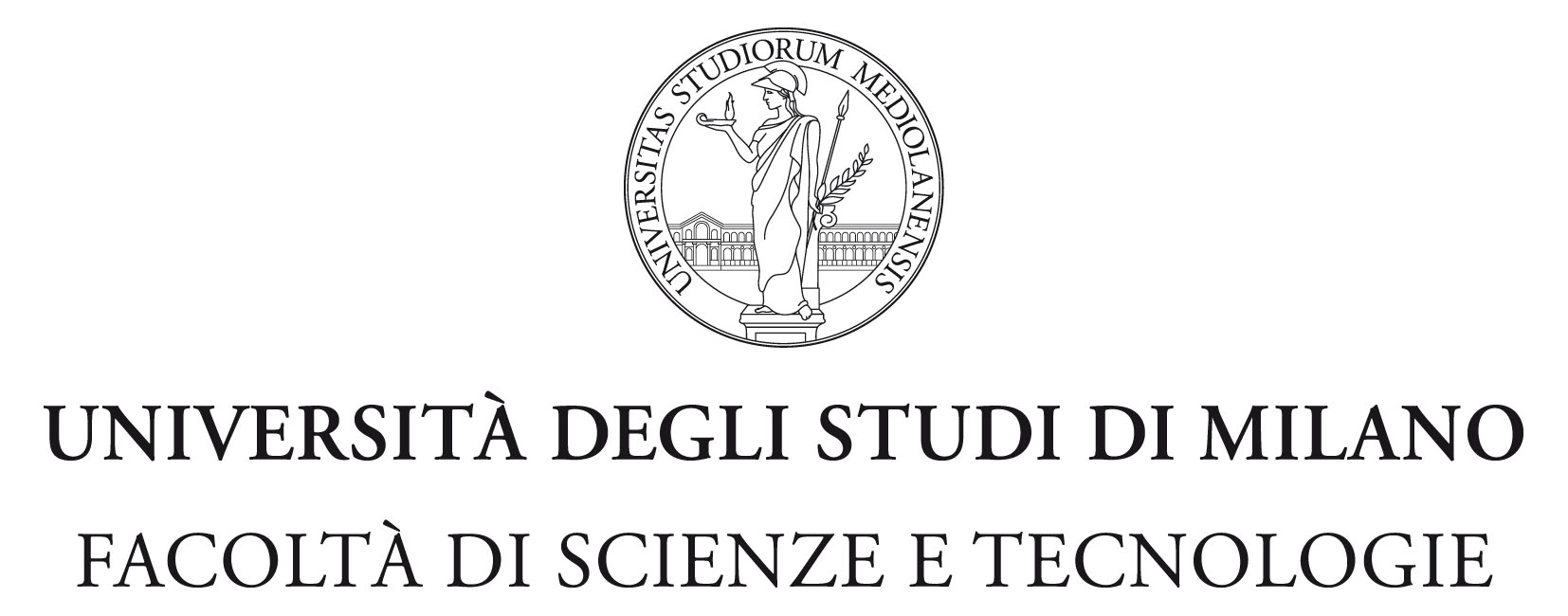}
\end{figure}

\large\emph{Corso di Laurea Magistrale in Fisica}

\end{center}

\vskip 1.5cm

\begin{center}

\large\textbf{Tesi di Laurea in Fisica}

\vskip 0.5cm

\LARGE\textbf{Threshold resummation in SCET vs. direct QCD:\\ a systematic comparison}
 
\end{center}
\vskip 1.5cm

\begin{large}

\begin{flushleft}
 \emph{Relatore interno}: Prof. S. Forte
\vskip 0.5cm
 \emph{Correlatore}: Prof. G. Ridolfi
\vskip 0.5 cm
\end{flushleft}

\begin{flushright}
Luca Rottoli

matr. 808317
\end{flushright}

\vskip 1.5 cm

\begin{center}
Anno Accademico 2012/2013
\end{center}

\end{large}

\end{titlepage}

\thispagestyle{empty}
\tableofcontents
\cleardoublepage

%\title{Threshold resummation in SCET vs. direct QCD: a systematic comparison}
%\author{Luca Rottoli}
%\maketitle

\chapter*{Introduction}

The discovery of a Higgs boson at the LHC brought enthusiasm to the high
energy physics community and made physicists even more aware that accurate
theoretical predictions are of the utmost importance in order to achieve a
deeper understanding of the fundamental laws of nature. The next years will
be dedicated to the study of the detailed properties of this new particle
and to extend the energy frontier looking for physics beyond the Standard Model.
For this reason, a detailed study of high-energy processes will be necessary in order
to understand whether we are observing Standard Model physics or new physics when
performing precision measurement in hadron colliders.

In order to obtain reliable predictions for the cross-sections it is 
necessary to compute all the relevant corrections. 
Higher order perturbative QCD calculations are necessary in order 
to obtain a detailed comprehension of hadronic final states properties.
In some cases, enhanced logarithmically contributions appear at any order 
in perturbation theory. It is thus necessary to go beyond fixed-order 
perturbation theory and sum these contributions to all orders in the
strong coupling constant $\alpha_s$: hence resummation, which basically consists
in a reorganization of the perturbative expansion by an all-order summation of classes of logs.
This organization has been carried out in a variety of ways 
in a standard ``direct'' QCD (dQCD) approach.

In recent years an alternative approach to resummation, based 
on soft-collinear effective theory (SCET) has been proposed.
In this latter approach the soft scale whose logarithms are resummed is a soft scale 
$\mu_s$. Several choice for the soft scale $\mu_s$ are possible.
In particular, one can choose the soft scale as a function of the dimensionless hadronic variable $\tau$.
This choice avoids problems related to the Landau pole in the strong coupling, 
which have to be faced in the traditional dQCD formalism.

An increasing interest in a deeper comprehension of the main 
similarities and differences between the two approaches has grown. In particular,
the analytic equivalence of the 
dQCD and SCET approaches has been explored at various level. However,
it would be interesting to investigate the phenomenological 
implications of this state of affairs.

In this Thesis we will perform a systematic comparison of soft-gluon resummation in SCET and in dQCD,
both from an analytical and a phenomenological point of view.
In particular, we will concentrate on Higgs boson production in gluon-gluon fusion
in a hadron-hadron collider. It is indeed well know that the main contribution to Higgs production 
in the gluon-gluon channel comes from the threshold region; for this reason,
it is interesting to compare the predictions of the two approaches in 
order to obtain a deeper comprehension of possibile differences in the resummation procedure.
In particular, we will show how a systematic approach can be more easily performed
by means of a saddle-point argument. We will present 
the results we have obtained with this strategy and we will discuss
some open issues which would be interesting to further investigate.

The Thesis is organized as follows. In Chapter~\ref{chapter:sinphys} we will present a very short 
review of QCD in order to introduce some notation which we will use in the following
Chapters. In Chapter~\ref{chapter:resummation} we will summarize some general features of soft-gluon
(or threshold) resummation for inclusive cross-sections. 
We will then apply soft-gluon resummation in the dQCD approach
to a specific process, the Higgs production at LHC in gluon-gluon fusion, in Chapter~\ref{chapter:phenres}. 
In the next Chapter, we will introduce the SCET approach to soft-gluon resummation,
focusing on Drell-Yan process and Higgs production. In Chapter~\ref{chapter:ancomp} we will then
compare soft-gluon resummation in QCD as performed in the standard 
perturbative dQCD formalism, to resummation based on SCET in the Becher-Neubert approach. 
In the final Chapter we will perform a systematical comparison of soft-gluon resummation 
in SCET and in dQCD. In particular, we will present results for Higgs boson production
in gluon-gluon fusion at the LHC.

\cleardoublepage
% \mainmatter

\pagenumbering{arabic}

\chapter{Strong interaction physics}\label{chapter:sinphys}

The following Chapter is dedicated to a brief overview of Quantum Chromodynamics (QCD), 
the modern theory of strong interactions. We intend this chapter
as a very short review in order to introduce some notations that will be useful in the
next Chapters. For a complete review, we refer the reader to \cite{pinkbook}.

\section{Quantum Chromodynamics}

Quantum Chromodynamics is a gauge theory based on the gauge group $SU(3)$, with Lagrangian 
\bea\label{eq:qcd_lagr}
\mathcal L_{\textrm{QCD}} = \sum_f \bar \psi_i^{(f)} (i \bar D_{ij} -m_f \delta_{ij}) \psi_j^{(f)}
-\frac14 F^{\mu\nu}_a F_{\mu\nu}^a,
\eea
where we have introduced the covariant derivative
\bea
D^\mu_{ij} \equiv \de^\mu \delta_{ij} + i g_s t^a_{ij} A^\mu_{a}
\eea
and the field strength
\bea
F_{\mu\nu}^a = \de_\mu A^a_{\nu}-\de_{\nu} A^a_{\mu} - g_s f_{abc} A^b_{\mu} A^c_{\nu}.
\eea
By looking at Eq.~(\ref{eq:qcd_lagr}) we notice that a single parameter, $g_s$, regulates the strength of the interaction. The fermionic fields,
called quarks, have all the same coupling to the gauge fields, i.e. QCD is flavour blind.
Eight self-interacting gauge fields, the gluons, have been introduced in order to 
preserve local gauge invariance.  

As we have stated above, the gauge group of QCD is $SU(N)$, with $N=3$. 
The indexes $i,\ j$ in Eq.~(\ref{eq:qcd_lagr}) run from 1 to $N$, whereas the group 
indexes $a,\ b, \ c$ run from $1$ to $N^2-1$, which is the dimension of the 
fundamental representation of $SU(N)$. The eight matrixes $t^a_{ij}$ are usually normalized as
\bea
\Tr (t_a t_b) = T_F \delta_{ab}
\eea 
with $T_F = \frac12$. They satisfy the Lie algebra
\bea
[t_a,t_b] = f_{ab}^{\ \ c} t_c,
\eea
where $f_{abc}$ are the real structure constants of the $SU(N)$ algebra, and are fully 
antisymmetric. The invariants $C_F$ and $C_A$ are defined by
\bea
\sum_a t^a_{ij} t^a_{kj} &= C_F \delta_{ij}\\
\sum_{cd} f^{acd}f^{bdc} &= C_A \delta^{ab},
\eea
where
\bea
C_F &= T_F\frac{N^2-1}{N}\\
C_A &= N.
\eea
Thus for QCD we obtain
\bea
C_F =\frac43, \qquad C_A = 3.
\eea

\begin{table*}[t]
\begin{center}
\begin{tabular}[c]{l|cccccc}
flavour              & $d$ & $u$ & $s$ & $c$ & $b$ & $t$ \\
  \midrule
name & down & up & strange & charm & bottom & top \\
mass & $\sim 2.5$ MeV & $\sim 5$ MeV & $0.1$ GeV & $1.3$ GeV & $4.2$ GeV & $173$ GeV \\
charge     & $-1/3$ & $2/3$ & $-1/3$ & $2/3$ & $-1/3$ & $2/3$  \\
\end{tabular}
\end{center}
\caption{Masses and charges of the quarks.}
\end{table*}

In order to quantize the classical lagrangian Eq.~(\ref{eq:qcd_lagr}) it is necessary to 
introduce gauge fixing and ghost terms. We refer the reader to any QFT textbook for details. 
 The physical vertices in Eq.~(\ref{eq:qcd_lagr})
include not only a gluon-quark-antiquark vertex, which is analogous to the electromagnetic 
vertex in QED, but also a 3-gluon and a 4-gluon vertex, which have no analogue in QED.

The theory has an additional exact $U(1)$ symmetry, which corresponds to the conservation 
of the barion number. Furthermore, it is well know that the QCD spectrum
presents an extra accidental symmetry, which corresponds to the conservation of
the isospin number.  
The masses of the up quark and of the down quark are very small with respect to
the lightest bound states:
\bea
m_u, \, m_d \ll m_\pi,\, m_N.
\eea
For all practical purposes, the two lightest quarks can be considered massless.
The theory has therefore an additional approximate global $U(2)$ flavour symmetry,
i.e. isospin. If we neglect the
mass of the strange quark, the symmetry can be enlarged to an approximate $U(3)$ 
symmetry, which is at the basis of the Gell-Mann quark model. 

Since the two chiral 
components of quark fields are independent in the massless limit, if the 
masses of the up and down quark are neglected the Lagrangian has the larger symmetry
\bea
SU_L(2) \times SU_R(2) \times U(1)_L \times U(1)_R,
\eea
or equivalently, considering vector and axial combinations, 
\bea
SU_A(2) \times SU_V(2) \times U(1)_A \times U(1)_V.
\eea
$U(1)_V$ and $SU(2)_V$ are good symmetry in nature, which correspond respectively 
to the conservation of the baryon number and isospin number. The axial symmetry is
spontaneously broken; in this case we expect to find the existence of four Goldstone
bosons in the QCD spectrum. However, we observe only
three Goldstone bosons, the pions. The absence of the fourth boson was known as the
$U(1)_A$ problem: $U(1)_A$ is not realized neither à la Goldstone, neither à la
Wigner-Weyl. The solution of the $U(1)_A$ problem was separately found by 
Jackiv and 't Hooft and relies on the highly non trivial topological vacuum structure
of QCD.

\section{The running coupling constant and the Landau pole}

We define the QCD coupling constant as
\bea
\alpha_s = \frac{g_s^2}{4\pi}.
\eea
The renormalization group equation reads
\bea
\mu^2 \frac{d}{d\mu^2}\alpha_s(\mu^2) = \beta(\alpha_s(\mu^2)),
\eea
where the $\beta$-function starts at order $\alpha_s^2 $ and has the expansion
\bea
\beta(\alpha_s) = -\alpha_s^2(\beta_0+\beta_1\alpha_s+\beta_2\alpha_s^2+\ldots).
\eea
One finds
\bea
\beta_0 = \frac{11 C_A-2n_f}{12 \pi},
\eea
which is positive provided $n_f \leq 16$. If the coupling constant is small 
it is possible to compute the $\beta$-function in perturbation theory. 
The sign in front of $\beta_0$ is crucial: it decides the slope of the coupling.
Since in QCD $\beta$ is negative at small $\alpha_s$, the running 
coupling decreases as $\mu $ increases.

A theory like QCD, where the running coupling vanishes asymptotically 
at high energies is called (ultraviolet) \emph{asymptotically free}. 
It has been proven that in 4 spacetime dimensions all and only non-abelian gauge
theories are asymptotically free.

We can write the solution for $\alpha_s$ at one loop (leading log approximation):
\bea\label{eq:rgeLL}
\alpha_s(\mu^2)=\frac{\alpha(Q^2)}{1+\beta_0 \alpha(Q^2)\log \frac{\mu^2}{Q^2}}
\eea
with the initial condition $\alpha_s(Q^2)$. Usually in QCD one defines 
\bea\label{eq:alpha_run}
\frac{1}{\alpha(Q^2)}= \beta_0 \log\frac{Q^2}{\Lambda_{QCD}^2}
\eea
so that we can trade the parameter $Q$ for the dimensional parameter 
$\Lambda_{QCD}$ and we can write
\bea\label{eq:alpha_Lambda}
\alpha(\mu^2) = \frac{1}{\beta_0 \log \frac{\mu^2}{\Lambda_{QCD}^2}}.
\eea
The scale $\Lambda_{QCD}$ is of the order of a few hundred MeV.
Eqs.~(\ref{eq:rgeLL})-(\ref{eq:alpha_Lambda}) clarify what we intend for asymptotic freedom:
the coupling constant decreases logarithmically with $\mu^2$. In the
leading log approximation the coupling constant has been replaced in a way
by the value of $\Lambda_{QCD}$. $\Lambda_{QCD}$ depends
on the particular definition of $\alpha_s$, not only on the defining scale
$Q$ but also on the renormalization scheme (beyond leading order). Moreover, it 
also depends on the active number $n_f$ of coupled flavours. 
%Up to the scale $\mu$
%only quarks with masses $m \ll \mu$ contribute to the running of $\alpha_s$.
In QED and QCD, the effects of heavy quarks are power suppressed 
and can be taken separately into account. 
This is very important, since all applications of perturbative QCD so far
apply to energies below $m_t$.
In conclusion, in QED and QCD, quarks with a mass higher than the scale $\mu$ do
not contribute to $n_f$ in the coefficients of the relevant $\beta$ function.

One notes the presence of a singularity in Eqs.~(\ref{eq:alpha_run},\ref{eq:alpha_Lambda})
called the Landau pole. The coupling constant $\as$ blows up at
\bea
\mu_L^2 = \Lambda_{QCD}^2 = Q^2 \exp \left(-\frac{1}{\beta_0 \as(\mu^2)} \right).
\eea
In this region, where $\mu \sim \Lambda_{QCD}$, QCD is non-perturbative.
The scale $\Lambda_{QCD}$ is typically of the order of some hundreds of MeV. 
The problem of the definition and the behaviour of the physical coupling
in the region around the perturbative Landau pole is an issue that lies 
outside the domain of perturbative QCD.

Due to the failure of perturbation theory, we cannot predict the
behavior of the coupling constant at low energy. We however observe that in
this regime QCD is strongly coupled.  Moreover, 
quarks can form bound states, which are the well-known hadrons. We have only
a phenomenological knowledge of the nonperturbative QCD. When a bound $q\bar q$ pair 
is forced to separate, the potential between the quarks increases. At some separation
length, it becomes convenient for the $q \bar q$ pair to split up in two $q\bar q$ pairs. 
The strong color attraction precludes the possibility of observing isolated quarks.
This experimentally checked effect is known as \emph{confinement} and
it is observed whenever a high-energy collision takes place. 
The additional quarks or gluons produced in the collisions dress themselves 
into hadrons and form jets of particle which are seen in the detectors.

The most important non-perturbative method at present is the technique 
of lattice simulations which has produced very valuable results on
confinement, phase transitions, bound states, and so on, 
and it is by now an established basic tool. Nevertheless, a full non-perturbative 
explanation of confinement has not been achieved yet. 

\section{The parton model}

Scattering experiments have been the main source of information 
on hadronic structure in the second half of the twentieth century. In a typical
scattering experiment a beam of high-energy leptons (usually electrons) is made
to collide with hadrons.
When the momentum transfer of the virtual photon 
which mediates the collision is sufficiently large, a system with a large number of 
hadrons is produced:
\bea
e^-(k) + N(p) \rightarrow e^-(k') + X.
\eea
This process is called deep inelastic scattering (DIS).

The differential cross-section for electron-nucleon scattering can be parametrized 
by structure functions $F_i(x_{Bj},Q^2)$, which are determined by experiment,
where 
\bea
Q^2 = -(k-k')^2=-q^2, \qquad x_{Bj}\equiv \frac{Q^2}{2 p\cdot  q}.
\eea
The experimental data show that in the limit $Q^2 \rightarrow \infty$, but for 
finite value of $x_{Bj}$
the structure functions obey approximate Bjorken scaling 
(which in reality is broken by logarithmic corrections that can be computed in QCD, as we 
will see):
\bea
F_i(Q^2,  x_{Bj}) \stackrel{Q^2\rightarrow \infty}{\longrightarrow} F_i(x_{Bj}).
\eea

Bjorken and Feynman proposed a simple model, the so-called 
\emph{parton model}, in order to explain this phenomenon. The basic assumption of this
model is that the proton is made up of a smaller point-like, non interacting
constituents, called partons. Among these constituents one can find the fermions 
which carry electric charge, namely the quarks (and antiquarks), and possibly other neutral
constituents. In the parton model, each parton is characterized by the fraction $x$ of the 
hadron's total momentum that it carries. Consequently, for each species $i$
of parton, there is a function $f_i(x)$ which express the probability 
that a parton $i$ carries a fraction $x$ of the total momentum of the nucleon. 
The functions $f_i(x)$ are called \emph{parton distribution functions} (PDF) 
and must be experimentally determined. 
%One of the most important property
%of PDFs is their universality: they do not depend upon the particular process 
%considered.

We naively expect the validity of the momentum sum rule, which expresses the
conservation of incoming total momentum:
\bea
\int_0^1 dx \sum_i x f_i^{(p)}(x) = 1.
\eea
We also expect the proton flavour to be conserved. For example, 
\bea
\int_0^1 dx \left(f_u^{(p)}(x) - f_{\bar u}^{(p)} (x) \right) =2
\qquad
\int_0^1 dx \left(f_d^{(p)}(x) - f_{\bar d}^{(p)} (x) \right) =1.
\eea
Measurements confirm these naive expectations. One discovers however that the
fraction of total momentum carried by valence and sea quarks is about one-half;
the remaining half is in fact carried by the neutral gluons, the quanta of
the strong nuclear force. 
%This means that the gluon PDF should be quite sizable.

In conclusion, the parton model enables the computation of several high energy processes with hadrons in
the initial state using only a simple set of assumptions: the hadrons are made of partons,
whose momenta are distributed according to PDFs. Finally, transverse momenta and
masses of the partons must be neglected.

\begin{figure}[htbp]
\begin{center}
\includegraphics[width=9cm]{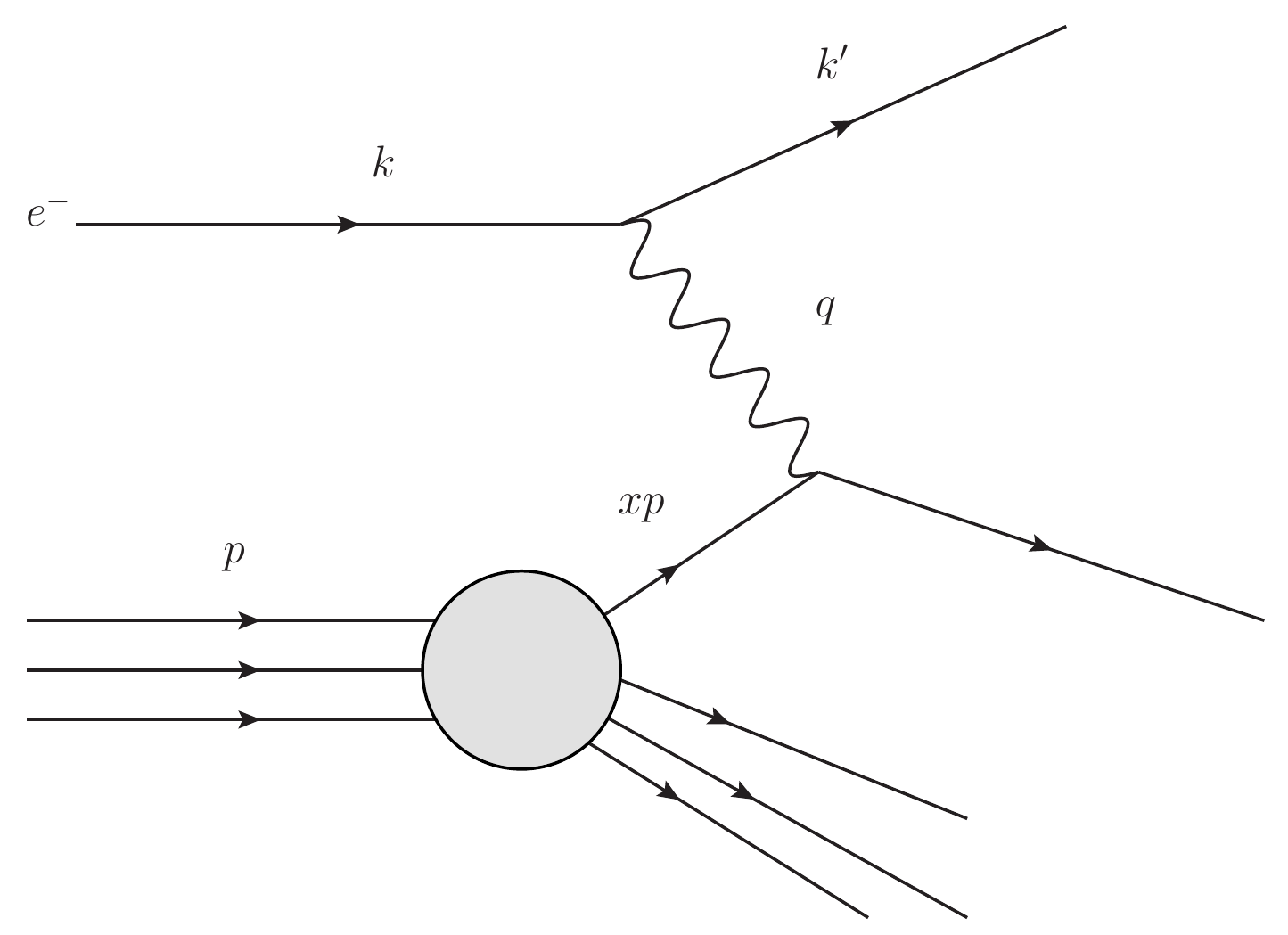}
\end{center}
\caption{Kinematics for the DIS process in the parton model framework.}\label{fig:DIS}
\end{figure}

We can now apply the parton model to inclusive DIS and verify that, despite its simplicity, 
it imposes strong constraints on the cross-section. The kinematics variables 
of the process are shown in Fig.~\ref{fig:DIS}. We defined the additional variables
\bea
s = (k+p)^2, \qquad 
y = \frac{p \cdot q}{k \cdot p}.
\eea
The observation of the outgoing electrons enables a determination of $s, \ x_{Bj}$ and 
$y$, which is the electron fractional energy loss in the laboratory frame 
of a fixed target experiment\footnote{In the centre-of-mass of the electron-quark 
system, $y=\frac{1-\cos \theta_{el}}{2}$ where $\theta_{el}$ is the electron 
scattering angle.}.

The Feynman rules of QED allow for a 
computation the partonic cross-section
\bea
\frac{d \hat \sigma}{d \hat y} = q_l^2 \frac{ \hat s}{Q^4} 2 \pi 
\alpha^2(1+(1-\hat y)^2)
\eea
where $q_l$ is the electric charge of the (anti)quark $l$ and
\bea
\hat p = x p, \qquad \hat s&= (k +\hat p)^2= 2 k \cdot p,  \qquad
\qquad (\hat p + q)^2=2 \hat p \cdot q -Q^2 =0,\nonumber \\
\hat y &= \frac{\hat p \cdot q}{k \cdot \hat p} = y,  \qquad x_{Bj}=\frac{Q^2}{2 p \cdot q}
= x \frac{Q^2}{2 \hat p \cdot q}=x.
\eea 
According to the parton model, the hadronic cross-section can be computed as
\bea
\frac{d \sigma}{d y} = \int dx \sum_l f_l(x) \frac{d \hat \sigma_l}{d y}.
\eea
We thus have 
\bea\label{eq:DIS_crs}
\frac{d \sigma}{d y dx_{Bj}} &= \sum_l f_l(x) \frac{d \hat \sigma}{d y}\nonumber \\
&= \frac{ 2 \pi \alpha^2 s x_{Bj}}{Q^4} (1 + (1-y)^2) 
\sum_l q_l^2 f_l(x_Bj).
\eea

Looking at Eq.~(\ref{eq:DIS_crs}) we learn that the parton model, in its simplicity,
makes remarkable predictions. It shows that at fixed $x_{Bj}$
and $y$ the cross-section scales with the energy $s$. Moreover, it fully predicts the
$y$-dependance of the cross-section, which is typical of vector interaction with
fermions (Callan-Gross relation). It is customary to write the cross-section 
Eq.~(\ref{eq:DIS_crs}) introducing the structure function $F_2(x)$,
which obeys Bjorken scaling:
\bea
\frac{d \sigma}{dy dx} = \frac{2 \pi \alpha^2 s}{Q^4} (1 + (1-y)^2) F_2(x) 
\qquad F_2(x) = \sum_{l} x q_l^2 f_l(x).
\eea
In particular, for electron scattering on proton,
\bea
F_2(x) = x \left(\frac49 u_p(x)+\frac19 d_p(x) \right),
\eea
whereas for electron scattering on a neutron
\bea
F_2^n(x) = x \left(\frac19 d_n(x)+\frac49 u_n(x) \right)
= x \left(\frac49 d_p(x)+\frac19 u_p(x) \right)
\eea
and therefore knowledge of
$F_2^n$ and $F_2^p$ allows a determination of $u_p$ and $d_p$ separately. 

\begin{figure}[htbp]
\begin{center}
\includegraphics[width=9cm]{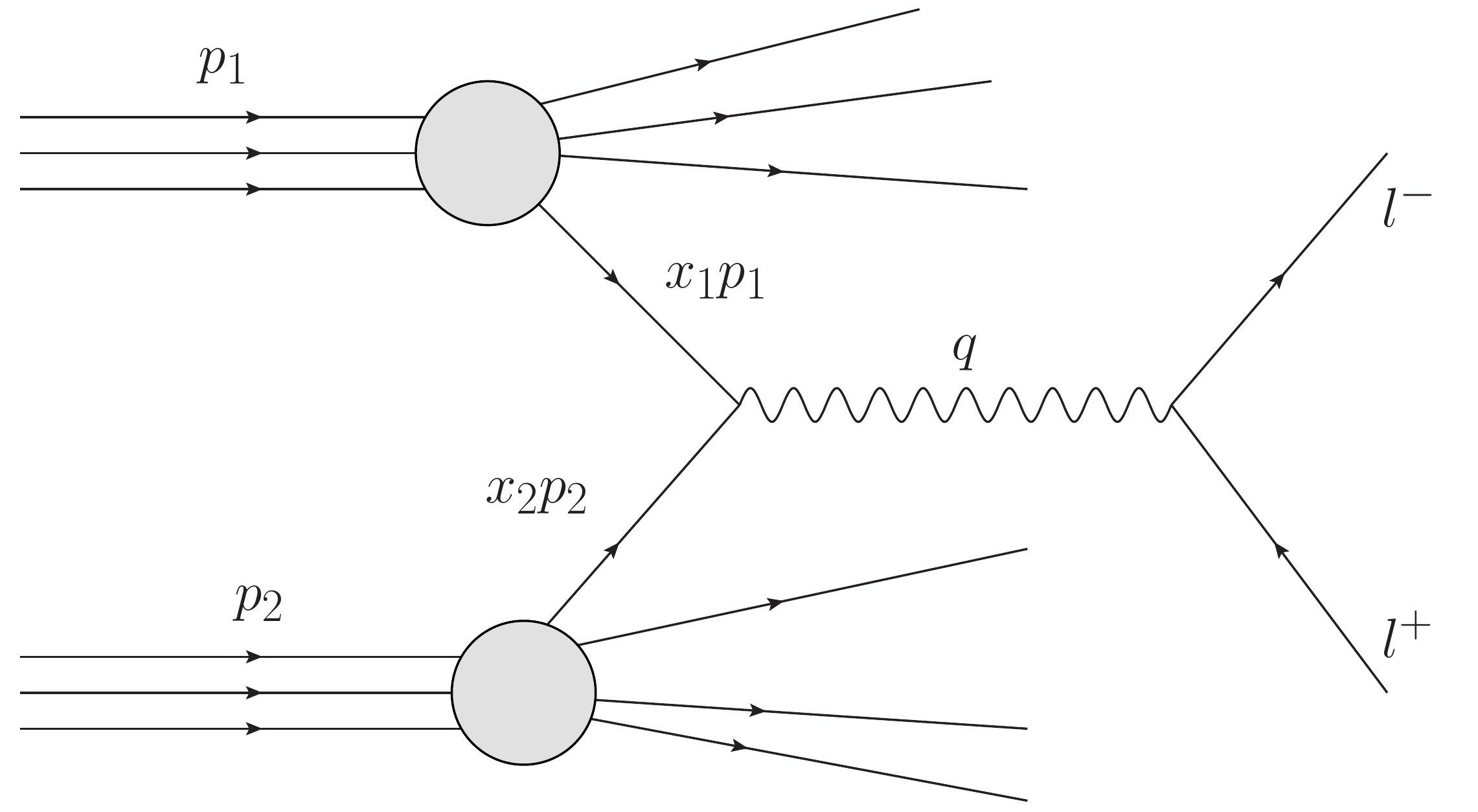}
\end{center}
\caption{Kinematics for the DY process in the parton model framework.}\label{fig:DY}
\end{figure}

The parton model can be used also for other processes. In the so-called Drell-Yan
process, a massive lepton pair emerges from a $q \bar q$ annihilation in a 
hadron-hadron collision. The Drell-Yan process is depicted in 
Fig.~\ref{fig:DY} in the parton model interpretation.
In this case, the partonic kinematic variables are
\bea
\hat p_i = x_i p_i, \quad i=1,2, \qquad 
s=(p_1+p_2)^2 = 2 p_1 \cdot p_2, \qquad Q^2 = q^2  .
\eea
The partonic cross-section is
\bea
\frac{d\hat \sigma^{DY}}{d Q^2}  = q_l^2 \frac{ 4 \pi \alpha}{9  Q^2} \delta(\hat s-Q^2).
\eea
where $\hat s= x_1 x_2 s$.
The hadronic cross-section in the parton model interpretation, where 
$Q^2$ is large, is then
\bea
\frac{d \sigma^{DY}}{d Q^2} = 
\sum_l \int dx_1 dx_2 \left( f_l^{(H_1)}(x_1)f_{\bar l}^{(H_2)}(x_2)
+ ( l \leftrightarrow \bar l)\right)\delta(x_1 x_2 s-Q^2) \sum_l q_l^2 \frac{4 \pi \alpha}{9 Q^2}.
\eea

In the parton model interpretation we can compute the cross-section for
several scattering processes, in which also gluons could enter in the initial state.
However not all hadronic processes can be computed in this way.
There is a rule of thumb which enables one to decide 
whether the process is a hard process or not, in the parton model context.
If the process is insensitive to 
the initial transverse momentum of the partons, which is of the order 
of typical hadronic scales, we can obtain cross-sections
in the parton model safely. The parton densities, in fact, do not carry any
information about the initial transverse momentum of the partons.

\begin{figure}[htbp]
\begin{center}
\includegraphics[width=15cm]{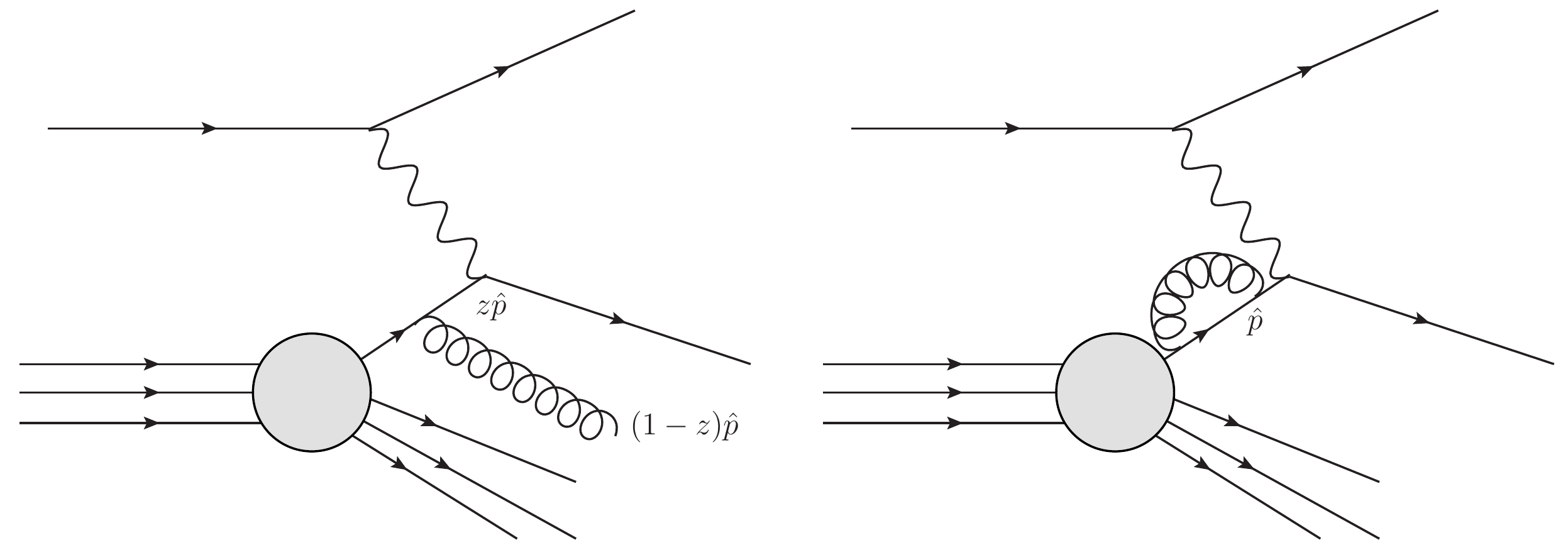}
\end{center}
\caption{Virtual and real radiative contribution to DIS.}\label{fig:DISvr}
\end{figure}

The parton model works well at the first order, but it is completely useless
at the next order correction, because of the appearance of collinear singularities.
For simplicity, let us consider again DIS.
To first order in the coupling it is necessary to consider the emission of
one real gluon and a virtual one (see Fig.~\ref{fig:DISvr}). 
If one adds real and virtual contributions, the partonic cross-section reads
\bea\label{eq:sigma_1}
\sigma^{(1)} = \frac{C_F \alpha_s}{2 \pi} \int dz 
\frac{d k_\perp^2}{k_\perp^2} \frac{1+z^2}{1-z} 
\left( \sigma^{(0)} ( z \hat p) - \sigma^{(0)} (\hat p) \right).
\eea
We see that there is a singularity at $z=1$ which cancels between real and virtual
corrections. This region corresponds to the soft limit. The cross-section 
Eq.~(\ref{eq:sigma_1}) thus does not have soft singularities. However, there is also a
collinear singularity in the limit $k_\perp \rightarrow 0$ for finite $z$. 
This singularity does not cancel.

The naive parton model therefore does not survive radiative corrections. As we will see
in the next Section, collinear divergences from initial state emissions are 
absorbed into parton distributions functions, similarly to what happens when
UV divergences are renormalized. 

\section{Improved parton model}

The presence of collinear divergences makes clear that the parton model is,
if not totally wrong, rather incomplete. 
The parton model today is viewed as an approximate consequence
of the leading order perturbative treatment of QCD, the modern theory of strong 
interactions.

In the real world, divergences do not exist. There are several remedies we could 
take in order to make the divergences go away, for example by introducing the quark masses,
or put a lower cutoff $\lambda$ in the transverse momentum integral. If we do so,
however, the cross-section would acquire a strong dependence upon low energy details,
such as quark masses or confinement effects. These dependences would in the end
spoil the parton model, since by assumptions these details should not count.

It is possible however to rescue the parton model if we make some improvements. 
We now fix some notation which will be useful in the following Chapters.
We define the Altarelli-Parisi splitting function
\bea
P_{qq}^{(0)}(z) = C_F \left( \frac{1+z^2}{1-z} \right)_+,
\eea
where we have introduced the \emph{plus prescription}. The plus 
distribution naturally arises from the cancellation of soft divergences; it is
defined as
\bea
\int_0^1 dz\ [f(z)]_+ g(z) = \int_0^1 dz\ f(z)[g(z)-g(1)]. 
\eea
We introduce a cutoff $\lambda$ and we rewrite Eq.~(\ref{eq:sigma_1}) as
\bea
\sigma^{(1)} = \frac{\alpha_s}{2 \pi} \log\frac{Q^2}{\lambda^2}
\int dz\ P_{qq}^{(0)} (z) \sigma^{(0)}(z \hat p). 
\eea
The partonic cross-section is found to be
\bea
\sigma(\hat p) = \sigma^{(0)}(\hat p) + \sigma^{(1)}(\hat p)=
\int dz \left(\delta(1-z) + \frac{\alpha_s}{2 \pi} \log \frac{Q^2}{\lambda^2}
P_{qq}^{(0)}(z) \right) \sigma^{(0)} (z \hat p). 
\eea
If we insert this formula in the parton model formula for the hadronic cross-section
we obtain
\bea
\sigma(p) = \int dydz\ f_q(y) \Gamma_{qq}(z,Q^2) \sigma^{(0)}(zyp)
\eea
where we have defined
\bea
 \Gamma_{qq}(z,Q^2) = \delta(1-z) + \frac{\alpha_s}{2 \pi} \log \frac{Q^2}{\lambda^2}
P_{qq}^{(0)}(z).
\eea
We then introduce an intermediate scale $\mu_F$ and we recombine the logarithms
using
\bea
\log \frac{Q^2}{\lambda^2} = \log\frac{Q^2}{\mu_F^2}+ \log \frac{\mu_F^2}{\lambda^2}
\eea
and we rewrite the hadronic cross-section (up to $\mathcal O(\alpha_s^2)$
corrections) as
\bea
\sigma(p) = \int dx\ f_q(x, \mu_F^2) 
\hat \sigma(xp, \mu^2)
\eea
where
\bea\label{eq:fmu}
f_q(x, \mu_F^2) = \int dy dz f_q(y) \Gamma_{qq}(z, Q^2) \delta(x-zy).
\eea

This means that the radiative corrections to the parton process can be 
absorbed redefining the parton densities, which acquire a scale dependence. 
This redefinition is universal and does not depend upon the hard process we are 
considering. In the improved parton model the physical cross section can be defined 
in terms of the new scale-dependent PDFs. We can write (we omit
the sum over flavours)
\bea\label{eq:qcd_impr}
\sigma = \int dx_1 dx_2 f_1^{(P_1)}(x_1,\mu^2) f_2^{(P_2)}(x_2, \mu_F^2)
\hat \sigma (x_1 x_2 s, \mu_F^2).
\eea

In the QCD-improved parton model a short distance cross-section $\hat \sigma$ appears,
which is obtained by subtracting the long distance part from the partonic cross-section.
In this way it is possible to rely on perturbation theory. The introduction of
the factorization scale allows to shift the divergent contribution into 
non-perturbative PDF. A choice of the
factorization scale $\mu_F$ similar to $Q$ avoids large logarithms in the short
distance cross-section. 

The argument we have presented was carried out only at leading order in 
perturbation theory. However, there are several arguments which show 
that Eq.~(\ref{eq:qcd_impr}) holds to all orders in perturbation theory.
This result is known as factorization theorem. In case of DIS, a solid proof of 
the theorem, which relies on a clever analytic continuation property of the 
DIS cross-section, exists. For production processes in hadronic collisions the
situation is more difficult. All-order arguments for factorization have been
given in Ref.~\cite{CSS}. Today the factorization theorem is widely 
accepted. 

One can improve the accuracy of the
hadronic cross-section by extracting from experiment PDFs at higher order and 
computing the short distance cross-section at the desired level of accuracy:
\bea
\hat \sigma = [\hat \sigma_0+a\hat  \sigma_1+ a^2 \sigma_2^2+\ldots]
\eea
with\footnote{The coupling constant $\alpha_s$
is evaluated at the renormalization scale $\mu_R$, which is in principle 
different from the factorization scale $\mu_F$. For simplicity, from now on we will consider
only one scale $\mu=\mu_F=\mu_R$.} $a=\alpha_s(\mu_R)/(2\pi)$.

\section{DGLAP equations}

We have seen how collinear singularities due to an initial state parton do not
cancel. Fortunately, it is possible to include into the PDFs 
initial state emission with $k_\perp$ below a given scale. The price to pay 
is the introduction of a factorization scale $\mu$, which separates the non-perturbative, low energy
dynamics from the perturbative hard cross section. As for the
renormalization scale, the dependence of cross-section on $\mu$ is due to the
fact that the perturbative expansion has been truncated. The $\mu$-dependence
becomes milder when including higher orders. 

The new scale-dependent PDFs contains uncalculable long distance effects. These effects
can be measured by using Eq.~(\ref{eq:qcd_impr}) with some reference hard process.
In this way it is possible to obtain the PDF at a given scale $\mu$. It is
however possible to obtain the $\mu$ dependence of PDFs by deriving Eq.~(\ref{eq:fmu}).
By doing so, the master equation of QCD, the Altarelli-Parisi (AP) 
equation (or Dokshitzer-Gribov-Lipatov-Altarelli-Parisi equation) is obtained:
\bea
\mu^2 \frac{\de f_q(x, \mu^2)}{\de \mu^2} = \int_x^1
\frac{dy}{y} f_q(y, \mu^2) P_{qq}^{(0)}\left( \frac{x}{y} \right).
\eea
Therefore, even if we cannot compute PDFs, we can predict how they evolve from one scale to another.
Since the splitting functions are universal, it is possible to measure PDFs 
in one process and use them as inputs for another one.

\begin{figure}[htbp]
\begin{center}
\includegraphics[width=9cm]{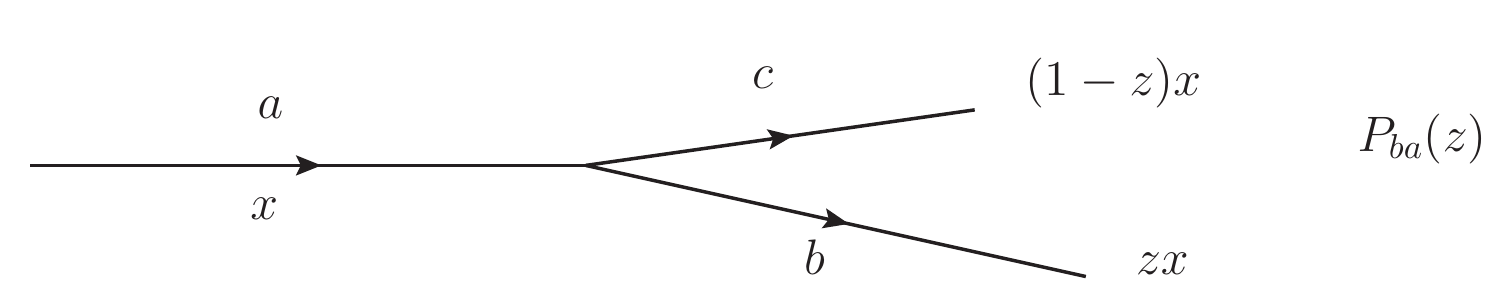}
\end{center}
\caption{Standard conventions for splitting functions.}\label{fig:splitting}
\end{figure}

It is necessary to take account of the different species of partons.
The standard notation for the splitting functions is shown in Fig.~\ref{fig:splitting}. 
The DGLAP equations are a system of coupled integro/differential equations which 
are usually written in compact notation as\footnote{The $\otimes$ product is 
defined as $f_1 \otimes f_2 \otimes \dots f_n(x) =
\int_0^1 dx_1 dx_2 \dots dx_n f_1(x_1) f_2(x_2) \dots
f-n(x_n) \delta(x-x_1 x_2 \dots x_n)$. }
\bea\label{eq:DGLAP}
\mu^2 \frac{\de f_i(x,\mu^2)}{\de \mu^2}=
\sum_j P_{ij} \otimes f_j(\mu^2).
\eea
with
\bea
P_{ij}(x) = \frac{\alpha_s}{2 \pi} P_{ij}^{(0)}+\left( \frac{\alpha_s}{2 \pi}\right)^2
P_{ij}^{(1)}+\ldots.
\eea
The splitting functions are now known up to $P_{ij}^{(2)}$. The determination of
$P_{ij}^{(2)}$ is an essential input in order to determinate
NNLO PDFs. In the following, we limit our discussions to $P_{ij}^{(0)}$. 
At leading order the splitting functions are
\bea
P_{qq}^{(0)}(z)&=P_{\bar q \bar q}^{(0)}(z) = C_F\left(\frac{1+z^2}{1-z} \right)_+,\\
P_{qg}^{(0)}(z)&=P_{\bar q g}^{(0)}(z) = T_F(z^2+(1-z)^2),\\
P_{gq}^{(0)}(z)&=P_{g \bar q}^{(0)}(z) = C_F \frac{1+(1-z)^2}{z},\\
P_{gg}^{(0)}(z) &= 2 C_A \left[ z \left(\frac{1}{1-z}\right)_+
+\frac{1-z}{z}+z(1-z)+\left(\frac{11}{12}-\frac{n_f}{6 C_A}\right)\delta(1-z)\right].
\eea
At higher orders components $P_{q_i q_j}$ with $i\neq j$, $P_{q_i \bar q_j}$ 
for any $i,j$ arise. 

We can take the difference of 
Eq.~(\ref{eq:DGLAP}) for a quark (antiquark) flavour $i$ with itself 
for a quark (antiquark) flavour $j$:
\bea
\mu^2 \frac{\de}{\de \mu^2}(f_i(\mu)-f_j(\mu))
= \sum_k (P_{ik} \otimes f_k(\mu)-P_{jk} \otimes f_k(\mu)).
\eea
Since the gluon contributions cancels out, we get
\bea
\mu^2 \frac{\de}{\de \mu^2}(f_i(\mu)-f_j(\mu))
= P_{qq} \otimes (f_i(\mu)-f_j(\mu)).
\eea
With $n_f$ flavours there are therefore $2 n_f-1$ independent combination of
PDFs which evolve independently, called non-singlet components. On the other hand,
we can take the sum of the DGLAP equations for all quark and antiquark 
flavours. In this way we get
\bea
\sum_{i\neq g} \frac{\de}{\de \mu^2} f_i(\mu) =
P_{qq} \otimes \sum_{i\neq g} f_i(\mu) + 2 n_f P_{ig} \otimes f_g(\mu). 
\eea
We can finally consider Eq.~(\ref{eq:DGLAP}) for the gluon:
\bea
\frac{\de}{\de \mu^2} f_g(\mu) = \sum_{i\neq g} P_{gi} \otimes f_i(\mu) + P_{gg}
\otimes f_{g}(\mu).
\eea
If we define 
\bea
S(\mu) = \sum_{i \neq g} f_i(\mu)
\eea
we have the following system of equations for the singlet component $S$
and the gluon:
\bea
\mu^2 \frac{\de}{\de \mu^2} f_{g}(\mu) &= P_{gq} \otimes S(\mu) + P_{gg} \otimes f_g(\mu)\\
\mu^2 \frac{\de}{\de \mu^2} S(\mu) &= P_{qq} \otimes S(\mu) + 2 n_f P_{ig} \otimes f_g(\mu).
\eea
We then learn that the singlet component mixes with the gluon density in
its evolution, contrary to the non-singlet components, which evolve independently. 
%Because of the coupled DGLAP evolution it is possible to acces the gluon PDF
%indirectly, through the way it changes the evolution of quark PDFs.

\begin{figure}[htbp]
\begin{center}
\includegraphics[width=12cm]{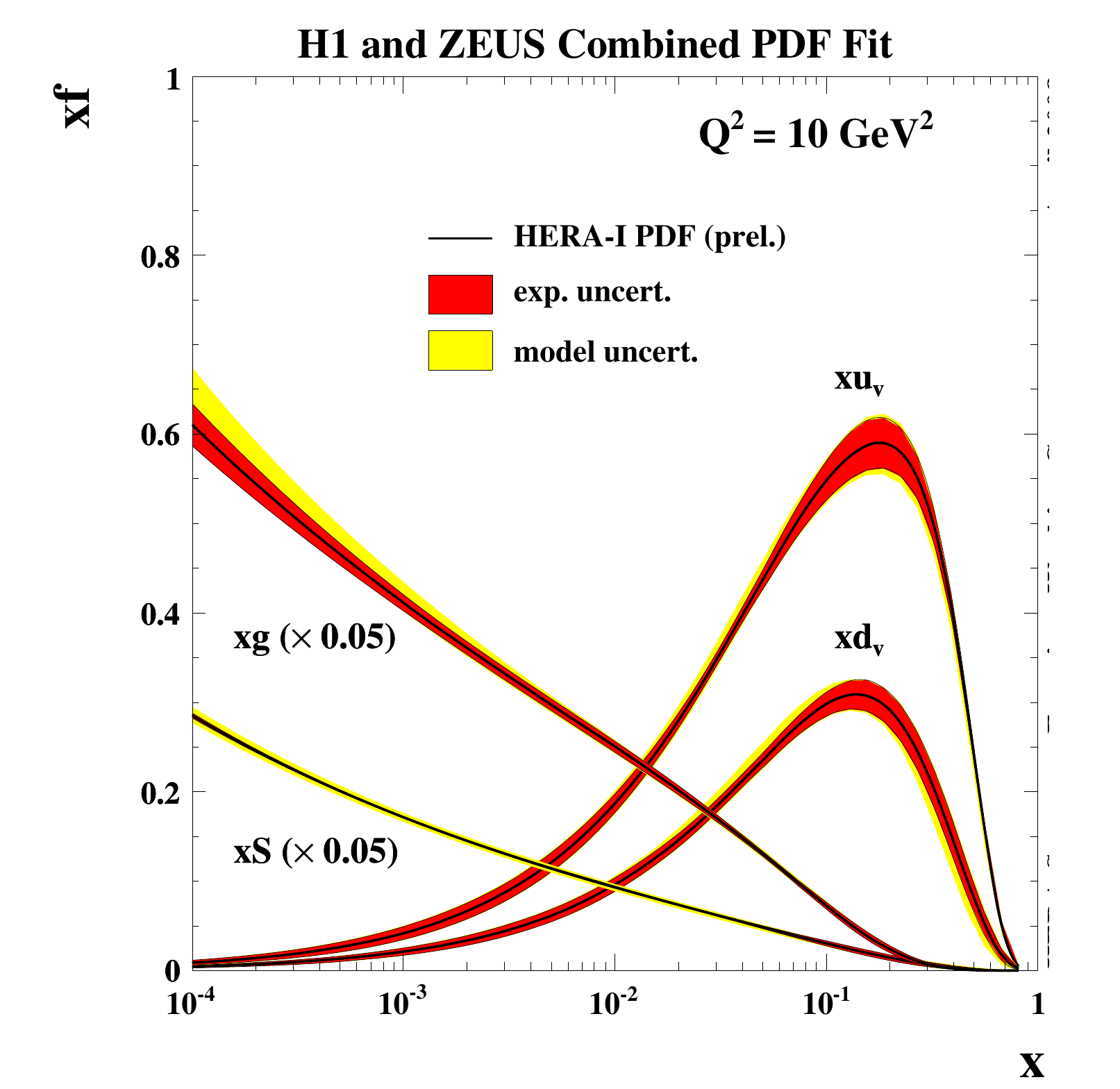}
\end{center}
\caption{Typical features of PDFs.}\label{fig:hera}
\end{figure}

In Fig.~\ref{fig:hera} it is possible to observe some typical features of parton distribution functions.
In particular, the gluon distribution is very large and dominates at small $x$.
The sea distributions grow at small $x$, too. The valence distributions present a 
peak around $x = 0.1-0.2$. Finally, the largest uncertainties are at very small, or very large $x$.

\chapter{Resummation}\label{chapter:resummation}

Higher order perturbative QCD calculations are necessary in order to obtain a 
detailed understanding of hadronic final states properties. In some cases, 
logarithmically enhanced contributions appear at all order in perturbation theory.
It is thus necessary to go beyond fixed-order perturbation theory and sum these
contributions to all orders in $\alpha_s$. The LHC provides 
very precise measurements of many processes, such as
Drell-Yan, Higgs or top production. Accurate theoretical predictions are necessary in order
to understand whether we are observing Standard Model physics or new physics signals.
There is therefore great interest in assessing the impact of threshold resummation
in order to obtain accurate phenomenological predictions.

In this Chapter we summarize some general features of soft-gluon
(or threshold) resummation for inclusive cross-sections. 
In Sect. \ref{sec:whyresumm} we introduce the idea of resummation classifying the
enhanced contributions which appear when physical observable are calculated. In Sects. 
\ref{sec:sgeffects} and \ref{sec:resexp} we present a simplified argument which sheds light on the
main features of soft resummation $-$ factorization and exponentiation. 
In Sect. \ref{sec:resproof} we present a proof of all-order resummation
which follows a renormalization group argument. Finally, in Sect.~\ref{sec:landaupole}
we introduce the reader to the well known problem of the Landau pole which appears
in the resummed expressions and in Sect.~\ref{sec:minimal} and \ref{sec:borel} 
we present two different solutions which remove it.

\section{Why resummation?}\label{sec:whyresumm}

When we compute a cross-section in dQCD we may run into 
several (logarithmically) enhanced contributions, which we can
classify into three main categories:
\begin{itemize}
\item logarithms of ultra-violet (UV) origin;
\item logarithms of collinear origin;
\item logarithms of soft origin.
\end{itemize}
We have seen how logarithms of collinear origin can be absorbed into scale-dependent PDFs
thanks to DGLAP equations; in the same way UV logarithms 
are absorbed into the running coupling constant $\alpha_s$. How can we deal with the last 
kind of logarithms?

Since any particle detector has a finite energy resolution, a generic 
physical cross-section is always inclusive over arbitrary soft particles which are
produced in the final state. Thanks to this inclusiveness, the radiation of 
undetected real gluons cancels exactly the divergences which 
appear when calculating perturbative contributions due to virtual gluons. 
However, despite the cancellation theorems of soft-gluon singularities, 
soft-gluon effects can still be large in kinematic configurations where 
high unbalance between real and virtual contributions persists. 
In this case, the convergence of fixed-order expansion is spoiled, and
calculations to all orders of perturbation theory are necessary in order to
achieve reliable predictions. An all-order exact treatment is hopeless,
and an improved perturbative expansion must be found: hence
resummation, which basically consists of a reorganization of the
perturbative expansion by an all-order summation of classes of logs. 

The idea of resummation should not sound completely new; in fact we have encountered 
another example of resummation when we have considered the solution of the
renormalization group equation for the running coupling constant $\alpha_s$
Eq.~(\ref{eq:alpha_Lambda}).
In the case of UV resummation we have to
deal with single logs. In the case of soft-gluon resummation
we have to deal with double logs, which originate because
soft logs are usually accompanied by collinear logs. Complications 
arise because of the non-abelian nature of QCD; gluons carry colour charge and 
multiple gluon emission is affected by dynamical correlations.

\section{Soft-gluon effects in QCD cross-sections}\label{sec:sgeffects}

In this Section and in the next Section, we will introduce soft-gluon resummation by following a simplified 
approach \cite{Catani:1996rb,Catani:1997vp} which nevertheless emphasizes the key ideas which enable a determination of
resummed expression.
We consider a generic process where a system 
of invariant mass $M$ is produced in a collision with centre of mass energy $s$.
At the first order in $\alpha_s$, 
the cross-section is affected by emission of virtual and real gluons, with
emission probabilities given by
\bea
\frac{d \omega_{1,v}(z)}{dz} &=  - 4 A_1 \alpha_s \delta(1-z) 
\int_0^{1-\xi}dz' \frac{\log(1-z')}{1-z'} \\
\frac{ d \omega_{1,r}(z)}{dz} &=4 A_1 \alpha_s  \frac{\log(1-z)}{1-z}
\Theta(1-z-\xi),
\eea 
where $1-z$ is the energy fraction carried by unobserved final state particles,
where $z=M^2/\hat s$, and $A_1$ is process-dependent ($A_1=C_F/\pi$ for the DY case and $A_1=C_A/\pi$ for
the Higgs case). As we have previously said, 
the combination of the bremsstrahlung spectrum and of the collinear spectrum leads to
double-logs. The virtual term has the same kinematics of the LO cross-section,
and thus contributes only on threshold, whereas the real contribution, which is
singular as well in $z=1$, is spread out in the whole interval $[\tau,1]$ where
$\tau$ is the energy fraction of the tagged final state, $\tau=M^2/s$ .  

In order to regularize the infrared divergences, we have introduced 
an unphysical cut-off $\xi$. Once the two contributions have been added, 
we can safely perform the physical limit $\xi \rightarrow 0$ and we thus obtain
a finite emission probability:
\bea\label{eq:singlegluonem}
 \frac{d \omega_1(z)}{dz} = 4 A_1 \alpha_s \left(\frac{\log(1-z)}{1-z}
\right)_+.
\eea
If we integrate over $z$ we finally obtain that the final leftover of infrared 
cancellation is therefore a double-logarithmic factor:
\bea
\int_\tau^1 dz \frac{d \omega_1(z)}{d z} =  2 A_1 \alpha_s
\log^2(1-\tau).
\eea
In the elastic limit $\tau \rightarrow 1$, when the final state carries a large 
fraction of the total energy, a perturbative expansion become meaningless because 
when $\alpha_s \log^2(1-\tau) \sim 1$ all the terms in the perturbative series are of 
the same order, even if the coupling constant is small.

We can generalize our argument and convince ourselves that a generic partonic 
coefficient function $C(z,\alpha_s)$ is enhanced in the elastic 
limit $z\rightarrow 1$. To this purpose, we consider a quark parton line
which emits $n$ gluons. For each emission, the energy decreases by a factor
$z=z_1 z_2 \dots z_n$ where $1-z_i$ is the energy fraction carried by the
$i$-th gluon. In the elastic limit, each gluon emission leads to an enhanced term
in the partonic coefficient. In the end, we end up with a tower of terms
which have the following form:
\bea\label{eq:towres}
\alpha_s^n \left(\frac{\log^k(1-z)}{1-z}\right)_+, \qquad 0 \leq k \leq a n -1,
\eea
where in general $a$ is process-dependent. In particular, $a=1$ for DIS and
$a=2$ for Drell-Yan and Higgs boson production. From now on, we will consider 
the latter case, and thus $a=2$. 
The higher-order soft contributions of Eq.~(\ref{eq:towres}) produced by
multiple soft radiation have to be resummed in order to achieve a reliable theoretical 
prediction.

One could ask when soft-gluon resummation is relevant for phenomenology. 
As we have seen, the inclusive cross-section (differential only in $M$)
for the production of a system of a 
large mass $M$ can be written in factorized form as
\bea\label{eq:cross-lum}
\frac{d \sigma}{d M^2} &=\sum_{ij} \int dx_i dx_j  f_i^{(1)}(x_i,\mu) f_j^{(2)}(x_j,\mu)
\times \hat \sigma_{ij} (M^2, \hat s, \mu, \alpha_s(\mu))\nonumber \\
&= \tau \sigma_0
 \sum_{ij} 
\int \frac{dx_i}{x_i} \frac{dx_j}{x_j} f_i^{(1)}(x_i,\mu) f_j^{(2)}(x_j,\mu)
\times C_{ij} (z,\mu,M^2,\alpha_s(\mu))\nonumber \\
&= \tau \sigma_0 \sum_{ij} \int_\tau^1 \frac{dz}{z} \mathscr L_{ij} \left(\frac{\tau}{z}\right)
C_{ij}(z, \mu,M^2, \alpha_s(\mu))
\eea
where we have defined the \emph{parton luminosity}
\bea\label{eq:partonlum}
\mathscr L_{ij} (x) = \int_x^1 \frac{dy}{y} 
f_i^{(1)}\left(\frac{x}{y},\mu\right) f_{j}^{(2)}(y,\mu)
\eea
and the sum runs over the flavour indexes $i,j$. We have introduced some conventional
notation that we will use in the next Chapters:
\bea\label{eq:zandtau}
\tau \equiv \frac{M^2}{s},\qquad 
z \equiv \frac{\tau}{x_i x_j}=\frac{M^2}{\hat s}
\eea
where $\hat s = x_i x_j s$; $\sqrt{s}$ is the hadronic center-of-mass energy 
whereas $\sqrt{\hat s}$ is the partonic center-of-mass energy. We can then identify
true threshold, $s\rightarrow M$, and partonic threshold, $x_i x_j s \rightarrow M^2$,
in the limits $\tau \rightarrow 1$ and $z \rightarrow 1$.
We have understood how the partonic coefficient functions are enhanced when $z$
get close to $1$; this means that soft-gluon resummation becomes relevant when 
the partonic process is close to threshold. Hence soft-gluon resummation is also
called threshold resummation. 

When a system of high mass $M^2$ is
produced, and thus $\tau$ tends to 1, the contribution to the cross-section
Eq.~(\ref{eq:cross-lum}) comes entirely from the threshold region. In this case,
soft resummation is obviously necessary. Among the systems where we encounter
such a behavior there are DIS, Drell-Yan processes where a very massive lepton pair
of invariant mass $M^2$ is produced, production of heavy $q \bar q$ pairs and 
pairs of jets with large transverse momentum. 

However, the relevance of resummation is determined by
partonic kinematics, which can be quite different from hadronic kinematics. Therefore
the effect of soft-gluon resummation can be relevant even relatively far
from the hadronic threshold, since the hadronic cross-sections are 
found by convoluting a partonic cross-section with a parton luminosity. 
In particular, the effect of soft-gluon resummation is found to be not negligible for 
Drell-Yan production at hadron colliders. 

Furthermore, it has to be noticed that the effect of soft resummation could be
significant even when $\tau$ is small and the process is very far from 
threshold, as for example Higgs boson production at the LHC. This region is
clearly perturbative; however, the size of the soft corrections depends in
a non-trivial way on the shape of the parton distribution functions. For this reason,
threshold resummation is expected to be relevant if the PDFs are peaked at small
$x$. In this case it turns out that, partly because of the partonic kinematics, partly 
because of the cross-section shape, most of the cross-section is governed by the 
logarithmically enhanced terms.

%The accuracy of an approximation to the Higgs production cross section at the
%LHC based on the dominance of threshold terms can be studied [30] by using
%the saddle point method to determine which is the region in N space that gives
%the bulk of the contribution to the cross section. It turns out that, despite 
%the fact that Higgs production at the LHC is far from the kinematic threshold, 
%partly because of the underlying partonic kinematics and partly because of the shape
%of the cross section, at the LHC with 8 TeV center-of-mass energy, logarithmically
%enhanced terms are still providing most of the cross section, though the situation
%gradually changes as the center-of-mass energy increases.

\section{Soft-gluon resummation and exponentiation in  \texorpdfstring{$N$-space}{N-space}}\label{sec:resexp}

%In this section we present a simplified discussion \cite{Catani:1996rb} of the resummation mechanism,
%which however enlightens the fundamental idea which permits to obtain 
%resummed expressions for physical observables.
For the sake of simplicity, we omit the sum over
parton subprocesses and we write perturbative QCD factorization as
\bea\label{eq:sigma_xspace}
\sigma(\tau, M^2) =\int_\tau^1 
\frac{dz}{z} \mathscr L\left( \frac{\tau}{z} \right)
C(z, M^2),
\eea 
where we have defined a dimensionless cross-section as
\bea
\sigma(\tau, M^2) = \frac{1}{\tau \sigma_0} \frac{d \sigma}{d M^2}
\eea
by imposing that at the Born level the coefficient function at LO is simply
$C(z,M^2)=\delta(1-z)$. Without loss of generality, we choose $\mu_F^2=\mu_R^2=M^2$.

The physical basis for soft-gluon resummation relies on factorization. 
In particular, it is possible to prove that 
in the soft limit the $n$-gluon probability can be written in a factorized form:
\bea
\frac{d \omega_n(z_1, \dots z_n)}{dz_1\dots dz_n}\simeq \frac{1}{n!}
\prod_{i=1}^n \frac{d \omega_1(z_i)}{dz_i}.
\eea
The proof, which is based on the \emph{eikonal approximation}, is rather easy in
QED, but is more cumbersome in a non-abelian theory like QCD, where the gluons
are not neutral but carry color charge instead.
The coefficient function can be written as
\bea\label{eq:Cxspace}
C(z,M^2)= \delta(1-z) + \sum_{n=1}^\infty \int_0^1 dz_1 \dots dz_n 
\frac{d \omega_n(z_1, \dots z_n)}{dz_1\dots dz_n} \delta(z-z_1\dots z_n).
\eea

The presence of $\delta(z-z_1\dots z_n)$ due to the longitudinal-momentum
conservation spoils the factorization in Eq.~(\ref{eq:Cxspace}). However, 
we can obtain a fully factorizable expression performing a \emph{Mellin 
transform} and moving to $N$-space. We define the Mellin transform of a 
function $f(x)$ defined in the range $(0,1)$ as
\bea
f(N) \equiv \mathcal M[f] (N) \equiv \int_0^1 dx\ x^{N-1} f(x).
\eea
The Mellin transform is a Laplace transform where a change of variables have 
been performed. If $x=e^{-\xi}$,
\bea
\int_0^1 dx\ x^{N-1} f(x) = \int_0^\infty d \xi\ e^{-\xi N} f(x(\xi)).
\eea 
The inverse Mellin transform is then
\bea
f(x) = \frac{1}{2 \pi i} \int_{c-i\infty}^{c+i\infty} dN \ x^{-N} g(N),
\eea
where $c$ is greater than the real part of the rightmost singularity. Under
Mellin transform, the convolution product diagonalizes:
\bea
(f \otimes g)(N) &= \int_0^1 dx\ x^{N-1} \int_x^1
\frac{dy}{y} f(y) g \left(\frac{x}{y} \right) =
\int_0^1 dx\ x^{N-1} \int_0^1 dy \int_0^1 dz f(y) g(z) \delta(x-yz)\nonumber \\
&= \int_0^1 dy \ y^{N-1} f(y) \int_0^1 dz z^{N-1} g(z) = f(N) g(N).
\eea
The cross-section Eq.~(\ref{eq:sigma_xspace}) can consequently be written 
in $N$-space as
\bea\label{eq:Mellinsigma}
\sigma(N, M^2) = \int_0^1 d\tau \ \tau^{N-1} \sigma(\tau, M^2)
= C(N, M^2) \mathscr L(N).
\eea
Upon Mellin transform, the $n$-particle longitudinal phase space factorizes:
\bea
\int_0^1 dz \ z^{N-1} \delta(z-z_1\dots z_n) = z_1^{N-1} \dots z_n^{N-1}.
\eea
This means that in $N$-space the coefficient function is fully factorized and 
can be recast in exponential form:
\bea\label{eq:CNQED}
C(N, M^2) =& 1 + \sum_{n=1}^\infty \int_0^1 dz_1 \dots dz_n \frac{d \omega_n(z_1, \dots z_n)}{dz_1\dots dz_n}
z_1^{N-1}\dots z_n^{N-1}\nonumber \\
=& 1 + \sum_{n=1}^\infty \frac{1}{n!} \int_0^1 dz_1 \frac{d \omega_1(z_1)}{dz_1} z_1^{N-1}
\dots \int_0^1 dz_n \frac{d \omega_1(z_n)}{dz_n}z_n^{N-1}\nonumber \\
=&  \exp\left[ \int_0^1 dz\ z^{N-1} \frac{d \omega_1(z)}{dz}\right]
\eea  

In the large $N$ limit, one can show that by inserting Eq.~(\ref{eq:singlegluonem}) 
in Eq.~(\ref{eq:CNQED}) the exponential reduces to the exponential of a double-log:
\bea\label{eq:CNLL}
C(N, M^2) =  \exp\left[
 4 A_1 \alpha_s\int_0^1 dz\ z^{N-1} \frac{z^{N-1}-1}{1-z}\right]=
 \exp\left[2 A_1\alpha_s(\log^2 N +\mathcal O(N)) \right].
\eea
In $N$-space the threshold region $z\sim 1$ corresponds to the large $N$ region,
and the tower of logs Eq.~(\ref{eq:towres}) converts into
\bea
\alpha_s^n \log^k \frac{1}{N}, \qquad 0 \leq k \leq 2n.
\eea
However, in Eq.~(\ref{eq:CNLL}), only the highest power $k=2n$ is resummed. 
The result we have obtained is said to hold at leading-logarithmic (LL) accuracy. 

In this simplified presentation we have not taken into account running 
coupling effects. The argument in this form would be complete in QED, where it
is called \emph{Sudakov} resummation. Nevertheless, it shows that 
the physical basis for soft resummation are the factorization of the multi-gluon
amplitudes, which follows from QCD dynamics, and the factorization of the phase space, which 
can take place only by working in $N$-space.

In order to correctly take into account the running of $\alpha_s$, one must consider
the longitudinal momenta of the gluons, which fix the energy scale of gluon emission.
Taking into account the running coupling effects,
the exponential Eq.~(\ref{eq:CNQED}) becomes
\bea\label{eq:ccoupli}
C(N, M^2) = \exp\left[\int_0^1 dz \ z^{N-1} 2 A_1 
\left( \frac{1}{1-z} \int_{M^2}^{(1-z)^2 M^2} \frac{d\mu^2}{\mu^2} 
\alpha_s(\mu^2)\right)_+ \right]
\eea
which reduces to the previous result when $\alpha_s$ is constant.
The expression, however, is ill-defined; by integrating in $z$ up to $z=1$ 
at some point the coupling constant has to be evaluated in the 
non-perturbative region and hits the Landau pole. This singularity has thus to be properly regolarized. The 
problem can be circumvented by expanding the integrand in powers of $\alpha_s$,
computing the Mellin transform of each term in the large-$N$ limit and finally 
resumming the series, obtaining a finite result. However, we will see that the Landau pole problem
will make a reappearance again.

Even if Eq.~\eqref{eq:ccoupli} includes the proper running coupling effects,
it still holds only at LL accuracy. One can prove that the most general expression for the
$N$-space resummed coefficient function is
\bea\label{eq:Cres}
C^{\textrm{res}}(N, M^2) = {\bar g}_0(\alpha_s) \exp \bar {\mathcal S}\left( M^2, \frac{M^2}{N^2} \right),
\eea
where the \emph{Sudakov form factor} is
\bea\label{eq:barS}
\bar {\mathcal S} \left(M^2, \frac{M^2}{N^2} \right) =
\int_0^1 dz \ z^{N-1} \left[\frac{1}{1-z} \int_{M^2}^{M^2(1-z)^2}
\frac{d \mu^2}{\mu^2} 2 A \left(\alpha_s(\mu^2)\right) + 
D\left( \alpha_s([1-z]^2 M^2)\right) \right]_+,
\eea
which reduces to Eq.~\eqref{eq:ccoupli} at LL accuracy.
The functions $\bar g_0(\alpha_s)$, $A(\alpha_s)$ and $D(\alpha_s)$ are
power series in $\alpha_s$, with $\bar g_0(\alpha_s) = 1 + \mathcal O(\alpha_s)$.
The functions $D(\alpha_s)$ and $\bar g_0(\alpha_s)$ are process-dependent,
whereas $A(\alpha_s)$ is order by order the coefficient of the soft singularities
in the Altarelli-Parisi splitting function for the relevant partonic process. 

\begin{table*}[t]
\begin{center}
\begin{tabular}{lccr}
  log approx.  & $g_i$ up to & $g_0$ up to order & accuracy: $\as^n \log^k\frac{1}{N}$\\
  \midrule
  LL     & $i=1$ & $(\alpha_s)^0$ & $k= 2n$ \\
  NLL    & $i=2$ & $(\alpha_s)^1$ & $2n-2 \leq k= 2n$ \\
  NNLL   & $i=3$ & $(\alpha_s)^2$ & $2n-4 \leq k= 2n$
  % \\  NNNLL  & 4-loop & 3-loop & 2-loop & $2n-5\le k\le 2n$ 
\end{tabular}
\caption{Orders of logarithmic approximations and accuracy of the
  predicted logarithms.}
\label{tab:count1}
\end{center}
\end{table*}

Using the notation of Ref.~\cite{Catani:1989aj}, the resummed coefficient function
can be written in the form
\bea\label{eq:cresg0expS}
C^{\textrm{res}}(N, M^2) &= g_0(\alpha_s) \exp \mathcal S \left(\bar \alpha
L, \bar \alpha \right) \\
\bar \alpha &\equiv 2 \beta_0 \alpha, \qquad L \equiv \log \frac{1}{N}, 
\eea 
where $g_0$ collects all the constant terms
and can be written as
\bea
g_0(\alpha_s) = 1 + \sum_{j=1}^\infty g_{0j} \alpha_s^j, 
\eea
and $\mathcal S$ has a logarithmic expansion
\bea
\mathcal S(\bar \alpha L, \bar \alpha) = \frac{1}{\bar \alpha} g_1(\bar \alpha L) +
g_2(\bar \alpha L) 
+ \bar \alpha g_3 (\bar \alpha L) + \bar \alpha^2 g_4(\bar \alpha L) + \ldots.
\eea
The functions $g_i$ can be obtained performing the integrals Eq.~\eqref{eq:cresg0expS}, and
they are determined by a small number of coefficients of the expansion of the 
functions $A$ and $D$, and 
are of order $g_1(\bar \alpha L) = \mathcal O(\as^2)$, 
$g_i(\bar \alpha L)= \mathcal O(\as)$ for $i>1$.
We note that the functions $g_0$ and $\mathcal S$ do not coincide with
the functions $\bar g_0$ and $\bar{\mathcal S}$ since by definition $\mathcal S$
does not contain not logarithmically enhanced terms and $g_0$ includes
non-logarithmic contribution both from $\bar g_0$ and from $\bar {\mathcal S}$.

If contributions up to $g_n$ are included in $\mathcal S$, $\log C^{\textrm{res}}(N,M^2)$
is determined up to $\mathcal O(\alpha_s^{k+(n-1)}L^k)$ subleading corrections.
In order to achieve the exact predictions of all the coefficients of the logarithms 
it is mandatory to include up to the relevant order the function $g_0$, 
because of its interference with the logarithmically enhanced contributions.
In particular, at the next$^k$-to-leading logarithmic (N$^k$LL) accuracy 
it is necessary to include functions up to $g_{k+1}$ and compute $g_0$ up to
order $\alpha_s^k$. At LL accuracy only the largest power of $\log\frac{1}{N}$
are correctly predicted; by adding one order in each of the functions $g_i$,
$g_0$ the NLL accuracy, which predicts two powers more, is achieved, and so on.
We show in Tab.~\ref{tab:count1} the order up to which the expansion of 
$g_0$ and $g_i$ should be included in order to get a given logarithmic accuracy
and the terms which are correctly predicted. Consider for example the result at NLL
accuracy. We can write
\bea
g_k(\lambda) = (2 \beta_0)^{2-k} \sum_{j=1}^\infty g_{kj} \left( \frac{\lambda}{2 \beta_0} \right)^j,
\qquad g_{11}=0.
\eea
If we expand $C^\textrm{res}$ in powers of $\as$, we obtain
\bea\label{eq:cnll}
C^\textrm{res} (N, \as) =& 1 + \as[g_{12} L^2 + g_{21} L + g_{01}]\nonumber \\
&+\as^2 \left[\frac{g_{12}^2}{2} L^4 + (g_{12}g_{21}+g_{13})L^3
+ \left( \frac{g_{21}^2}{2}+g_{22} + g_{12} g_{01}\right) L^2 + \mathcal O(L)\right]\nonumber \\
& + \mathcal O(\as^3).
\eea
As expected, the order $\as$ and the coefficients of the powers $k=4,3,2$ of $L$ at 
order $\as^2$ depend only on the NLL contributions $g_1$, $g_2$ and $g_{01}$.

\begin{table*}[t]
\begin{center}
\begin{tabular}{lcccr}
  log approx.  & $A(\as)$ & $D(\as)$ & $\bar g_0(\as)$ & accuracy: $\as^n \log^k\frac{1}{N}$\\
  \midrule
  LL     & 1-loop & --- & tree-level & $k= 2n$ \\
\addlinespace[0.8\defaultaddspace]
  NLL*   & 2-loop & 1-loop & tree-level & $2n-1\le k\le 2n$ \\
  NLL    & 2-loop & 1-loop & 1-loop & $2n-2\le k\le 2n$ \\
\addlinespace[0.8\defaultaddspace]
  NNLL*  & 3-loop & 2-loop & 1-loop & $2n-3\le k\le 2n$ \\
  NNLL   & 3-loop & 2-loop & 2-loop & $2n-4\le k\le 2n$
  % \\  NNNLL  & 4-loop & 3-loop & 2-loop & $2n-5\le k\le 2n$ 
\end{tabular}
\caption{Orders of logarithmic approximations and accuracy of the
  predicted logarithms for the N$^k$LL and the N$^k$LL$^*$ counting. The last columns refers to the
  coefficient function.}
\label{tab:count2}
\end{center}
\end{table*}

We emphasize that the functions $g_i$ depend on a few coefficients of the expansion of the 
functions $A$ and $D$; $g_1$ depends on the 
coefficient $A_1$ previously introduced, and $g_2$ on a 
couple of other coefficients. In particular, the function $A$ is 
fully determined by the Altarelli-Parisi anomalous dimension. It is therefore 
necessary to compute the N$^p$LO anomalous dimension in order to compute
$g_{p+1}$, which is necessary for the result with N$^p$LL accuracy. On the other hand, the 
process-dependent functions can be fixed by matching the expansion of the
resummed coefficient function in $\as$ with a fixed order computation.
In conclusion, in order to obtain LL, NLL, NNLL accuracy it is necessary to
know the GLAP anomalous dimension respectively at 1, 2, 3 loop, whereas the fixed order
expressions at NLO and NNLO are needed for NLL and NNLL accuracy.

It has been pointed out that if one exponentiates $g_0$ (or equivalently $\bar g_0$) and
then performs the power counting at the level of exponent it 
may be more natural to
include one order less in $g_0$. However, this decreases the logarithmic accuracy of the 
coefficient function by half of a logarithmic order; for example, the coefficient of
$L^2$ at order $\as^2$ in Eq.~\eqref{eq:cnll} is not correctly predicted without $g_{01}$. 
This is sometimes called N$^k$LL$^*$ 
accuracy. For phenomenology it is often more convenient to examine the accuracy 
of the predicted logarithms by considering the functions $A(\alpha_s)$,
$D(\alpha_s)$ and $\bar g_0(\alpha_s)$ in Eq.~(\ref{eq:barS}).
A summary for the first orders both for the N$^k$LL and the N$^k$LL$^*$ counting
is given in Tab.~\ref{tab:count2}.

Correct predictions
for phenomenology with N$^p$LO+N$^k$LL accuracy are obtained by combining the resummed 
coefficient function expanded in powers of $\as$ with the fixed-order coefficient
function and then subtracting the double-counting terms:
\bea
C^{\textrm{N$^p$LO}}_{\textrm{N$^k$LL}}(N, \as) =
\sum_{j=0}^p \as^j C^{(j)}(N) + C^\textrm{res}_{\textrm{N$^k$LL}} (N, \as) 
-\sum_{j=0}^p \frac{\as^j}{j!} \left[ \frac{d^j C^\textrm{res}_{\textrm{N$^k$LL}}(N, \as)}{d \as^j}\right]_{\as=0}.
\eea

\section{Soft gluon resummation and renormalization group approach}\label{sec:resproof}

In the previous Sections we have present a simplified discussion of the resummation mechanism,
which however has shed light on the fundamental ideas which allow for a determination of
resummed expressions for physical observables.

Exponentiation formulae have been obtained first at LO 
and subsequently extended at NLO by recurring to eikonal \cite{Catani:1989aj} or 
factorization \cite{Sterman:1986aj} techniques. The exponentiation of soft logs related to gluon emission has
been then proved at all orders in \cite{Contopanagos:1996nh,Forte:2002aj}.
In all these approaches soft-gluon resummation is performed in $N$-space,
after the Mellin transformation of the cross-section which factorizes into
the product of a partonic cross-section and a parton luminosity Eq.~(\ref{eq:Mellinsigma}). 
More recently threshold resummation was based on direct diagrammatic analysis
in Refs.~\cite{Laenen:2008gt,Laenen:2010uz}. In the latter approach factorization and resummation are performed without
the need of a Mellin transform by recurring to path-integral methods. 

A complete review of these proofs is beyond the scope of this Thesis.
In this Section we present a simple proof of all-order soft-logarithms 
exponentiation in QCD by following the argument presented in \cite{Forte:2002aj}.
The result is essentially based on kinematics arguments and renormalization-group
techniques. 

For the sake of simplicity, we will consider a DIS-like 
process. We consider the total cross-section $\sigma(x,Q^2)$ in the vicinity of the
kinematic boundary $x=1$. Let $\sigma(N,Q^2)$ be its Mellin transform which
can be written in factorized form as
\bea\label{eq:fact_sigma}
\sigma(N,Q^2) = C\left(N, \frac{Q^2}{\mu^2},\alpha_s(\mu^2)\right)F(N,\mu^2)
\eea 
where $F(N,\mu^2)$ are Mellin moments of parton densities $F(x,\mu^2)$. 
In the case of Drell-Yan process or Higgs production Eq.~\eqref{eq:fact_sigma}
should be appropriately modified in order to take into account that there are
two partons in the initial state.
Since resummation takes the form of an exponentiation,
it is convenient to introduce the physical anomalous dimension,
which is defined as the log-derivative of the cross-section 
\bea\label{eq:ad_physical}
Q^2 \frac{\de \sigma(N, Q^2)}{\de Q^2} =
\gamma (N, \as(Q^2),) \sigma(N, Q^2). 
\eea
We can thus write the cross-section as
\bea
\sigma(N, Q^2) = \exp[E(N,Q_0^2,Q^2)] \sigma(N, Q_0^2)
\eea
with
\bea
E(N,Q_0^2,Q^2) = \int_{Q_0^2}^{Q^2} \frac{dk^2}{k^2} \gamma(N, \alpha_s(k^2)).
\eea
We observe that by comparison of Eq.~(\ref{eq:ad_physical}) and Eq.~(\ref{eq:fact_sigma})
\bea
\gamma(N, \as(Q^2)) = \frac{\de}{\de \log Q^2} \log  C\left(N, \frac{Q^2}{\mu^2},\alpha_s(\mu^2)\right).
\eea

The only assumption we make is the validity of the standard factorization Eq.~(\ref{eq:fact_sigma}). 
This means that the coefficient function can be multiplicatively renormalized. 
We can therefore remove all the divergences from the so-called bare coefficient function
$C^{(0)}(N, Q^2, \alpha_0, \epsilon)$ by introducing a process-independent factor $Z_C$
\bea
C\left(N,\frac{Q^2}{\mu^2}, \as(\mu^2), \epsilon \right) = Z_C (N, \as(\mu^2), \epsilon)
C^{(0)}(N, Q^2, \alpha_0, \epsilon)    
\eea
such that the renormalized coefficient function is finite at $\epsilon=0$ since the
multiple poles in $Z_C$ cancel out the infinities of the bare coefficient function. 
We have introduced a renormalized coupling constant $\as(\mu^2)$ which is related to
the bare coupling constant $\alpha_0$ by the implicit equation
\bea
\alpha_0(\mu^2, \as(\mu^2),\epsilon) = \mu^{2 \epsilon} \as(\mu^2) Z_\alpha (\as(\mu^2), \epsilon). 
\eea
Both $Z_\alpha$ and $Z_C$ are computable in perturbation theory and are $Q^2$-independent
with a convenient choice of the factorization scheme (such as the $\overline{\textrm{MS}}$ scheme).
Therefore the physical anomalous dimension is, according to Eq.~(\ref{eq:ad_physical})
\bea\label{eq:degamma0}
\gamma &= \frac{\de}{\de \log Q^2} \log C^{(0)} (N, Q^2, \alpha_0, \epsilon)\nonumber \\
&= - \epsilon \alpha_0 \frac{\de}{\de \alpha_0} \log C^{(0)} (N, Q^2, \alpha_0, \epsilon). 
\eea
When we have derived Eq.~(\ref{eq:degamma0}) we have taken into account the fact that for 
dimensional reason the dependence on $Q^2$ of the bare coefficient function is through
the combination $\alpha_0 Q^{-2 \epsilon }$. 

In order to obtain a resummed expression for the physical anomalous dimension it is 
necessary to understand the structure of the bare coefficient function in $N$-space. 
The bare coefficient function and its Mellin transform can be expanded in powers of the
bare coupling constant $\alpha_0$:
\bea
C^{(0)} (x, Q^2, \alpha_0, \epsilon) &= \sum_{n=0}^\infty \alpha_0^n
C_n^{(0)} (x, Q^2, \epsilon)\\
C^{(0)} (N, Q^2, \alpha_0, \epsilon) &= \sum_{n=0}^\infty \alpha_0^n
C_n^{(0)} (N, Q^2, \epsilon).
\eea

The main result of \cite{Forte:2002aj}, which we will not prove here, consists in 
showing first that the $N$-dependence of the bare coefficient function is determined by 
the kinematic structure of the $k$-particle phase space at tree-level, and secondly that
loop integrations do not affect this result.
One first proves that
\bea
C_n^{(0)}(x, Q^2, \epsilon) =
(Q^2)^{-n \epsilon} \left[ 
C_{n0}^{(0)} (\epsilon) \delta(1-x) + \sum_{k=1}^n \frac{C_{nk}^{(0)} (\epsilon)}{\Gamma(-ak\epsilon)}
(1-x)^{-1-ak\epsilon}\right]+ \mathcal O[(1-x)^0]
\eea
where we have denoted by $O[(1-x)^0]$ terms which are not divergent as $x\rightarrow 1$
when $\epsilon$ tends to 0, and $a=1$ for DIS, where $Q^2$ is the virtuality of the
exchanged boson, and $a=2$ for DY or Higgs boson production processes, where the hard scale 
$Q$ is the invariant mass of the final state. Since 
\bea
\int_0^1 dx \ x^{N-1} (1-x)^{-1 -ak\epsilon} =
\frac{\Gamma(N) \Gamma(-ak\epsilon)}{\Gamma(N-ak\epsilon)} = \Gamma(-ak\epsilon) 
N^{a k \epsilon} + \mathcal O(1/N),
\eea 
whereas the Mellin transform of any function of $x$ which is not divergent as $x\rightarrow 1$
vanishes in the large $N$ limit, one has
\bea\label{eq:CnN}
C_n^{(0)} (N, Q^2, \epsilon) = \sum_{k=0}^n C_{nk}^{(0)} (\epsilon ) 
(Q^2)^{-(n-k)\epsilon} \left(\frac{Q^2}{N^a}\right)^{-k\epsilon} + \mathcal O(1/N).  
\eea

Furthermore, one can see that the coefficients $C_{nk}^{(0)}$ have a pole of order $2n$ in $\epsilon$. 
These poles are related to infrared singularities, and despite the fact that they cancel
in the coefficient function, their interference with the powers of $\epsilon$ in the series expansion of
$N^{-\epsilon}$ leads to powers of $\log N$ in $\sigma(N,Q^2)$ (which are related to powers of
$\log(1-x)$ under Mellin transform). The presence of soft logs
is therefore a consequence of an incomplete cancellation of real and virtual contributions to 
$C_n^{(0)}$ in the soft limit.

We can conclude that when $N\rightarrow \infty$, the regularized cross-section
depends on $N$ only through integer powers of a dimensionful variable $(Q/N^a)^{-\epsilon}$. 
The fact that the coupling constant in $d$ dimension has the dimensions of $Q^{2\epsilon}$
relates the dependence of $(Q/N^a)^{-\epsilon}$ to the running of the coupling 
through the renormalization group. We will see how this fact is sufficient in order to 
prove all-order resummation. 

From Eq.~(\ref{eq:CnN}) and Eq.~(\ref{eq:degamma0}) we get
\bea
\gamma(N, Q^2, \alpha_0, \epsilon) &= \sum_{i=1}^\infty \alpha_0^i \sum_{j=0}^i
\gamma_{ij}(\epsilon) (Q^2)^{-(i-j)\epsilon} \left(\frac{Q^2}{N^a}\right)^{-j \epsilon}
+\mathcal O(1/N)\nonumber  \\
&= \sum_{i=1}^{\infty}\sum_{j=0}^i \gamma_{ij} (\epsilon) [(Q^2)^{-\epsilon} \alpha_0]^{i-j}
\left[ \left( \frac{Q^2}{N^a} \right)^{-\epsilon} \alpha_0 \right]^j + \mathcal O(1/N),
\eea
which is a power series with $N$-independent coefficients. It is possibile to obtain a 
renormalized expression of the physical anomalous dimension by expressing the bare coupling 
in terms of the renormalized bare coupling. It is convenient to introduce the function
\bea
\bar \alpha_0 \left( \frac{Q^2}{\mu^2}, \as(\mu^2), \epsilon\right) &=
Q^{-2 \epsilon} \alpha_0(\mu^2, \as(\mu^2), \epsilon)\nonumber  \\
&= \left( \frac{Q^2}{\mu^2} \right)^{- \epsilon} \as (\mu^2) Z_{\as} (\as(\mu^2), \epsilon). 
\eea
which is invariant under renormalization group transform:
\bea
\frac{d \bar \alpha_0}{d \log \mu^2} =0. 
\eea
In particular, when $\mu=Q$, 
\bea
\bar \alpha_0 \left( \frac{Q^2}{\mu^2}, \as(\mu^2), \epsilon\right) =
\bar \alpha_0(1, \as(Q^2), \epsilon)=
\as(Q^2) Z_{\as}(\as(Q^2), \epsilon). 
\eea
We can therefore write the renormalized anomalous dimension as
\bea\label{eq:gammaN2}
\gamma\left(N,\frac{Q^2}{\mu^2}, \as(\mu^2), \epsilon\right)
&= \sum_{i=1}^\infty \sum_{j=0}^i \gamma_{ij} (\epsilon) 
[ \bar \alpha_0 (1, \as(Q^2), \epsilon)]^{i-j} 
\left[\bar \alpha_0 \left(1, \as\left(\frac{Q^2}{N^a}\right), \epsilon\right) \right]^j + \mathcal O(1/N) \nonumber\\
&=
\sum_{m=1}^\infty\sum_{n=0}^m \gamma_{mn}^R (\epsilon) \as^{m-n} (Q^2) \as^n \left( \frac{Q^2}{N^a} \right) + \mathcal O(1/N)
\eea
where the coefficients $\gamma_{mn}^R$ depends on $\gamma_{ij}(\epsilon)$ and $Z_{\as}$. 
The fact that the physical anomalous dimension is finite as $\epsilon\rightarrow 0$ however does not imply that
it admits an expansion of the same form of Eq.~(\ref{eq:gammaN2}) since the 
coefficients $\gamma_{mn}^R(\epsilon)$ might not be finite in the $\epsilon \rightarrow 0$ limit. 
In order to obtain an expression for the four-dimensional physical anomalous dimensions in terms of 
finite quantities it is convenient to separate the $N$-independents terms in 
Eq.~(\ref{eq:gammaN2}) ($n=0$) from the $N$-dependent terms ($n>0$) by defining
\bea
\hat \gamma^{(l)} \left( \as(Q^2), \as\left( \frac{Q^2}{N^a} \right), \epsilon\right) &\equiv
\sum_{i=1}^\infty \sum_{j=1}^i \gamma_{ij} (\epsilon) [ \bar \alpha_0(1, \as(Q^2), \epsilon)] ^{i-j}
\left[ \bar \alpha_0 \left(1, \as\left( \frac{Q^2}{N^a} \right) , \epsilon\right) \right]^j\nonumber \\
&= \sum_{m=0}^\infty \sum_{n=1}^\infty \gamma_{m+n\, n}^R (\epsilon) \as^m(Q^2) \as^n \left( \frac{Q^2}{\bar N^a} \right) 
\eea
and
\bea
\hat \gamma^{(c)} (\as(Q^2), \epsilon) &\equiv \sum_{i=1}^\infty 
\gamma_{i0} (\epsilon) [ \bar \alpha_0 (1, \as(Q^2), \epsilon)]^i\nonumber \\
&= \sum_{m=1}^\infty \gamma_{m0}^R(\epsilon) \as^m(Q^2)
\eea
such that
\bea\label{eq:gammaN3}
\gamma\left(N,\frac{Q^2}{\mu^2}, \as(\mu^2), \epsilon\right) =
\hat \gamma^{(l)}  \left( \as(Q^2), \as\left( \frac{Q^2}{N^a} \right), \epsilon\right)
+\hat \gamma^{(c)} (\as(Q^2), \epsilon) + \mathcal O(1/N).
\eea

Eq.~(\ref{eq:gammaN3}) implies that, though $\hat \gamma^{(l)}$ and $\hat \gamma^{(c)}$
might not separately be finite in the $\epsilon \rightarrow 0$, they can be made finite by adding an
appropriate counterterm $Z_\gamma$:
\bea
\gamma^{(l)} \left( \as(Q^2), \as\left( \frac{Q^2}{N^a} \right), \epsilon\right) &=
\hat \gamma^{(l)} \left( \as(Q^2), \as\left( \frac{Q^2}{N^a} \right), \epsilon\right) 
- Z_\gamma(\as(Q^2), \epsilon),\\
\gamma^{(c)} (\as(Q^2), \epsilon) &= \hat  \gamma^{(c)} (\as(Q^2), \epsilon)  + Z_\gamma(\as(Q^2), \epsilon)
\eea
where now $\gamma^{(l)}$ and $\gamma^{(c)}$ have a finite $\epsilon \rightarrow 0$ limit. 
The counterterm has to be $N$-independent since $\hat \gamma^{(c)}$ is $N$-independent.
A particularly convenient choice is 
\bea\label{eq:Zcount}
Z_\gamma (\as(Q^2), \epsilon) = \hat \gamma^{(l)} (\as(Q^2), \as(Q^2), \epsilon).
\eea
With this choice $\gamma^{(l)}$ vanishes for $N=1$ (and therefore is finite), but thanks to the
$N$-independence of the counterterm this holds for any $N$. This means that a different 
choice of the counterterm consists in a redefinition of the finite part, which is therefore 
a mere reshuffle of the finite $N$-independent term in $\gamma^{(l)}$ and $\gamma^{(c)}$. 
With the peculiar choice Eq.~(\ref{eq:Zcount}) $\gamma^{(l)}$ is purely logarithmic;
in fact it vanishes at $N=1$, where $\log N=0$. With this choice
\bea
\gamma^{(l)} \left( \as(Q^2), \as\left( \frac{Q^2}{N^a} \right), \epsilon\right) &=
\hat \gamma^{(l)} \left( \as(Q^2), \as\left( \frac{Q^2}{N^a} \right), \epsilon\right) 
- \hat \gamma^{(l)} (\as(Q^2), \as(Q^2), \epsilon),\\
\gamma^{(c)} (\as(Q^2), \epsilon) &= \hat  \gamma^{(c)} (\as(Q^2), \epsilon)  + \hat \gamma^{(l)} (\as(Q^2), \as(Q^2), \epsilon),
\eea
and the physical anomalous dimension is thus
\bea
\gamma\left(N,\frac{Q^2}{\mu^2}, \as(\mu^2), \epsilon\right) &= 
\gamma^{(l)} \left( \as(Q^2), \as\left( \frac{Q^2}{N^a} \right), \epsilon\right)+
\gamma^{(c)} (\as(Q^2), \epsilon)+ 	\mathcal O(1/N)\nonumber \\
&= \gamma^{(l)}  \left( \as(Q^2), \as\left( \frac{Q^2}{N^a} \right), \epsilon\right)+\mathcal O(N^0).
\eea

Now both $\gamma^{(l)}$ and $\gamma^{(c)}$ are finite in the $\epsilon\rightarrow 0$ limit. 
The function $\gamma^{(l)}$ provides an expression of the resummed
physical anomalous dimension up to non-logarithmic in the large $N$ limit.
The choice Eq.~(\ref{eq:Zcount}) let us notice that
\bea
\gamma^{(l)}  \left( \as(Q^2), \as\left( \frac{Q^2}{N^a} \right), \epsilon\right) 
=\int_1^{N^a} \frac{dn}{n} g  \left( \as(Q^2), \as\left( \frac{Q^2}{n} \right), \epsilon\right) 
\eea
where we have defined
\bea
g(\as(Q^2), \as(\mu^2), \epsilon) \equiv & -\frac{\de}{\de  \log \mu^2} \hat \gamma^{(l)} 
(\as(Q^2), \as(\mu^2), \epsilon) \nonumber  \\
&= - \beta^{(d)} (\as(\mu^2), \epsilon) \frac{\de}{\de \as(\mu^2)} \hat \gamma^{(l)} 
(\as(Q^2), \as(\mu^2), \epsilon),
\eea
where $\beta^{(d)}$ is the $d$-dimensional beta function, defined as
\bea
\frac{\de}{\de \log \mu^2} \as(\mu^2)  &\equiv \beta^{(d)} (\as(\mu^2), \epsilon)\nonumber \\
&= -\epsilon \as(\mu^2) + \beta(\as (\mu^2)). 
\eea
It follows that
\bea
g(\as (Q^2), \as(\mu^2), \epsilon) =
\sum_{m=0}^\infty \sum_{n=1}^\infty g_{mn} (\epsilon) \as^m(Q^2) \as^n(\mu^2).
\eea

We have therefore obtained that the resummed physical anomalous
dimension is
\bea\label{eq:gammares}
\gamma^{\textrm{res}}(N, \as(k^2)) &= \int_1^{N^a} \frac{dn}{n} g\left( \as(k^2), \as\left( \frac{k^2}{n} \right)\right) +
\mathcal O(N^0)\\
g\left( \as(k^2), \as\left( \frac{k^2}{n} \right)\right) &=
\sum_{m=0}^\infty \sum_{n=1}^\infty g_{mn} \as^m(k^2) \as^n \left(\frac{k^2}{n} \right); \qquad g_{11}=0.\label{eq:gmn}
\eea
The coefficients $g_{mn}$ can be obtained by comparing the resummed result with the fixed-order
expressions. In the N$^{k-1}$LL approximation it is necessary to know all the
$\frac{k(k+1)}{2}$ coefficients $g_{mn}$ with $n+m \leq k$, and thus \cite{Bolzoni:2005xn} a 
full N$^{2k-1}$LO fixed-order calculation is needed.  

The result obtained coincides with the all-order resummation obtained in 
\cite{Contopanagos:1996nh} but it is less predictive, since in \cite{Contopanagos:1996nh}
one assumes the validity of a factorization formula more restrictive than 
Eq.~(\ref{eq:fact_sigma}). In the latter approach in fact the coefficient function
is written in the large $N$ limit as
\bea
C\left( N, \frac{Q^2}{\mu^2}, \as(\mu^2)\right) 
=C^{(l)} \left( \frac{Q^2}{\mu^2 N^a}, \as(\mu^2)\right) C^{(c)} \left( \frac{Q^2}{\mu^2},
\as(\mu^2)\right),
\eea
which implies that the bare coefficient function is therefore
\bea\label{eq:barefactplus}
C^{(0)}(N, Q^2, \alpha_0, \epsilon) = C^{(0,l)} 
\left( \frac{Q^2}{N^a}, \alpha_0, \epsilon\right) C^{(0,c)} (Q^2,\alpha_0, \epsilon).
\eea
It follows that the the physical anomalous dimension has the form
\bea
\gamma\left(N, \frac{Q^2}{\mu^2}, \as(\mu^2), \epsilon \right) =
\gamma^{(l)} \left( \frac{Q^2}{\mu^2 N^a}, \as(\mu^2), \epsilon \right) +
\gamma^{(c)} \left( \frac{Q^2}{\mu^2}, \as(\mu^2), \epsilon \right),
\eea
where $\gamma^{(l)}$ depends on $\as\left(\frac{Q^2}{N^a}\right)$ only. If one follows 
the lines of the proof above, one obtains that $g=g(\as(\mu^2), \epsilon)$,
i.e. only the terms with $m=0$ in Eq.~(\ref{eq:gmn}) are left. Therefore the
knowledge of the first $k$ coefficients (which corresponds to a N$^k$LO calculation)
is sufficient in order to obtain the N$^{k-1}$LL approximation. 
Eq.~(\ref{eq:barefactplus}) is satisfied if and only if the coefficients 
$C_n^{(0)}(\epsilon)$ have the following structure in the large $N$ limit:
\bea\label{eq:barefactplus2}
C_n^{(0)}(\epsilon) = \sum_{k=0}^n F_k(\epsilon) G_{n-k} (\epsilon) 
(Q^2)^{-(n-k) \epsilon} \left( \frac{Q^2}{N^a}\right)^{-k\epsilon}. 
\eea
It was shown in \cite{Collins:1989gx} that the validity of the factorization
Eq.~(\ref{eq:barefactplus}) holds for different process at various perturbative orders.
%More recently the validity of Eq.~(\ref{eq:barefactplus}) has been proved at NNLL \cite{Moch:2005ba}.
However, an all-order proof of the validity of Eq.~(\ref{eq:barefactplus2}) is not yet available. 

We can finally cast the results Eq.~(\ref{eq:gammares},\ref{eq:gmn}) in $x$-space,
thanks to the relation between leading $\log \frac{1}{N}$ and leading $\log(1-x)$ resummation:
\bea
\gamma^{\textrm{res}} (N, \as(k^2)) &= \int_0^1 dx \ x^{N-1} P^\textrm{res}(x, \as(k^2)) + \mathcal O(N^0)\\
P^\textrm{res} (x, \as(k^2)) &\equiv \left[ \frac{\hat g(\as(k^2), \as(k^2(1-x)^a))}{1-x}\right]_+\\
\hat g(\as(k^2), \as(\mu^2)) &= -\sum_{p=0}^\infty \frac{\Delta^{(p)} (1) a^p}{p!}
\frac{d^p}{d \log^p \mu^2} g(\as(k^2), \as(\mu^2)),
\eea 
where we have defined
\bea
\Delta(p) \equiv \frac{1}{\Gamma(p)}.
\eea
It is therefore possible to compute the resummed evolution factor $K^\textrm{res}(N; Q_0^2, Q^2)$ 
defined by
\bea\label{eq:kres}
K^\textrm{res}(N; Q_0^2,Q^2) &= \exp [E^\textrm{res}(N, Q^2_0, Q^2)]\nonumber \\
&= \exp \int_{Q_0^2}^{Q^2} \frac{dk^2}{k^2} \gamma^\textrm{res}(N, \as(k^2)).
\eea

It is possibile to obtain resummed expression for the coefficient function and anomalous 
dimension in any factorization scheme from Eq.~\eqref{eq:kres}. In particular, it is 
shown in Ref.~\cite{Forte:2002aj} that the resummed coefficient function can be brought in 
the form of Eq.~\eqref{eq:cresg0expS}.

\section{The Landau pole}\label{sec:landaupole}

We can rewrite \cite{Forte:2006mi} the resummed expression of the physical anomalous dimension Eq.~(\ref{eq:gammares}) as
\bea
\gamma^\textrm{res}(N,\as(Q^2)) = \int_1^N \frac{dn}{n}
\sum_{k=1}^\infty g_k \as^k \left(\frac{Q^2}{n}\right) + \mathcal O(N^0) 
\eea
where the constants $g_k$ can be determined by matching with fixed-order expression
and we have neglected $N$-independent or power-suppressed terms. Without loss of generality, 
we have taken $a=1$. The resummed expression of the physical anomalous dimension
is necessary in order to obtain the resummed Mellin transform of the cross section
$\sigma(N, Q^2)$. The physical cross section $\sigma(x,Q^2)$
can finally be calculated by inverting the Mellin transform. 
The resummed result Eq.~(\ref{eq:gammares}) depends on $N$ through the
rescaled coupling  $\as \left(\frac{Q^2}{N}\right)$. This means that
$\gamma^\textrm{res}(N,\as(Q^2))$ and equivalently $\sigma(N,Q^2)$ 
have a branch cut along the real positive axis for $N>N_L$,
where $N_L$ is the location of the Landau pole in $N$-space:
\bea
N_L = \frac{Q^2}{\Lambda^2}.
\eea
As a consequence,
$\sigma(N,Q^2) $ is not the Mellin transform of $\sigma(x,Q^2)$; in fact a Mellin transform
always has a convergence abscissa, i.e. it is an analytic function in the half-plane
$\Re(N) >c$, where $c$ is a real constant. 

For the sake of simplicity, let us consider the resummed physical anomalous dimension
at LL accuracy:
\bea
\gamma_{LL} (N, \as(Q^2)) &= g_1 \int_1^N \frac{dn}{n} \as \left( \frac{Q^2}{n} \right)\nonumber \\
&= - \frac{g_1}{\beta_0} \log \left( 1+\beta_0 \as(Q^2)\log \frac{1}{N} \right)
\eea
where we have used the LL expression of $\as$ Eq.~(\ref{eq:rgeLL}). We notice that
the the LL expression of $\gamma(N, \as(Q^2))$ has a branch cut for $N \geq N_L$,
whith
\bea
N_L \equiv \exp\left( \frac{1}{\beta_0 \as(Q^2)}\right).
\eea
On the other hand, one can consider any finite truncation of the series expansion
\bea
\gamma_{LL} (N, \as(Q^2)) = \sum_{k=0}^\infty \as^k(Q^2) \gamma_k(N),
\eea
which is free of singularities for $N$ large enough. 
The resummed expression in $x$ space can be constructed as
\bea
P_{LL}(x, \as(Q^2)) = \sum_{k=0}^\infty \as^k P_k(x),
\eea
where the coefficients are computed as the inverse Mellin transform of the series expansion
in $N$ space:
\bea
 P_k(x) = \mathcal M^{-1} [\gamma_k(N)] (x).
\eea
In particular, one obtains
\bea\label{eq:seriesmtLL}
P_{LL}(x, \as(Q^2)) =
-\lim_{k\rightarrow \infty} \frac{g_1}{\beta_0} \sum_{k=1}^K 
\frac{(-1)^{k+1}}{k} \beta_0^k \as^k (Q^2) \times
\frac{1}{2 \pi i} \int_{\bar N-i \infty}^{\bar N +i \infty} dN\ x^{-N} \log^k \frac{1}{N}, \qquad \bar N >0.
\eea

The series Eq.~(\ref{eq:seriesmtLL}), however, does not converge. If it did converge,
it would be possible to interchange the sum and the integral, but the sum
\bea
\frac{(-1)^{k+1}}{k} \beta_0^k \as^k (Q^2)\log \frac{1}{N}
\eea
is convergent for
\bea
\left| \beta_0 \as(Q^2) \log \frac{1}{N} \right| < 1,
\eea
whereas the integral involves values of $N$ which are not in this range.
However, if the inverse Mellin transform is taken at finite
logarithmic accuracy, the perturbative series converges.
In the LL case,
\bea
\frac{1}{2 \pi i} \int_{\bar N-i \infty}^{\bar N +i \infty} dN\ x^{-N} \log^k \frac{1}{N}
= k \left[ \frac{\log^{k-1} (1-x)}{1-x}\right]_+ + \textrm{NLL}, 
\eea
and therefore
\bea\label{eq:Pllx}
P_{LLx}(x, \as(Q^2)) &= -g_1 \left[\frac{1}{1-x} \frac{\as(Q^2)}{1+\beta_0 \as(Q^2) \log(1-x)} \right]_+\nonumber \\
&= -g_1 \left[\frac{\as (Q^2 (1-x))}{1-x} \right]_+,
\eea
where the scale $Q^2$ is replaced by $Q^2(1-x)$. Eq.~(\ref{eq:Pllx}) is convergent for 
\bea
x < x_L \equiv 1-\frac{\Lambda^2}{Q^2}.
\eea

Similar arguments apply to the expression of the resummed coefficient function
Eq.~(\ref{eq:Cres}); in particular, the series
\bea\label{eq:creszseries}
C^\textrm{res}(z, \as) = \sum_{k=0} \as^k C^\textrm{res}_k(z)
\eea 
acquires a nonzero convergence radius if the term-by-term inverse Mellin transform of
\bea\label{eq:cresNseries}
C^\textrm{res}(N, \as) = \sum_{k=0} \as^k C^\textrm{res}_k(N)
\eea 
is performed at the relevant logarithmic level. Nevertheless, the convolution integral
Eq.~(\ref{eq:sigma_xspace}) extends to the region above the Landau pole where the series
diverges. It is therefore necessary a definition of what these divergent series mean. 
We now introduce two prescriptions which provide resummed expressions to 
which the divergent series is asymptotic.

\section{Minimal prescription}\label{sec:minimal}

\begin{figure}[htbp]
\begin{center}
\includegraphics[width=8cm]{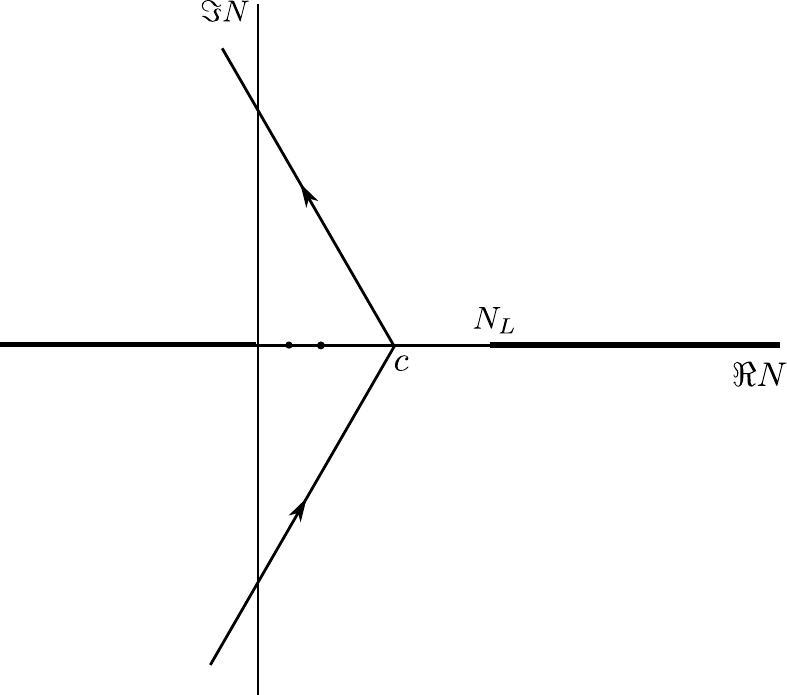}
\end{center}
\caption{Minimal prescription path.}\label{fig:mppath}
\end{figure}

A simple solution in order to handle the Landau pole problem was first presented in Ref.~\cite{Catani:1996yz}.
The resummed hadronic cross-section could formally be written as a power expansion
\bea\label{eq:seriessigman}
\frac{1}{\tau\sigma_0}\frac{d\sigma}{dM^2} = \frac{1}{2 \pi i} \sum_{k=0}^\infty \as^k \int_{c-i \infty}^{c+i\infty}
dN\ \tau^{-N} \mathscr L(N) C_k^\textrm{res}(N)
\eea 
with well-defined coefficients computed as the exact Mellin transform of the coefficients of
the series in $z$ space Eq.~(\ref{eq:creszseries}).
However, as we have seen, the series Eq.~(\ref{eq:seriessigman}) is not
convergent because of the presence of the Landau pole. Equivalently, 
it is not possibile to construct the hadronic cross-section as the inverse Mellin
transform of the product of the $N$-space parton luminosity and coefficient function
since in the integral
\bea
\frac{1}{\tau\sigma_0}\frac{d\sigma}{dM^2} = \frac{1}{2 \pi i} \int_{c-i \infty}^{c+i\infty}
dN\ \tau^{-N} \mathscr L(N) C(N,\as)
\eea 
$c$ does not exist because of the branch cut in the $N$-space. The minimal prescription (MP)
consists of defining the resummed hadronic cross-section as 
\bea\label{eq:MP}
\frac{1}{\tau\sigma_0} \frac{d \sigma^\textrm{MP}}{dM^2}= \frac{1}{2 \pi i} \int_\textrm{MP}
dN\ \tau^{-N} \mathscr L(N) C^\textrm{res}(N,\as)
\eea
where the integration path is chosen in such a way that all the singularities, with the exception
of the branch cut, are to the left of the integration contour. The numerical convergence of the
integral is guaranteed by modifying the slope of the path as shown in Fig.~\ref{fig:mppath}.
The MP is well defined for all $\tau$ and is exact for invertible functions.
In Ref.~\cite{Catani:1996yz} it was shown that the expansion Eq.~(\ref{eq:seriessigman}) converges asymptotically to the MP
formula Eq.~(\ref{eq:MP}) and that if one truncates the series expansion Eq.~(\ref{eq:seriessigman})
at the order at which its terms are at a minimum, the difference between the truncated expansion and 
the full MP formula is suppressed by a more than power suppression factor
\bea\label{eq:suppr}
e^{-\frac{HQ(1-\tau)}{\Lambda}}
\eea
where $H$ is a slowly varying positive function. 

The MP, however, is not a convolution. In fact, if one defines the MP resummed
coefficient function as
\bea
C^\textrm{MP} (z, \as)= \frac{1}{2 \pi i} \int_\textrm{MP} dN\ z^{-N} C^\textrm{res}(N, \as)
\eea
and write $\mathscr L(N)$ as the Mellin transform of $\mathscr L(z)$, the 
resummed hadronic cross-section results in
\bea
\frac{1}{\tau\sigma_0} \frac{d \sigma^\textrm{MP}}{dM^2} =
\int_\tau^\infty \frac{dz}{z} \mathscr L \left( \frac{\tau}{z}\right) C^\textrm{MP}(z, \as).
\eea
Because of the branch cut, in fact, $ C^\textrm{MP}(z, \as)$ does not vanish when
$z>1$ and therefore the integral extends to $\infty$. However, the 
contribution from the region $z>1$ is exponentially suppressed in 
$\frac{\Lambda}{M}$, see Eq.~(\ref{eq:suppr}).  
In conclusion, the MP has good properties but presents some problems. Firstly, the hadronic
cross-section receives contributes from a unphysical region; secondly, it was found in Ref.~\cite{Bonvini:2010tp} that
MP partonic cross-section strongly oscillates in the prossimity of $z=1$ and therefore
problems in the numerical implementation arise. In order to solve this last problem
different solutions have been suggested. 
In Ref.~\cite{Catani:1996yz} the problem is handled by adding and subtracting 
the results of the minimal prescription evaluated with a toy luminosity. 
In Ref.~\cite{Bolzoni:2006ky} the use parton distributions whose Mellin
transform can be computed exactly at the initial scale has been proposed. This solution 
however restricts the choice of PDFs and therefore is not a suitable solution for
precision phenomenology. 
In Ref.~\cite{Bonvini:2010tp} the parton luminosities is expanded 
on a basis of Chebyshev polynomials, whose Mellin transform can be computed analytically.

\section{Borel prescription}\label{sec:borel}

In Refs.~\cite{Forte:2006mi,Abbate:2007qv} an alternative prescription, based on Borel summation of divergent series,
was developed. In this Section we will follow the simpler presentation given in Refs.~\cite{Bonvini:2012sh,Bonvini:2010tp}.
We consider the resummed coefficient function Eq.~(\ref{eq:cresg0expS}) 
which we write as
\bea
C^\textrm{res}(N, \as) = h_0(\as) + \tilde \Sigma(\bar \alpha L, \as)
\eea
where $h_0(\as)$ contains only terms constants in $N$
and $\Sigma$, which includes only logarithmically enhanced contributions,
can be expanded in series as
\bea\label{eq:Sigmatilde}
\tilde \Sigma(\bar \alpha L, \as) = \sum_{k=1}^\infty h_l (\as) (\bar \alpha L)^k.
\eea
The coefficient function in $z$ space can be written as the series
\bea\label{eq:origdiv}
C^\textrm{res}(z, \as) = \sum_{k=0}^\infty h_k (\as) \bar \alpha^k c_k(z),
\eea
where we have included the constant term in the sum and $c_k(z)$
are the inverse Mellin transform of $\log^k\frac{1}{N}$, which can be written
in one of the following alternative forms:
\bea
\mathcal M\left[\log^k\frac{1}{N}\right]&=\delta_{k0} \delta(1-z)
+ \left(\frac{d^k}{d \xi^k} \frac{\log^{\xi-1}\frac{1}{z}}{\Gamma(\xi)}
\Big|_{\xi=0} \right)_+\\
&= \delta_{k0} \delta(1-z) + \frac{k!}{2 \pi i} \left(
\oint \frac{d \xi}{\xi^{k+1}} \frac{\log^{\xi-1}\frac{1}{z}}{\Gamma(\xi)}\right)_+.
\eea
In the last form the 
path of integration must inclose $\xi=0$. 
Therefore by interchanging the sum and the integral the resummed coefficient
function in $z$ space becomes
\bea\label{eq:Cresfactdiv}
C^\textrm{res}(z, \as) = \frac{1}{2 \pi i} \oint
\frac{d \xi}{\xi} \left( \frac{\left[\log^{\xi-1} \frac{1}{z}\right]_+}{\Gamma(\xi)}
+\delta(1-z) \right) \sum_{k=0}^\infty h_k(\as) \left( \frac{\bar \alpha}{\xi}\right)^k k!.
\eea

The series in Eq.~(\ref{eq:Cresfactdiv}) is factorially divergent. However, we can sum 
the series using the Borel method:
\bea\label{eq:seriesnocnv}
C^\textrm{res} &= \frac{1}{2 \pi i} \oint \frac{d \xi}{\xi} \left( 
\frac{\left[\log^{\xi-1} \frac{1}{z}\right]_+}{\Gamma(\xi)} + \delta(1-z)\right) 
\int_0^\infty \d\omega \ e^{-\omega} \sum_{k=0}^\infty h_k (\as) 
\left( \frac{\bar \alpha \omega}{\xi} \right)^k\nonumber \\
&=  \frac{1}{2 \pi i} \oint \frac{d \xi}{\xi} \left( 
\frac{\left[\log^{\xi-1} \frac{1}{z}\right]_+}{\Gamma(\xi)} + \delta(1-z) \right)
\int_0^\infty \d\omega \ e^{-\omega} \sum_{k=0}^\infty h_0 + 
\tilde \Sigma\left(\frac{\bar \alpha \omega}{\xi}, \as \right)\nonumber \\
&=  \frac{1}{2 \pi i} \oint \frac{d \xi}{\xi} \left( 
\frac{\left[\log^{\xi-1} \frac{1}{z}\right]_+}{\Gamma(\xi)} + \delta(1-z) \right)
\int_0^\infty \d\omega \ e^{-\omega} \sum_{k=0}^\infty \Sigma\left(\frac{\bar \alpha \omega}{\xi}, \as \right)
\eea 
where we have used the definition Eq.~(\ref{eq:Sigmatilde}) and we have defined the function
$\Sigma(b, \as) = h_0 (\as) + \tilde \Sigma(b, \as) $. We observe that the branch cut
$-\infty < \bar \alpha L \leq -1$ of the function $\Sigma $ has been mapped in terms of 
the new variable $\xi$ onto the range $-\bar \alpha \omega \leq \bar \alpha \xi \leq 0 $ 
in the complex $\xi$ plane. Therefore the $\xi $ integration path, which has to encircle
the $\xi$ pole order by order, is any closed curve encircling the cut. 
We further observe that the $\omega$ integral move the lower branch point to $-\infty$.
This means that the inverse Borel of Eq.~(\ref{eq:seriesnocnv}) does not 
exists because of the factorial growth of the function $1/\Gamma(\xi)$ as $\xi \rightarrow \infty$.

We now formulate the Borel prescription.
It is convenient to recast the integral Eq.~(\ref{eq:seriesnocnv}) into a 
different form by changing variable in the $\omega$ integral defining 
$\omega' = \bar\alpha \omega $:
\bea
C^\textrm{res} &=  \frac{1}{2 \pi i} \oint \frac{d \xi}{\xi} \left( 
\frac{\left[\log^{\xi-1} \frac{1}{z}\right]_+}{\Gamma(\xi)} + \delta(1-z) \right)
\int_0^\infty \frac{\d\omega}{\bar \alpha} \ e^{-\frac{\omega}{\bar \alpha}}
\sum_{k=0}^\infty \Sigma\left(\frac{\omega}{\xi}, \as \right). 
\eea 
This integral is divergent; to make it convergent, we put a upper cutoff at some finite value
$W$ of the Borel variable $\omega$:
\bea\label{eq:BP}
C^\textrm{BP} (z, \as, W) = \frac{1}{2 \pi i} \oint \frac{d \xi}{\xi} \left( 
\frac{\left[\log^{\xi-1} \frac{1}{z}\right]_+}{\Gamma(\xi)} + \delta(1-z) \right)
\int_0^W \frac{\d\omega}{\bar \alpha} \ e^{-\frac{\omega}{\bar \alpha}}
\sum_{k=0}^\infty \Sigma\left(\frac{\omega}{\xi}, \as \right).
\eea

In Refs.~\cite{Forte:2006mi,Abbate:2007qv} it is proved that the divergent series Eq.~(\ref{eq:BP}) is asymptoptic 
to the original divergent series Eq.~(\ref{eq:origdiv}) and that the neglected terms due to the cutoff
are higher twist:
\bea
\exp \frac{W}{\bar \alpha}= \left( \frac{\Lambda^2}{M^2}\right)^{W/2}
\eea
where $\Lambda$ is the Landau pole scale. This means that the full and cutoff results 
differ for any truncation by a twist $2+W$ (for Drell-Yan or Higgs production). Since
the first subleading twist is twist 4, the parameter $W$ can be chosen freely in the
range $W \geq 2$. Different choices of the parameter $W$ correspond to 
equivalent results which differ by power suppressed terms. It is therefore possibile 
to use $W$ as a parameter to estimate ambiguities in the resummation procedure. 
In Ref.~\cite{Bonvini:2010tp} the ``minimal'' choice $W=2$ was used.
Finally, it is important to stress the fact that the resummed expression is at parton level
and therefore does not spoil the convolution structure as the MP does.

\chapter{Phenomenology of threshold resummation in dQCD}\label{chapter:phenres}

In this Chapter we will apply soft-gluon resummation in the dQCD approach
to a specific process, Higgs production at LHC. 
In particular, we will concentrate on the $gg\rightarrow H$ production mechanism, 
which is the dominant mechanism for SM Higgs boson production at LHC. 
It is well know \cite{Grazzini:2010zc} that the 
main contribution to Higgs production in the gluon-gluon channel comes from the
threshold region; for this reason, resummation arguments can been used in order to estimate
higher order approximations of the cross-section \cite{Ball:2013bra}. 
In Sect.~\ref{sect:higgsLHC} we will briefly describe the main production modes of a SM 
Higgs boson at the LHC.
We will discuss in more detail Higgs production in gluon-gluon fusion at fixed perturbative 
order in Sect.~\ref{sect:ggH}. We will finally see the effects of soft-gluon resummation
on this process in Sect.~\ref{sect:higgsressqcd}. 

\section{Higgs production at the LHC}\label{sect:higgsLHC}

\begin{figure}[htbp]
\begin{center}
\includegraphics[width=10cm]{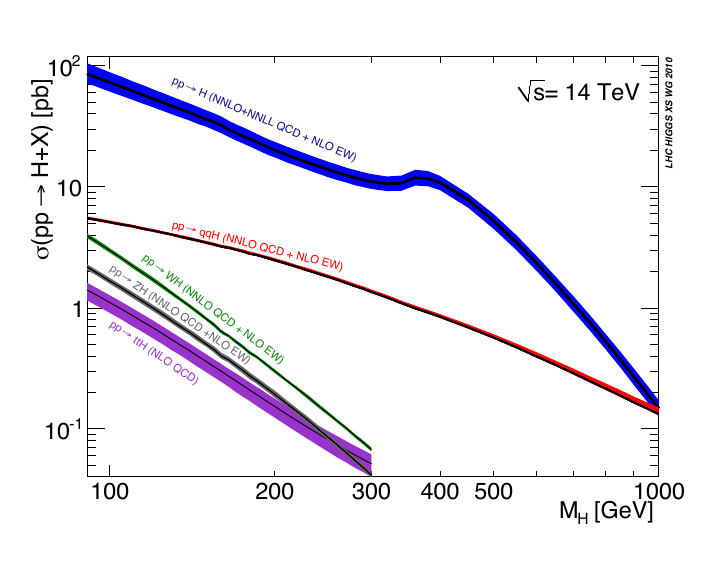}
\end{center}
\caption{Higgs production cross-sections in the Standard Model,
including parametric and systematic theoretical errors, for the LHC 
at c.m. energy 14 TeV.}\label{fig:higgs_cross_tot}
\end{figure}

In this section we briefly describe the main Higgs production mechanisms
at a high energy collider, such the LHC proton-proton collider. 
There are only a few production mechanisms which are relevant
for the production of a Higgs boson with $m_H \sim 125$ GeV at the LHC. The cross-sections
for such processes are represented in Fig.~\ref{fig:higgs_cross_tot}. In order of decreasing importance the
measurable production modes are:
\begin{enumerate}[(a)]
\item gluon-gluon fusion: $gg\rightarrow H$,
\item vector boson fusion (VBF): $qq\rightarrow qqH$, via $W^+W^-, \ ZZ \rightarrow H$,
\item associated production with vector bosons: $q\bar q\rightarrow WH, \ ZH $,
\item associated production with top quarks: $gg, q \bar q \rightarrow Ht\bar t$.
\end{enumerate}

\begin{figure}[htbp]
\centering
\subfloat[][$gg\rightarrow H$.]
   {\includegraphics[width=.2\columnwidth]{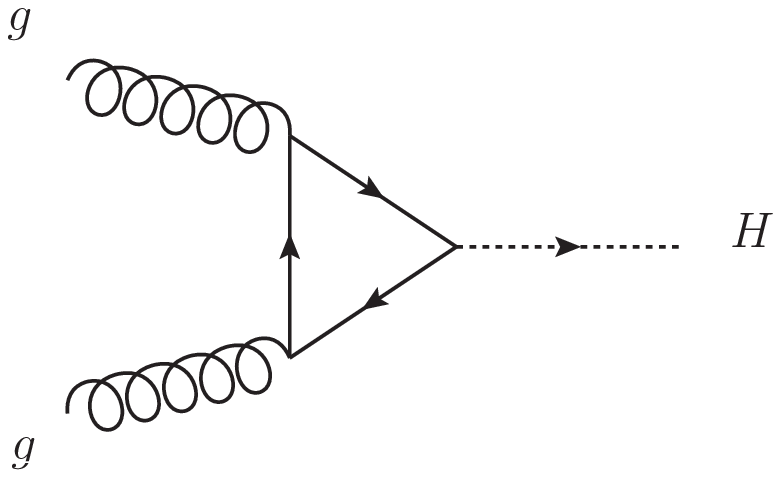}} \quad
\subfloat[][\emph{VBF}.]
   {\includegraphics[width=.2\columnwidth]{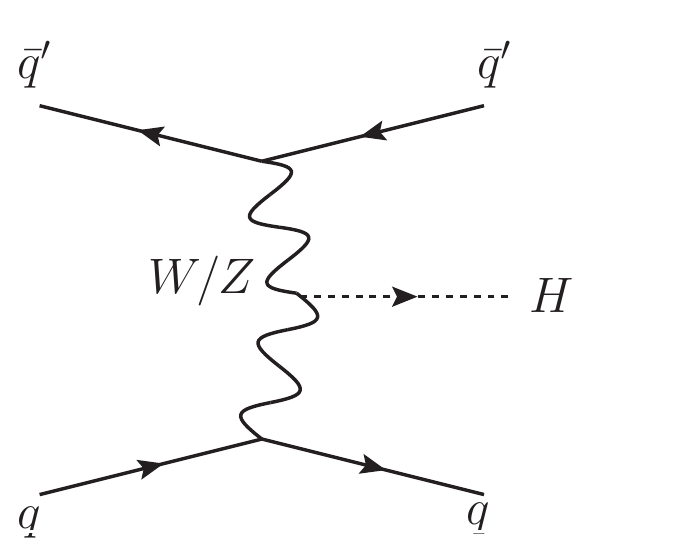}} \quad
\subfloat[][\emph{VH}.]
   {\includegraphics[width=.2\columnwidth]{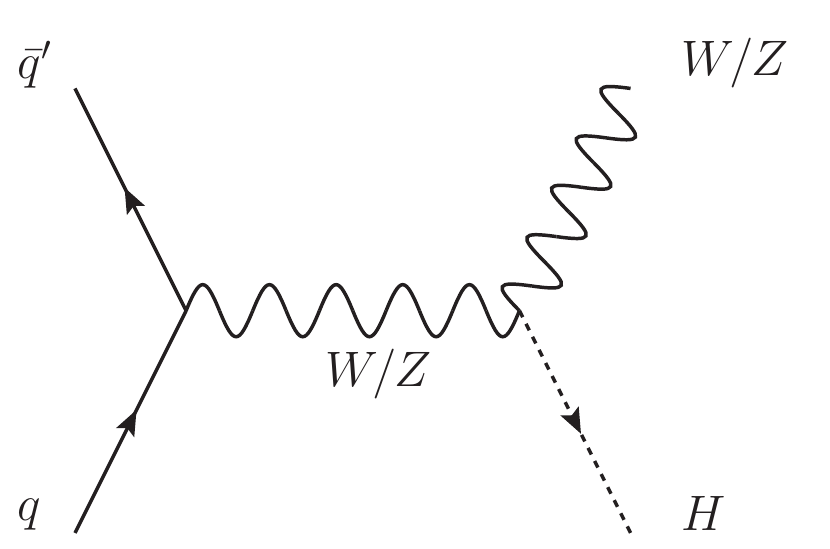}} \quad
\subfloat[][$t\bar t \rightarrow H$.]
   {\includegraphics[width=.2\columnwidth]{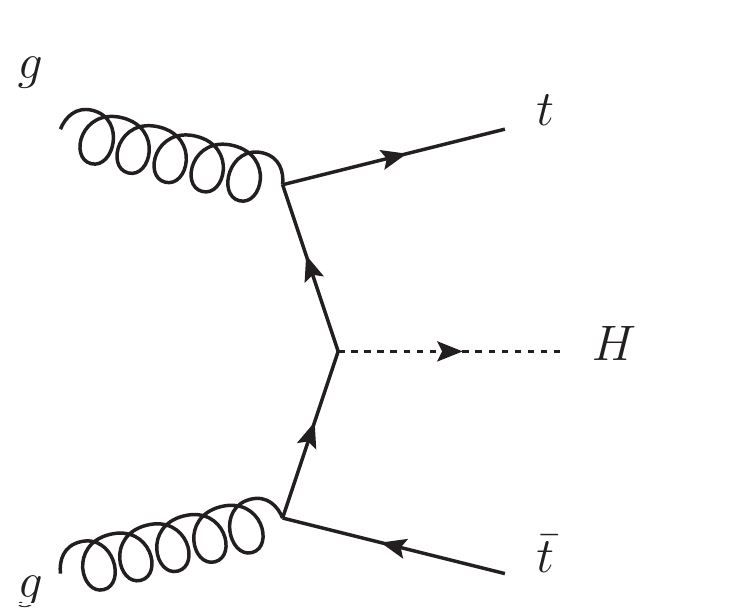}}\\
\caption{Higgs production mechanisms at the LHC.}
\label{fig:higgs_prod}
\end{figure}

The leading order Feynman diagrams for these mechanisms are shown in Fig.~\ref{fig:higgs_prod}. 
The main contribution is the gluon-gluon fusion process, via a loop of 
heavy quarks (we recall that Yukawa Higgs couplings are proportional to masses).
We will describe Higgs production in this channel more accurately in the next Section.

The next contribution, vector-boson fusion, has been calculated 
at NNLO QCD with the inclusion of NLO EW corrections and it presents a well-converging 
perturbative expansion.
The third largest cross-section, the associated production 
with a massive vector boson, has been calculated at NNLO QCD + NLO EW. The 
perturbative expansion has also in this case good properties of convergence.
The fourth and less frequent production mechanism is the associated production 
with a couple of top quarks. At the moment this process is known less accurately.
It has been calculated at NLO QCD and has therefore larger uncertainties. 
However, since the cross-section for this specific process increases 
rapidly with energy, there is an increasing interest in this latter process in
prospect of the LHC upgrade at 14 TeV. 

\section{Higgs production in gluon-gluon fusion at fixed perturbative order}\label{sect:ggH}

The Higgs production in gluon-gluon fusion is the dominant Higgs production
mechanism in the SM. The perturbative expansion of the cross-section is 
slowly convergent;
it has been calculated with full-dependence
on the top and bottom quarks at NLO and at NNLO in the heavy-top limit. In 
Refs.~\cite{Marzani:2008ih,Marzani:2008az} a computation of
the cross-section with finite $m_t$ mass in the limit of high
partonic c.m.s. energy was presented. A recent
approximate determination of the N$^3$LO was presented in Ref.~\cite{Ball:2013bra}.
The cross-section for the threshold production 
of the Higgs boson at hadron-colliders 
at N$^3$LO has been recently performed in Ref.~\cite{Anastasiou:2014vaa}. A complete
N$^3$LO in the heavy-top limit is on the way. Mixed QCD+EW corrections have been calculated in
Ref.~\cite{Anastasiou:2008tj}.
%The fixed-order predictions have been improved by adding the resummation of the soft contribution at NNLL accuracy.

\begin{figure}[tbp]
\begin{center}
\includegraphics[width=10cm]{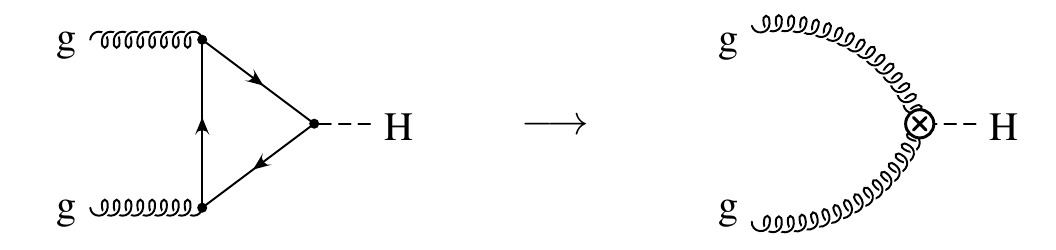}
\end{center}
\caption{Effective vertex in the $m_t \rightarrow \infty$ limit.}\label{fig:eff_vertex}
\end{figure}

The NLO corrections to this process, which arises mainly from the
radiation of soft and collinear gluons, are very large. Gluon radiation 
gives the dominant contribution in the soft limit, where $\hat s \sim m_{H}^2$.
This leading contribution, however, does not resolve the top quark loop in the 
large $m_t$ limit, i.e. when $m_H \ll 2 m_t$. Therefore, it is possibile to
calculate the NLO corrections with a good level of accuracy in the limit
$m_t \rightarrow \infty$, which greatly reduces the complexity of the calculation
since it is reduced by one order of loops. In the infinite top-quark mass limit
the one loop $ggH$ vertex is reduced to a tree level effective vertex (see Fig. \ref{fig:eff_vertex})
which can be derived by the effective lagrangian
\bea
\mathcal L = -\frac{1}{4 v} W G^a_{\mu\nu}G^{a,\mu\nu}H, \qquad v =\frac{1}{(2 G_F^2)^{1/4}}.
\eea
The Wilson coefficient $W$
% which contains the 
%residual logarithmic dependence on the top quark, 
in the $\overline{\textrm{MS}}$ scheme is given by \cite{Chetyrkin:1997un,Harlander:2001is}
\bea
W = - \frac{\as}{3 \pi} \left[1+ \frac{11}{4} \frac{\as}{\pi} +
\left[\frac{2777}{288} + \frac{19}{16}\log \frac{\mu^2}{m_t^2} + n_f 
\left(-\frac{67}{96} + \frac{1}{3} \log \frac{\mu^2}{m_t^2} \right) \right] 
\left( \frac{\as}{\pi} \right)^2+ \mathcal O(\as^3)\right].
\eea

\begin{figure}[tbp]
\begin{center}
\includegraphics[width=11cm]{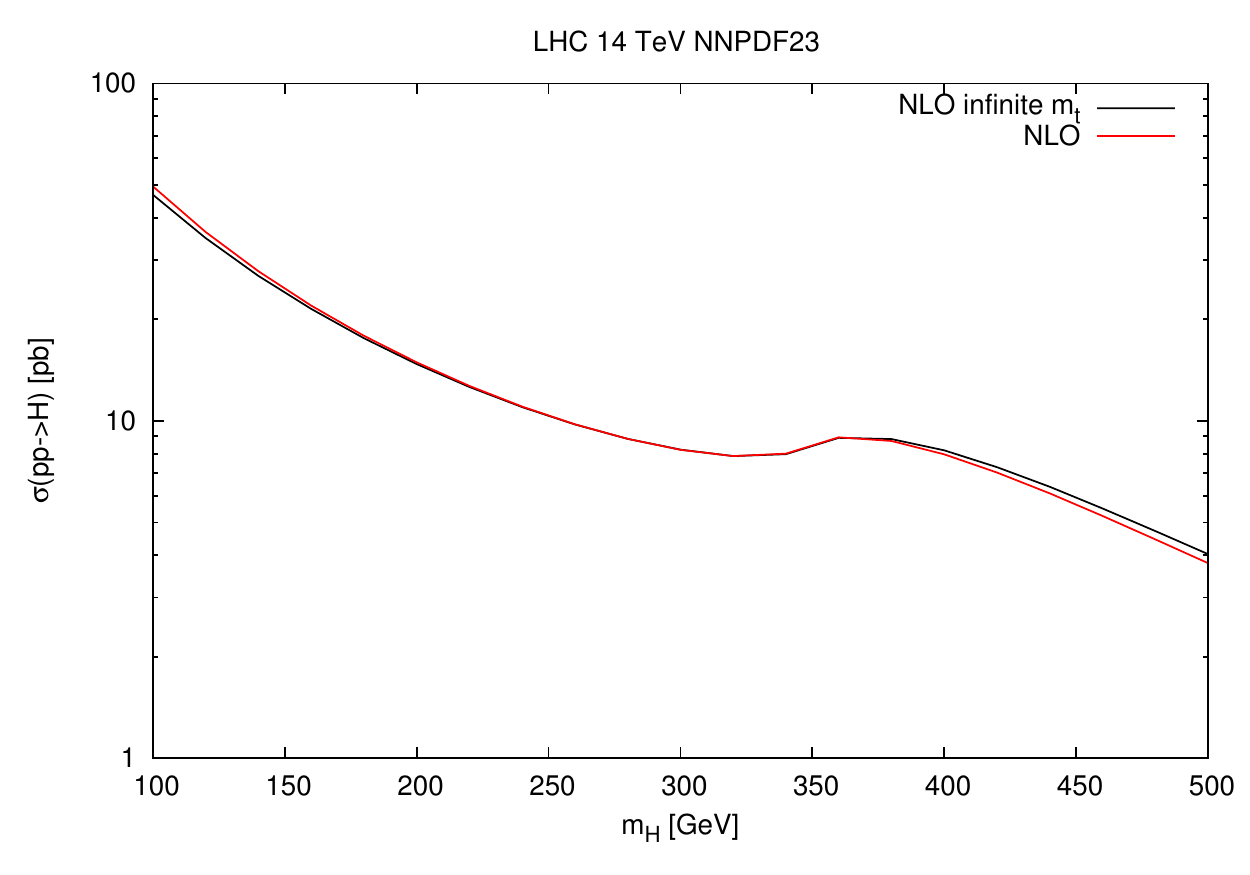}
\end{center}
\caption{Comparison between the NLO exact cross-section and the
cross-section calculated in the infinite $m_t$ limit. The curves have 
been obtained with \texttt{ggHiggs} and using NNPDF23 NNLO PDF set.}\label{fig:topinf}
\end{figure}

In the infinite $m_t$ limit it is possibile to calculate the NLO and NNLO 
QCD corrections as corrections to the effective vertex. Since the NLO correction
has been calculated with full $m_t$ dependence, it has been possibile to test 
how accurate is the infinite $m_t$ approximation at NLO. The approximate and the
exact results shows a remarkable agreement at the level of the total cross-section
as one can see in Fig.~\ref{fig:topinf} even for a heavy Higgs. 

At NNLO the total cross-section requires the calculation of two-loop diagrams
instead of the original three-loop diagrams.
%The full NNLO correction has been computed in Ref. .
The cross-section has been calculated in the soft limit in Ref.~\cite{Harlander:2001is}. The
partonic cross-section $\hat \sigma_{ij}$ has the perturbative 
expansion
\bea
\hat \sigma_{ij}^{(n)} = \sum_{n=0}^\infty \left(\frac{\as}{\pi}\right)^n 
\hat \sigma_{ij}^{(n)}.
\eea
In the soft limit ($z\rightarrow 1$) it is possibile to write the
$n$-th term of the expansion by expanding about $z=1$ and after 
separating off distributions from regular terms as
\bea
\hat \sigma_{ij}^{(n)} = 
\overbrace{a^{(n)} \delta(1-z) + \sum_{k=0}^{2n-1} b_k^{(n)}
\left[ \frac{\log^k(1-z)}{1-z} \right]_+}^{\textrm{purely soft terms}} + 
\overbrace{\sum_{l=0}^\infty\sum_{k=0}^{2n-1}
c_{lk}^{(n)} (1-z)^l \log^k (1-z)}^{\textrm{collinear + hard terms}}.
\eea 
The NNLO cross-section is obtained by calculating the perturbative 
coefficient $a^{(2)},\ b^{(2)},\ c^{(2)}_{lk}$ for $l\geq 0$
and $k=0,\ldots,3$. The result obtained in this limit 
has been confirmed by the full calculation of Ref.~\cite{Anastasiou:2002yz}. 

The total cross section
for the production of a Higgs boson in hadronic collisions at 
center-of-mass energy $\sqrt{s}$ can be written as
\bea
\sigma(\tau, M^2) &= \sum_{ij} \int_0^1 dx_1 \int_0^1 dx_2 \int_0^1 
dz f_i(x_1) f_j (x_2) \hat \sigma_{ij} (z) \delta \left( z -\frac{\tau}{x_1 x_2}\right)\\
&= \tau \sigma_0 \as^2 \sum_{ij} \int_\tau^1 \frac{dz}{z} \mathscr L_{ij} 
\left(\frac{\tau}{z} \right) C_{ij} (z)
\eea
where
\bea
\tau &= \frac{M^2}{s},\\
\sigma_0 &= \frac{G_F}{288 \pi \sqrt{2}} \left| \sum_q A(x_q) \right|^2, \qquad 
x_q = \frac{4 m_q^2}{m_H^2},\\
A(x) &= \frac32 x [1+(1-x) f(x)], \qquad f(x) = 
\begin{cases} \arcsin^2 \frac{1}{\sqrt{x}} & x \geq 1\\
-\frac{1}{4} \left[ \log \frac{1+\sqrt{1-x}}{1-\sqrt{1-x}}-i \pi\right]^2 & x <1
\end{cases},
\eea
$\mathscr L_{ij}$ is the parton luminosity Eq.~(\ref{eq:partonlum}) and the
coefficient functions $C_{ij}$ are related to the partonic cross-section as
\bea
\hat \sigma_{ij} (z) = \sigma_0 z \as^2 C_{ij}(z, \as)
\eea
and have a perturbative expansion in $\as$:
\bea
C_{ij} = C_{ij}^{(0)} + \as C_{ij}^{(1)} + \as^2C_{ij}^{(2)} + \ldots .
\eea
At LO only $C_{gg}$ contributes; the coefficient functions at NLO in the
large-$m_t$ limit can be found in Refs.~\cite{Dawson:1990zj,Dawson:1993qf},
whereas the full NNLO correction can be found in Ref.~\cite{Anastasiou:2002yz}.

\begin{figure}[htbp]
\centering
   {\includegraphics[width=.8\columnwidth]{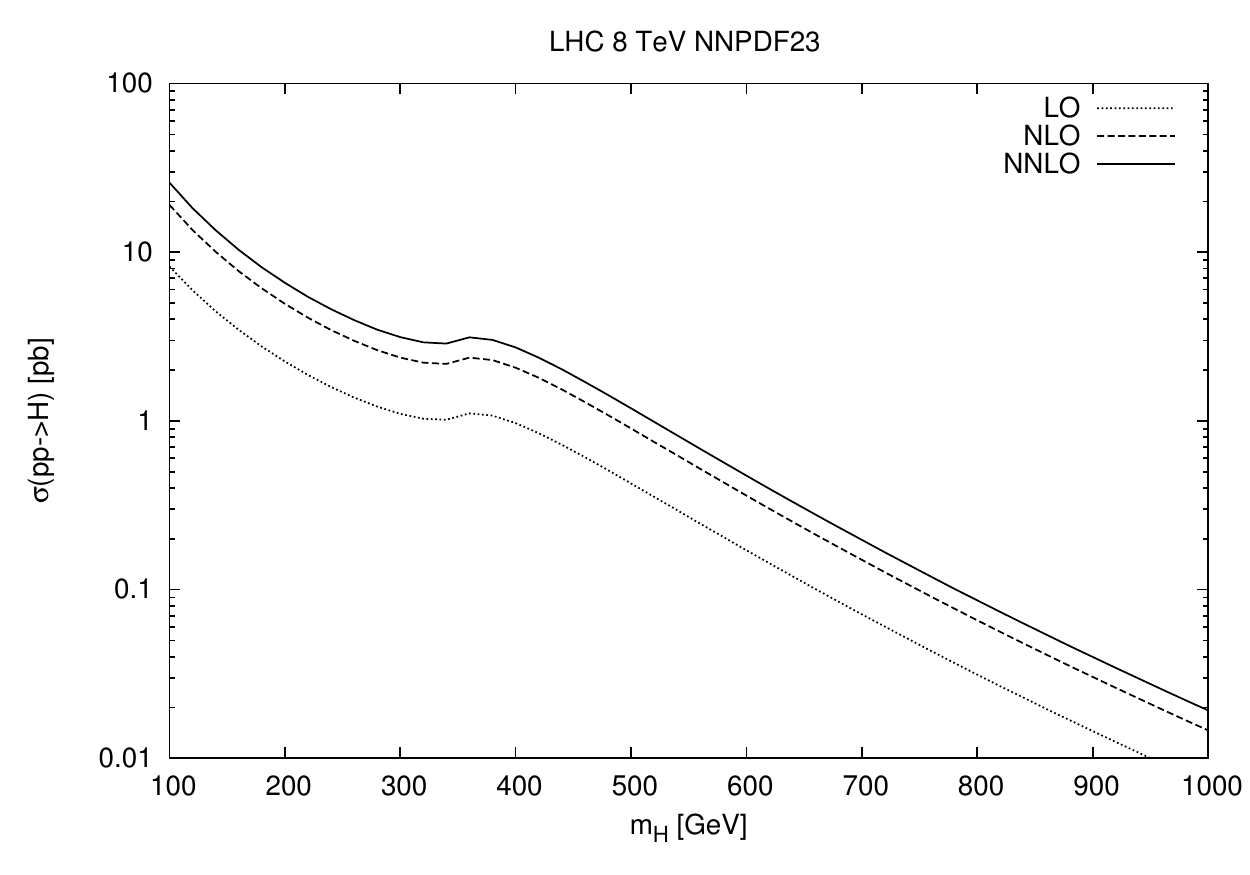}}\\
\caption{Total cross-section for $gg\rightarrow H$ as a function of the
Higgs mass $m_H$ at LO, NLO and NNLO at the LHC, setting 
$\mu_F = \mu_R = m_H$, computed with \texttt{ggHiggs} in the large-$m_t$
approximation in the LHC 8 TeV kinematics.}\label{fig:gghigsfixed}
\end{figure}

We show in Fig. \ref{fig:gghigsfixed} the cross-section for the Higgs production in 
gluon-gluon fusion calculated at LO, NLO and NNLO with \texttt{ggHiggs} \cite{ggHiggs} in the 
large-$m_t$ limit. All the curves have been obtained using the NNPDF2.3 NNLO PDF set
and and $\as(m_Z) =0.0117$ in order to show the perturbative behavior of the
hard partonic cross-section. We 
observe that the cross-section has a slowly convergent expansion.

\section{Higgs resummed results}\label{sect:higgsressqcd}

The NNLO cross section has been further improved in Ref.~\cite{Catani:2003zt} by
resumming up to NNLL accuracy.
In this Section we present a phenomenological study of the impact of threshold 
resummation on the inclusive cross-section for the Higgs boson production 
process. Once
the resummed coefficient functions have been computed, their convolution with
PDFs provide us (after matching consistently with fixed-order results) with the 
resummed hadronic observables.  

As we have seen in the previous Chapters, in dQCD resummation is performed in 
$N$ space to finite logarithmic accuracy. After performing the Mellin inversion 
with one of the aforementioned prescriptions, retaining terms to all log orders in 
$1-z$ the physical hadronic cross-section can be finally obtained. 
The most general expression for the resummed coefficient in the $N$ space is, as we have seen,
\bea
C^\textrm{res} (N, \as) = g_0 (\as) \exp \left[ \frac{1}{\bar \alpha} 
g_1 \left(\bar \alpha L\right)+ 
g_2 \left(\bar \alpha L\right)+
\bar \alpha g_3 \left(\bar \alpha L\right) + \ldots\right].
\eea
where $L = \log \frac{1}{N}$.
The explicit expressions of the coefficients $g_i $ can be found in Appendix~\ref{app:analy}. The
double-counting terms which have to be subtracted when matching with the fixed-order 
result can be found as the Taylor expansion in
powers of $\as$ of $g_0(\as) \exp \mathcal S(\lambda, \bar \alpha)$:
\bea
g_0(\as) \exp \mathcal S(\lambda, \bar \alpha) &= (1+ \as g_{01} + \as g_{02}^2 + \ldots)
e^{\as \mathcal  S_1 + \as^2 \mathcal S_2+ \ldots}\nonumber \\
&= 1+ (\mathcal S_1 + g_{01}) \as +
\left( \frac{\mathcal S_1^2}{2} + \mathcal S_2 + \mathcal S_1 g_{01} + g_{02} \right) \as^2 
+\mathcal O(\as^3).
\eea

\begin{figure}[htbp]
\centering
   {\includegraphics[width=.75\columnwidth]{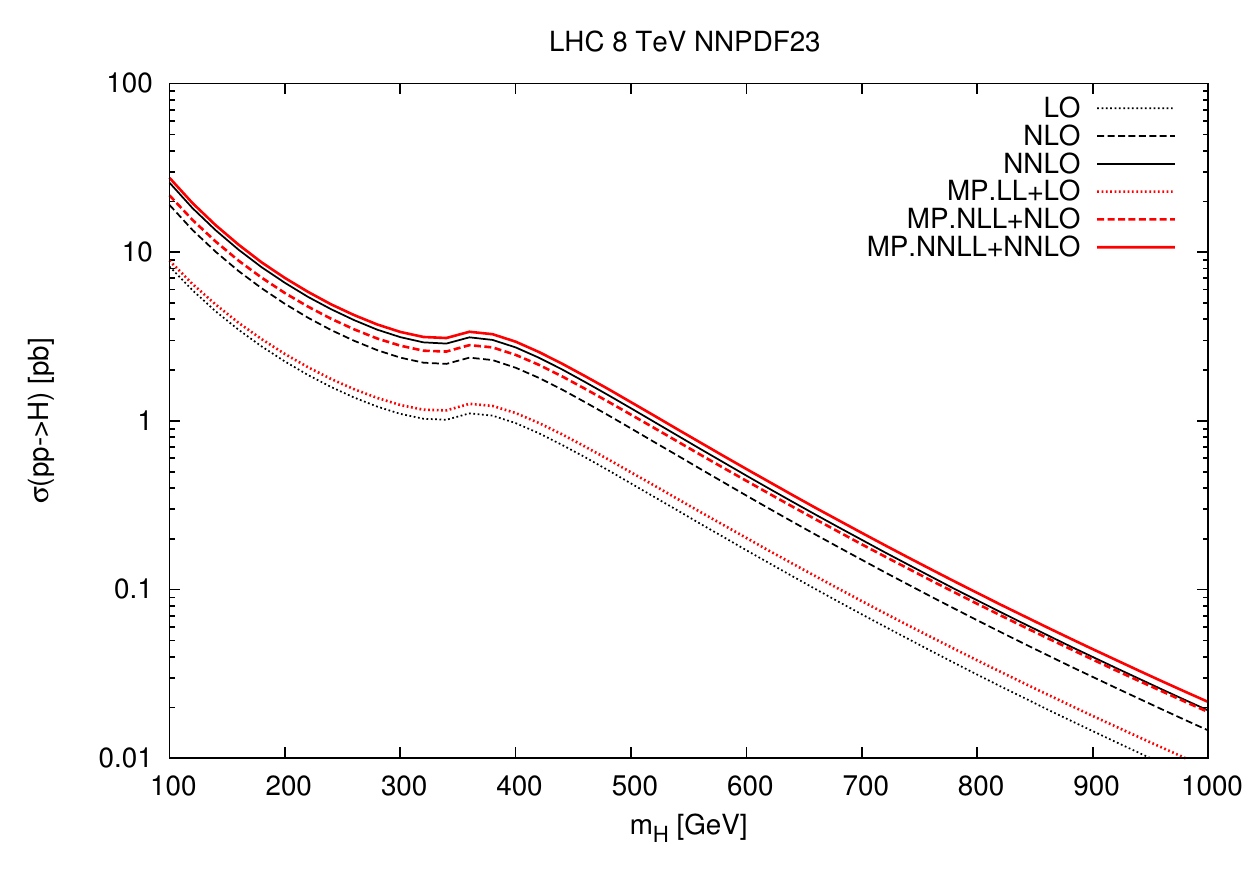}} 
\caption{Resummed cross-section as a function of the
Higgs mass $m_H$ at LL, NLL and NNLL at the LHC. The curves have been calculated with 
NNPDF2.3 NNLO PDF set, with $\mu_F=\mu_R=m_H$.}\label{fig:gghiggsres}
\end{figure}

In Fig.~\ref{fig:gghiggsres} we show the results for the inclusive invariant mass distribution
for Higgs boson production. The resummed result have been obtained with the minimal prescription 
described in Sect.~\ref{sec:minimal} and then matched with the fixed-order results obtained with 
\texttt{ggHiggs}. Also in this case we have used the NNPDF2.3 NNLO PDF set. We have 
used a code by M. Bonvini which resums up to NNLL accuracy and is consistently matched with 
fixed-order corrections up to NNLO.

\begin{figure}[htbp]
\centering
   {\includegraphics[width=.75\columnwidth]{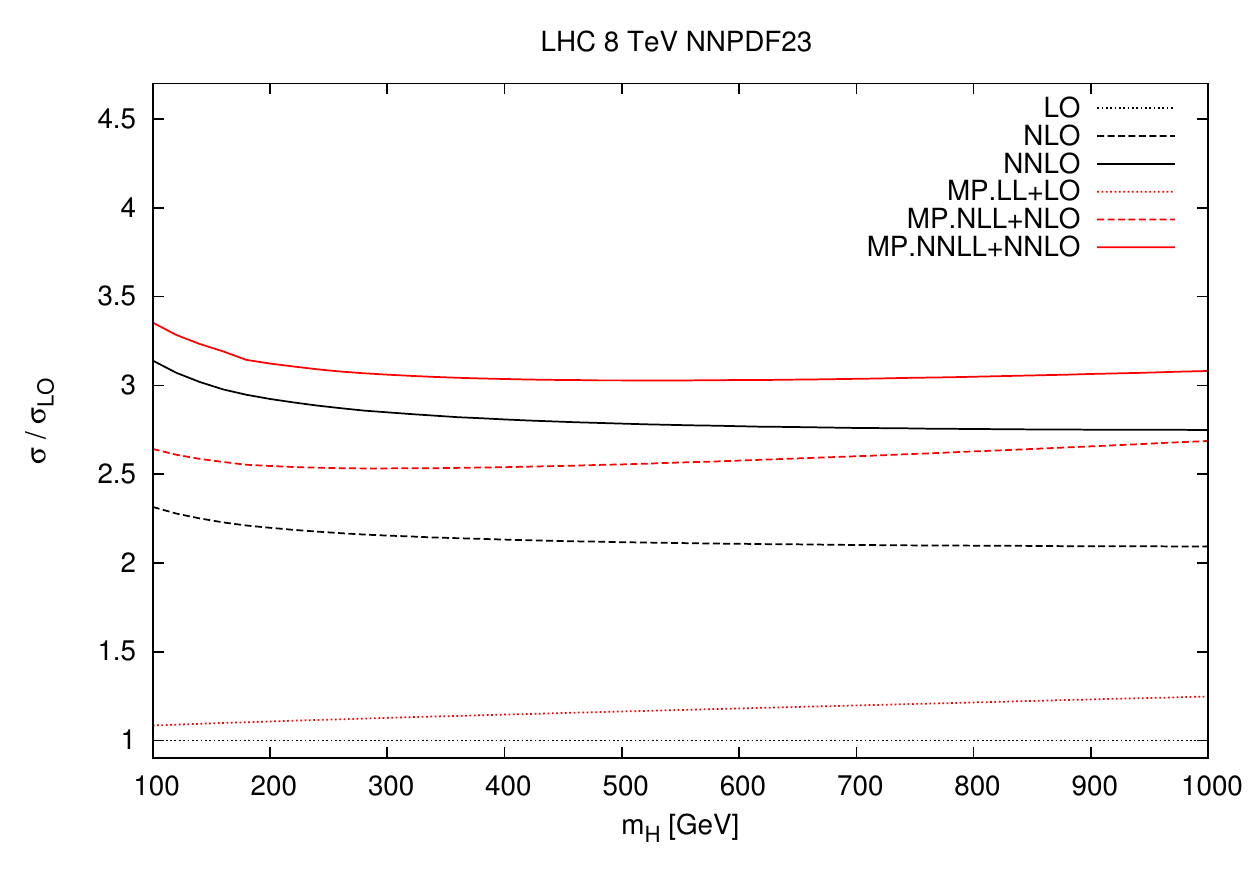}}
\caption{$K$-factor for the resummed cross-section as a function of the
Higgs mass $m_H$ at LL, NLL and NNLL at the LHC. The curves have been calculated with 
NNPDF2.3 NNLO PDF set, with $\mu_F=\mu_R=m_H$. }\label{fig:gghiggskfact}
\end{figure}

In order to obtain a better assesment of the effect of threshold resummation it 
is useful to introduce the $K$-factor defined as 
\bea\label{eq:kfactor}
K(\tau, M^2) = \frac{ \sigma(\tau, M^2)}{\sigma^\textrm{LO}(\tau, M^2)}.
\eea
We show in Fig.~\ref{fig:gghiggskfact} the $K$-factor for the fixed-order and the
resummed cross-section as a function of the Higgs mass. We note once again that the
NLO correction to the total cross-section is very large and it amounts to 
a correction by $100\%$ of the LO cross-section, and that only at NNLO the perturbative series
starts to show a better behaviour. We observe that the effects of soft-gluon resummation 
are small at LL. We have checked that this fact is due to subleading terms,
which are known to be relavant in Higgs boson production since the process is far
from threshold. On the other hand, the difference between 
NNLL and NNLO is about one half of the difference between NNLO and NLO.  We 
observe that NLL correction is not very distant from the NNLO curve, and it gets
closer as the Higgs mass grows.

\chapter{Resummation in the SCET approach}\label{chapt:scetapp}

In recent years an alternative approach to resummation, based on soft-collinear
effective theory\footnote{In this Thesis we are not interested in a discussion of 
soft-collinear effective theory and we will limit our discussion to its
implications on phenomenology. An introduction to SCET is available 
online at \href{http://physicslearning.colorado.edu/tasi/tasi_2013/notes/june13/Bauer1.pdf}
{\texttt{http://physicslearning.colorado.edu/tasi/tasi\_2013/notes/june13/Bauer1.pdf}}.} (SCET) has been proposed. Soft-collinear 
effective theory techniques provide a powerful way of performing resummation 
in the standard Mellin space \cite{Manohar:2003vb}. However, since effective 
theory deals with hadronic degrees of freedom, resummed expressions can be 
derived in a SCET approach in terms of hadronic-related quantities, as it was pointed
out in Ref.~\cite{Becher:2006nr}. In this latter approach the scale whose logarithms are resummed 
is a soft scale $\mu_s$ fixed in terms of the hadronic momentum scale $M(1-\tau)$.  
The Becher-Neubert (BN) choice of scale avoids the Landau pole problem, since the divergences 
related to the need to integrate over the parton kinematics disappear. 
The approach of Ref.~\cite{Becher:2006nr} (that we will refer to as the SCET approach) has been
subsequently extended to various physical processes, such as deep inelastic scattering \cite{Becher:2006mr}, 
Drell-Yan \cite{Becher:2007ty}, and Higgs production \cite{Ahrens:2008nc}. 

However, the hadronic scale choice leads to puzzling consequences, since the
partonic coefficient function depends on hadron-level physics, a statement which is unclear 
in the standard dQCD approach. 
As a consequence of the BN scale choice the partonic coefficient function depends on
$\tau$ through the convolution variable \textit{and} through the soft scale $\mu_s$.
Hence the resummed result does not factorize under Mellin transform into the 
product of the coefficient function and a parton luminosity as in Eq.~(\ref{eq:Mellinsigma}).
For this reason, as we will see in the next Chapter, it is not possible to 
compare directly the SCET approach to the standard dQCD approach at the parton level. 

In this Chapter we study the emergence of an effective 
physical scale that characterizes the soft emission using the approach to threshold resummation 
based on effective field theory. We follow the argument of Ref.~\cite{Becher:2007ty},
which considers as an example Drell-Yan production. We will then see \cite{Ahrens:2008nc} how it is 
possible to apply the same argument 
to Higgs-boson production at hadron colliders in Sect.~\ref{sect:higgsprodscet}. In Sect.~\ref{sect:bnscale} we 
will discuss in more detail the BN scale choice and its consequences. Finally,
we will present predictions for the total resummed cross-section in case of
Higgs boson production in gluon-gluon fusion in hadron collider in Sect.~\ref{sect:phenscet}.  

\section{Resummation in the Drell-Yan process}

\subsection{Fixed-order calculation of the Drell-Yan process}

We consider the production of a lepton pair of invariant mass $M$ at a 
hadron collider with centre-of-mass energy $\sqrt{s}$. In particular,
we limit our discussion for simplicity to the process
\bea
N_1+N_2 \rightarrow \gamma^* + X \rightarrow l^- + l^+ + X.
\eea
We consider the cross-section 
\bea
\frac{d^2 \sigma}{dM^2 dY} = \frac{4 \pi \alpha^2}{3 N_c M^2 s} 
\sum_{ij} \int dx_1 dx_2 \ C_{ij}(x_1,x_2, s, M, \mu_f) f_{i/N_1} (x_1, \mu_f)
f_{j/N_2} (x_2, \mu_f)
\eea
differential on $M^2=q^2$ and $Y=\frac{1}{2} \log \frac{1^0+q^3}{q^0-q^3}$,
where $q$ is the virtuality of the photon and $Y$ is the rapidity of the 
lepton pair in the centre-of-mass frame. The hard-scattering kernels
$C_{ij}$ have a power expansion in $\as$. At LO 
only the $q \bar q$, $\bar q q$ channels contribute, while at 
NLO also the $qg,\ gq,\ \bar q g,\ g\bar q $ channels contribute.    

It is useful to define new kernels $C_{ij}$ in terms of the variable $z$ Eq.~(\ref{eq:zandtau})
and the quantity
\bea
y= \frac{\frac{x_1}{x_2}e^{-2 Y}-z}{(1-z)(\left(1+\frac{x_1}{x_2} e^{-2 Y}\right)}
\eea
with $0\leq y \leq 1$, $\tau \leq z \leq 1$. In particular,
\bea
\tilde C_{ij}(x_1,x_2, s, M, \mu_f) =
\left| \frac{dz dy}{dx_1 dx_2} \right| \frac{ C_{ij}(z,y, s, M, \mu_f) }{
[[1-y(1-z)][1-(1-y)(1-z)]]}.
\eea
The hard-scattering kernels at NLO can be found 
in Ref.~\cite{Anastasiou:2003yy} and at NNLO in Refs.~\cite{Anastasiou:2003yy,Anastasiou:2003ds}.
The coefficient functions $C_{ij}$ contains terms which 
are singular at the partonic threshold $z=1$. In particular, $C_{ij}$ contains 
logarithms of the form
\bea
\log \frac{M^2(1-z)^2}{z^2 \mu_f^2}
\eea
which suggests that two mass scales are relevant in the DY process: a hard scale
$\mu_h \sim M$ and a soft scale $\mu_s\sim M(1-z)/\sqrt{z}$, which characterize
respectively the invariant mass of the lepton pair and the 
energy of the remnant $X$ which is produced in the collision. As we have seen in
Chapter~\ref{chapter:resummation}, the goal of resummation is to resum these logarithms, which become
large when a large separation between the hard and the 
soft scale exists (i.e. when $z\rightarrow 1$), to all order in
perturbation theory.

The leading singular terms in the partonic threshold in the hard scattering kernels are 
contained in $C_{q \bar q}$. In the $z \rightarrow 1$ limit it is possible to 
rearrange the expression such that the singular terms are always accompanied by 
$\delta$-distributions in the variable $y$:
\bea
C_{q \bar q} = \frac{\delta(y) + \delta(1-y)}{2} e_q^2 C(z, M, \mu_f) + C_{q \bar q}^\textrm{subl}.
\eea
The complete expression of the leading singular terms can be found
in Appendix of Ref.~\cite{Becher:2007ty} and it can be derived from the 
hard scattering kernels~\cite{Anastasiou:2003yy,Anastasiou:2003ds}. As we will see,
the SCET approach provides a method to resum these leading singular terms to all orders in
perturbation theory.

Upon integration over $y$ and rapidity $Y$, the total cross-section,
differential only in $M$, is obtained:
\bea\label{eq:sigmaDYscet1}
\frac{d \sigma^\textrm{threshold}}{d M^2} 
&= \frac{4 \pi \alpha^2}{3 N_c M^2 s}\sum_{q} e_q^2 
\int \frac{dx_1}{x_1} \frac{dx_2}{x_2} C(z, M, \mu_f) [f_{q/N_1} (x_1, \mu_f) 
f_{\bar q/N_2} (x_2, \mu_f) + (q \leftrightarrow \bar q)],
\eea
where the integration region is bounded from the condition $x_1 x_2 \geq \tau$. We
rewrite Eq.~(\ref{eq:sigmaDYscet1}) as
\bea
\frac{d \sigma^\textrm{threshold}}{d M^2} 
&= \frac{4 \pi \alpha^2}{3 N_c M^2 s}
\int_{\tau}^1 \frac{dz}{z} C(z, M, \mu_f) \mathscr L \left( \frac{\tau}{z}, \mu_f \right)
\eea
where we have introduced the parton luminosity
\bea
\mathscr L(y, \mu_f)= \sum_{q} e_q^2 \int_q^1 \frac{dx}{x} 
\left[ f_{q/N_1} (x, \mu_f) f_{\bar q/N_2}\left( \frac{y}{x}, \mu_f \right) + (q \leftrightarrow \bar q)\right].
\eea

\subsection{Factorization formula for the hard-scattering kernel}

The singular part of the coefficient function $ C(z, M, \mu_f)$ can be written in a factorized form as
\bea\label{eq:factC}
 C(z, M, \mu_f) = H(M, \mu_f) S(\sqrt{\hat s} (1-z), \mu_f),
\eea
where $H$ and $S$ are known respectively as hard function and soft function and will 
be defined below. This factorization
formula has been known for a long time, but it was rederived in Ref.~\cite{Becher:2007ty} 
using soft collinear effective theory methods. The hard and soft functions 
in this way are related to functions $-$ Wilson coefficients $-$ in effective theory, which 
obey renormalization group (RG) equation. Since the calculation of 
$H$ and $S$ at a given order is simpler than the calculation of the Drell-Yan cross-section
at the same order, Eq.~(\ref{eq:factC}) is an approximation to the
cross-section which reduces the amount of calculational work like the resummed coefficient
function in dQCD. Moreover,
it is possibile to perform threshold resummation directly in $z$ space by solving
the RG equations satisfied by $H$ and $S$.

The starting point of the derivation of the factorization formula is the standard formula
for the Drell-Yan cross-section
\bea
d \sigma = \frac{4 \pi \alpha^2}{3 s q^2} \frac{d^4 q}{( 2 \pi)^4} 
\int d^4 x e^{-i q \cdot x} \langle N_1(p_1) N_2 (p_2) | (-g_{\mu\nu})J^{\mu \dagger}(x)
J^\nu (0) | N_1 (p_1) N_2 (p_2) \rangle, 
\eea
where $J^\mu = \sum_{q} e_q \bar q \gamma^\mu q$ is the electromagnetic current. 
The factorization formula follows from a sequence of matching steps. The
product of current is matched onto operators in SCET; the matching lets a 
correspondent Wilson coefficient arise. In the case of DY, the matching proceeds in
two steps. In the first step a matching with a first version 
of SCET, which contains soft degrees of freedom along with two types of 
hard-collinear fields, is performed; in the second step this effective theory is matched 
onto a SCET version where the soft modes are integrated out and 
fields of lower virtuality replace the hard-collinear modes. 
The correspondent Wilson coefficients are the hard function $H$ and
the soft function $S$. The remaining low-energy matrix element is then identified with
the parton luminosity. We will present here the results of this matching procedure; 
we refer the interested reader to Ref.~\cite{Becher:2007ty} and references therein for further details.

\begin{figure}[htbp]
\centering
   {\includegraphics[width=.5\columnwidth]{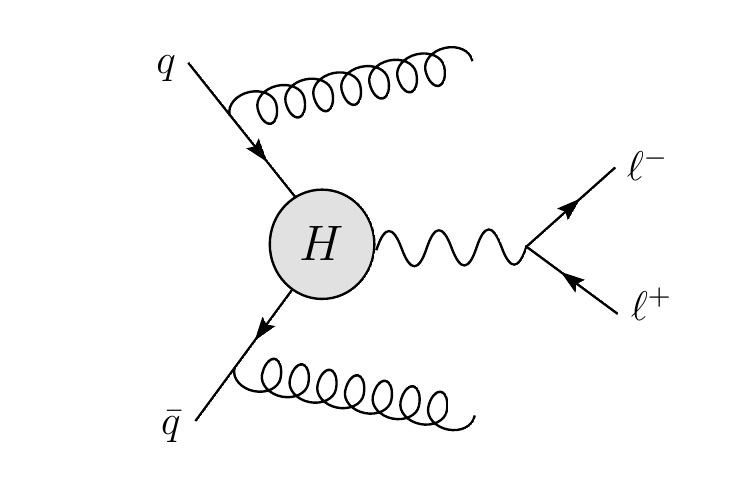}}
\caption{Radiation from the initial-state partons in 
Drell-Yan process at parton level. Figure from Ref.~\cite{Becher:2007ty}.}\label{fig:Hscet}
\end{figure}

The physical basis of the factorization formula Eq.~(\ref{eq:factC}) is made clearer
by an intuitive argument, which we will now sketch. We consider the kinematics of the 
DY process at the parton level, which is shown in Fig.~\ref{fig:Hscet}. Near the 
partonic threshold, where the energy of the partonic centre-of-mass 
is just above the invariant mass of the lepton pair, the initial-state 
partons radiate multiple soft gluons. This soft radiation can be
exponentiated and it can be described by objects called Wilson lines,
which are phase factors represented by a path ordered exponential of gauge fields.
To leading power, the incoming partons are left on the mass-shell; 
it is therefore possible to describe the DY pair by an on-shell quark form factor,
which can be identified with the hard function $H$. In particular, the 
hard function $H$ is given by
\bea
H(M, \mu_f) = |C_V (-M^2-i\epsilon, \mu_f)|^2
\eea
where an $i \epsilon$ prescription is required since the Wilson coefficient $C_V$
has a branch cut along the positive $q^2$ axis. The coefficient $C_V$ is 
known at 2 loop, and its expression can be found in Appendix~\ref{app:analy}. In Ref.~\cite{Becher:2007ty} it is finally shown that
the soft function can be identified as
\bea\label{eq:softWDY}
S(\sqrt{\hat s}(1-z), \mu_f) = \sqrt{\hat s}W_\textrm{DY}(\sqrt{\hat s}(1-z), \mu_f),
\eea
where the Wilson loop $W_\textrm{DY}$ is an object which describes the properties of the hadronic 
final state and whose perturbative expression in position space has been obtained in Ref.~\cite{Belitsky:1998tc}.

\subsection{Resummation in momentum space}

The resummation of large logarithms in the threshold region $z \rightarrow 1$ 
is performed in the SCET approach by solving the renormalization group 
equation which the hard and the soft function, now characterized as
field-theoretic objects, obey. 

The matching coefficient $C_V$ obeys the evolution equation 
\bea
\frac{d}{ d \log \mu} C_V (-M^2-i \epsilon, \mu) = \left[ \Gamma_\textrm{cusp}(\as) 
\left( \log \frac{M^2}{\mu^2}- i \pi \right) + \gamma^V (\as) \right]
C_V (-M^2 - i \epsilon, \mu),
\eea
where the term in $\Gamma_\textrm{cusp}$ is associated with Sudakov double
logarithms, whereas $\gamma^V$ is associated with single logarithms. This equation can 
be solved exactly; the solution is
\bea\label{eq:resultcv}
C_V (-M^2, \mu_f ) = \exp [2 S (\mu_h, \mu_f) - a_{\gamma^V}(\mu_h, \mu_f) 
+ i \pi a_\Gamma(\mu_h, \mu_f)]\left( \frac{M^2}{\mu_h} \right)^{- a_{\Gamma}(\mu_h, \mu_f)}
C_V (-M^2, \mu_h)
\eea
where a hard matching scale $\mu_h \sim M$, at which $C_V$ is calculated using 
fixed-order perturbation theory, has been introduced. 
The Sudakov exponent $S$ and the exponent $a_\Gamma$ are
\bea
S(\nu, \mu) = 
-\int_{\as(\nu)}^{\as(\mu)}d \alpha \frac{\Gamma_\textrm{cusp}(\alpha)}{\beta(\alpha)}
\int_{\as(\nu)}^\alpha \frac{d \alpha'}{\beta(\alpha')}, \qquad
a_\Gamma(\nu,\mu) = - \int_{\as(\nu)}^{\as(\mu)}d \alpha \frac{\Gamma_\textrm{cusp}(\alpha)}{\beta(\alpha)}.
\eea
with an analogous expression for $a_{\gamma^V}$.

The Wilson loop, which is related to the soft function by Eq.~(\ref{eq:softWDY}), 
obeys the following integro-differential equation:
\bea\label{eq:wsloopeq}
\frac{d W_\textrm{DY}(\omega, \mu)}{d \log \mu} =& 
-\left[ 4 \Gamma_\textrm{cusp} (\as) \log \frac{\omega}{\mu} + 2 \gamma^W (\as) \right]
W_\textrm{DY} (\omega, \mu)\nonumber  \\
&-4 \Gamma_{\textrm{cusp}} (\as) \int_0^\omega d \omega'\frac{W_\textrm{DY}(\omega', \mu)
-W_\textrm{DY}(\omega,\mu)}{\omega-\omega'},
\eea
where 
\bea
\gamma^W = 2 \gamma^\phi + \gamma^V,
\eea
where $\gamma^\phi$ can be obtained by expanding the Altarelli-Parisi splitting function 
as
\bea
P_{qq}(z) = \frac{2 \Gamma_\textrm{cusp}(\as)}{[1-z]_+}
+ 2 \gamma^\phi (\as) \delta(1-z)+ \ldots\ .
\eea
The solution of Eq.~(\ref{eq:wsloopeq}) is
\bea\label{eq:resultwdy}
W_\textrm{DY} (\omega, \mu_f) =
\exp [ -4 S (\mu_s, \mu_f) + 2 a_{\gamma^W} (\mu_s, \mu_f) ] 
\tilde s_\textrm{DY} (\de_\eta, \mu_s) 
\frac{1}{\omega} \left(\frac{\omega}{\mu_s}\right)^{2 \eta} \frac{e^{-2 \gamma_E \eta}}{\Gamma(2 \eta)},
\eea
where we have introduced an auxiliary parameter $\eta= 2 a_\Gamma(\mu_s, \mu_f)$.
The result Eq.~(\ref{eq:resultwdy}) is well defined if $\eta>0$. The result 
is analytically continued to negative values of $\eta$ (which is typically the case in 
DY-like processes) by means of the identity
\bea\label{eq:starreg}
\int_0^\Omega d \omega \frac{f(\omega)}{\omega^{1-2 \eta}} =
\int_0^\Omega d \omega \frac{f(\omega)-f(0)}{\omega^{1-2 \eta}}
+\frac{f(0)}{2 \eta} \Omega^{2 \eta}
\eea
which is valid if $-\frac12 < \eta < 0$; other subtractions are required for 
$\eta < \frac12$. 
The function $\tilde s_\textrm{DY}$ is given by the Laplace transform
\bea
\tilde s_\textrm{DY} (L, \mu_s) 
= \int_0^\infty d \omega \ e^{-s \omega} W_\textrm{DY} (\omega, \mu_s), 
\qquad s = \frac{1}{e^{\gamma_E} \mu_s e^{L/2}}.
\eea

The resummed expression for the hard scattering coefficient Eq.~(\ref{eq:factC})
is given by the product of the solutions of the renormalization group equations
Eq.~(\ref{eq:resultcv}) and Eq.~(\ref{eq:resultwdy}). The result can be written as
\bea\label{eq:resummdyscet}
C(z, M, \mu_f) =& |C_V(-M^2, \mu_h)|^2 U(M, \mu_h, \mu_s, \mu_f)\nonumber \\
& \times \frac{z^{-\eta}}{(1-z)^{1-2 \eta}} \tilde s_\textrm{DY} 
\left( \log \frac{M^2 (1-z)^2}{\mu_s^2 z} + \de_\eta, \mu_s \right) 
\frac{e^{-2 \gamma_E \eta}}{\Gamma(2 \eta)}
\eea
where we have defined
\bea
U(M, \mu_h, \mu_s, \mu_f) = \left(\frac{M^2}{\mu_h^2} \right)^{-2 a_\Gamma(\mu_h, \mu_s)}
\exp[4 S(\mu_h, \mu_s)-2 a_{\gamma^V} (\mu_h, \mu_s) + 4 a_{\gamma^\phi}(\mu_s,\mu_f)].
\eea
The resulting expression is, from a formal point of view, independent of the 
hard scale $\mu_h$ and the soft scale $\mu_s$; however, in phenomenological 
applications, one has to truncate the perturbative expansions of $C_V$, 
$\tilde s_\textrm{DY}$ and the anomalous dimensions $\gamma^V$ and $\gamma^\phi$. 
Furthermore, the evolution function, which is an exponential, contains all the 
order in $\as$, whereas the hard and the soft function are truncated at fixed 
perturbative order. For this reason a residual, non trivial dependence on the matching scales 
cannot be avoided when phenomenological analysis are performed.

\section{Resummation for the Higgs production process}\label{sect:higgsprodscet}

In this Section we will show how soft collinear effective theory 
provides a method to resum logarithmically enhanced contribution which 
arise in the Higgs production in the gluon-gluon fusion process. As for
the DY process, the SCET expressions are cast in momentum space 
and are free of Landau pole. 

\subsection{Fixed-order expressions and factorization formula for Higgs production}

We have discussed in some details Higgs production process in gluon-gluon 
fusion at fixed-order in Sect.~\ref{sect:ggH}. We have seen that
the total cross-section for the production of a Higgs boson of mass $m_H$ in hadronic collisions
at centre-of-mass energy $\sqrt{s}$ can be written as
\bea\label{eq:sigmaHiggsscet1}
\sigma(\tau, M^2) = \tau \sigma_0 \as^2(\mu_f^2)\sum_{ij} \int \frac{dz}{z} 
C_{ij}(z, m_t, m_H, \mu_f) \mathscr L_{ij} \left( \frac{\tau}{z}, \mu_f \right), 
\eea
where
\bea
\mathscr L_{ij} (y, \mu) = \int_y^1 \frac{dx}{x} f_{i/N_1} (x, \mu) f_{j/N_2}\left( \frac{y}{x}, \mu \right)
\eea
are the parton luminosity, $\tau= \frac{m_H^2}{s}$, and $C_{ij}$ are hard scattering kernels.  
The hard scattering kernels contains terms which are singular in the
partonic threshold limit $z \rightarrow 1$, where $\hat s \sim m_H^2$. These
terms are contained in the hard-scattering kernel $C_{gg}$, which can be written 
therefore as
\bea
C_{gg} (z, m_t, m_H, \mu_f) = C(z,m_t, m_H, \mu_f)+ C_{gg}^\textrm{reg} (z, m_t, m_H, \mu_f),
\eea
where $C_{gg}^\textrm{reg}$ contains only terms which are not singular in $z=1$.

It is possible to write the coefficient function $C(z,m_t, m_H, \mu_f)$ in a 
factorized form, using effective field theory methods in the same way as it was done for the factorized 
scattering kernel in DY production. As for the DY production, a sequence of matching
step is made in order to obtain a factorized formula. One first integrates out the 
top quark by matching the Standard Model with six quarks flavour onto a 
five-flavour theory. In this way a Wilson coefficient $C_t$ arises. The five-flavour 
theory is then matched on SCET in analogy with what was done in DY production; 
in this way the hard function $H$ and the soft function $S$ arise. The remaining 
effective-theory matrix element is then identified with the convolution of the parton
distribution functions. The resulting coefficient function can be written as
\bea
C(z,m_t, m_H, \mu_f) = [C_t(m_t^2,\mu_f^2) ]^2 H(m_H^2, \mu_f^2) 
S(\hat s (1-z)^2, \mu_f)
\eea
In this way the cross-section is approximated by a factorized formula.

\subsection{Resummation in momentum space}

The hard and soft function $H$ and $S$ are related to Wilson coefficients in SCET.
In particular, the hard function can be written as
\bea
H(m_H^2, \mu^2) = |C_S(-m_H^2-i\epsilon, \mu^2)|^2,
\eea
where $C_S$ has an expansion in powers of $\as(\mu^2)$. The soft function is 
related to the Wilson loop $W_\textrm{Higgs}$, which coincides at
two loop order with the Wilson loop $W_\textrm{DY}$ after the replacement
$C_F\rightarrow C_A$. In particular,
\bea
S(\hat s (1-z)^2, \mu_f^2) =& \sqrt{\hat s} W_\textrm{Higgs} (\hat s (1-z)^2,\mu^2_f)\nonumber \\
=& \sqrt{\hat s} W_\textrm{DY} (\hat s (1-z)^2,\mu^2_f)\Big|_{C_F\rightarrow C_A} + 
\mathcal O(\as^3).
\eea
The Wilson coefficient $C_t$, $C_S$ obey RG equations whose solutions provide a 
method to resum logarithmically enhanced contributions in momentum space. 
In particular, the Wilson coefficient $C_t$ at scale $\mu_f$ is
\bea
C_t (m_t^2, \mu_f^2) = \frac{\beta(\as(\mu_f^2))/\as^2(\mu^2_f))}{\beta(\as(\mu_t^2))/\as^2(\mu^2_t))}
C_t(m_t^2,\mu_t^2),
\eea 
with $\mu_t \sim m_t$, which is the scale at which the top quark is integrated out and
$C_S$ can be written as
\bea
C_S(-m_H^2-i\epsilon, \mu^2_f) =&\exp\left[2 S(\mu_h^2, \mu_f^2)-a_\Gamma(\mu_h^2, \mu_f^2)
\log \frac{-m_H^2-i \epsilon}{\mu_h^2} -a_{\gamma^S}(\mu_H^2, \mu_f^2) \right]\nonumber \\
&\times C_S(-m_H^2 -i\epsilon, \mu_h^2),
\eea
where a hard matching scale $\mu_h$ has been introduced. The solution above depends on
the cusp anomalous dimension $\Gamma^A_\textrm{cusp}$, which controls the double-log
evolution, and the anomalous dimension $\gamma^S$, which controls the single-log evolution.

The Wilson loop $W_\textrm{Higgs}$ obeys an integro-differential equation 
analogous to Eq.~(\ref{eq:wsloopeq}), whose solution is
\bea
\omega W_\textrm{Higgs} (\omega^2, \mu_f^2) =
\exp[-4 S (\mu_s^2, \mu_f^2)+ 2 a_{\gamma^W}(\mu_s^2, \mu_f^2)]
\tilde s_\textrm{Higgs} (\de_\eta, \mu_s^2) 
\left(\frac{\omega^2}{\mu_s^2} \right)^\eta \frac{e^{-2 \gamma_E \eta}}{\Gamma(2 \eta)},
\eea
where $\eta = 2 a_\Gamma(\mu_s^2, \mu_f^2)$. The anomalous dimension $\gamma^W$ 
is related to $\gamma^S$ from the equation
\bea
\gamma^W = \frac{\beta(\as)}{\as} + \as^2 \frac{d}{d \as} \frac{\beta(\as)}{\as^2} 
+ \gamma^S + 2 \gamma^B,
\eea
where $\gamma^B$ is the coefficient of the 
$\delta(1-z)$ term in $P_{gg}(z)$. The function $\tilde s_\textrm{Higgs} $ 
is related to the corresponding function $\tilde s_\textrm{DY} $  by a simple
replacement of color factors. The explicit expressions for the relevant coefficients can
be found in Appendix~\ref{app:analy}. 

Putting everything together allows one to write the resummed hard-scattering 
kernel as 
\bea\label{eq:cfhiggsscet}
C(z, m_t, m_H, \mu_f) =& [C_t(m_t^2, \mu_t^2)]^2 |C_S(-m_H^2-i \epsilon, \mu_h^2)|^2
U(m_H, \mu_t, \mu_h, \mu_s, \mu_f)\nonumber \\
&\times \frac{z^{-\eta}}{(1-z)^{1-2 \eta}} \tilde s_\textrm{Higgs} 
\left( \log \frac{m_H^2 (1-z)^2}{\mu_s^2 z}+\de_\eta, \mu_s^2 \right)\frac{e^{2 \gamma_E \eta}}{\Gamma(2 \eta)}
\eea
where $U$ is the evolution factor defined is
\bea
U(m_h, \mu_t, \mu_h, \mu_s, \mu_f) =& \frac{\as^2(\mu_s^2)}{\as^2(\mu_f^2)}
\left[ \frac{\beta(\as(\mu_s^2))/\as^2(\mu_s^2)}{\beta(\as(\mu_t^2))/\as^2(\mu_t^2)} \right]^2
\left|\left(\frac{-m_H^2 - i \epsilon}{\mu_h^2} \right)^{-2 a_\Gamma(\mu_h^2,\mu_s^2)} \right|\nonumber \\
&\times |\exp[4 S(\mu_h^2,\mu_s^2)-2 a_{\gamma^S} (\mu_h^2, \mu_s^2)+ 4 a_{\gamma^B}(\mu_s^2,\mu_f^2)]|.
\eea
As we emphasized in case of DY production, the resummed hard-scattering coefficient is 
formally independent of the matching scales $\mu_s$, $\mu_h$ and $\mu_t$, 
but a residual scale dependence remains due to the truncation of perturbation theory.

The total cross-section for the Higgs-boson production is obtained by integrating the
resummed coefficient convoluted with the gluon-gluon luminosity $\mathscr L_{gg}$. It is 
finally necessary to match the result with the fixed-order computation. In the
momentum-space approach the double-counting terms are avoided by the simple 
subtraction
\bea
\sigma = \sigma^\textrm{resummed}\Big|_{\mu_t, \mu_h, \mu_s, \mu_f} 
+ \left(\sigma^\textrm{fixed order}\Big| _{\mu_f} -\sigma^\textrm{resummed}\Big|_{\mu_t=\mu_h=\mu_s=\mu_f} \right),
\eea
since by construction the last term, where all the evolution scales are evaluated 
at $\mu_f$, contains only the fixed order truncation of the resummed result.

\section{The Becher-Neubert scale choice}\label{sect:bnscale}

In the standard dQCD approach soft gluon emission is characterized by an
energy scale of order $M(1-z)$, where $z$ is the energy fraction of the
observed final state. Therefore, the energy available for unobserved gluon
radiation is much smaller than $M$ in the partonic threshold limit $z \rightarrow 1$. 
As we have previously discussed in Chapter \ref{chapter:resummation}, the resummation of $\log(1-z)$ 
has important phenomenological applications; when a partonic scale is chosen
threshold resummation could be important even relatively far from the hadronic threshold,
because of the convolution structure of the integral Eq.~(\ref{eq:sigmaHiggsscet1}) . 

In the SCET expressions one resums both of $1-z$ and logs of $M/\mu_s$, where $M$ is a hard scale whereas 
$\mu_s$ is a soft scale which can be chosen in different ways.
In the SCET approach hence the choice of the soft scale $\mu_s$ determines what is being resummed.
Different choices are possibile. When $\mu_s\sim M$ nothing is resummed and by construction
of the SCET expressions fixed-order result in the soft limit is reproduced. The 
natural partonic choice, $\mu_s \sim M(1-z)$, corresponds to the resummation of logs of $1-z$. 

In this last case the resummed SCET coefficient function can be directly compared with 
the perturbative coefficient function $C_\textrm{dQCD}(z,M^2) $ obtained as the 
leading-log truncation of the inverse Mellin
of the $N$-space coefficient function $C_\textrm{dQCD}(N,M^2)$. A comparison can hence
be done only order by order. On the other hand, 
the SCET coefficient function in momentum space is formally defined for $\mu_s=M(1-z)$. 
It was shown in Ref.~\cite{Bonvini:2012az} that, away from the endpoint $z=1$, all the logarithmically  
enhanced terms in the partonic cross-section are reproduced order by order with the partonic scale
choice. However, the SCET expression with $\mu_s = M(1-z)$ is ill-defined when $z \rightarrow 1$.
Moreover, since $\eta$ depends on $z$, it is not possibile to use Eq.~(\ref{eq:starreg}) to regulate the 
behaviour of the resummed coefficient function at the endpoint. A 
possible way out, which consists in a cutoff of the convolution integral at $z = \bar z  < 1$,
was proposed in Ref.~\cite{Beneke:2011mq}. 

Alternatively, one can chose the soft scale as a function of the hadronic variable $\tau$,
which avoids the Landau problem pole.
This choice was proposed by Becher and Neubert in
Refs.~\cite{Becher:2006nr,Becher:2006mr,Becher:2007ty,Ahrens:2008nc}. This particular choice has
some puzzling consequences if compared to the standard perturbative approach, since 
the analytical structure of the integral
\bea\label{eq:convintscet}
\sigma_\textrm{SCET}(\tau, M^2) &= \int_\tau^1
\frac{dz}{z} \mathscr L \left( \frac{\tau}{z} \right) C_\textrm{SCET} 
(z,M^2,\mu_s^2)
\eea
changes. In fact, if the soft scale is chosen as
a function of the partonic variable $z$ Eq.~(\ref{eq:convintscet}) factorizes upon Mellin transform. This 
is a sufficient and a necessary condition for parton radiation to respect longitudinal 
momentum conservation. However, if $\mu_s$ depends on $\tau$, the convolution structure 
is spoiled, and upon Mellin transform the cross-section no longer factorizes
and violates longitudinal momentum conservation. In particular, this choice
violates the QCD factorization theorem since the partonic coefficient function depends on 
the hadronic variable $\tau$, while it should not.

The choice of soft scale proposed by Becher and Neubert is
\bea\label{eq:softscale}
\mu_s \sim M(1-\tau).
\eea
However, when $\tau$ is small (as for Higgs production at the LHC), with this particular choice $\mu_s \sim M$, which 
is an hard scale. In this case, as we have discussed, the SCET approach reproduces 
the soft-limit of the fixed order result. In the small $\tau$ case Becher and Neubert suggest 
a slightly more general choice of scale. The soft scale Eq.~(\ref{eq:softscale}) is generally rescaled by
a function of $\tau$ which does not vanish in the hadronic endpoint and which is determined 
by minimization of perturbative contributions of $\tilde s$ to the cross-section.
In particular, two criteria are proposed:
\begin{itemize}
\item Starting from a high scale, one determines the value of $\mu_s$ at which the one-loop
correction drops below $15\%$.
\item one choose the value of $\mu_s$ for which the one-loop correction is minimized.
\end{itemize}

In the Drell-Yan case the two resulting scales are chosen as 
\bea
\mu_I = \frac{M(1-\tau)}{1 + 7 \tau} \qquad \mu_{II} = \frac{M(1-\tau)}{\sqrt{1 + 150 \tau}},
\eea
whereas for the Higgs production two different functional forms are chosen, which 
well interpolates the numerical results obtained from the two criteria proposed. 
The soft scale is varied between the two scales $\mu_I$ and $\mu_{II}$
and, somewhat arbitrarily, the average of the two scales is taken as the default choice.
We will investigate further the consequences of the BN scale choice in Chapter~\ref{chapter:ancomp} when we
will make a systematical comparison of the dQCD and the SCET approach.

\section{Phenomenology}\label{sect:phenscet}

We have discussed how SCET provides an alternative way to compute resummed 
observables. The results depend on various scales from which the SCET predictions 
are formally independent, but we have seen how a residual dependence cannot be avoided 
if phenomenological study are performed.  
 
In this Section we will apply soft-gluon resummation in the SCET approach to a specific process,
namely the Higgs boson production at the LHC. We will present the prediction for the total
cross-section and the $K$-factor at the resummed level. All the result have been 
computed setting the matching scales $\mu_h=\mu_t=m_H$, the factorization scale $\mu_f=m_H$
and choosing $\mu_s$ as the average value of the two scales $\mu_I$ and $\mu_{II}$, which are chosen according to
the Becher-Neubert criteria. In Ref.~\cite{Ahrens:2008nc} a more detailed study\footnote{Actually, the results of 
Ref.~\cite{Ahrens:2008nc} include RG improvement up to NNNLL$^*$. In our study, however, we switched from
the starred counting used in Ref.~\cite{Ahrens:2008nc} to the un-starred counting.} has been performed, that
includes an estimation of the theoretical uncertainties and the resummation of 
contributions of the form $(C_A \pi \as)^n$, which arise in the analytic 
continuation of the gluon form factor $C_S$. The possibility of the resummation of this enhanced terms, 
related to the Sudakov logarithms, was observed long time ago \cite{Parisi:1979xd}. In Ref.~\cite{Ahrens:2008nc} 
$\pi^2$ resummation was claimed to lead to large corrections to resummed Higgs cross-section.
However, since it is a separate subject, we will not consider it in our phenomenological 
study. 

\begin{figure}[htbp]
\centering
   {\includegraphics[width=.75\columnwidth]{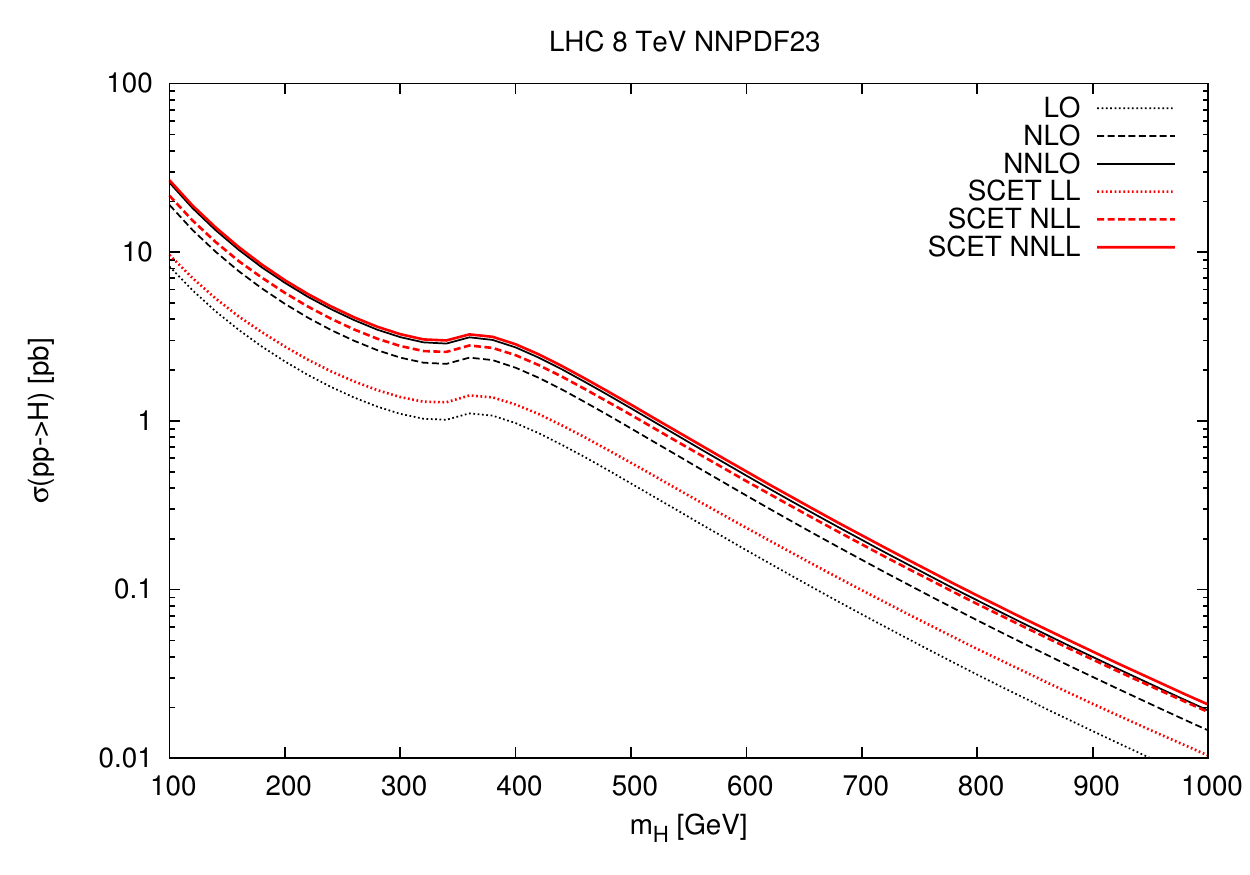}} 
\caption{Resummed cross-section as a function of the
Higgs mass $m_H$ at LL, NLL and NNLL at the LHC. The curves have been calculated with NNPDF2.3 NNLO PDF set,
with  $\mu_h=\mu_t=m_H=\mu_f$.}\label{fig:SCEThiggsres}
\end{figure}

\begin{figure}[htbp]
\centering
   {\includegraphics[width=.75\columnwidth]{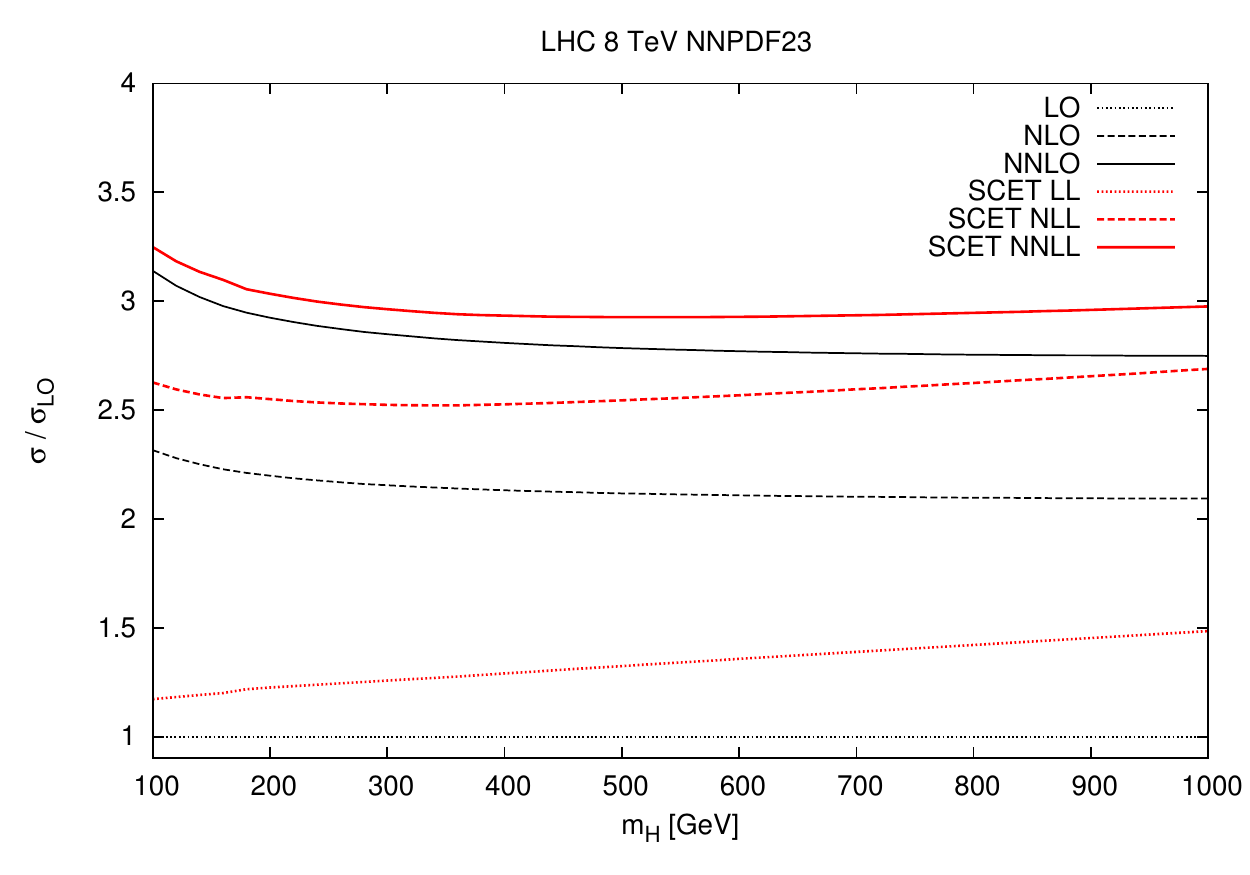}}
\caption{$K$-factor for the resummed cross-section as a function of the
Higgs mass $m_H$ at LL, NLL and NNLL at the LHC. The curves have been calculated with NNPDF2.3 NNLO PDF set,
with  $\mu_h=\mu_t=m_H=\mu_f$.}\label{fig:SCEThiggskfact}
\end{figure}

We show in Fig.~\ref{fig:SCEThiggsres} the resummed cross-section as a function of the
Higgs mass $m_H$ at LL, NLL, NNLL compared with the fixed-order results obtained with
\texttt{ggHiggs}. All the curves have been obtained in the large-$m_t$ limit with the
NNPDF2.3 NNLO PDF set and $\as(m_Z) =0.0117$. The predictions have been obtained with our own code which
we have checked to produce similar results
to the public code \texttt{RGHiggs} \cite{RGHiggs}. The effect 
of soft-gluon resummation is made clearer in Fig.~\ref{fig:SCEThiggskfact} where 
we show the predictions for the $K$-factor Eq.~(\ref{eq:kfactor}). The effect of threshold resummation is relevant
already at LL, where it leads to an increase of the cross-section by 20$\%$ up to
$ 50\%$ for a heavy Higgs. The effect is relevant at NLL, where the resummed
cross-section gets close to the NNLO result. At NNLL accuracy the effect is slightly 
reduced and it corresponds to an increase of the cross-section by 10$\%$ at most.

\begin{figure}[htbp]
\centering
   {\includegraphics[width=.75\columnwidth]{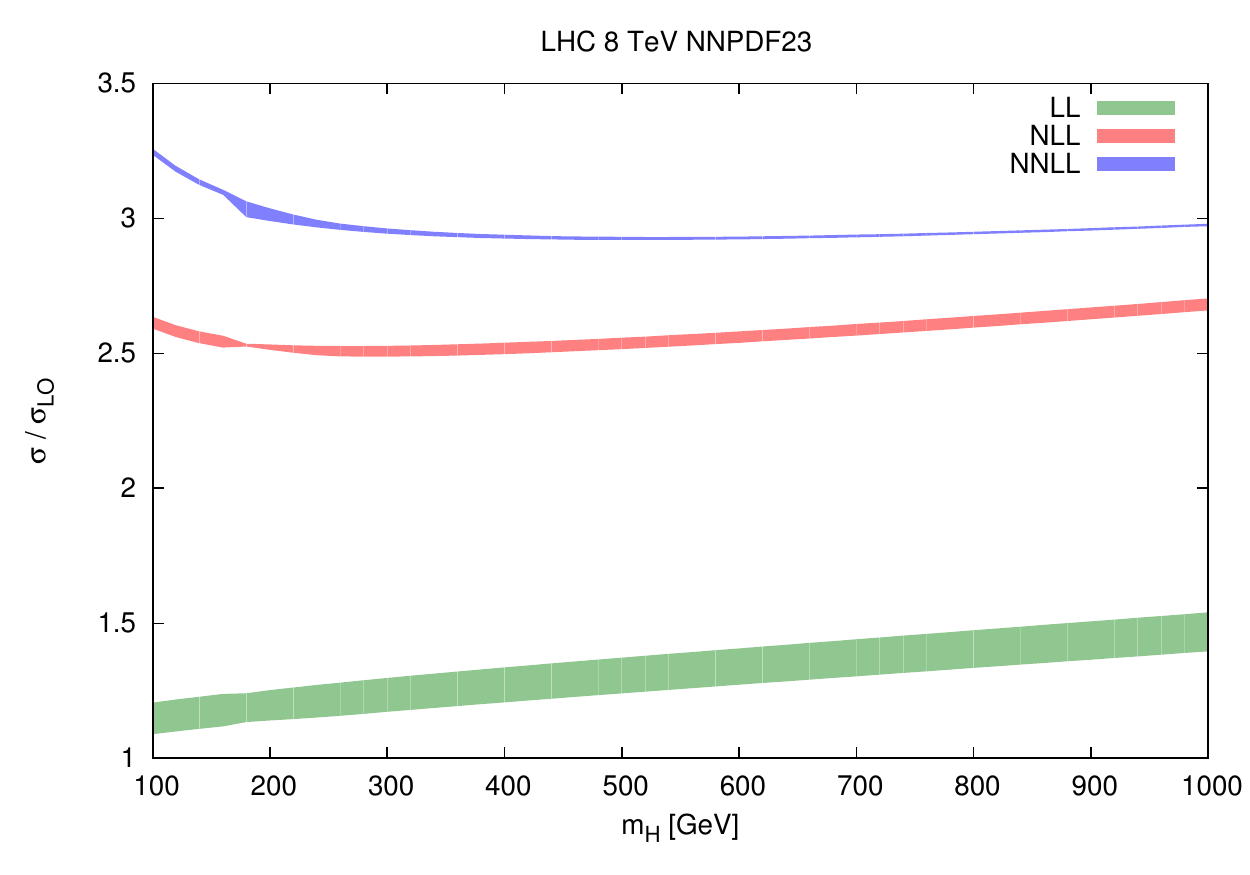}}
\caption{Dependence on $\mu_s$ of the $K$-factor for the resummed cross-section as a function of the
Higgs mass $m_H$ at LL, NLL and NNLL at the LHC. The curves have been calculated with NNPDF2.3 NNLO PDF set,
with  $\mu_h=\mu_t=m_H=\mu_f$.}\label{fig:SCEThiggsmusdep}
\end{figure}

Finally, we show in Fig.~\ref{fig:SCEThiggsmusdep} the dependence of the $K$-factor for the resummed cross-section on the
soft scale $\mu_s$, which varies between the two values $\mu_s^I$ and $\mu_s^{II}$.
As one would expect, the width of the band gets thinner the higher the
logarithmic accuracy is. 

\chapter{Resummation in SCET and in dQCD: an analytic comparison}\label{chapter:ancomp}

In this Chapter we compare soft-gluon resummation in QCD as performed in the standard 
perturbative QCD formalism, to
resummation based on SCET in the Becher-Neubert approach. 

In the past years an increasing interest in a deeper understanding of the main
similarities and differences in the two methods has grown. Both in Refs.~\cite{Becher:2006mr,Becher:2007ty,Ahrens:2008nc}
and in Refs.~\cite{Bonvini:2012az,Bonvini:2013td} the analytic equivalence of the dQCD and SCET approaches has been
explored at various levels. More recently, in Ref.~\cite{Sterman:2013nya}, the relationship between
the two formalisms has been investigated by focusing on their common basis 
which relies on soft-gluon factorization properties and then by analyzing how
the two formalisms differ in the choice of scales and in the derivation of
physical observables.

We will follow the argument presented in Refs.~\cite{Bonvini:2012az,Bonvini:2013td} which concentrates on Drell-Yan 
production and that can be straightforwardly generalized to 
Higgs boson production. 
In Sect.~\ref{sec:firstprob} we will briefly summarize some features 
of threshold resummation in both the approaches and we will discuss some of the problems which
arise in performing such a task in the two formalisms. In Sect.~\ref{sect:dynot} we will fix some
notation and in Sect~\ref{sect:masterformula} we will derive a master formula which relates dQCD and SCET.
We will finally discuss the logarithmic accuracy of the SCET result with the 
BN scale choice in Sect.~\ref{sect:compbn}.

\section{Threshold resummation}\label{sec:firstprob}

In the previous Chapters we have learned how the resummation of logarithmically
enhanced contributions arises naturally when one computes cross-sections
in perturbative QCD. In this Section we summarize some of the features
we have already encountered in order to have a better understanding of 
similarities and differences in the dQCD and in the SCET approach.

Standard dQCD resummation is traditionally performed in Mellin space since 
the truncation of resummed results in momentum space to any finite logarithmic 
accuracy induces terms which violate the conservation of the longitudinal momentum.
As a consequence, it is not possible to perform dQCD resummation in momentum space since 
this would lead to divergent hadronic cross-sections. In order to obtain finite physical 
hadronic cross-section one has therefore to perform resummation in $N$ space to finite 
logarithmic accuracy and subsequently to invert the Mellin transform retaining terms to all 
logarithmic orders in $1-z$ (which correspond under Mellin transform to logs of $N$). 

We have however learned how the running of the coupling introduces new difficulties
while performing threshold resummation. In particular, the resummed coefficient function
in $z$ space, viewed as a series in $\as$ Eq.~(\ref{eq:creszseries}), where each term
is constructed as the exact inverse Mellin transform of Eq.~(\ref{eq:cresNseries}), 
does not converge. This divergence is a direct consequence of the presence of the
Landau pole in $\as$. The fact that the series Eq.~(\ref{eq:creszseries})
acquires a nonzero radius of convergences if the Mellin inversion is performed to
finite logarithmic accuracy does not help; since upon inverse Mellin transform
the hard scale $Q^2$ is replaced with a scale $Q^2(1-z)^a$ (with $a=2$ for DY and Higgs
production) related to soft-gluon emission
the convolution integral Eq.~(\ref{eq:sigma_xspace}) always intercepts the region 
$z\rightarrow 1$ where the running coupling gets over the Landau pole. Fortunately,
this non-perturbative divergence can be removed by adding suitable subleading
terms: in the MP the particular integration path chosen when performing the Mellin 
inversion corresponds to the addition of more-than-power suppresed terms, while in
the BP the divergent series is made Borel-summable by adding a higher twist term.

On the other hand, approaches based on soft-collinear effective theory are valid
alternatives to dQCD and provide a powerful tool in order to obtain resummed observables.
In Ref.~\cite{Becher:2006nr} it was suggested that in the SCET approach
resummed expressions in terms of the hadronic kinematic variables could be derived; this allows therefore 
the resummation of $\log(1-\tau)$, where $\tau$ is an (hadronic) dimensionless variable.
The divergence due to the presence of the Landau pole can hence be avoided.
In fact in the resummed SCET expressions one can choose the soft 
scale $\mu_s$ in terms of the kinematic
variables of the hadronic process. With this scale choice
resummed expression for different physical processes have been obtained. However, 
it is not immediate to understand
the implications of the BN scale choice in the traditional formalism
because one of the main features of dQCD, perturbative factorization $-$
the independence of the partonic cross-section from the hadronic kinematics variables $-$
is lost.

\section{Resummation of the Drell-Yan process in SCET and dQCD}\label{sect:dynot}

In this Section we consider for definiteness inclusive Drell-Yan production at a hadron collider;
we will however see how the same argument can be applied with 
minimal modification to Higgs boson production in gluon-gluon fusion. In particular, 
we concentrate on the invariant mass distribution $\frac{d\sigma_\textrm{DY}}{d M^2}$,
where $M$ is the invariant mass of the DY pair. We define the dimensionless cross-section
\bea
\sigma(\tau, M^2) = \frac{1}{\tau \sigma_0} \frac{d\sigma_\textrm{DY}}{d M^2}
\eea
where $\sigma_0$ is the leading order partonic cross section. We can write the invariant 
mass distribution in the schematic form
\bea
\sigma(\tau, M^2) = \int_\tau^1 \frac{dz}{z} C(z, M^2) \mathscr L\left( \frac{\tau}{z} \right).
\eea
From now on, without significant loss of generality, we shall always choose
$\mu_F = \mu_R = M$.

\subsection{Perturbative resummation in dQCD}

For convenience of the reader, in this Section we briefly summarize how soft resummation is performed in 
dQCD. We have discussed at length in Chapter~\ref{chapter:resummation} how dQCD resummation is performed in $N$ space, by considering 
the Mellin transform of the cross-section
\bea
\sigma(N, M^2) = \int_0^1 d\tau \ \tau^{N-1} \sigma(\tau, M^2) = C(N,M^2) \mathscr L(N).
\eea
The resummed coefficient function in $N$ space has the form Eq.~(\ref{eq:Cres}) which we report here:
\bea
C^\textrm{res}_\textrm{dQCD}(N,M^2) = \bar g_0(\as) \exp \bar S \left(M^2, \frac{M^2}{N^2} \right)
\eea
where
\bea
\bar S \left(M^2, \frac{M^2}{N^2} \right)
= \int_0^1 dz\ z^{N-1} \left[\frac{1}{1-z}\int_{M^2}^{M^2(1-z)^2}
\frac{d \mu^2}{\mu^2} 2 A (\as(\mu^2)) + D(\as([1-z]^2M^2))\right]_+.
\eea
The functions $\bar g_0(\as)$, $A(\as)$ and $D(\as)$ are power series 
in $\as$, with $\bar g_0(0)=1$ and $A(0)=D(0)=0$. We have discussed in Chapter~\ref{chapter:resummation}
how the inclusion of extra terms in $\bar g_0$ is mandatory in order to improve the accuracy 
beyond NLL; the inclusion of an extra term in $\bar g_0$ increases the logarithmic accuracy 
of the coefficient function by half a logarithmic order.

\subsection{Resummation in the SCET approach}

In this Section we collect the relevant expressions for resummed quantities in the 
SCET approach of Refs.~\cite{Becher:2006nr,Becher:2006mr,Becher:2007ty,Ahrens:2008nc}.
In order to compare the two approaches it is useful to 
rewrite the resummed coefficient function Eq.~(\ref{eq:resummdyscet}) as
\bea\label{eq:resummdyscetonescale}
C_\textrm{SCET}^\textrm{res} = H(M^2) U(M^2, \mu_s^2) S(z, M^2, \mu_s^2),
\eea
where the energy scales $\mu_h$ and $\mu_f$ which appear in Eq.~(\ref{eq:resummdyscet}) are taken for simplicity 
equal to the hard scale $M^2$. In Eq.~(\ref{eq:resummdyscetonescale}) we recognize the hard function $H(M^2)$,
which arises from the matching at the hard scale $\mu_h=M$, and which have an expansion
in powers of $\as$; the soft function 
\bea
S(z, M^2, \mu_s^2) = \tilde s_\textrm{DY}  
\left(\log\frac{M^2}{\mu_s^2}+\frac{\de}{\de_\eta},\mu_s\right)
\frac{1}{1-z} \left(\frac{1-z}{\sqrt{z}}\right)^{2 \eta} \frac{e^{-2 \gamma_E \eta}}{\Gamma(2 \eta)}
\eea
where
\bea
\eta = \int_{M^2}^{\mu_s^2} \frac{d \mu^2}{\mu^2} \Gamma_\textrm{cusp}(\as(\mu^2));
\qquad \Gamma_\textrm{cusp} (\as) = A(\as),
\eea
and $\tilde s_\textrm{DY}$ has a perturbative expansion in powers of $\as(\mu^2)$. We
finally have
\bea
U(M^2, \mu_s^2) = \exp\left\{-\int_{M^2}^{\mu_s^2} \frac{d \mu^2}{\mu^2} \left[
\Gamma_\textrm{cusp} (\as(\mu^2)) \log \frac{\mu^2}{M^2}-\gamma_W(\as(\mu^2))
\right]
\right\}
\eea
which accounts RG evolution from the hard scale $\mu_h=M$ to the soft scale $\mu_s$. Observe that
$\Gamma_\textrm{cusp}(\as)= A(\as)$.
We note that the choice of the soft scale $\mu_s$ determines the form of the
logs, i.e. it determines what is being resummed.

We show in Tab.~\ref{tab:count} the logarithmic accuracy of the SCET and the dQCD results and 
its dependence on the perturbative order of the relevant functions. In Refs.~\cite{Becher:2006mr,Becher:2007ty,Ahrens:2008nc}
the counting is performed at the level of exponents, which corresponds to the 
N$^k$LL$^*$ accuracy.  

\begin{table*}[t]
\begin{center}
\begin{tabular}{lcccr}
  dQCD:  & $A(\as)$ & $D(\as)$ & $\bar g_0(\as)$ & accuracy: $\as^n \ln^kN$\\
  SCET: & $\Gamma_{\rm cusp}(\as)$ & $\gamma_W(\as)$ & $H$, $\tilde s_{\rm DY}$
  & accuracy: $\as^n \ln^k(\mu_s/M)$\\
  \midrule
  LL     & 1-loop & --- & tree-level & $k= 2n$ \\
\addlinespace[0.8\defaultaddspace]
  NLL*   & 2-loop & 1-loop & tree-level & $2n-1\le k\le 2n$ \\
  NLL    & 2-loop & 1-loop & 1-loop & $2n-2\le k\le 2n$ \\
\addlinespace[0.8\defaultaddspace]
  NNLL*  & 3-loop & 2-loop & 1-loop & $2n-3\le k\le 2n$ \\
  NNLL   & 3-loop & 2-loop & 2-loop & $2n-4\le k\le 2n$
  % \\  NNNLL  & 4-loop & 3-loop & 2-loop & $2n-5\le k\le 2n$ 
\end{tabular}
\caption{Orders of logarithmic approximations and accuracy of the
  predicted logarithms in dQCD and SCET. The last columns refers to the content of the
  coefficient function.}
\label{tab:count}
\end{center}
\end{table*}

\section{Comparison at NNLL}\label{sect:masterformula}

In this Section we will derive a master formula which relates the SCET and the dQCD 
result for a generic choice of the soft scale $\mu_s$. An analytic comparison 
between the two approaches can be performed in $N$ space, where dQCD 
admits a perturbative expansion in powers of the strong coupling $\as$. 

\subsection{Drell-Yan production}

The NNLL resummed expression in dQCD is given by Eq.~(\ref{eq:Cres}), 
where
\bea
A(\as) &= A_1 \as + A_2 \as^2 +A_3\as^3+\mathcal O(\as^4),\\
D(\as) &= D_1 \as + D_2 \as^2 + \mathcal O(\as^3),\\
\bar g_0 (\as) &= 1 + \bar g_{01}\as + \bar g_{02} \as^2+\mathcal O(\as^3) .
\eea 
The analytical expressions for the coefficients $A_i$, $D_i$ and $\bar g_{0i}$ can be found in Appendix~\ref{app:analy}.
In Ref.~\cite{Bonvini:2012az} it was proven that 
\bea
\bar S_\textrm{dQCD}\left(M^2, \frac{M^2}{\bar N^2} \right) =&
\int_{M^2}^{\frac{M^2}{\bar N^2}} \frac{d\mu^2}{\mu^2}
\left[A(\as(\mu^2)) \left(\log \frac{1}{\bar N^2} -
\log \frac{\mu^2}{M^2} \right) + \hat D_2 \as^2 (\mu^2) \right]\nonumber \\
&+ 2 \zeta_2 \frac{C_F }{\pi }\as(M^2)
\eea
where $\bar N = N e^{\gamma_E}$ and 
\bea
\hat D (\as) = \frac{1}{2} D(\as) - 2 \zeta_2 \frac{C_F}{\pi} \beta_0 \as^2 
= \hat D_2 \as^2 + \mathcal O(\as^3). 
\eea

The dQCD result can be therefore rewritten in the convenient form
\bea\label{eq:cqcdhat}
C_\textrm{dQCD} (N, M^2) =
\hat g_0 (\as(M^2)) \exp \hat S_\textrm{dQCD}
\left( M^2, \frac{M^2}{\bar N^2} \right)
\eea
where
\bea
\hat g_0 (\as) &= \bar g_0(\as) 
\exp\left[2 \zeta_2 A(\as) + \frac{8}{3} \zeta_3 \beta_0 
\frac{C_F}{\pi}\as^2\right],\\
\hat S_\textrm{dQCD} \left(M^2, \frac{M^2}{\bar N^2} \right) &=
\int_{M^2}^{M^2/\bar N^2} \frac{d\mu^2}{\mu^2} 
\left[A(\as(\mu^2)) \log \frac{M^2}{\mu^2 \bar N^2}+ \hat D(\as(\mu^2)) \right].
\eea

We now turn to the SCET expression at NNLL. In order to compare the SCET expression
to the perturbative QCD result, one has to perform a Mellin transform 
with respect to $z$. It is very important to observe that this Mellin transform has to 
be computed \textit{at fixed} $\mu_s$. Since all the $z$ dependence is contained in the 
soft function $S(z,M^2,\mu_s^2)$ one has
\bea\label{eq:exactmt}
\mathcal M[S(z, M^2,\mu_s^2)] = \tilde s_\textrm{DY} \left(\log \frac{M^2}{\mu_s^2} +\frac{\de}{\de \eta}, \mu_s \right)
\frac{\Gamma(N-\eta) \Gamma(2 \eta)}{\Gamma(N+\eta)}\frac{e^{-2 \gamma_E \eta}}{\Gamma(2 \eta)}.
\eea
In the large-$N$ limit this expression turns into 
\bea
\mathcal M[S(z, M^2,\mu_s^2)] = 
\tilde s_\textrm{DY}  \left(\log \frac{M^2}{\mu_s^2} +\frac{\de}{\de \eta}, \mu_s \right)
\bar N^{-2 \eta} + \mathcal O \left(\frac{1}{N} \right).
\eea
The analytical comparison between dQCD and SCET of Refs.~\cite{Bonvini:2012az,Bonvini:2013td} has been performed in this 
limit. However, we will investigate in next Chapter if these
neglected terms will have any phenomenological effects. 

The large-$N$ limit of the SCET expression can be finally written as 
\bea\label{eq:resummscethats}
C_\textrm{SCET} (N, M^2, \mu_s^2) = \hat H(M^2) 
E\left( \frac{M^2}{\bar N^2}, \mu_s^2 \right) 
\exp \hat S_\textrm{SCET}(N, M^2, \mu_s^2),
\eea
where
\bea
\hat H(M^2) &= H(M^2) \exp \left[\frac{\zeta_2}{2} \frac{C_F}{\pi} \as(M^2)\right],\\
E\left( \frac{M^2}{\bar N^2},\mu_s^2 \right)&= \tilde s_\textrm{DY} 
\left(\log \frac{M^2}{\mu_s^2 \bar N^2},\mu_s^2 \right)
\exp\left[-\frac{\zeta_2}{2} \frac{C_F}{\pi} \as(\mu_s^2) \right], \label{eq:estilde}\\
\hat S_{\textrm{SCET}}\left(N, M^2,\mu_s^2\right) 
&= \int_{M^2}^{\mu_s^2} \frac{d \mu^2}{\mu^2} 
\left[\Gamma_{\textrm{cusp}}(\alpha_s (\mu^2)) \log \frac{M^2}{\mu^2 \bar N^2} + \hat \gamma_W (\alpha_s(\mu^2)) \right]
\eea
and
\bea
\hat \gamma_W(\as) = \gamma_W (\as) - \frac{\zeta_2}{2}{C_F}{\pi}\beta_0 \as^2.
\eea
In particular, one has that $\hat \gamma_W(\as) = \hat D(\as)$ at this order. 

The ratio $C_r (N, M^2, \mu_s^2)$ between the dQCD and the SCET expressions is defined from
\bea
C_\textrm{dQCD} (N, M^2) = C_r (N, M^2, \mu_s^2) C_\textrm{SCET} (N, M^2, \mu_s^2).
\eea
We find therefore that
\bea
C_r (N, M^2, \mu_s^2) =
\frac{\hat g_0(\as(M^2))}{\hat H(M^2) E \left(\frac{M^2}{\bar N^2}, \mu_s^2 \right)}
\exp \hat S \left( \mu_s^2, \frac{M^2}{\bar N^2} \right)
\eea
with
\bea
\hat S \left(\mu_s^2, \frac{M^2}{\bar N^2} \right) =
\int_{\mu_s^2}^{\frac{M^2}{\bar N^2}} \frac{d \mu^2}{\mu^2} 
\left[A(\as(\mu^2)) \log \frac{M^2}{\mu^2 \bar N^2}+ \hat D (\as(\mu^2)) \right].
\eea
One sees that
\bea
\frac{\hat g_0 (\as(M^2))}{\hat H(M^2) E(M^2, M^2)} = 1+ \mathcal O(\as^3 (M^2))
\eea
so that to NNLL accuracy
\bea\label{eq:crerehats}
C_r(N, M^2, \mu_s^2) = \frac{E(M^2, M^2)}{E\left( \frac{M^2}{\bar N^2}, \mu_s^2 \right)}
\exp \hat S \left(\mu_s^2, \frac{M^2}{\bar N^2} \right).
\eea
Using Eq.~(\ref{eq:estilde}) and the 2-loop expression of $\tilde s_\textrm{DY}$ one has
\bea
E\left( \frac{M^2}{\bar N^2}, \mu_s^2 \right) =
1 +E_1(L) \as(\mu_s^2) + E_2 (L) \as^2 (\mu_s^2) + \mathcal O(\as^3),
\eea
where
\bea
E_1(L) =& \frac{A_1}{2} L^2,\\
E_2(L) =& \frac{A_1^2}{8}L^4 - \frac{L^3}{3} \frac{A_1}{2} \beta_0 
+\frac{L^2}{2} A_2 + L \hat D_2\nonumber \\
&+ \frac{C_A C_F}{\pi^2} \left[\frac{602}{324} + \frac{67}{144} \zeta_2 
-\frac{3}{4} \zeta_2^2 - \frac{11}{72} \zeta_3 \right]\nonumber \\
&+ \frac{C_F n_f}{\pi^2} \left[-\frac{41}{162} - \frac{5}{72} \zeta_2 +
\frac{\zeta_3}{36} \right],
\eea
with 
\bea\label{eq:Lmus}
L \equiv \log \frac{M^2}{\mu_s^2 \bar N^2}.
\eea
In particular, $L=0$ if the two arguments of $E$ are equal to each other. 

We observe that $C_r(N, M^2, \mu_s^2)=1$ for $\mu_s= M/\bar N$, up to 
subleading (NNNLL$^*$) terms. Therefore with this particular scale choice 
the SCET result reproduces the dQCD result to NNLL accuracy. 

\subsection{Higgs boson production}

In the case of Higgs boson production an equivalent master formula which relates
dQCD and SCET exists. The differences between the two process are
limited to a rearrangement of color factors and different values of the 
process-dependent coefficient. In particular, the Higgs resummed coefficients in dQCD are
related to the Drell-Yan coefficient by
\bea
A_k^\textrm{Higgs} = \frac{C_A}{C_F} A_k^\textrm{DY},\qquad 
D_k^\textrm{Higgs} = \frac{C_A}{C_F} D_k^\textrm{DY}.
\eea
On the other hand, the function $\bar g_0$ in the Higgs case is different
from the function $\bar g_0$ in the Drell-Yan case; the first coefficients
of its perturbative expansion can be found in Appendix~\ref{app:analy}. 

A similar argument also holds for the SCET expression in this case; the values
of $\tilde s_\textrm{Higgs}$, $\Gamma_\textrm{cusp}$ and $\gamma_W$ are related to
the analogous DY functions by a color factor rearrangement. However, the perturbative 
expansion of $H(M^2)$ is instead different from the DY one. 

The dQCD result in case of Higgs production can be written as
\bea
C_\textrm{dQCD} (N, M^2) =
\hat g_0 (\as(M^2)) \exp \hat S_\textrm{dQCD}
\left( M^2, \frac{M^2}{\bar N^2} \right)
\eea
where
\bea
\hat g_0 (\as) &= \bar g_0(\as) 
\exp\left[2 \zeta_2 A(\as) + \frac{8}{3} \zeta_3 \beta_0 
\frac{C_A}{\pi}\as^2\right],\\
\hat S_\textrm{dQCD} \left(M^2, \frac{M^2}{\bar N^2} \right) &=
\int_{M^2}^{M^2/\bar N^2} \frac{d\mu^2}{\mu^2} 
\left[A(\as(\mu^2)) \log \frac{M^2}{\mu^2 \bar N^2}+ \hat D(\as(\mu^2)) \right]\\
\hat D (\as) &= \frac{1}{2} D(\as) - 2 \zeta_2 \frac{C_A}{\pi} \beta_0 \as^2 
= \hat D_2 \as^2 + \mathcal O(\as^3). 
\eea
whereas the SCET result is 
\bea
C_\textrm{SCET} (N, M^2, \mu_s^2) = \hat H(M^2) 
E\left( \frac{M^2}{\bar N^2}, \mu_s^2 \right) 
\exp \hat S_\textrm{SCET}(N, M^2, \mu_s^2),
\eea
where
\bea
\hat H(M^2) &= H(M^2) \exp \left[\frac{\zeta_2}{2} \frac{C_A}{\pi} \as(M^2)\right],\\
E\left( \frac{M^2}{\bar N^2},\mu_s^2 \right)&= \tilde s_\textrm{Higgs} 
\left(\log \frac{M^2}{\mu_s^2 \bar N^2},\mu_s^2 \right)
\exp\left[-\frac{\zeta_2}{2} \frac{C_A}{\pi} \as(\mu_s^2) \right],\\
\hat S_{\textrm{SCET}}\left(N, M^2,\mu_s^2\right) 
&= \int_{M^2}^{\mu_s^2} \frac{d \mu^2}{\mu^2} 
\left[\Gamma_{\textrm{cusp}}(\alpha_s (\mu^2)) \log \frac{M^2}{\mu^2 \bar N^2} + \hat \gamma_W (\alpha_s(\mu^2)) \right]
\eea
and
\bea
\hat \gamma_W(\as) = \gamma_W (\as) - \frac{\zeta_2}{2}{C_A}{\pi}\beta_0 \as^2.
\eea
In particular, also in this case one has that $\hat \gamma_W(\as) = \hat D(\as)$ at this order. 

We have checked that also in this case
\bea
\frac{\hat g_0 (\as(M^2))}{\hat H(M^2) E(M^2, M^2)} = 1+ \mathcal O(\as^3 (M^2))
\eea
so that to NNLL accuracy Eq.~(\ref{eq:crerehats}) holds, where
in this case
\bea
E\left( \frac{M^2}{\bar N^2}, \mu_s^2 \right) =
1 +E_1(L) \as(\mu_s^2) + E_2 (L) \as^2 (\mu_s^2) + \mathcal O(\as^3),
\eea
where
\bea
E_1(L) =& \frac{A_1}{2} L^2,\\
E_2(L) =& \frac{A_1^2}{8}L^4 - \frac{L^3}{3} \frac{A_1}{2} \beta_0 
+\frac{L^2}{2} A_2 + L \hat D_2 \nonumber \\
&+ \frac{C_A^2}{\pi^2} \left[\frac{602}{324} + \frac{67}{144} \zeta_2 
-\frac{3}{4} \zeta_2^2 - \frac{11}{72} \zeta_3 \right] \nonumber \\
&+ \frac{C_A n_f}{\pi^2} \left[-\frac{41}{162} - \frac{5}{72} \zeta_2 +
\frac{\zeta_3}{36} \right],
\eea
and $L$ given by Eq.~(\ref{eq:Lmus}).

\subsection{Master formula}\label{sect:mformulas}

We are interested in studying $C_r$ for different scale choices, and in
particular with the Becher-Neubert scale choice. It may result 
useful to recast $C_r$ in
a particularly transparent form suitable for analytic comparison by 
exponentiating the ratio 
\bea
\frac{E(M^2, M^2)}{E \left(\frac{M^2}{\bar N^2}, \mu_s^2 \right)}.
\eea
This manipulation introduces terms of order 
$\as^3$ or higher in the ratio, which does not spoil the NNLL accuracy. 
In this way we can find a final form of $C_r$ which represents the irreducible
difference between dQCD and SCET at NNLL accuracy. 
One has
\bea
\log \frac{E(M^2, M^2)}{E \left(\frac{M^2}{\bar N^2}, \mu_s^2 \right)} 
=& \as(\mu_s^2) A_1 \frac{L^2}{2}\nonumber \\
&+ \as^2 (\mu_s^2) \left[\beta_0 \frac{A_1}{2} \frac{L^3}{3}
-A_2 \frac{L^2}{2} - \hat D_2 L \right]
+ \mathcal O(\as^3).
\eea
By using
\bea
\frac{L^{k+1}}{k+1} = \int_{\mu_s^2}^{M^2/\bar N^2}
\frac{d \mu^2}{\mu^2} \log^k \frac{M^2}{\mu^2 \bar N^2}, 
\eea
and taking the running of $\as$ into account, we get to
\bea\label{eq:crfinal}
C_r(N,M^2, \mu_s^2) 
= \exp \int_{\mu_s^2}^{M^2/\bar N^2} \frac{d \mu^2}{\mu^2} \log \frac{M^2}{\mu^2 \bar N^2}
\left[A(\as(\mu^2))- A_1 \as(\mu^2) - A_2 \as^2 (\mu^2) \right]
\eea
which is the final NNLL master dQCD-SCET comparison formula of Ref.~\cite{Bonvini:2013td}. We observe that,
thanks to the exponentiation, the exponent in Eq.~(\ref{eq:crfinal}) is of order $\as^3$.

We remark however that a full numerical comparison between the SCET and dQCD results 
should rather be performed at first place by using the exact Mellin Eq.~(\ref{eq:exactmt}) and only 
at second place considering the large-$N$ approximation which leads to Eqs.~(\ref{eq:crerehats}) and (\ref{eq:crfinal}). 
Furthermore, one should also consider the effect of subleading terms which change
between the expression Eq.~(\ref{eq:Cres}) and Eq.~(\ref{eq:cqcdhat}) (dQCD) and 
the Mellin of Eq.~(\ref{eq:resummdyscetonescale}) without analytical manipulation and
Eq.~(\ref{eq:resummscethats}) (SCET). In the next Chapter we 
will perform such a systematic comparison which will enable us to understand the 
phenomenological effects of the analytical manipulations which lead to Eq.~(\ref{eq:resummscethats}).

\section{Comparison in hadronic space}\label{sect:compbn}

We now move to the BN scale choice and we compare the dQCD and the SCET result in this
case. The BN approach is based on the choice
\bea
\mu_s \sim M(1-\tau) 
\eea
which, as discussed in Chapter \ref{chapt:scetapp}, provides a solution to the Landau pole problem. 
In the following argument we are interested on the logarithmics accuracy of the SCET
result and therefore we will discuss the choice $\mu_s=M(1-\tau)$ and we will 
not consider the more general choice $\mu_s = M(1-\tau) g(\tau)$ discussed in Sect.~\ref{sect:bnscale}.
Since $\mu_s$ depends on a hadronic variable, a partonic comparison cannot be performed,
and one must carry out the comparison at the level of hadronic cross-sections. 
We therefore consider
\bea
\sigma_\textrm{dQCD}(\tau, M^2) &= \int_\tau^1 
\frac{dz}{z} \mathscr L \left( \frac{\tau}{z} \right) C_\textrm{QCD} (z, M^2),\\
\sigma_\textrm{SCET}(\tau, M^2) &= \int_\tau^1
\frac{dz}{z} \mathscr L \left( \frac{\tau}{z} \right) C_\textrm{SCET} 
(z,M^2,M^2(1-\tau)^2)
\eea
where $C_\textrm{dQCD} (z, M^2)$ and $C_\textrm{SCET} (z, M^2,\mu_s^2)$ are the
inverse Mellin transform of the $N$-space coefficient function Eqs.~(\ref{eq:cqcdhat}) and (\ref{eq:resummscethats}) and 
$C_\textrm{SCET} (z, M^2,\mu_s^2)$ has to be calculated as a Mellin transform at fixed $\mu_s$.
$C_\textrm{dQCD} (z, M^2)$ should be understood as the order-by-order Mellin inversion
at a (arbitrarily high) finite order, since it is given by a divergent series in
$\as(M^2)$. The dQCD cross-section can be written as
\bea\label{eqdqcdscetcr}
\sigma_\textrm{dQCD} (\tau, M^2) = \int_\tau^1 \frac{dz}{z} \sigma_\textrm{SCET}
\left( \frac{\tau}{z},M^2 \right) C_r (z, M^2, M^2 (1-\tau)^2 )
\eea
where $C_r  (z, M^2, M^2 (1-\tau)^2 )$ is the inverse Mellin transform of $ C_r (N, M^2, M^2 (1-\tau)^2 )$
carried out at fixed $\mu_s$ and evaluated at $\mu_s = M(1-\tau)$. 

$C_r(N, M^2, \mu_s^2)$ can be rewritten as 
\bea
C_r(N, M^2, \mu_s^2) &= \frac{E(M^2,M^2)}{E(\mu_s^2,\mu_s^2)}
\frac{E(\mu_s^2,\mu_s^2)}{E\left(\frac{M^2}{\bar N^2}, \mu_s^2 \right)}
\exp \hat S \left( \mu_s^2, \frac{M^2}{\bar N^2} \right)\nonumber \\
&= \frac{E(M^2,M^2)}{E(\mu_s^2,\mu_s^2)} 
\left[1 + F_r\left(\as(\mu_s^2), \log \frac{M^2}{\mu_s^2\bar N^2} \right) \right],
\eea
where $F_r\left(\as(\mu_s^2), \log \frac{M^2}{\mu_s^2\bar N^2} \right)$ is 
of order $\as^3$. Its inverse Mellin transform is \cite{Bonvini:2012az}
\bea
C_r (z, M^2, \mu_s^2 ) = \frac{E(M^2,M^2)}{E(\mu_s^2,\mu_s^2)}
\left[\delta(1-z) +
F_r \left( \as(\mu^2_s), 2 \frac{\de }{\de \xi}  \right)
\frac{(1-\tau)^{-\xi} \log^{\xi-1}\frac{1}{z}}{e^{\gamma_E \xi}\Gamma(\xi)
}\Big|_{\xi=0}\right]
\eea
Plugging this last expression into Eq.~(\ref{eqdqcdscetcr}) we obtain
\bea
\sigma_\textrm{dQCD} (\tau, M^2) 
= \frac{E(\mu_s^2, \mu_s^2)}{E\left(\frac{M^2}{\bar N^2}, \mu_s^2 \right)}
\left[\sigma_\textrm{SCET}(\tau,M^2) + F_r\left(\as(\mu_s^2),2 \frac{\de}{\de \xi} \right)
\Sigma (\tau, M^2, \xi) \Big|_{\xi=0} \right]
\eea
where
\bea
\Sigma (\tau, M^2, \xi) &= \frac{(1-\tau)^{-\xi}}{e^{\gamma_E \xi}\Gamma(\xi)}
\int_{\tau}^1 \frac{dz}{z} \sigma_\textrm{SCET}\left( \frac{\tau}{z}, M^2 \right)
\log^{\xi-1}\frac{1}{z} \nonumber \\
&= \sum_{k=0}^\infty c_k (\xi) \frac{d^k \sigma_\textrm{SCET} (\tau, M^2) }{d \log^k(1-\tau)}
[1+ \mathcal O(1-\tau)]
\eea
for $\mu_s = M(1-\tau)$. All the logarithmic enhancement are now contained in
the ratio 
\bea
\frac{E(M^2,M^2)}{E(\mu_s^,\mu_s^2)}&=
1+ E_2(0) [\as^2 (M^2) - \as^2 (\mu_s^2)]+ \cdots\nonumber \\
&= 1 + 4 E_2 (0) \beta_0 \as^3 (M^2) \log (1-\tau)+ \mathcal O(\as^4).
\eea

Therefore the leading difference between dQCD and SCET expressions is
\bea
\sigma_\textrm{dQCD}(\tau, M^2) - \sigma_\textrm{SCET}(\tau, M^2) =
\sigma_\textrm{SCET}(\tau, M^2) [\as^3 4 \beta_0 E_2(0) \log (1-\tau)+ \ldots]
\eea
where the ellipses stands for terms of relative order $\mathcal O(\as^3)$,
without logarithmic enhancement, or terms of relative order $\mathcal O(\as^4)$.

The logarithmic counting is now based on counting powers of $\log (1-\tau)$.
Since the SCET expression violates factorization, the difference 
between dQCD and SCET results is not universal and it depends on parton
luminosity through $\sigma_\textrm{SCET}$. The cross-section $\sigma_\textrm{SCET}$
contains in general leading log terms of form
\bea
\sigma_\textrm{SCET} \sim \as^k \log^{2k+p}(1-\tau)
\eea
where terms of order $\as^k \log^{2k}(1-\tau)$ comes from the coefficient function 
while terms of order $\as^k \log^{p}(1-\tau)$ are due to the parton luminosity.

If we set $p=0$, which correspond to assume that the parton luminosity does not lead 
any logarithmic enhancement, the difference between dQCD and SCET is
\bea
\sigma_\textrm{dQCD}-\sigma_\textrm{SCET} \sim \as^{h} \log^{2h-5}(1-\tau); \qquad h=k+3
\eea
which corresponds to a NNNLL$^*$ correction. 

In conclusion, SCET and dQCD results differ by a luminosity-dependent term. The BN
scale choice avoids the Landau pole problem but the price to pay is the introduction 
of logarithmically suppressed non-universal terms, in contrast with MP or BP which 
introduces respectively non universal, but more than power-suppressed terms or
power-suppressed universal terms. Moreover, the parton distribution are expected to contain
logarithmically enhanced contributions, which may lead the terms introduced by
the BN scale choice to become leading or even super-leading. 

Finally, when $\tau$ is far from threshold one has $\mu_s=M(1-\tau)\sim M$ and 
therefore $C_r$ is actually leading log. It was observed in Ref.~\cite{Bonvini:2012az} that this problem
could be partially removed through generalizations of the scale $\mu_s$ 
which lead to values of $\mu_s$ far from the hard scale $M$; these choices however
do not change the log counting. 

In order to obtain a full understanding of this state of affairs and to investigate the
phenomenological implications of the results discussed in this Chapter it seems necessary to 
appeal to numerical methods. In the next Chapter we will discuss our strategies 
and we will perform a phenomenological comparison of the two approaches.

\chapter{Resummation in SCET and in dQCD: a phenomenological comparison}

In this Chapter we will perform a systematic comparison of soft-gluon resummation 
in SCET and in dQCD. In particular, we will present results for Higgs boson production
in gluon-gluon fusion in a hadron-hadron collider. 

As we have understood, phenomenological predictions in the two
formalisms differ significantly in their implementation. The naivest way to 
compare the two approaches would basically consist of a $\tau$-space comparison
of the numerical results. With this results at hand, however, we 
would not gain any particular understanding of possibile sources of 
differences between SCET and dQCD. 
On the other hand, we have seen in the previous Chapter how an analytic comparison 
can be performed in $N$-space and we have found master formulas which relate
the two formalisms. For this reason, it would be of great interest to 
perform a comparison in $\tau$ space but armed with the analytic insight of 
Chapter~\ref{chapter:ancomp}. We will show how this can be done by means of a saddle-point method. 
Furthermore, we will show that a saddle-point argument allows the quantification
of factorization violation effects in SCET. We will present the results we have 
obtained with this strategy and we will discuss some open issues which 
would be interesting to investigate further.

\section{Comparison of the resummed results}

In this Section we will compare the numerical predictions in SCET and in dQCD 
by looking at the total cross-section and at the $K$-factor which we have 
computed numerically. At first, we will show the predictions for the matched results.
However, matching with fixed-order calculations reduces the differences between 
the resummed results; moreover, the analytic comparison of Chapter~\ref{chapter:ancomp} involves the
pure resummed coefficient functions. For this reason, we will move to 
resummed predictions and we will not consider the matching with the fixed-order. 

\begin{figure}[htbp]
\centering
   {\includegraphics[width=.75\columnwidth]{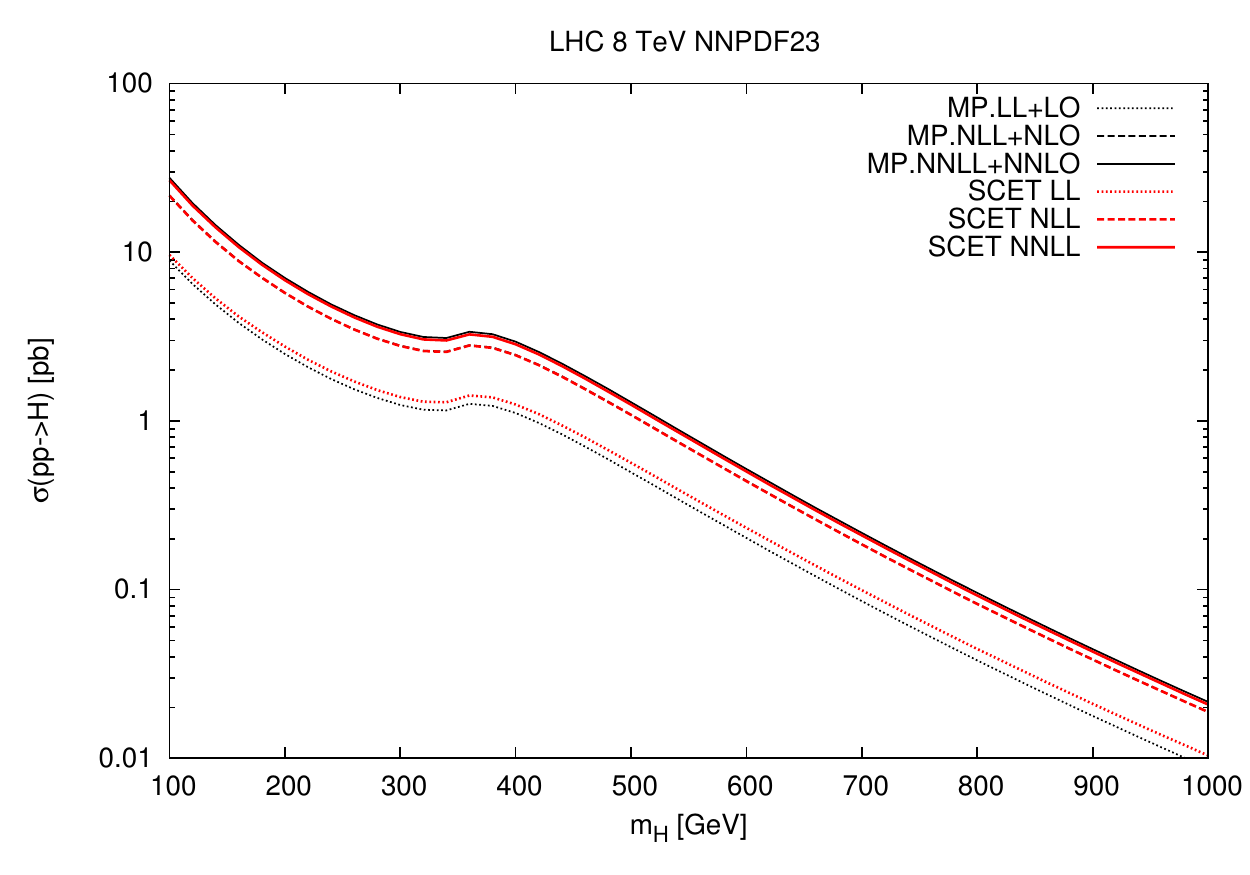}}
\caption{Predictions in SCET and in dQCD for the resummed cross-section as a function of the
Higgs mass $m_H$ at LL, NLL and NNLL at the LHC.}\label{fig:SCETvsgghiggsTotal}
\end{figure}

\begin{figure}[htbp]
\centering
   {\includegraphics[width=.75\columnwidth]{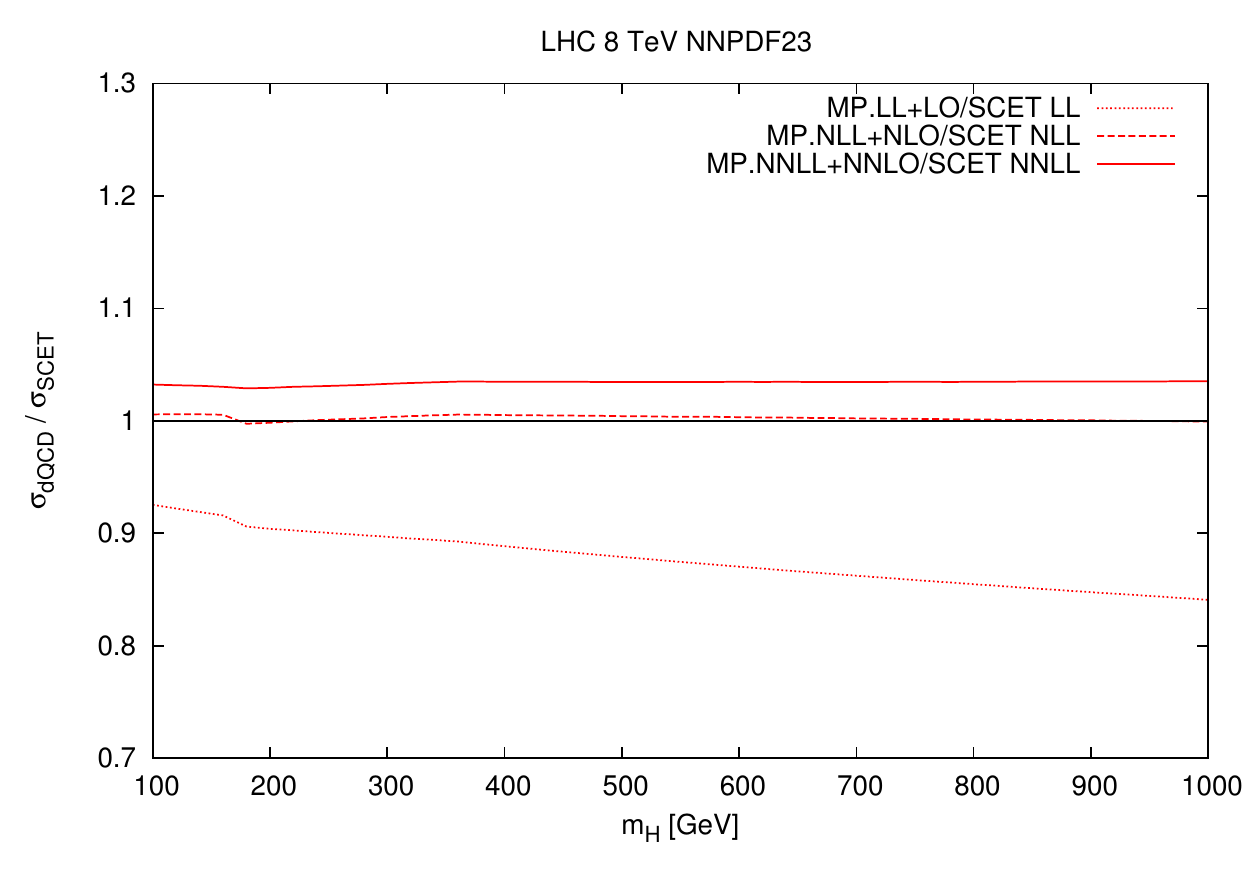}}
\caption{Ratio of the dQCD prediction to the SCET prediction for the total cross-section, matched with 
fixed-order results, as a function of the
Higgs mass $m_H$ at LL, NLL and NNLL at the LHC.}\label{fig:SCETvsgghiggs}
\end{figure}

In Fig.~\ref{fig:SCETvsgghiggsTotal} we show the predictions for the
total cross-section for Higgs boson production at LL, NLL and NNLL, with 
consistent matching with fixed-order results. The 
curves have been calculated with NNPDF23 NNLO PDF set in the large-$m_t$ limit
and $\as(m_Z) =0.0117$. 
We observe that at LL accuracy the SCET result is larger than the dQCD result,
but the situation changes as the logarithmic accuracy grows, and at NNLL accuracy 
the dQCD prediction is larger than the SCET prediction. The predictions for the
total cross-section in the two formalism can made more transparent by comparing their
ratio. In Fig.~\ref{fig:SCETvsgghiggs} we show the ratio of the dQCD prediction 
to the SCET prediction. We observe that at LL accuracy the dQCD prediction is 
about $10\%$ lower than the dQCD one; at NLL accuracy the predictions are comparable, 
and at NNLL accuracy the SCET result is $3-4\%$ lower than the dQCD result.

\begin{figure}[htbp]
\centering
   {\includegraphics[width=.75\columnwidth]{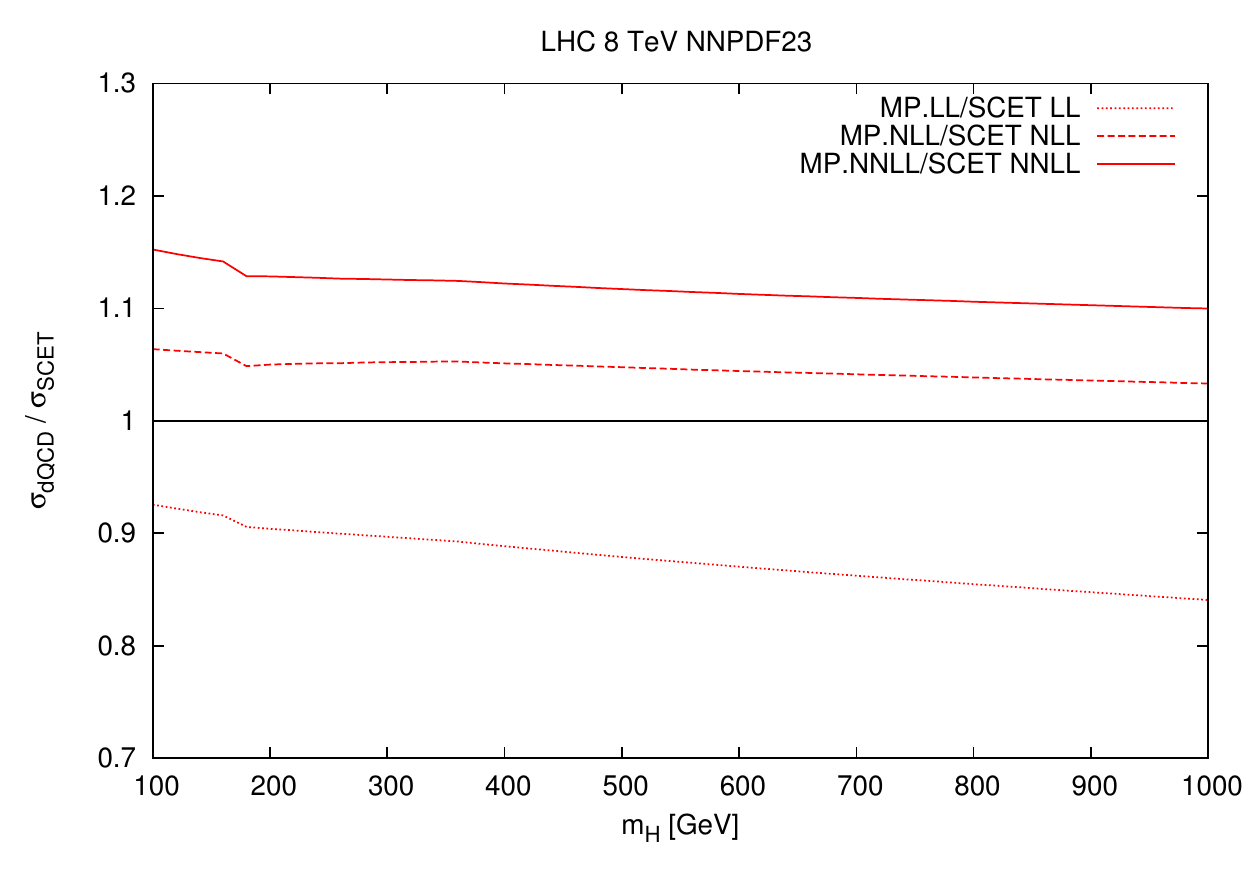}}
\caption{Ratio of the dQCD prediction to the SCET prediction for the total cross-section as a function of the
Higgs mass $m_H$ at LL, NLL and NNLL at the LHC, without matching with the fixed-order results.}\label{fig:SCETvsgghiggsnmomatch}
\end{figure}

We show in Fig.~\ref{fig:SCETvsgghiggsnmomatch} the ratio of the resummed predictions
without matching with fixed-order results. We observe that in this case 
the dQCD prediction at NLL accuracy  is about $5\%$ larger than the SCET prediction
and at NNLL accuracy the dQCD result is up to $15\%$ larger for a light Higgs 
and more than $10\%$ larger for a heavy Higgs.

From now on, we will not consider anymore the matched results and we will consider the
resummed prediction in both formalism. The results obtained give a first feeling 
for the size of resummation in the two formalisms. However, a more detailed comparison 
will be carried out in the following Sections.

\clearpage

\section{The resummation region for \texorpdfstring{$gg\rightarrow H$}{gg->H} production mechanism}

In the previous Section we have shown a first comparison of the predictions 
for the cross-section in dQCD and in SCET. However, we learn very little 
from the previous results, which simply state how much the two approaches
differ in their numerical predictions. For this reason, it would be interesting to
find a new strategy which would let us study the differences between dQCD and 
SCET in more detail. 

In this Section we will show that a saddle-point argument
is particularly suited for this goal and it presents several advantages. First, through a saddle-point argument it is possibile
to estimate the region which is effected the most by resummation. Second, 
it is possible to perform a phenomenological comparison with hadronic variables
and $N$-space expressions. Third, the approximate result obtained is very close to
the exact result but requires a minimal amount of computational work. Fourth,
by means of the saddle-point method it is possibile to estimate the effects of factorization violation 
in SCET.

We will show in Sect.~\ref{sec:psaddle}, by means of 
Mellin-space argument, how it is possible to determine the region 
of $N$ which provides the dominant contribution to the cross-section. 
We will discuss how this argument can be applied in the SCET approach in
Sect.~\ref{sect:scetsaddle}. We will then apply this argument to Higgs production in gluon-gluon fusion
and we will then show in Sect.~\ref{sect:saddleres} the goodness of our approximation by 
comparing our previous predictions with the results obtained 
with the saddle point argument. We will finally address the problem of 
the violation of factorization in SCET and we will try to quantify it in Sect.~\ref{sect:factviol}.

\subsection{Saddle point argument}\label{sec:psaddle}

In Ref.~\cite{Bonvini:2010tp,Bonvini:2012an} it was shown that the
logarithmic contribution to the coefficient function is sizable or 
even dominant for $N\gtrsim 2$
and it rapidly deviates from it as $N$ gets smaller.
This means that the region where logarithmic effects are important is 
wider than the region where $\as \log^2 N \sim 1$. In particular, 
in this region, where logarithmically enhanced terms
behave in a perturbative way, they may lead to a substantial contribution. 

As we have seen in Chapter~\ref{chapter:resummation}, since the hadronic cross-sections are calculated 
as the convolution of the partonic cross-section with a parton luminosity,
the effect of soft-gluon resummation can be relevant far from the 
hadronic threshold. This can be made quantitative using a saddle-point argument
in $N$-space.
More precisely, in Refs.~\cite{Bonvini:2010tp,Bonvini:2012an} it was shown that for a given process and for any
given value of the hadronic ratio $\tau$ the cross-section receives the 
dominant contribution from a narrow range of $N$, the conjugate to 
$\tau$ upon Mellin transform. In particular, it was shown 
that the position of the saddle of the Mellin inversion integral is 
mostly determined by the parton luminosity and that it is not particularly sensitive to the non-perturbative
shape of PDFs whereas it is mainly determined by the small-$x$ (low-$N$) behaviour of the 
relevant Altarelli-Parisi splitting functions. In particular, if the splitting function has a fast
small-$x$ growth, then the average center of mass energy gets smaller and therefore
the resummation can be relevant far from the hadronic threshold.
In this case, threshold resummation is controlled 
by perturbative physics. 
We will see that that 
since the saddle point is mostly determined by luminosity it may be quite large,
even if $\tau \ll 1$. The region in which resummation is relevant is therefore 
extended to small values of $\tau$.

We consider a suitable quantity $\sigma$ for a hadronic process,
related in a simple way to a cross-section or a distribution, characterized
by a scale $M^2$ and a centre-of-mass energy $s=M^2/\tau$ which 
has the property of factorizing as
\bea\label{eq:sigmasaddle1}
\sigma(\tau, M^2) = \int_\tau^1 \frac{dz}{z} \mathscr L(z) 
C\left( \frac{\tau}{z},\as(M^2)\right)
\eea
in terms of a partonic coefficient function $C$ and a 
parton luminosity
\bea
\mathscr L(z,\mu^2) = \int_z^1 \frac{dx}{x} f_1(x,\mu^2) 
f_2\left( \frac{z}{x},\mu^2\right), 
\eea
which in turn depends on parton distributions $f_i(x_i,\mu^2)$. 

In general, $\sigma$ receives different contributions from each of the
parton channels which contribute in the process. This is however inessential
for our discussion. 
We will concentrate on the dominant channel. 

In Eq.~(\ref{eq:sigmasaddle1}) we recognize the partonic threshold, 
i.e. the region where the partonic coefficient function is
enhanced and soft-gluon resummation is relevant, as the region where
$\hat s=x_1 x_2 s$ (where $x_1=x$ and $x_2=z/x$) is close to $M^2$.  
Since all the values of $x_1$ and $x_2$ between $\tau$ and $1$ 
are accessible, the relevance of resummation depends on the
dominant region in the convolution integral, which can be determined
using a saddle point argument. 

We consider the Mellin transform of $\sigma(\tau,M^2)$ 
\bea\label{eq:saddlemellin}
\sigma(N, M^2) = \int_0^1 d\tau\ \tau^{N-1} \sigma(\tau, M^2), 
\eea
with inverse 
\bea
\sigma(\tau, M^2) = \frac{1}{2 \pi i} \int_{c -i \infty}^{c + i \infty}
dN\ \tau^{-N} \sigma(N, M^2) =
\frac{1}{2 \pi i} \int_{c -i \infty}^{c + i \infty}dN\ e^{E(\tau, N; M^2)}
\eea
where we have defined 
\bea
E(\tau, N; M^2) \equiv N \log \frac{1}{\tau} + \log \sigma (N,M^2).
\eea
The function $E(\tau, N; M^2)$ always has a minimum on the real 
positive $N$ axis at some $N_0$, since $\sigma (N,M^2)$ is 
a decreasing function of $N$ (the area below $\tau^{N-1} \sigma (\tau,M^2)$
decreases as $N$ increases). Therefore the inversion integral is
dominated by the region $N\sim N_0$, and it can be approximated 
by a saddle-point argument looking at the expansion of $E(\tau, N; M^2)$ around $N_0$.

$E(\tau,N)$ has a minimum on the real $N$ axis where $N=N_0(\tau)$, whith
\bea
E'(\tau, N_0(\tau)) = \log \frac{1}{\tau} + \frac{\sigma'(N_0(\tau),M^2)}{\sigma(N_0(\tau),M^2)}=0,
\eea
where we have denoted with a prime the differentiation with respect to $N$. 
Hence the inversion integral can be approximated by
\bea\label{eq:saddleappr}
\sigma(\tau, M^2) &\approx \int_{c-i\infty}^{c+i\infty} dN \
e^{E(\tau, N_0(\tau))+ \frac{E''(\tau,N_0(\tau))}{2}(N-N_0(\tau))^2}\nonumber \\
&= \frac{1}{\sqrt{2 \pi }} \frac{e^{E(\tau,N_0(\tau))}}{\sqrt{E''(\tau, N_0(\tau))}}
\eea
which has been calculated by changing the variable $N=N_0+it$ and
performing the gaussian integral.

It was shown in Refs.~\cite{Bonvini:2010tp,Bonvini:2012an} that the position of the saddle point 
for the fixed order cross-section in case of DY process and Higgs process 
is mostly determined by the PDFs and it is very much larger 
than the one calculated with only the coefficient function. 
As a consequence, the convolution with
PDFs greatly enhances the impact of resummation and it
extends it to a wider kinematic region.

The saddle-point strategy is particularly useful because it allows a transparent comparison
of the SCET and the dQCD expression at the parton level. As we know, the SCET expression
depends through $\mu_s$ on a hadronic scale $\tau$. Since the 
SCET expression is no longer in the form of a convolution product, and therefore it
does not factorize into a parton luminosity and a partonic cross-section, 
it is not immediate to perform a partonic comparison. However, through a saddle-point 
argument we can obtained an approximate factorized expression both for 
dQCD and SCET. By considering the first order of the saddle-point approximation we obtain
\bea
\sigma(\tau, M^2) &\approx \frac{1}{\sqrt{2 \pi}}e^{E(\tau,N_0(\tau))}\nonumber\\
&= \frac{1}{\sqrt{2 \pi}} \tau^{-N_0(\tau)}  \mathscr L(N_0(\tau), M^2) C(N_0(\tau), M^2).
\eea
In this way the cross-section is factorized into a luminosity and a 
coefficient function, evaluated at the saddle-point $N_0$. In the next Section 
we will discuss how the saddle-point argument can be used in SCET; 
however, it is clear that if one can obtain an analogous expression for the SCET case
a partonic comparison could be possible.

\subsection{The saddle point in SCET}\label{sect:scetsaddle}

In the previous Section we have seen how the saddle point depends on
the Mellin transform of the cross-section $\sigma (\tau,M^2)$ 
Eq.~(\ref{eq:saddlemellin}). In Mellin space the cross-section 
calculated in dQCD factorizes in the product of the parton luminosity
and a coefficient function:
\bea
\sigma(N, M^2) = \mathscr L(N, M^2) C(N, M^2).
\eea

This is not the case if one considers the Mellin transform of 
the SCET expression in the BN approach. In fact, since the
BN choice of $\mu_s$ depends on the hadronic variable $\tau$,
the cross-section $\sigma(\tau,M^2)$ is not a convolution:
\bea
\sigma^\textrm{SCET}(\tau,M^2) = 
\int_\tau^1 \frac{dz}{z} \mathscr L(z) C \left( \frac{\tau}{z}, 
M^2, \mu_s^2(\tau)\right)
\eea
It is however possible to compute the Mellin transform of 
$\sigma^\textrm{SCET}(\tau,M^2)$ \textit{at fixed} $\mu_s$ 
and then consider 
\bea
\sigma^\textrm{SCET}(N, M^2;\tau) = \mathscr L(N, M^2) C(N, M^2,\mu_s^2(\tau)),
\eea
where with an abuse of notation we have denoted the 
the cross-section in $N$ space as $\sigma^\textrm{SCET}$, even if
it is not the Mellin transform of $\sigma^\textrm{SCET}(\tau,M^2)$.
The cross-section can be calculated as
\bea\label{eq:saddlescet}
\sigma^\textrm{SCET}(\tau, M^2) = \frac{1}{2 \pi i} \int_{c -i \infty}^{c + i \infty}
dN\ \tau^{-N} \sigma^\textrm{SCET}(N, M^2;\tau) =
\frac{1}{2 \pi i} \int_{c -i \infty}^{c + i \infty}dN\ e^{E^\textrm{SCET}(\tau, N; M^2)}
\eea
keeping in mind that $\sigma^\textrm{SCET}(N, M^2;\tau)$ is not the Mellin 
transform of $\sigma^\textrm{SCET}(\tau, M^2)$ if $\mu_s$ depends on $\tau$.
Following the steps of the previous Section it is possible to 
look for the saddle-point of the integral Eq.~(\ref{eq:saddlescet}).
In this case $E^\textrm{SCET}(\tau,N)$ will have a minimum on the real $N$ axis 
where $N=\bar N_0(\tau)$, whith
\bea
{E^\textrm{SCET}}'(\tau, \bar N_0(\tau)) = 
\log \frac{1}{\tau} + \frac{{\sigma^\textrm{SCET}}'(\bar N_0(\tau),M^2;\tau)}{\sigma^\textrm{SCET}(\bar N_0(\tau),M^2;\tau)}=0,
\eea
where in principle $N_0(\tau)\neq \bar N_0(\tau)$. 

We can now compare the saddle-point approximation of the cross-section in the
two approaches, which at first order read:
\bea\label{eq:sadappdqcd}
\sigma^\textrm{dQCD}(\tau, M^2) &\approx \frac{1}{\sqrt{2 \pi}} \tau^{-N_0(\tau)}  \mathscr L(N_0(\tau), M^2) C^\textrm{dQCD}(N_0(\tau), M^2)\\
\sigma^\textrm{SCET}(\tau, M^2) &\approx \frac{1}{\sqrt{2 \pi}} \tau^{-\bar N_0(\tau)}
 \mathscr L(\bar N_0(\tau), M^2) C^\textrm{SCET}(\bar N_0(\tau), M^2,\mu_s^2(\tau))
 \label{eq:sadappscet}.
\eea
If $N_0=\bar N_0$ we conclude that 
\bea
\frac{\sigma^\textrm{dQCD}(M^2,\tau)}{\sigma^\textrm{SCET}(M^2,\tau,\mu_s^2(\tau))} \simeq 
\frac{C^\textrm{dQCD}(M^2,\tau)}{C^\textrm{SCET}(M^2,\tau,\mu_s^2(\tau))}
= C_r(N_0(\tau), M^2, \mu_s^2(\tau)).
\eea
It is therefore possible a comparison of the two approaches trough partonic expressions. 

However, it is not obvious that the two saddle point should be equal.
In fact we expect a very small difference between $N_0$ and $\bar N_0$ because the coefficient function in 
SCET and in dQCD are two different functions; however, a potentially bigger 
difference between the two saddle-point descends from the non-trivial dependence
of $C^\textrm{SCET}$ on the hadronic variable $\tau$. 
By comparing the resulting expressions it is hence possible to investigate 
the factorization violation of the SCET result.  
The study of the saddle-point positions in dQCD and in SCET therefore allows one to quantify 
the violation of factorization in the SCET formalism. Since the dependence 
of the saddle-point is mostly determined by PDFs, we will find
this violation small.

\subsection{Results}\label{sect:saddleres}

In this Section we show the results obtained in the specific case of 
Higgs production in gluon-gluon fusion. In this particular case the 
quantity $\sigma(\tau,M^2)$ of Eq.~(\ref{eq:sigmasaddle1}) is 
\bea
\sigma(\tau, M^2) = \frac{1}{\tau \sigma_0} \frac{d \sigma_{gg\rightarrow H}}{dM^2}
(\tau, M^2).
\eea
We concentrate on the $gg$ channel, which is dominant.

\begin{figure}[htbp]
\centering
   {\includegraphics[width=.49\columnwidth]{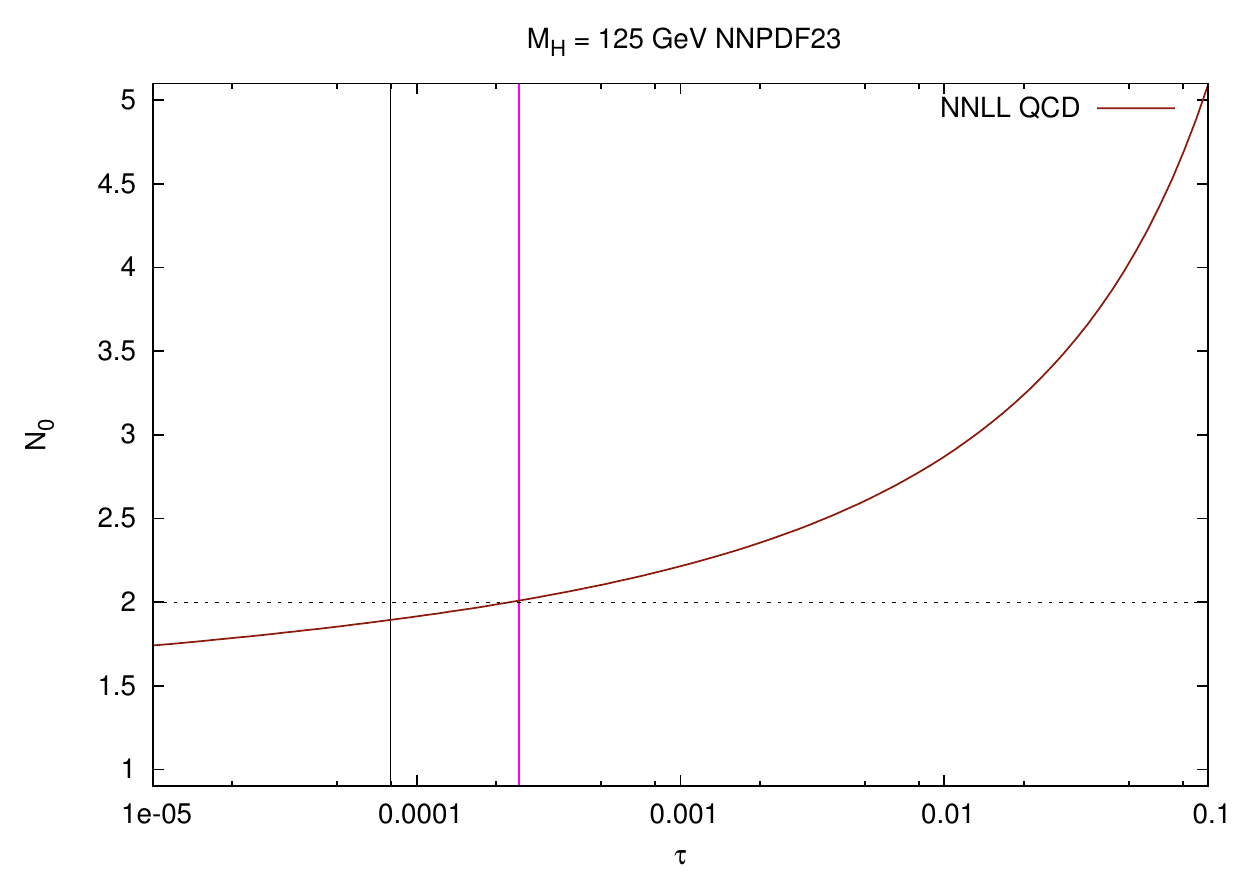}}
   {\includegraphics[width=.49\columnwidth]{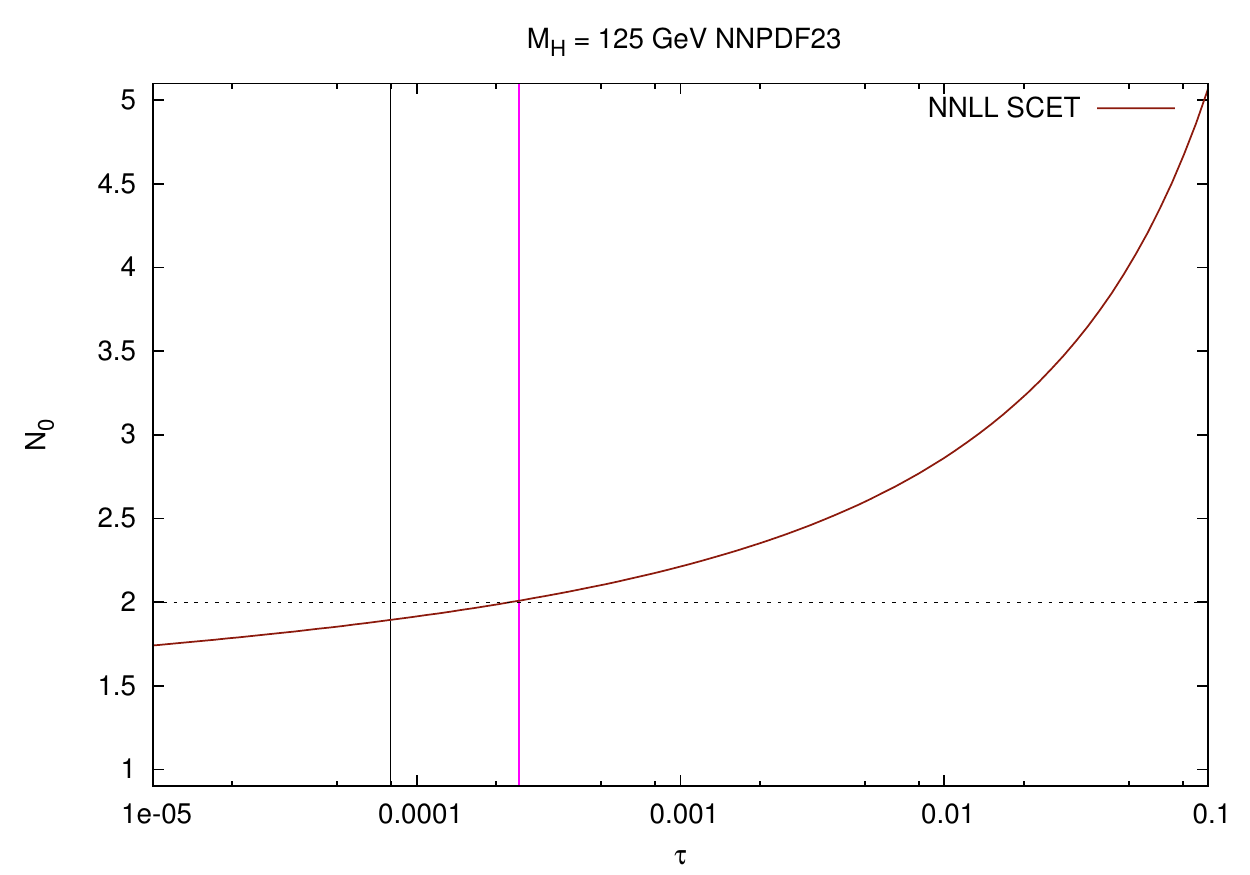}}
\caption{The position of the saddle-point $N_0$ for the Mellin inversion integral as a 
function of $\tau$ with $m_H=125$ GeV using NNPDF2.3 NNLO parton distributions.
Left: dQCD; right: SCET. The vertical black line marks the saddle point region when $\sqrt{s}= 14$ TeV
whereas the vertical purple line marks the saddle point region when $\sqrt{s}= 8$ TeV.
Left: dQCD; right: SCET.}\label{fig:saddlepoint}
\end{figure}

\begin{figure}[htbp]
\centering
   {\includegraphics[width=.49\columnwidth]{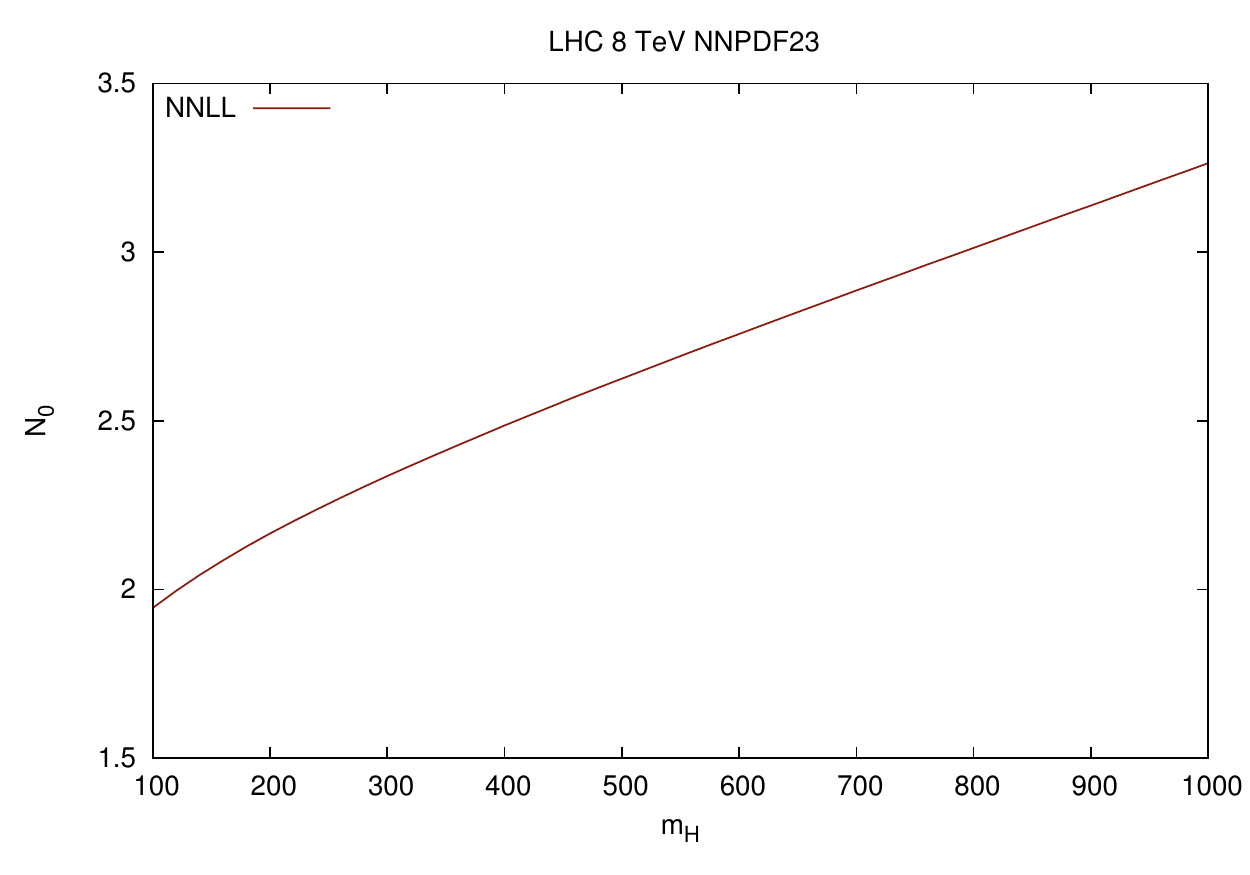}}
   {\includegraphics[width=.49\columnwidth]{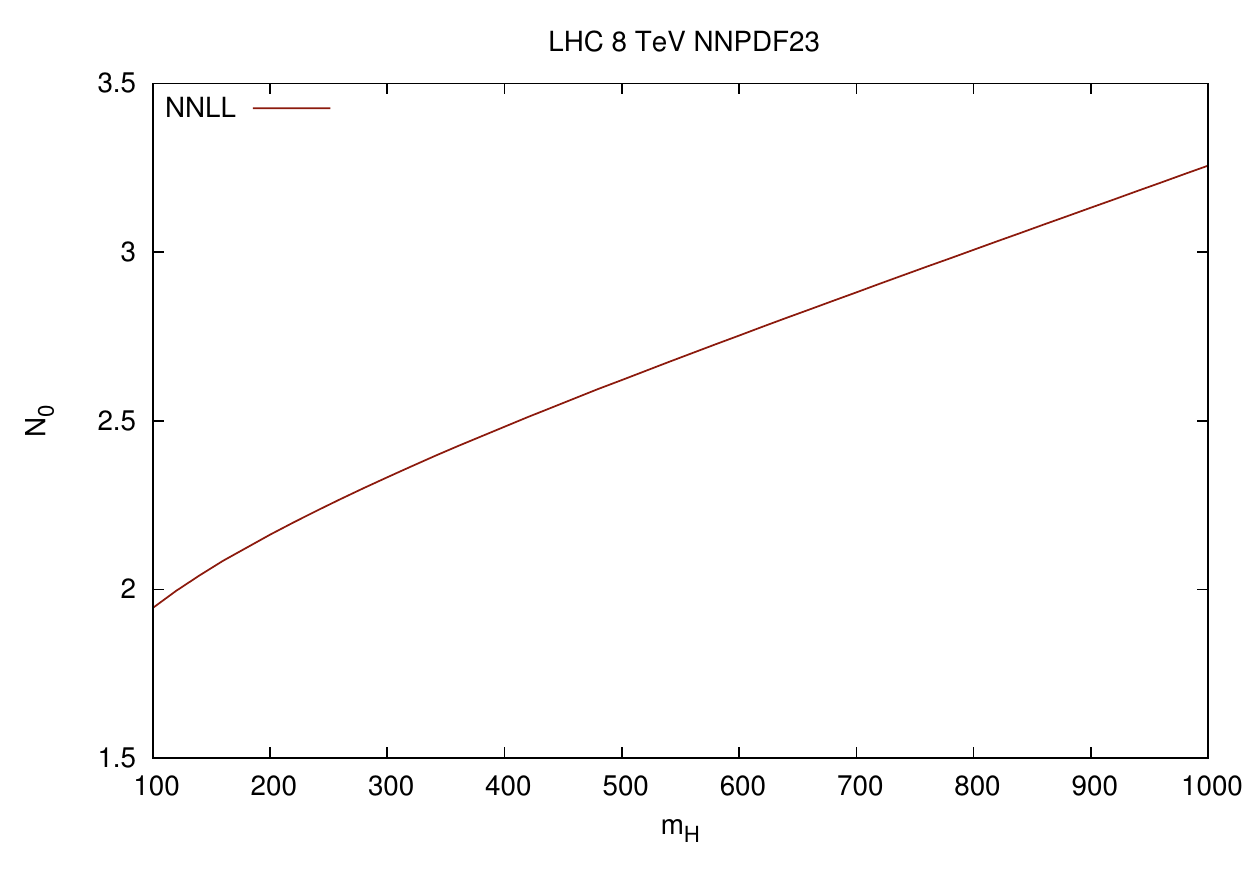}}
\caption{The position of the saddle-point $N_0$ for the Mellin inversion integral as a 
function of $m_H$ with $\sqrt{s}=8$ TeV using NNPDF2.3 NNLO parton distributions.
Left: dQCD; right: SCET.}\label{fig:saddlepointmu}
\end{figure}

We determine the position of the saddle-point $N_0$ for the production 
of a Higgs boson of invariant mass $m_H=125$ GeV in a $pp$ collider,
with a parton luminosity determined using NNPDF2.3 NNLO parton distributions
with $\as(m_Z) = 0.117$. 
In Fig.~\ref{fig:saddlepoint} we show the saddle point determined using the resummed coefficient
function Eq.~(\ref{eq:Cres}) in dQCD and Eq.~(\ref{eq:cfhiggsscet}) in SCET with
$\mu_h=\mu_t=\mu_f=m_H$ and choosing $\mu_s$ as the average value of the two scales $\mu_I$ and $\mu_{II}$
according to BN criteria. We will discuss below the consequences of such a choice.
The saddle point 
has been determined in both plots using the NNLL resummed coefficient.
The vertical black line
denotes the saddle-point region for a Higgs boson of $125$ GeV produced 
in the LHC 14 TeV kinematics, where $\tau \sim 0.0001$, whereas the
vertical purple line denotes the saddle-point region for a 
$125$ GeV Higgs boson produced in the LHC 8 TeV kinematics
where $\tau \sim 0.00025$.
We find that in these particular kinematics settings the saddle point is
close to the point $N=2$, in good agreement with what was found using
fixed-order partonic coefficient functions in Ref.~\cite{Bonvini:2012an}.
We show in Fig.~\ref{fig:saddlepointmu} the position of the saddle-point as a function of $m_H$ in the 
LHC 8 TeV kinematics.

\clearpage

\begin{figure}[htbp]
\centering
   {\includegraphics[width=0.75\columnwidth]{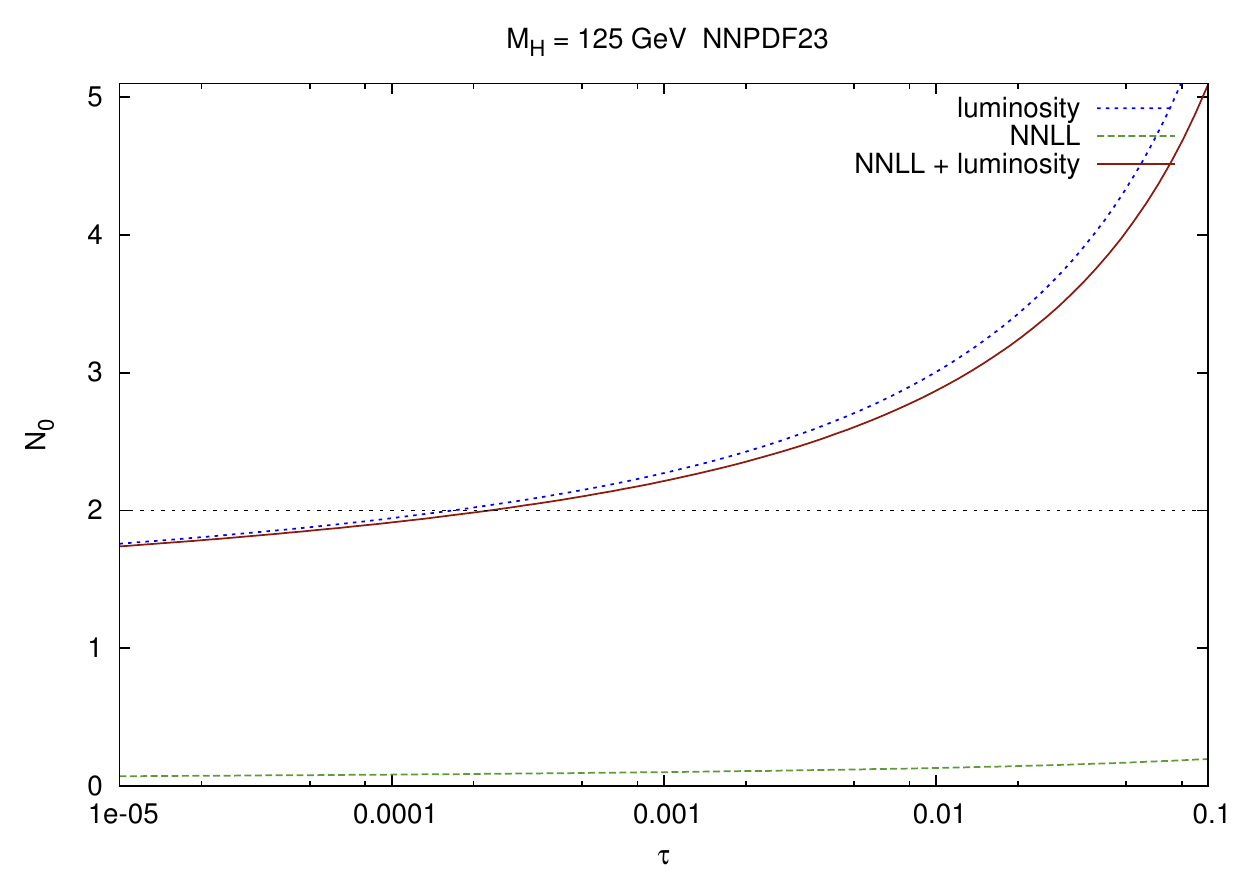}}
\caption{The position of the saddle-point $N_0$ for the Mellin inversion integral as a 
function of $\tau$ with $m_H=125$ GeV using NNPDF2.3 NNLO parton distributions.
Green dashed curve: position of the saddle point omitting the parton luminosity;
blue dashed curve: position of the saddle point omitting the partonic coefficient function.}\label{fig:saddlepointlum}
\end{figure}

In Fig.~\ref{fig:saddlepointlum} we show the position of the saddle obtained
omitting the parton luminosity and omitting the coefficient function, in case of dQCD.
As expected, the position of $N_0$ depends mostly on the
behaviour of the parton distribution functions.

\clearpage

\begin{figure}[htbp]
\centering
   {\includegraphics[width=0.75\columnwidth]{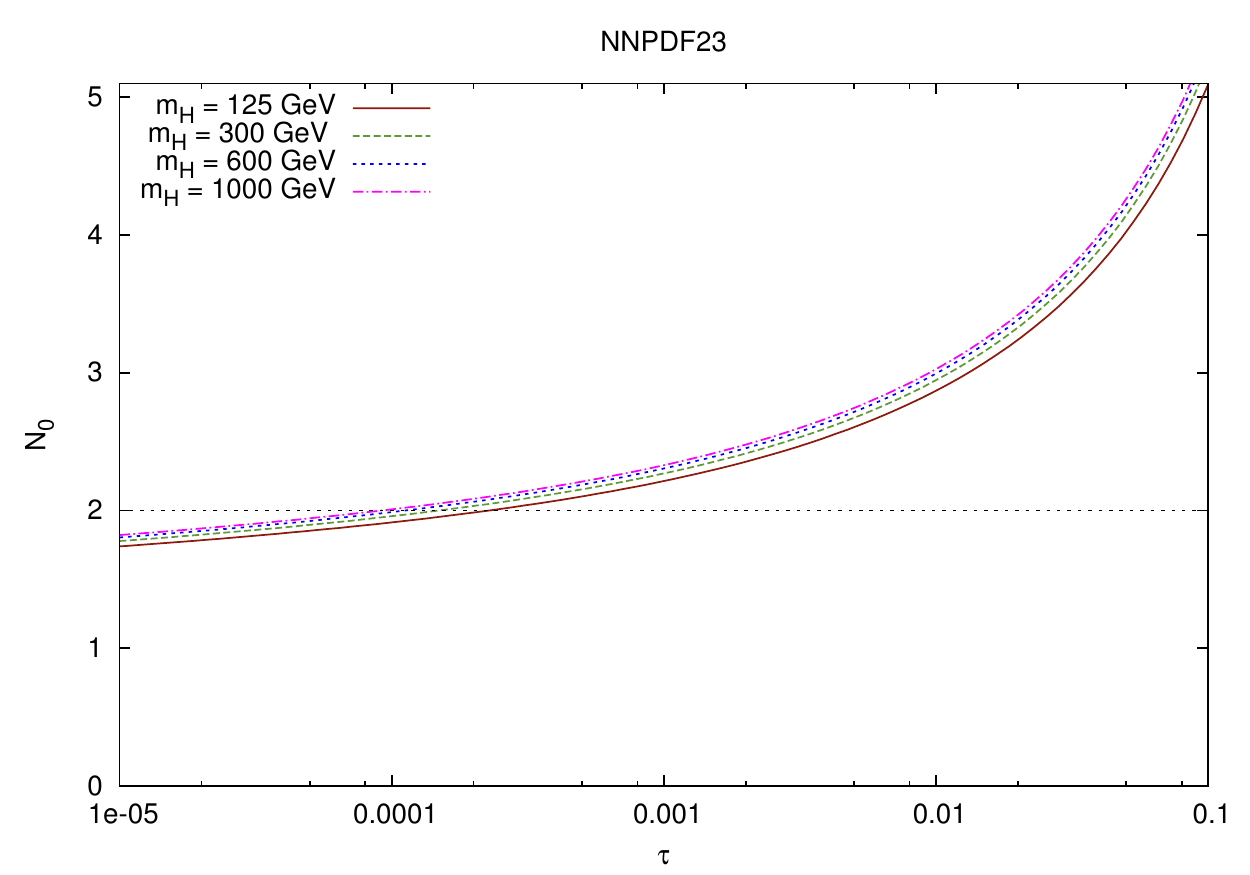}}
\caption{The position of the saddle-point $N_0$ for the Mellin inversion integral as a 
function of $\tau$ using NNPDF2.3 NNLO parton distributions,
with different values of $m_H$.}\label{fig:saddlepointmass}
\end{figure}

We note that the saddle-point depends weakly on the mass of the Higgs boson.
In particular, we show in Fig.~\ref{fig:saddlepointmass} the saddle-point as a function of $\tau$ 
with different values of the Higgs mass from 125 GeV to 1 TeV. The saddle-point 
has been calculated using the dQCD resummed coefficient with NNLL accuracy.

\begin{figure}[htbp]
\centering
   {\includegraphics[width=.75\columnwidth]{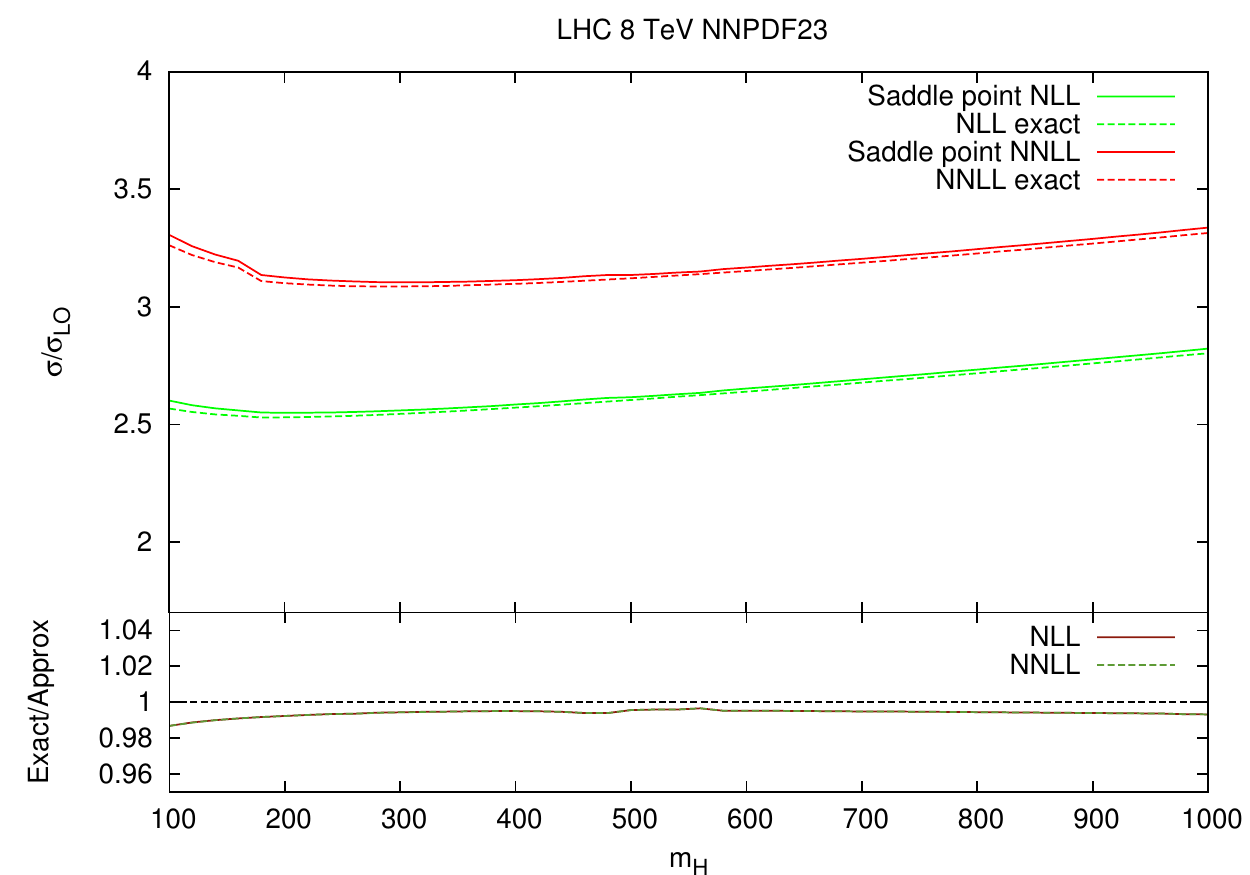}}
   {\includegraphics[width=.75\columnwidth]{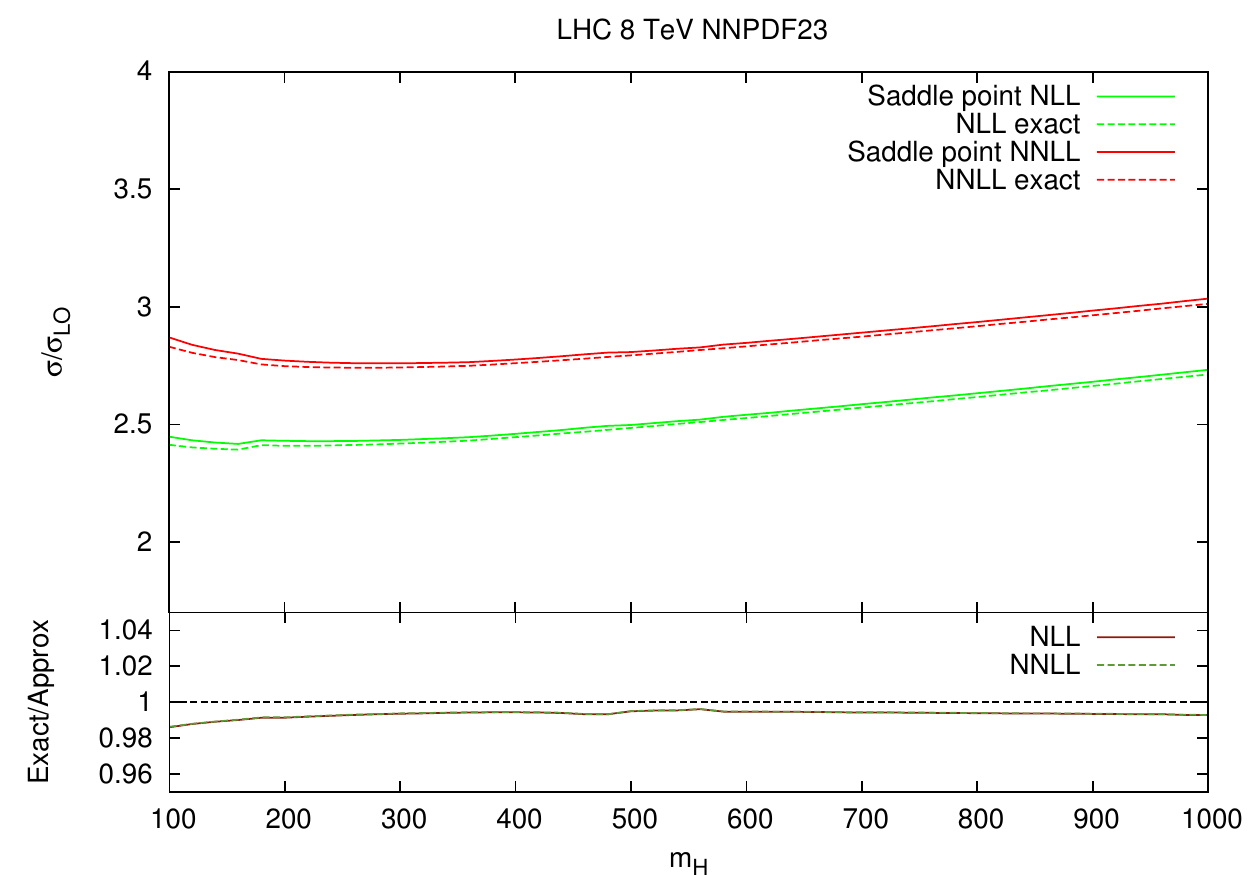}}
\caption{Upper plot: the exact $K$-factor for the resummed cross-section as a function of the
Higgs mass $m_H$ at NLL and NNLL compared with the $K$-factor calculated with the saddle point 
approximation Eq.~(\ref{eq:saddleappr}) using NNPDF2.3 NNLO parton distributions, with $\sqrt{s} = 8$ TeV. 
Small lower plots: ratio between the exact and the approximate $K$-factor. Up: dQCD;
low: SCET.
}\label{fig:saddlepointapprox}
\end{figure}

In Figs.~\ref{fig:saddlepointapprox} we compare the
results obtained for the resummed cross-section 
in dQCD and in SCET with the approximation Eq.~(\ref{eq:saddleappr}). 
We observe that thanks to the saddle-point argument it is possibile
to obtain an approximation of the resummed cross-section with accuracy at
the percent level on a wide mass range, both at NLL and NNLL.
We observe that the saddle-point approximation has the same level of
accuracy both for the SCET and dQCD; in both cases the value obtained
with Eq.~(\ref{eq:saddleappr}) slightly overestimates the exact value.

\clearpage

\subsection{Factorization violation in SCET}\label{sect:factviol}

In this Section we quantify the effect of violation of factorization in SCET
by comparing the position of the saddle point of the SCET result to the 
position of the saddle point in dQCD. Since we have seen that the position of
$N_0$ saddle depends very strongly on luminosity, we expect small effects. 

\begin{figure}[htbp]
\centering
   {\includegraphics[width=.75\columnwidth]{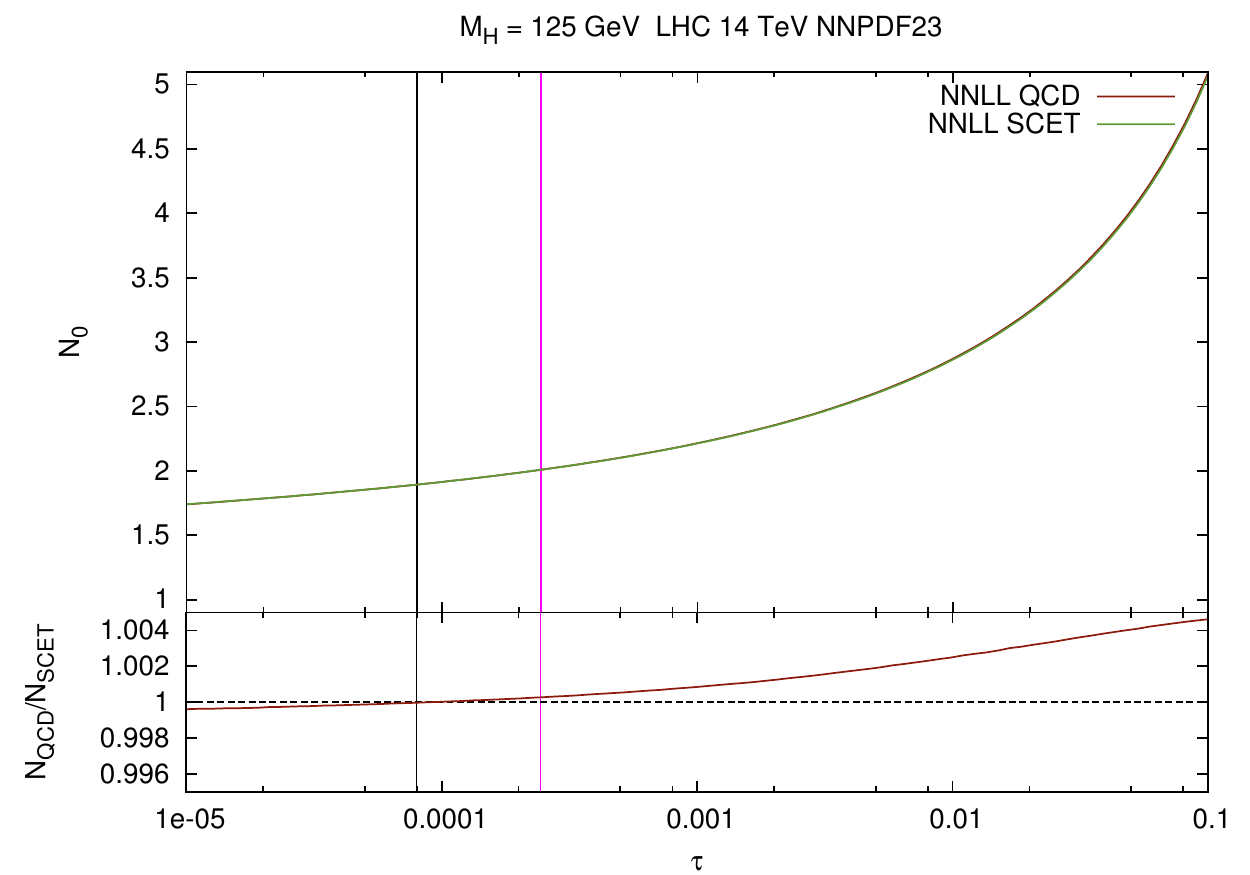}}
\caption{Upper plot: the position of the saddle-point $N_0$ for the Mellin inversion integral as a 
function of $\tau$ with $m_H=125$ GeV using NNPDF2.3 NNLO parton distributions
for dQCD and for SCET. Lower plot: ratio between the saddle point $N_0^\textrm{dQCD}$ and
$\bar N_0^\textrm{SCET}$.
The vertical black line marks the saddle point region when $\sqrt{s}= 14$ TeV
whereas the vertical purple line marks the saddle point region when $\sqrt{s}= 8$ TeV.
}\label{fig:saddlepointfact1}
\end{figure}

In Fig.~\ref{fig:saddlepointfact1} we show a comparison of the 
position of the saddle point for the Mellin inversion integral as a function 
of $\tau$ and a Higgs boson mass of $125$ GeV. We use the NNPDF2.3 NNLO PDF set and 
we compare the predictions obtained using the NNLL coefficient function. We observe 
that the saddle points differ by less than $1\permil$ for a large range of the variable
$\tau$ and reach the level of $1\permil$ only for values of $\tau$ much larger than
the interesting values at LHC. We have checked that the violation of factorization 
is negligible in a large $\tau$ range even for the production of a Higgs with a mass of the TeV order.

\begin{figure}[htbp]
\centering
   {\includegraphics[width=.75\columnwidth]{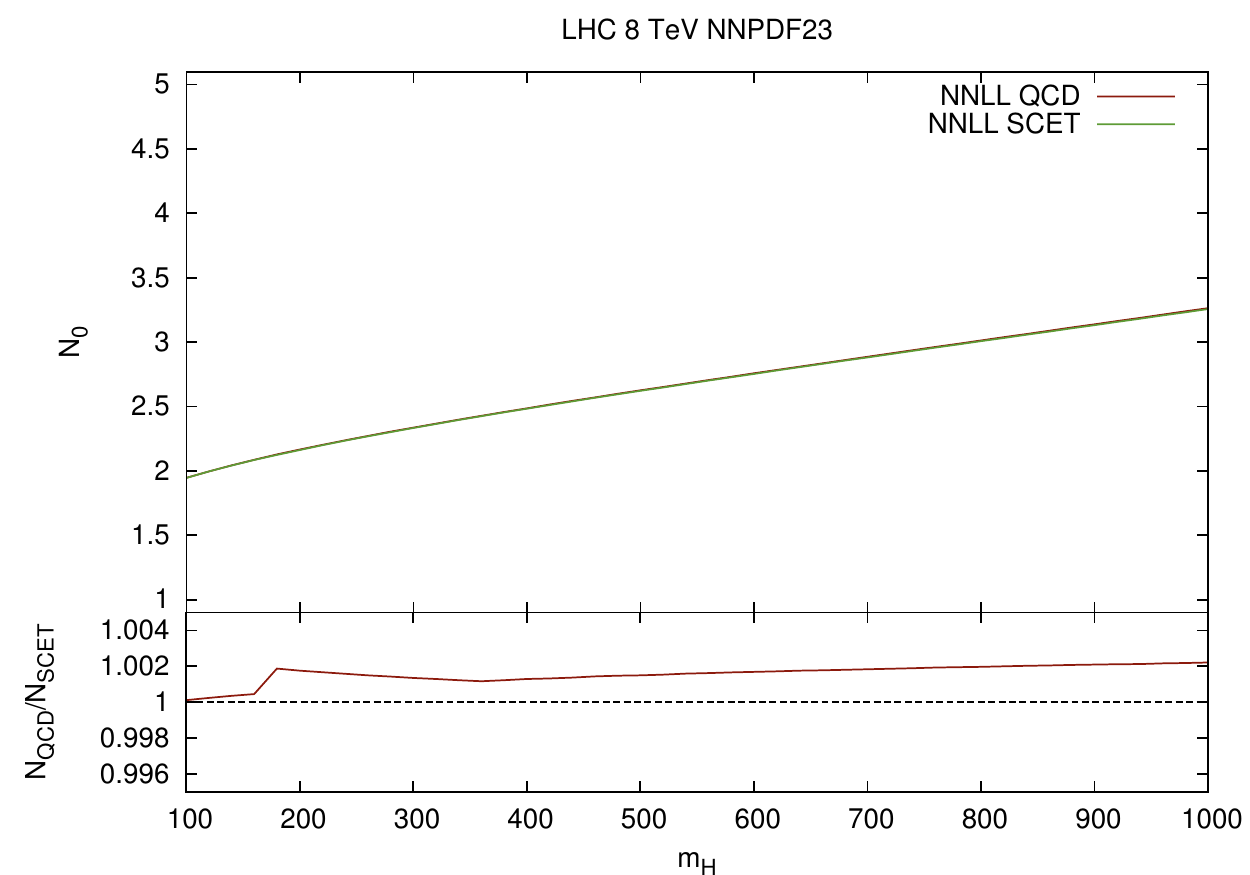}}
\caption{Upper plot: The position of the saddle-point $N_0$ for the Mellin inversion integral as a 
function of $m_H$ with $\sqrt{s}=8$ TeV using NNPDF2.3 NNLO parton distributions
for dQCD and for SCET. Lower plot: ratio between the saddle point $N_0^\textrm{dQCD}$ and
$\bar N_0^\textrm{SCET}$.}\label{fig:saddlepointfact2}
\end{figure}

In Fig.~\ref{fig:saddlepointfact2} we show the position of the saddle point for SCET and dQCD
as a function of the Higgs mass and with the center of mass energy equal to $8$ TeV. 
We observe that also in this case the differences between the values of the saddle 
point is at the per thousand level for all the mass range considered. 

We conclude that no significant difference exists between the position of the 
saddle point in the two approaches and that therefore the effects of the violation
of factorization in SCET are indeed small.

\clearpage

\section{Higgs resummed results: \texorpdfstring{$N$-space}{N-space} comparison}

In the previous Section we have found a way to 
perform a phenomenological comparison between SCET and dQCD by using N-space expressions
with hadronic variables. The study of $N$-space quantities can hence be 
used in order to obtain a better assesment of the differences between the two
approaches. We are particularly interested in the study of the ratio between
the dQCD resummed coefficient function and the SCET resummed coefficient function. 
We have observed that, thanks to the saddle-point argument, the difference between the
predictions for the total cross-section can be related in an extremely simple 
way to $C_r(N,M^2,\mu_s^2)$.

By comparing the values of $C_r(N,M^2,\mu_s^2)$ to the ratio of the
dQCD on the SCET cross-section we are therefore able on one hand to validate 
the saddle-point strategy and on the other hand to verify once more that the 
factorization violation is to all effects small. Moreover, we are able to quantify the 
consequences of the analytic manipulations of $C_r(N,M^2,\mu_s^2)$
and we are able to inspect the differences between SCET and dQCD at various levels. 

\begin{figure}[htbp]
\centering
   {\includegraphics[width=.49\columnwidth]{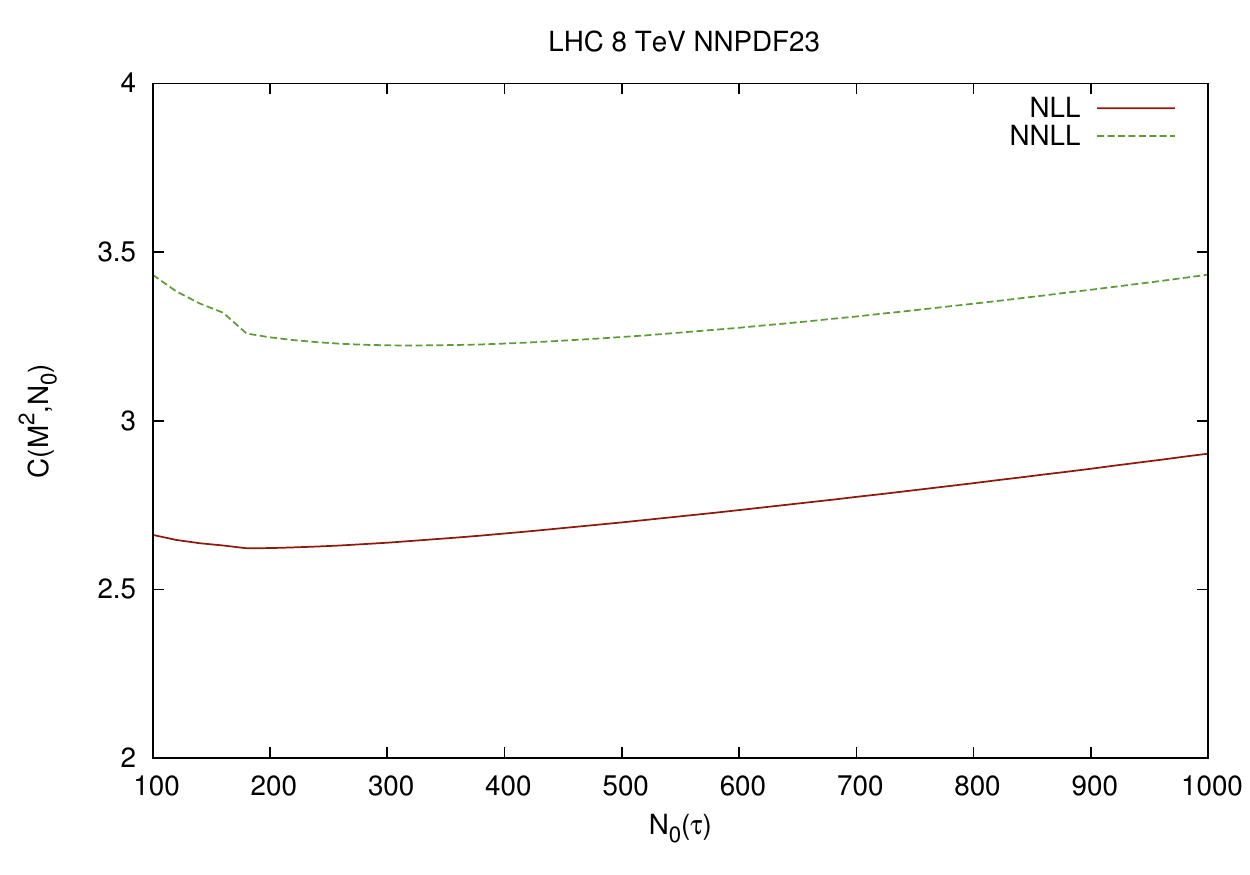}}
   {\includegraphics[width=.49\columnwidth]{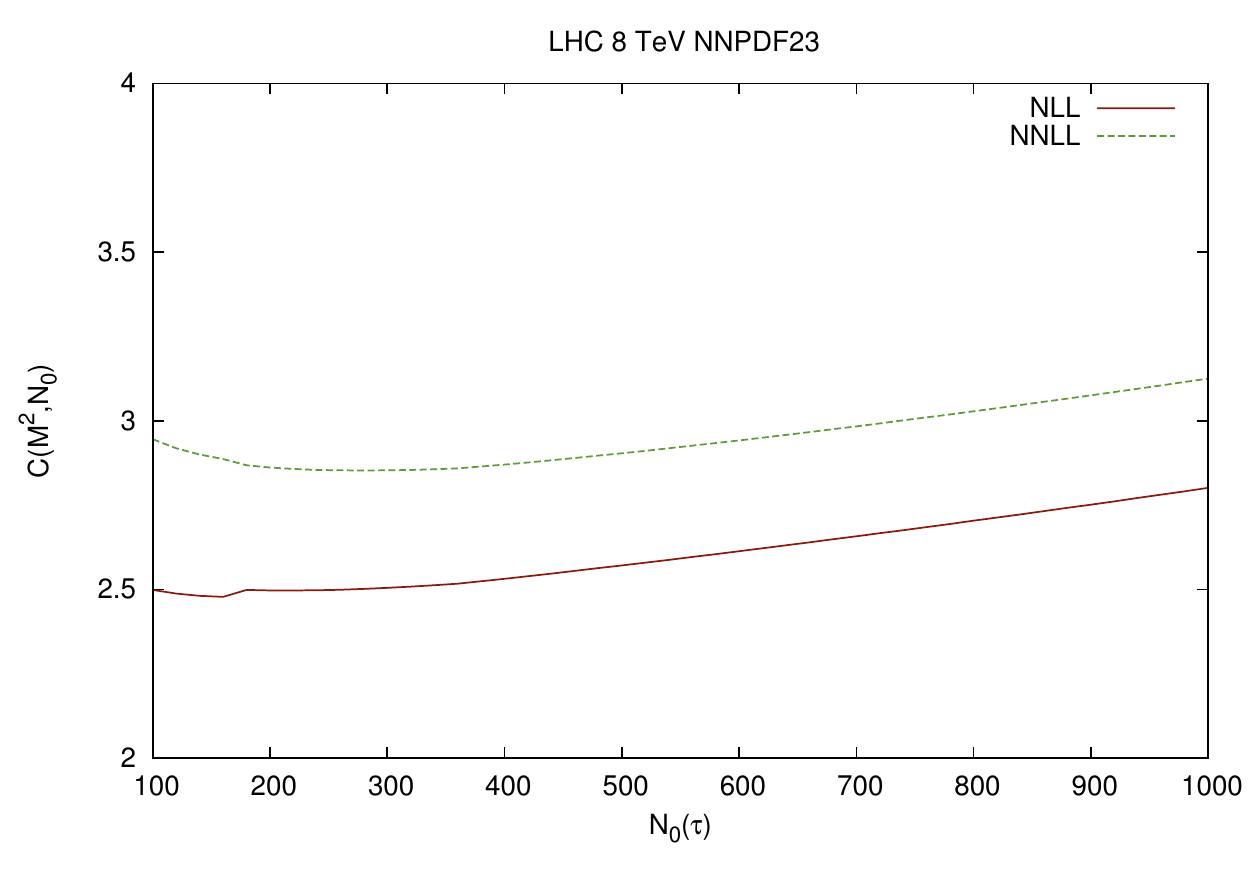}}
\caption{The partonic coefficient function evaluated in the saddle point 
$N_0$ as a function of the Higgs mass $m_H$ at NLL and NNLL accuracy. 
Left: dQCD; right: SCET.}\label{fig:parcoefffunn0}
\end{figure}

We show in Fig.~\ref{fig:parcoefffunn0} the partonic coefficients functions
both for SCET and for dQCD, evaluated at $N_0(\tau)$, as a function of the
Higgs mass $m_H$ at NLL and at NNLL accuracy.  We observe that both at NLL and at 
NNLL accuracy the dCQD coefficient functions are bigger than the SCET 
coefficient functions; however, we observe that whereas the difference is small
at NLL accuracy, it gets above the $10\%$ level at NNLL accuracy. For this numerical 
evaluation we have used the dQCD expression Eq.~(\ref{eq:Cres}) and the exact $N$-space Mellin transform of Eq.~(\ref{eq:cfhiggsscet}) 
with $\mu_t=\mu_h=\mu_f=m_H$ and $\mu_s$ equal to the average value of the two scales
$\mu_I$ and $\mu_{II}$, which are chosen according to the Becher-Neubert criteria.

\begin{figure}[htbp]
\centering
   {\includegraphics[width=.6\columnwidth]{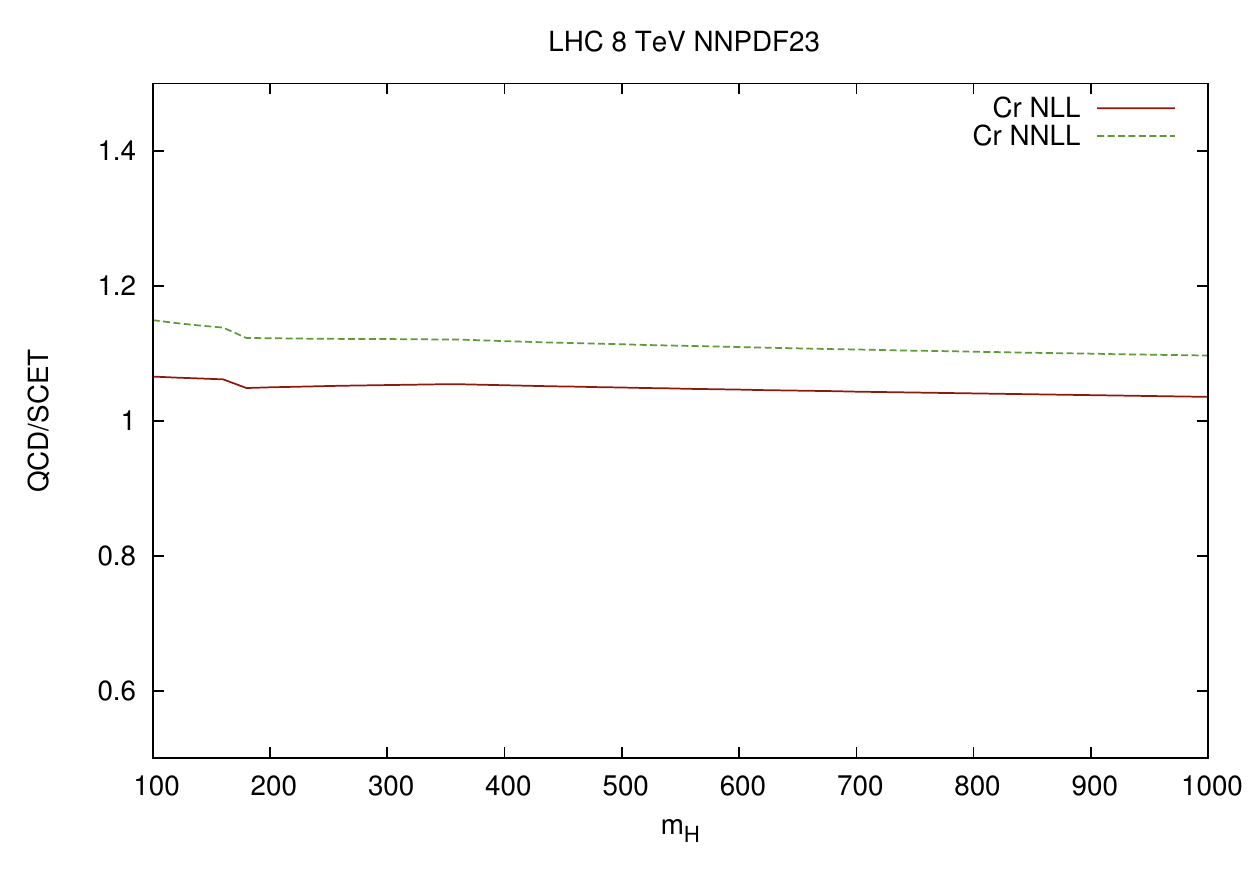}}
\caption{$C_r(N_0(\tau),M^2,\mu_s^2)$ as a function of the 
Higgs mass $m_H$ at NLL and NNLL accuracy.}
\label{fig:crhmass}
\end{figure}

We show in Fig.~\ref{fig:crhmass} the value of $C_r(N_0(\tau),M^2,\mu_s^2)$ as a function of the 
Higgs mass $m_H$ at NLL accuracy and at NNLL accuracy. The value of 
$C_r(N_0(\tau),M^2,\mu_s^2)$ is evaluated as the ratio of the partonic coefficient
function depicted in Fig.~\ref{fig:parcoefffunn0}. We will study how the value of $C_r$ is modified by the
manipulations of Chapter~\ref{chapter:ancomp} in Sect.~\ref{sect:cancomp}.
We observe that the value of $C_r$ is about 1.05 at NLL accuracy  and is 
about 1.15 at NNLL accuracy, i.e. the effect of resummation is higher in dQCD than in SCET.

\begin{figure}[h!]
\centering
   {\includegraphics[width=.6\columnwidth]{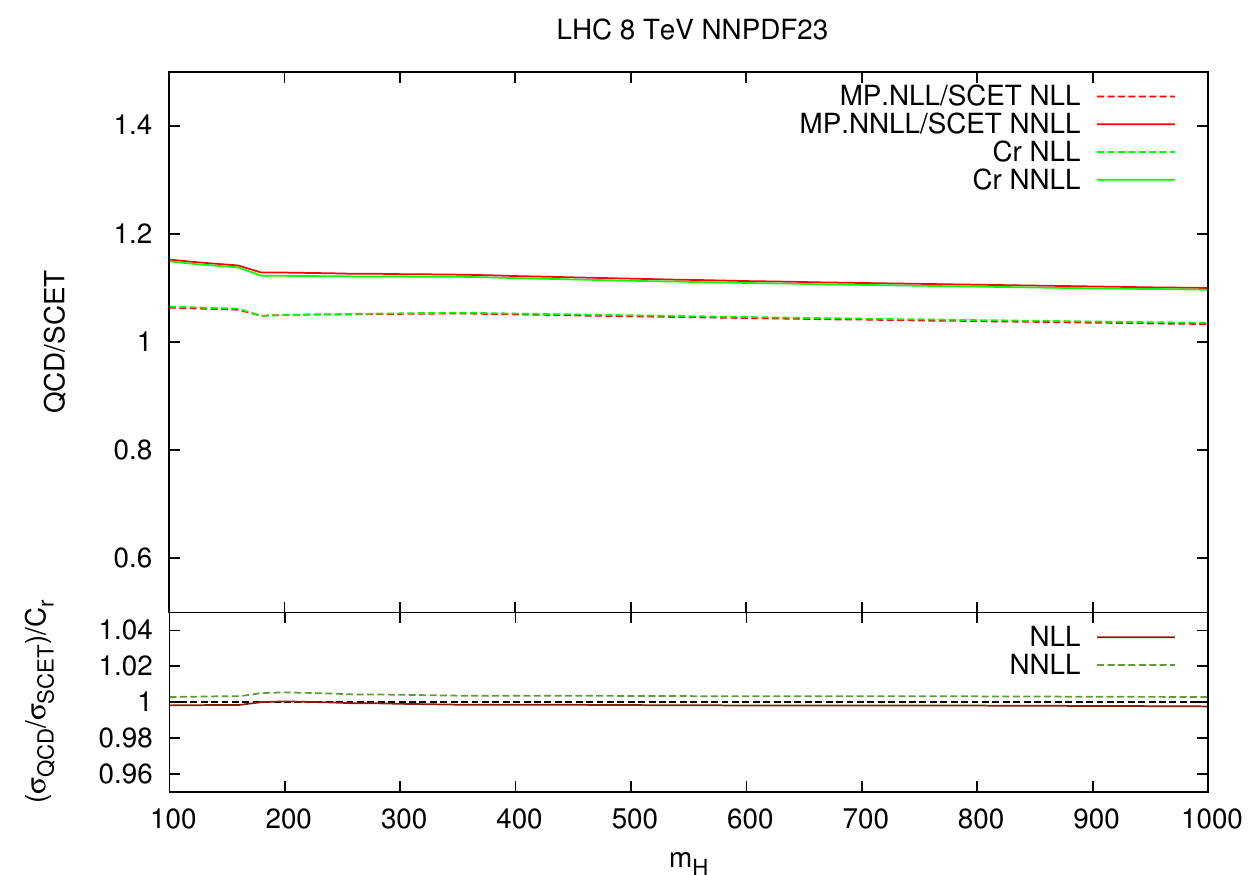}}
\caption{Upper plot: the ratio of the dQCD cross-section to the SCET cross-section 
compared with the value of $C_r(N_0(\tau),M^2,\mu_s^2)$ as a function of the 
Higgs mass $m_H$ at NLL accuracy and at NNLL accuracy. Lower plot:
ratio of the two curves at NLL accuracy and at NNLL accuracy.}
\label{fig:crsigmacomp}
\end{figure}

We show in Fig.~\ref{fig:crsigmacomp} the ratio of the dQCD cross-section to the SCET cross-section 
compared with the value of $C_r(N_0(\tau),M^2,\mu_s^2)$ as a function of the 
Higgs mass $m_H$ at NLL accuracy and at NNLL accuracy. In both cases we have 
chosen $\mu_s$ as the average value of the two soft scales proposed by 
Becher and Neubert in Ref.~\cite{Ahrens:2008nc}. 
In the lower plot we show the ratio of the two values. We observe that the 
results are in remarkable agreement both at NLL and at NNLL. 

\clearpage

\section{Detailed comparison}

In the previous Section we have compared the numerical predictions obtained in the 
SCET and in the dQCD approaches. We have observed that at NNLL accuracy the 
results differ at $15\%$ level and we have traced this difference to the
value of $C_r(N,M^2,\mu_s^2)$ evaluated at the saddle point $N_0$. Now we 
would like to understand the origin of this difference. This will be done in 
two steps.

First, we would like to concentrate on the analytic comparison of Chapter~\ref{chapter:ancomp}. 
After some analytical manipulations, we have found the final NNLL master dQCD-SCET comparison
formula Eq.~(\ref{eq:crfinal}). However, we would like to better understand the consequences of those
analytical manipulation. 

Second, we would like to investigate the phenomenological consequences of the 
BN scale choice. Since the soft scale $\mu_s$ determines what is being resummed, 
it is important to understand the effect of the factors which lead 
to values of $\mu_s$ far from the hard scale thus determining the resummation enhancement.
We address these problems in this Section.

\subsection{Connection with the analytical comparison}\label{sect:cancomp}

In Sect.~\ref{sect:mformulas} we have obtained a master formula which relates SCET and dQCD. In 
particular, we have defined the ratio $C_r(N,M^2,\mu_s^2)$ as
\bea
C_\textrm{dQCD} (N, M^2) = C_r (N, M^2, \mu_s^2) C_\textrm{SCET} (N, M^2, \mu_s^2).
\eea
We have shown in the previous Section the values of $C_r(N,M^2,\mu_s^2)$
computed as the exact ratio of the dQCD and the SCET expression at NNLL, without
any analytical manipulation, and we have found that the value of 
$C_r$ is about 1.15. 

However, all the master formulas of Chapter~\ref{chapter:ancomp}, which
relate the SCET and dQCD results, were obtained using the large-$N$ limit of 
the SCET expression. In particular, the exact Mellin transform of the soft function 
\bea
\mathcal M[S(z, M^2,\mu_s^2)] = \tilde s \left(\log \frac{M^2}{\mu_s^2} +\frac{\de}{\de \eta}, \mu_s \right)
\frac{\Gamma(N-\eta) \Gamma(2 \eta)}{\Gamma(N+\eta)}\frac{e^{-2 \gamma_E \eta}}{\Gamma(2 \eta)}
\eea
in the large-$N$ limit becomes
\bea
\mathcal M[S(z, M^2,\mu_s^2)] = 
\tilde s  \left(\log \frac{M^2}{\mu_s^2} +\frac{\de}{\de \eta}, \mu_s \right)
\bar N^{-2 \eta} + \mathcal O \left(\frac{1}{N} \right).
\eea 

We have seen in Sect.~\ref{sect:saddleres} that the value of the saddle point $N_0$ lies between 
2 and 3. Therefore one may ask whether the large-$N$ limit is a good approximation or not. 
We will see that the large-$N$ limit does indeed make a difference. 

We will finally show the value of $C_r(N,M^2,\mu_s^2)$  Eq.~(\ref{eq:crfinal}) , which is the 
final NNLL master dQCD-SCET comparison formula of Ref.~\cite{Bonvini:2013td}. This corresponds to the irreducible difference 
between the two approaches provided that SCET and dQCD are treated exactly in the 
same way up to NNNLL$^*$ terms. However, since SCET does not exponentiate all the
resummed terms, which are contained both in the evolution function $U$ and in
soft function $S$, we expect a difference between this last master formula and the
value of $C_r$ evaluated as the ratio of the dQCD resummed coefficient function Eq.~(\ref{eq:Cres})
and the Mellin transform of the SCET resummed coefficient function
Eq.~(\ref{eq:cfhiggsscet}) in the large-$N$ limit.

\begin{figure}[htbp]
\centering
	{\includegraphics[width=.55\columnwidth]{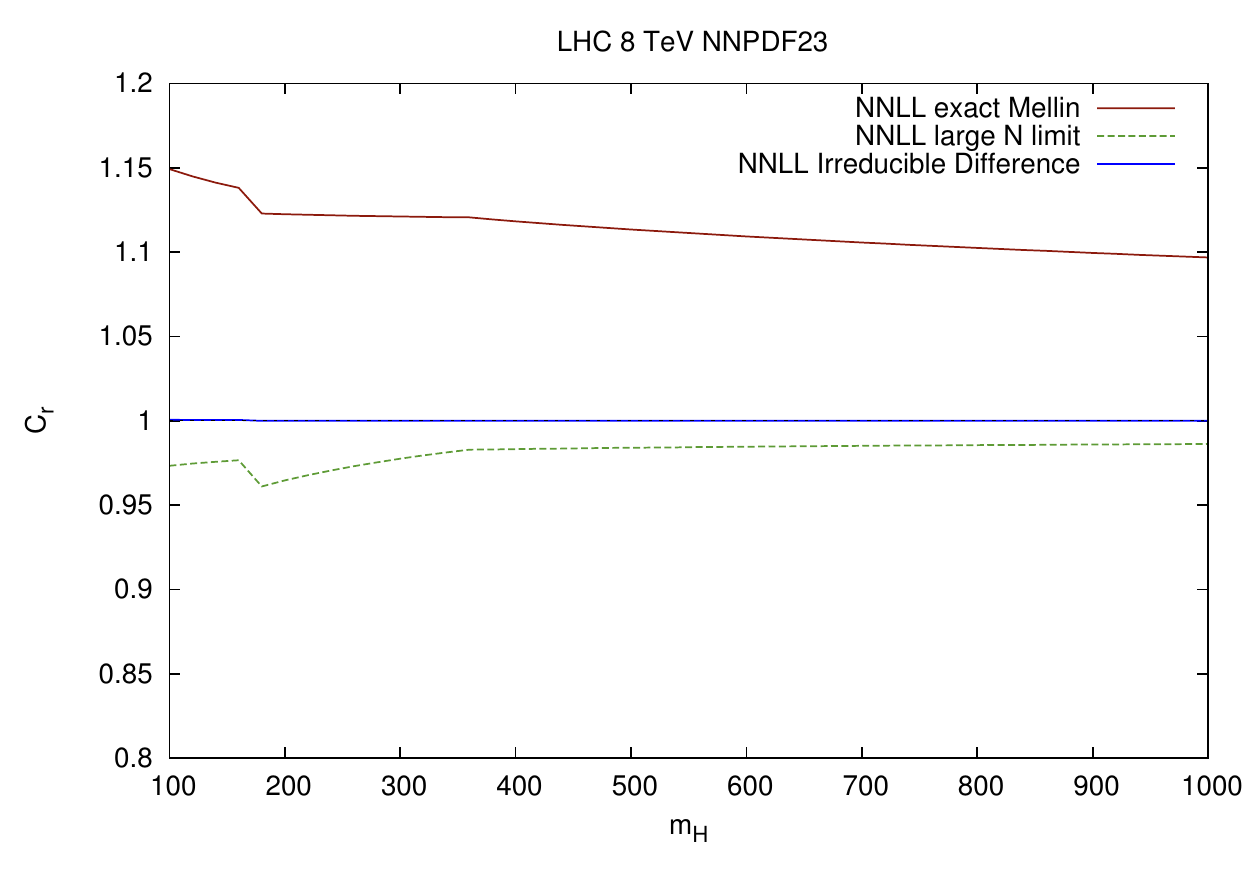}}
\caption{$C_r(N_0(\tau),M^2,\mu_s^2)$ as a function of the 
Higgs mass $m_H$ in different approximations.}
\label{fig:crallmh}
\end{figure}

\begin{figure}[htbp]
\centering
   {\includegraphics[width=.55\columnwidth]{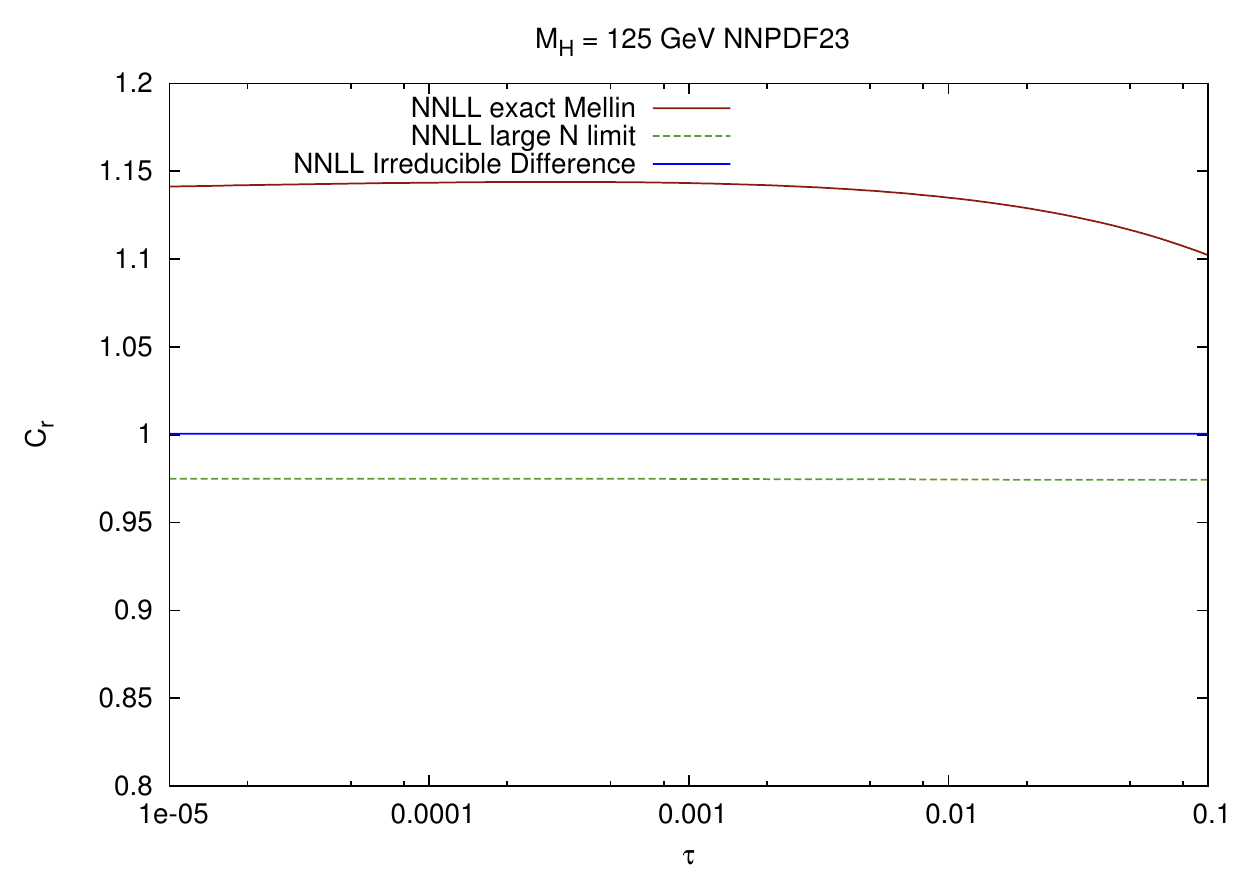}}
\caption{$C_r(N_0(\tau),M^2,\mu_s^2)$ as a function of $\tau$ in different approximations.}
\label{fig:cralltau}
\end{figure}

Our results are collected in Fig.~\ref{fig:crallmh} and in Fig.~\ref{fig:cralltau}. 
In particular, we show in Fig.~\ref{fig:crallmh}
the value of $C_r(N_0,M^2,\mu_s^2)$ as a function of the Higgs mass $m_H$ and center-of-mass energy of
$8$ TeV and in Fig.~\ref{fig:cralltau} the value of $C_r(N_0,M^2,\mu_s^2)$ as a function of $\tau$ with $m_H=125$ GeV.
We observe that in the large-$N$ limit the value of $C_r$ is about 0.98, which 
corresponds to a difference of almost $20\%$ with respect to the value of $C_r$ computed 
with the exact Mellin transform of the SCET coefficient function. The value of 
$C_r$ computed as Eq.~(\ref{eq:crfinal}) is almost equal to 1, the difference being well below $1\%$ percent. 
It appears from Fig.~\ref{fig:crallmh} that there is a small dependence on mass for the large-$N$ 
and for the final $C_r$, whereas the exact Mellin $C_r$ gets closer to 1 as the 
mass grows. This behaviour can be easily understood since the larger is the Higgs mass, 
the bigger is the saddle-point $N_0$, the closer is the value of the exact Mellin to the 
large-$N$ $C_r$. One observes the same dependence, this time on $\tau$, in Fig.~\ref{fig:cralltau}.

We have seen that the main difference between SCET and dQCD is due to the 
fact that the two approaches treat differently the subleading terms. 
If we move back to $z$ space, we can compare the soft function $S$ in the
original formalism, namely
\bea
S(z, M^2, \mu_s^2) = \tilde s
\left(\log\frac{M^2}{\mu_s^2}+\frac{\partial}{\partial\eta},\mu_s\right)
\frac{1}{1-z} \left(\frac{1-z}{\sqrt{z}}\right)^{2 \eta} \frac{e^{-2 \gamma_E \eta}}{\Gamma(2 \eta)}
\eea
to the soft function in the large $N$ limit
\bea
S(z, M^2, \mu_s^2) = \tilde s
\left(\log\frac{M^2}{\mu_s^2}+\frac{\partial}{\partial\eta},\mu_s\right)
(-\log z)^{-1+2 \eta} \frac{e^{-2 \gamma_E \eta}}{\Gamma(2 \eta)}.
\eea
In particular, we observe that in the first case the action of the $\eta$ derivative 
produces logs of $z$, whereas in the second case it produces logs of $\log(z)$. Note that
\bea
(-\log z)^{-1+2 \eta}= \sqrt{z} \frac{z^{-\eta}}{(1-z)^{1-2 \eta}} \left[ 1 +\mathcal O \left(( 1-z)^2\right)\right].
\eea
Therefore, different classes of logs are resummed. 
These choices are equivalent from the point of view of resummation, as they differ by
subleading terms. Nevertheless, the difference in their effects on the prediction is not negligible. 

\begin{figure}[htbp]
\centering
	{\includegraphics[width=.6\columnwidth]{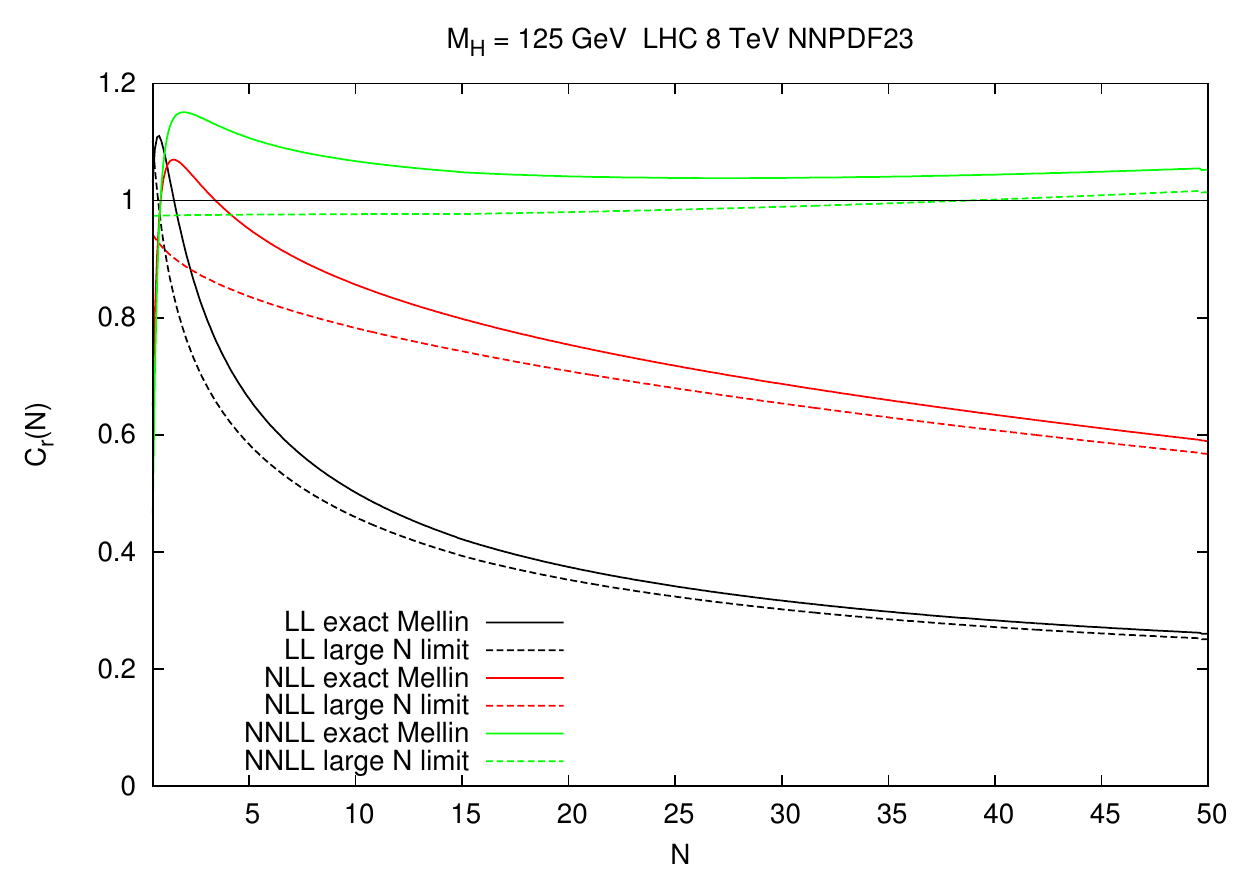}}
\caption{Exact $C_r(N,M^2,\mu_s^2=M^2/\bar N^2)$ and large-$N$ $C_r(N,M^2,\mu_s^2=M^2/\bar N^2)$
as a function of $N$ at LL, NLL and NNLL accuracy. }
\label{fig:crn}
\end{figure}

In order to assess the impact of the subleading terms, we show in Fig.~\ref{fig:crn} the value of $C_r(N,M^2,\mu_s=M/\bar N)$,
computed as the exact ratio of the dQCD and the SCET expression, without
any analytical manipulation, at LL, NLL, NNLL, compared to the value of 
$C_r(N,M^2,\mu_s=M/\bar N)$ in the large-$N$ limit at the same logarithmic accuracy, as a function of $N$.
We note that the result is close to 1 only at NNLL, where the effect of the
subleading terms is smaller. However, we note that for small $N$ the 
difference between curves with different logarithmic accuracy is comparable, or even smaller,
than the difference between the curves evaluated at the same logarithmic accuracy 
but which differ by $\mathcal O(1/N)$ terms.

\subsection{The Becher-Neubert scale choice}

The Becher-Neubert scale choice is
\bea
\mu_s = M(1-\tau)g(\tau)
\eea
where the function $g(\tau)$ is chosen in order to 
minimize the contributions of $\tilde s$ to the cross-section according to 
the two criteria discussed in Sect.~\ref{sect:bnscale}. If $\tau$ is small, as for 
Higgs production at the LHC, the naive choice 
$\mu_s = M(1-\tau)$ would not enhance the cross-section, by construction 
of SCET. 
The value of the 
factors hence determines the impact of 
soft-gluon resummation in the SCET formalism by rescaling the soft scale to a value which is 
smaller than the hard scale $M$. 
In this Section we inspect the phenomenological
consequences of such a choice. 

\begin{figure}[htbp]
\centering
   {\includegraphics[width=.6\columnwidth]{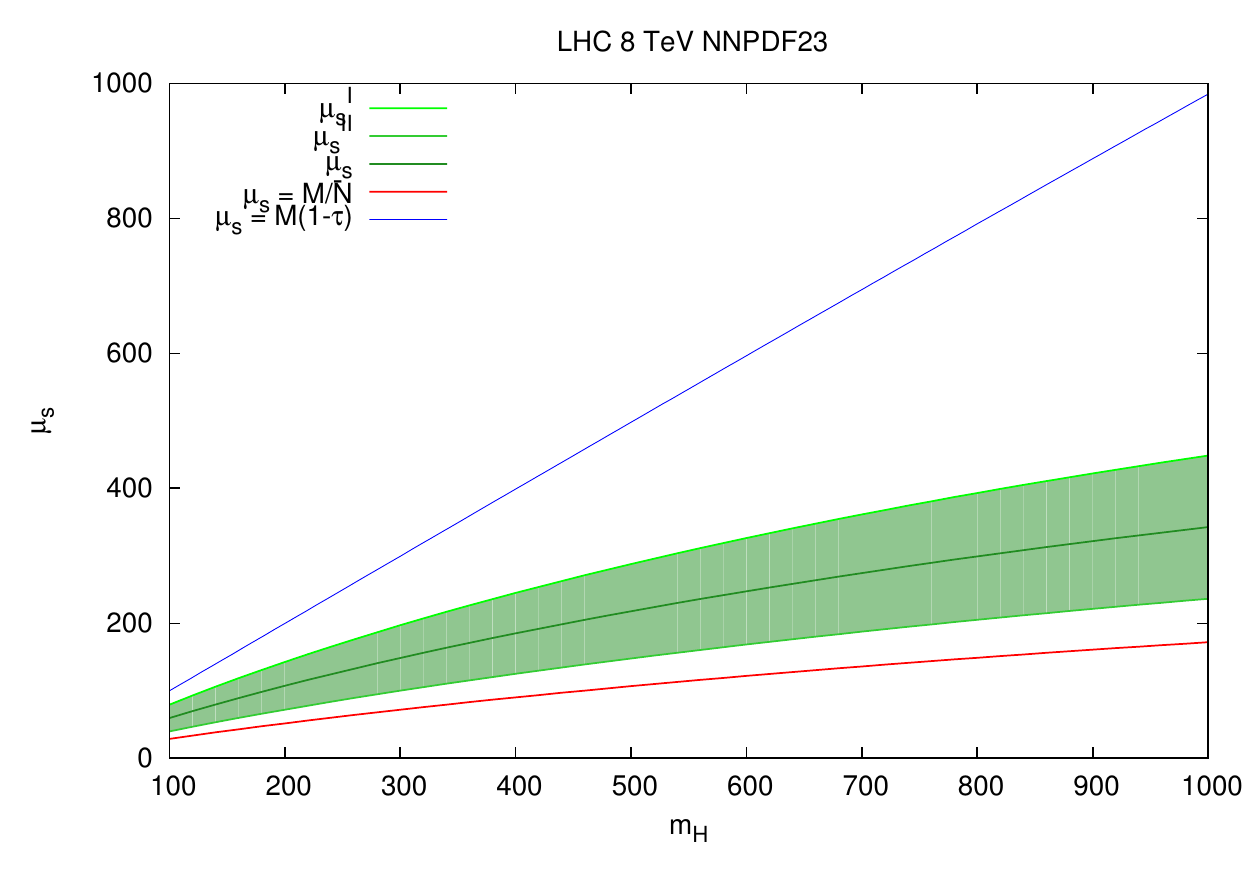}}
\caption{Values of different choices of the soft scale $\mu_s$
as a function of the Higgs mass $m_H$, with centre-of-mass energy of 8 TeV.}
\label{fig:muvar1}
\end{figure}

\begin{figure}[htbp]
\centering
   {\includegraphics[width=.6\columnwidth]{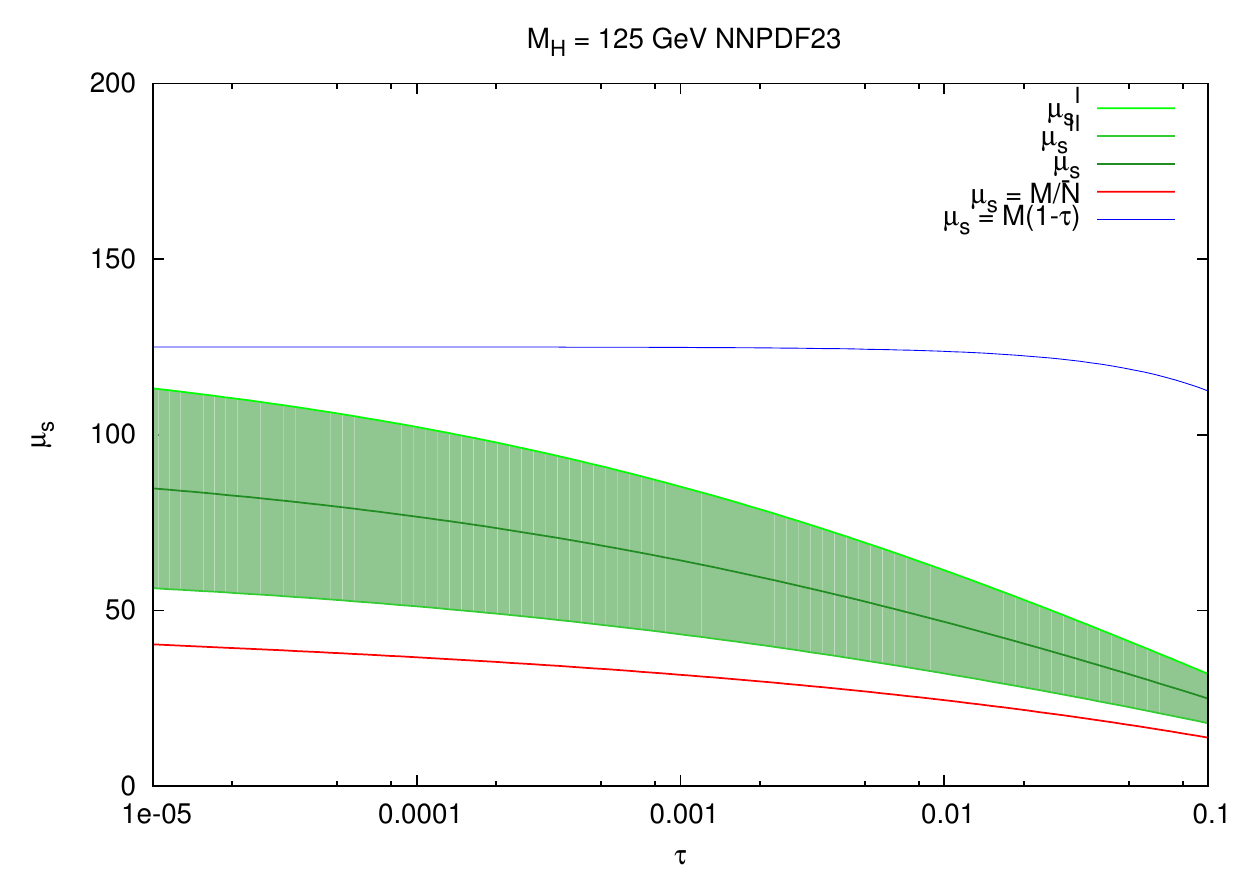}}
\caption{Values of different choices of the soft scale $\mu_s$
as a function of $\tau$, with $m_H=125$ GeV.}
\label{fig:muvar2}
\end{figure}

We know that with the choice $\mu_s=M/\bar N$ the SCET results in the large $N$ limit reproduce
the dQCD results. Therefore it is natural to compare the BN soft scale with the 
scale $\mu_s=M/\bar N_0$. In Fig.~\ref{fig:muvar1} we show the values of different choices of the soft scale $\mu_s$
as a function of the Higgs mass $m_H$, with centre-of-mass energy of 8 TeV. We observe
that the soft scale is almost equal to the Higgs mass $m_H$ if the naive choice 
$\mu_s=m_H$ is made. However, if $\mu_s$ is chosen according to the two criteria
proposed by Becher and Neubert, we find that the value of the soft scale gets 
closer to the value $\mu_s=M/\bar N_0$ depicted in red in Fig.~\ref{fig:muvar1}. 
We show in Fig.~\ref{fig:muvar2} the values of the same choices of the soft scale as a function of $\tau$ 
and with $m_H=125$ GeV. We observe that the naive choice leads to a almost constant value 
of $\mu_s$ except for large values of $\tau$. The Becher-Neubert choice on the other 
hand lies closer to the choice $\mu_s=M/\bar N_0$.

\begin{figure}[htbp]
\centering
   {\includegraphics[width=.6\columnwidth]{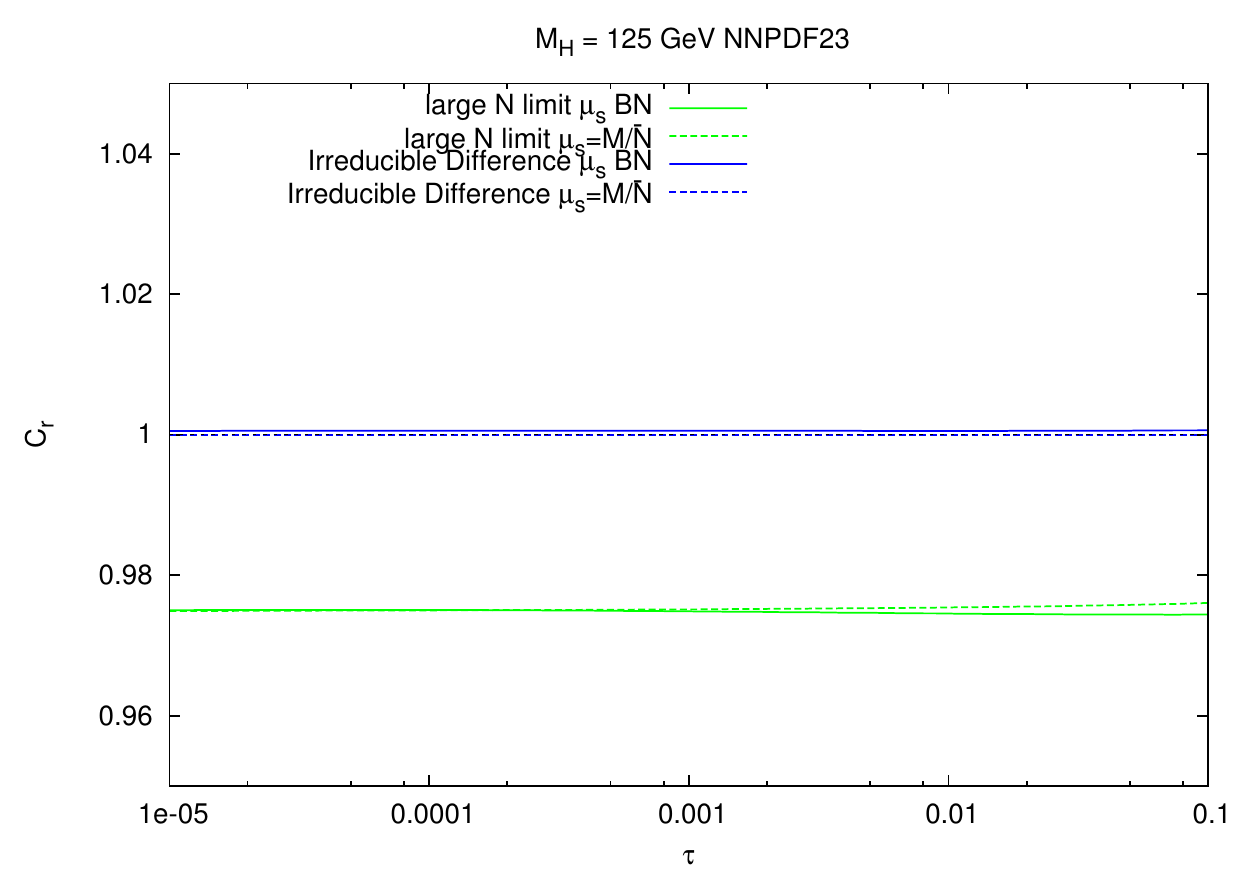}}
   \caption{Value of the large-$N$ limit $C_r(N_0(\tau),M^2,\mu_s^2) $ 
and of the final $C_r(N_0(\tau),M^2,\mu_s^2) $ computed with the BN choice of $\mu_s$
and with the choice $\mu_s=M/\bar N_0$ as a function of $\tau$.}
   \label{fig:muvar3}
\end{figure}

We expect therefore a small difference between the results obtained with the 
BN scale choice and the results obtained with the choice $\mu_s=M/\bar N_0$. 
In Fig.~\ref{fig:muvar3} we compare the value of the large-$N$ limit $C_r(N_0(\tau),M^2,\mu_s^2) $ 
and of the final $C_r(N_0(\tau),M^2,\mu_s^2) $ computed with the BN choice of $\mu_s$
and with the choice $\mu_s=M/\bar N_0$ as a function of $\tau$. We show results at NNLL accuracy 
evaluated with $m_H=125$ GeV. We observe that the difference is indeed very small. 

It is natural to ask why the BN choice is not dissimilar from the choice  $\mu_s=M/\bar N_0$.
A simple qualitative argument provides the answer to this question.
The Mellin transform of the soft function reads
\bea\label{eq:stildeoneloop}
\mathcal M[S(z, M^2, \mu_s^2)] =
\left[ 1 +\frac{C_A}{2 \pi} \as(\mu_s^2) \left( \log^2 \frac{M^2}{\mu_s^2 \bar N^2}
+\frac{\pi^2}{6}\right) \right] \bar N^{-2 \eta} + \mathcal O \left( \frac{1}{N} \right).
\eea
According to the BN criteria, $\mu_s$ is chosen in order to minimize the contributions of
$\tilde s$ to the cross-section. Since the integral is very well approximated by its value
in the saddle point, we conclude by looking at Eq.~(\ref{eq:stildeoneloop}) that a value of $\mu_s\sim M/\bar N_0$
does indeed minimize the first loop collection of $\tilde s$. Some difference 
persists because of the shift induced by the constant $\frac{\pi^2}{6}$ 
and, more subtly, by the fact that in the SCET approach one has to consider
the exact Mellin of the SCET moment-space coefficient function.

\subsection{Summary and outlook}

In Sect.~\ref{sect:factviol} we have shown that the factorization violation effects is SCET are to all 
practical purposes negligible for a large range of $\tau$. Thanks to 
a saddle point argument, we have hence been able to relate the difference between 
the SCET and dQCD predictions to the value of $C_r(N, M^2, \mu_s^2)$ in the saddle point $N_0$.
We have therefore 
performed a detailed comparison of soft gluon resummation in SCET and in dQCD. 

We have first 
concentrated on the analytical structure of the SCET and the dQCD result and we have 
assessed the impact of the subleading terms with the choice $\mu_s = M/\bar N$,
with which SCET reproduces the standard dQCD results. In particular, we have shown that 
for small $N$ the difference between the curves evaluated at the same logarithmic accuracy,
but which differ by $\mathcal O(1/N)$ terms (which we will call subdominant terms), is significant,
and is greater than the difference between predictions at different logarithmic accuracy.
The situation changes as $N$ grows, where the effect of the subleading terms is
greater than the effect of the subdominant terms up to NNLL accuracy. 

However, in SCET one resums logs of $\mu_s$ which is chosen according to BN criteria.
We have therefore compared the BN scale choice to the choice $\mu_s = M/\bar N$.
We have seen that the difference is small and that the BN choice is similar to the
choice $\mu_s=M/\bar N_0$. However, we would like 
to obtain a better understanding of the factors which reduce the value of the soft 
scale $\mu_s$. In particular, we would like to understand how the choice of $\mu_s$
according to BN criteria changes as the mass grows. In this way the value of the
saddle point would be in the large-$N$ region, which we have not studied in detail.

In conclusion, we have found that the main difference between SCET and dQCD resummed result
is due to subdominant terms. Therefore a significant ambiguity in the resummation 
procedure exists. One may conclude therefore that resummation cannot be used. 
However, a possible way out of the ambiguity due to the choice of the subleading terms consists in
the comparison of the results which differ by vanishing terms in the large-$N$ limit
to the fixed order result.  We mean to further clarify this issues
with further analytical analysis as a prosecution of this Thesis.

\cleardoublepage
\phantomsection
\addcontentsline{toc}{chapter}{Conclusions}
\chapter*{Conclusions}
\markboth{Conclusions}{}

\fancyhf{}
\fancyhead[RE]{\nouppercase{\textbf{\textsf{\leftmark}}}}
\fancyhead[LE,RO]{\bfseries\thepage}
\renewcommand{\refname}{Conclusions} % Al posto di Referenze puoi mettere quello che preferisci
								   % viene cambiato il nome solamente nel paragrafo, ma non
								   % nell'indice. 
  % questo comando aggiunge all'indice la sezione 
										  % della bibliografia

In this Thesis we have presented a systematic comparison of 
soft-gluon resummation in the more standard dQCD approach and in a SCET approach. 
We have focused on Higgs boson production in gluon-gluon fusion.
We now summarize our main results.

From the theoretical point of view, we have verified that the results of
Refs.~\cite{Bonvini:2012az,Bonvini:2013td} can be straightforwardly extended to Higgs boson production in gluon-gluon fusion.
We have then tackled the challenge of performing a phenomenological comparison
between the dQCD and the SCET resummed result on various levels.

We have developed a strategy which allows for a comparison
between the SCET and dQCD expressions by means of partonic expressions, for which 
an analytical comparison can be performed in an easier way, despite the fact that 
the SCET result depend on hadronic variables. 
This comparison can be performed provided the fact that the effect of violation 
of factorization in SCET is small. Our results have indeed confirmed that
this effect is negligible. Furthermore, our strategy allows to obtain 
resummed cross-section with a minimal amount of computational work and with 
very good accuracy. 

We have reduced the difference between the SCET and the dQCD resummed results
to the value of the ratio between the dQCD and the SCET resummed coefficient function
$C_r(N,M^2,\mu_s^2)$ evaluated in the saddle point $N_0(\tau)$ of the 
Mellin inversion integral. We have shown how the value of $C_r$ depends
on various approximations and we have studied the resulting effects on the 
SCET and the dQCD predictions.
In particular, we have concluded that the main difference
between the two approaches is related to subleading terms in $N$ space.
However, we have studied a process for which resummation is
perturbative and for which the effect of these subleading terms has
a considerable phenomenological effect. 

We have investigated the consequences of the Becher-Neubert scale choice and we have
performed a comparison between this scale choice and the scale $M/\bar N$ which 
is the scale which is naturally resummed in dQCD. By a qualitative analisys, we
have understood how the BN scale choice is similar to 
the saddle point-driven scale $M/\bar N_0$.

In our work we have inspected some issues which we would like to
further investigate. In particular, we have observed that SCET and dQCD
resummation formulae agree in the large-$N$ or $x\rightarrow 1$ soft limit, but 
disagree by subleading terms. Therefore, there are two possibilities: either resummation
cannot be used because of these large ambiguities, or else a comparison with the 
fixed order result could be a possible way to resolve the ambiguity in the 
resummation procedure. A deeper analytical insight
is needed in order to obtain a more complete
comprehension of the resummation procedure in the two formalisms. 
To this purpose, it would be worthwhile to perform a more detailed study of 
the ambiguities of the resummation by using different prescriptions in 
the traditional dQCD approach. The Borel 
resummation prescription could therefore result particularly useful, 
in particular as a mean to asses the different treatments of the subleading terms.

Finally, it would be interesting to perform an analogous systematic comparison
of dQCD and SCET predictions focusing on Drell-Yan process in hadron-hadron collider.
It would be then possible to discuss the relevance of resummation
in SCET and dQCD for the production of light mass states ($W,\ Z$ production)
and for the production of heavy dileptons, with masses in the TeV region.

\appendix
\chapter{Analytical expressions}\label{app:analy}

In this Appendix we collect various analytical expressions which complete the 
discussion of the resummation procedure.

\section{Soft-gluon resummation formulae in dQCD}

The general structure of the resummed coefficient function for Drell-Yan and Higgs
production is 
\bea
C^{\textrm{res}}(N, M^2) &= g_0(\alpha_s) \exp \mathcal S \left(\bar \alpha
L, \bar \alpha \right) \\
\bar \alpha &\equiv 2 \beta_0 \alpha, \qquad L \equiv \log \frac{1}{N}, 
\eea 
where $g_0$ collects all the constant terms
and has the expansion
\bea
g_0(\alpha_s) = 1 + \sum_{j=1}^\infty g_{0j} \alpha_s^j, 
\eea
and the Sudakov form factor has the logarithmic expansion
\bea
\mathcal S(\bar \alpha L, \bar \alpha) = \frac{1}{\bar \alpha} g_1(\bar \alpha L) +
g_2(\bar \alpha L) 
+ \bar \alpha g_3 (\bar \alpha L) + \bar \alpha^2 g_4(\bar \alpha L) + \ldots.
\eea

In this Section we use the expansion coefficient of the $\beta$ function defined from
the renormalization-group equation
\bea
\mu^2 \frac{d}{d \mu^2} \as(\mu^2) = \beta(\as(\mu^2))
\eea
where
\bea
\beta(\as) &= - \as^2 (\beta_0 + \beta_1 +\beta_2 \as^2 + \ldots)\nonumber \\
&= -\beta_0 \as^2 (1 + b_1 \as + b_2 \as^2 + \ldots)
\eea
and $\beta_k = b_k\beta_0 $ for $k \geq 1$. So far the first four 
coefficients are known \cite{Vermaseren:1997fq,Czakon:2004bu}. The first three coefficients are
\begin{subequations}
\bea
  \beta_0 &= \frac{11 C_A - 4\,T_F\, n_f}{12\pi} = \frac{33-2n_f}{12\pi}\label{eq:beta0}\\
  \beta_1 &= \frac{17\,C_A^2-(10\,C_A+6\,C_F)\,T_F\,n_f}{24\pi^2}
  = \frac{153-19\,n_f}{24\pi^2}\\
  \beta_2 &= \frac{1}{(4\pi)^3} \bigg[
  \frac{2857}{54}\,C_A^3 + \left( 2\, C_F^2 - \frac{205}{9}\,C_F\, C_A
    - \frac{1415}{27}\,C_A^2 \right) T_F\, n_f\nonumber\\
  &\qquad\qquad + \left( \frac{44}{9}\,C_F + \frac{158}{27}\,C_A \right) T_F^2\, n_f^2 \bigg]
  \nonumber\\
  & = \frac{1}{128 \pi^3}\left[ 2857-\frac{5033}{9}\,n_f 
    +\frac{325}{27}\,n_f^2 \right].
\eea
\end{subequations}

\subsection{Drell-Yan process}

In this Section we give the explicit expressions of the functions $g_i$ which appear
in the resummed Drell-Yan cross-section. We have \cite{Moch:2005ba}
\begin{subequations}
\bea\label{eq:gi}
g_1(\lambda) =&\, \frac{2A_1}{\beta_0} \left[ (1+\lambda)\log(1+\lambda) -\lambda \right] \\
g_2(\lambda) =&\, \frac{A_2}{\beta_0^2} \left[ \lambda - \log(1+\lambda) \right]
   + \frac{A_1}{\beta_0} \left[ \log(1+\lambda) \left( \log\frac{M^2}{\mu_R^2} -2\gamma_E \right) -\lambda \log\frac{\mu_F^2}{\mu_R^2} \right] \nonumber\\
       &\,+ \frac{A_1b_1}{\beta_0^2} \left[ \frac{1}{2}\log^2(1+\lambda) +\log(1+\lambda) -\lambda \right]\\
g_3(\lambda) =&\, \frac{1}{4\beta_0^3} \left( A_3 - A_1  b_2 + A_1  b_1^2 - A_2  b_1 \right) \frac{\lambda^2}{1+\lambda} \nonumber\\
&\, + \frac{A_1 b_1^2}{2\beta_0^3} \; \frac{\log(1+\lambda)}{1+\lambda} \left[ 1 + \frac{1}{2} \log(1+\lambda)\right]
      + \frac{ A_1  b_2 - A_1  b_1^2 }{2\beta_0^3} \; \log(1+\lambda) \nonumber\\
&\, +  \left( \frac{A_1  b_1}{\beta_0^2}  \gamma_E 
          + \frac{A_2  b_1 }{2\beta_0^3}
          \right) \left[ \frac{\lambda}{1+\lambda} 
- \frac{\log(1+\lambda)}{1+\lambda} \right]
\nonumber\\
&\, - \left(
            \frac{A_1  b_2 }{2\beta_0^3}
          + \frac{A_1}{\beta_0}   (\gamma_E^2 + \zeta_2) 
          + \frac{A_2}{\beta_0^2}  \gamma_E
          - \frac{D_2}{4\beta_0^2} 
         \right) 
           \frac{\lambda}{1+\lambda} 
\nonumber\\
&\,
       +   \left[
         \left(
            \frac{A_1}{\beta_0}  \gamma_E
          + \frac{A_2 - A_1  b_1 }{2\beta_0^2}
          \right) 
           \frac{\lambda}{1+\lambda} 
       +  \frac{A_1  b_1}{2\beta_0^2} \;
          \frac{\log(1+\lambda)}{1+\lambda}
          \right] \log\frac{M^2}{\mu_R^2}
\nonumber\\
&\,
       -  \frac{A_2}{2\beta_0^2} \, \lambda \, \log\frac{\mu_F^2}{\mu_R^2}
       + \frac{A_1}{4\beta_0} \left[
         \lambda\, \log^2\frac{\mu_F^2}{\mu_R^2}
          - \frac{\lambda}{1+\lambda} \, \log^2\frac{M^2}{\mu_R^2}
          \right].
\eea
\end{subequations}
The full expression of the coefficient $g_{0k}$ is given by \cite{Moch:2005ky,Bonvini:2012sh}
\begin{subequations}
\begin{align}
g_{01} =&\, \frac{C_F}{\pi}\left[ 4\zeta_2 -4 + 2\gamma_E^2 
+\left(\frac{3}{2}-2\gamma_E\right)\log\frac{M^2}{\mu_F^2} \right]\\
g_{02} =&\, \frac{C_F}{16\pi^2} \Bigg\{
          C_F \left( \frac{511}{4} - 198\,\zeta_2 - 60\,\zeta_3 
          + \frac{552}{5}\,\zeta_2^2 
              - 128\,\gamma_E^2 + 128\,\gamma_E^2 \zeta_2 
              + 32\,\gamma_E^4 \right) \nonumber\\
       &\qquad + C_A \bigg( - \frac{1535}{12} + \frac{376}{3}\,\zeta_2 
                  + {604 \over 9}\,\zeta_3 
                  - \frac{92}{5}\,\zeta_2^2 + \frac{1616}{27}\,\gamma_E
                  - 56\,\gamma_E \zeta_3\nonumber\\
       &\qquad\qquad\qquad+ {536 \over 9}\,\gamma_E^2
                  - 16\,\gamma_E^2 \zeta_2 + {176 \over 9}\,\gamma_E^3 \bigg)
\nonumber\\
&\qquad + n_f \left( {127 \over 6} - {64 \over 3}\,\zeta_2 
+ {8 \over 9}\,\zeta_3 
                - {224 \over 27}\,\gamma_E - {80 \over 9}\,\gamma_E^2 
                - {32 \over 9}\,\gamma_E^3 \right) \nonumber\\
&\qquad + \log^2\frac{M^2}{\mu_F^2} \left[ C_F \left(32\gamma_E^2-48\gamma_E+18\right)
 +C_A \left(\frac{44}{3}\gamma_E-11\right) +n_f\left(2-\frac{8}{3}\gamma_E\right) \right]
\nonumber\\
&\qquad + \log\frac{M^2}{\mu_F^2} 
\Bigg[ C_F \left(48\zeta_3 +72\zeta_2 -93 -128\gamma_E \zeta_2 +128\gamma_E + 48\gamma_E^2 -64\gamma_E^3 \right) \nonumber\\
       &\qquad\qquad\qquad +C_A \left(\frac{193}{3} -24\zeta_3 -\frac{88}{3}\zeta_2 +16\gamma_E\zeta_2 -\frac{536}{9}\gamma_E -\frac{88}{3}\gamma_E^2 \right) \nonumber\\ 
       &\qquad\qquad\qquad +n_f \left(\frac{16}{3}\zeta_2 -\frac{34}{3} +\frac{80}{9}\gamma_E +\frac{16}{3}\gamma_E^2 \right) \Bigg]
       \Bigg\} \nonumber\\
       &\,- \frac{\beta_0 C_F}{\pi}\left[ 4\zeta_2 -4 + 2\gamma_E^2 +\left(\frac{3}{2}-2\gamma_E\right)\log\frac{M^2}{\mu_F^2} \right] \log\frac{\mu_F^2}{\mu_R^2} \; .
\end{align}
\end{subequations}
For completeness, we list the coefficients $\bar g_{0k}$ \cite{Bonvini:2013td}:
\begin{subequations}
\bea
\bar g_{01} =& \frac{C_F}{\pi} (2 \zeta_2 -4)\\
\bar g_{02} =& \frac{C_F}{16 \pi^2} \Bigg[C_A \left(-\frac{12}{5} \zeta_2^2 + \frac{592}{9} \zeta_2 
+28 \zeta_3 - \frac{1535}{12} \right)\nonumber \\ 
&+C_F \left(\frac{72}{5} \zeta_2^2 - 70 \zeta_2 
-60 \zeta_3 + \frac{511}{4}\right) \nonumber \\
&+n_f \left(8 \zeta_3 -\frac{112}{9} \zeta_2 +\frac{127}{6} \right) \Bigg]. 
\eea
\end{subequations}
The coefficients appearing in the previous functions are
\bea
A_1 &= \frac{C_F}{\pi}\nonumber \\
A_2 &= \frac{C_F}{2\pi^2}
\left[ C_A \left(\frac{67}{18} - \frac{\pi^2}{6}\right) - \frac{10}{9}T_F\, n_f \right]\nonumber \\
A_3 &= \frac{C_F}{4\pi^3} \left[ C_A^2 
\left( \frac{245}{24} - \frac{67}{9}\,\zeta_2
+ \frac{11}{6}\,\zeta_3 + \frac{11}{5}\,\zeta_2^2 \right) 
+ \left( -  \frac{55}{24}  + 2\,\zeta_3 \right)C_F \,n_f  \right. 
\nonumber\\ & 
\left. \mbox{} \qquad
+ \left( - \frac{209}{108} + \frac{10}{9}\,\zeta_2 -\frac{7}{3}\zeta_3 \right)
C_A \,n_f - \frac{1}{27} \, n_f^2 \right]
\nonumber\\
D_2 &= \frac{C_F}{16\pi^2}
\left[ C_A\left(-\frac{1616}{27}+\frac{88}{9}\pi^2+56\zeta_3\right) 
+ \left(\frac{224}{27}-\frac{16}{9}\pi^2\right) n_f \right].
\eea

\subsection{Higgs production}

The Higgs resummation coefficient are simply related to those of Drell-Yan 
process and can be obtained by \cite{Moch:2005ba}
\bea
A_k^\textrm{Higgs} = \frac{C_A}{C_F} A_k^\textrm{DY},\qquad 
D_k^\textrm{Higgs} = \frac{C_A}{C_F} D_k^\textrm{DY}.
\eea
The coefficients $g_{0k}$, which collect the constant terms, can be found in Ref.~\cite{Catani:2003zt} with 
full scale dependence. They are:
\begin{subequations}
\bea
g_{01} =& \frac{C_A}{\pi} \left[ 4\zeta_2 + 2\gamma_E^2 + \frac23\pi\beta_0\log\frac{\mu_R^2}{\mu_F^2} - 2\gamma_E \log\frac{M^2}{\mu_F^2}\right]
+ \frac{11}{2\pi} \\
g_{02} =&  \frac{1}{\pi^2} \Bigg\{\delta g_{02} + \gamma_E \left(\frac{101}{3} - \frac{14}{9}n_f - \frac{63}{2} \zeta_3 \right)
          + \gamma_E^2 \left(\frac{133}{2}-\frac{5}{3}n_f+\frac{63}{2} \zeta_3 \right) 
          + 4\gamma_E^3\pi b_0 + 18 \gamma_E^4\nonumber\\
        & + \left(\frac{133}{2}-\frac{5}{3}n_f \right)\zeta_2 + \frac{261}{2} \zeta_4 
          + \left( 22-\frac{4}{3} n_f\right) \zeta_3\nonumber\\
        &  + \left[ \left( -\frac{165}{4}+\frac{5}{2} n_f\right) \gamma_E + 18 \gamma_E^2
          + 18 \zeta_2 \right] \log^2\frac{M^2}{\mu_F^2}\nonumber\\
        &  + 18 \pi \gamma_E b_0  \log \frac{M^2}{\mu_F^2}\log \frac{M^2}{\mu_R^2}
           -18 \pi b_0 (\gamma_E^2+\zeta_2)\log^2 \frac{M^2}{\mu_R^2}\nonumber\\
        &  + \left[ 12\left( \pi b_0-3 \gamma_E\right) \gamma_E^2 + \left(-\frac{133}{2} +\frac{5}{3} n_f
          -63 \zeta_2 \right) \gamma_E + 12 \pi b_0 \zeta_2 -72 \zeta_3 \right] \log \frac{M^2}{\mu_F^2} \Bigg\}\nonumber\\
\delta g_{02} =& \frac{11399}{144} + \frac{133}{2} \zeta_2-\frac{9}{20}\zeta_2^2 -\frac{165}{4}\zeta_3\nonumber\\
               & + \left(\frac{19}{8}+\frac{2}{3} n_f \right) \log \frac{M^2}{m_t^2} + n_f \left(
                   -\frac{1189}{144}-\frac{5}{3} \zeta_2 +\frac{5}{6} \zeta_3\right) 
                 + 3 b_0^2 \pi^2 \log^2 \frac{\mu_F^2}{\mu_R^2}- 18 \zeta_2 \log^2 \frac{M^2}{\mu_F^2}\nonumber\\
			   & + \left( \frac{169}{4} + \frac{171}{2} \zeta_3 -\frac{19}{6} n_f + 12 \pi b_0 \zeta_2\right)\log \frac{M^2}{\mu_F^2}\nonumber\\
			   & + \left(- \frac{465}{8} + \frac{13}{3} n_f - 18 \pi b_0 \zeta_2 \right) \log \frac{M^2}{\mu_R^2} .
\eea
\end{subequations}
The function $\bar g_0(\as)$ is related to $g_0(\as)$ by \cite{Ball:2013bra}
\bea\label{eq:g0bardefapp}
\bar g_0(\as)=g_0(\as) 
\exp\left[-\sum_{n=1}^\infty \as^n \sum_{k=0}^nb_{n,k}
\frac{\Gamma^{(k+1)}(1)}{k+1}\right].
\eea
where (omitting scale dependence)
\begin{subequations}
\bea
b_{1,1} &= \frac{4C_A}\pi, \\
b_{1,0} &= 0,\\
b_{2,2} &=\frac{1}{\pi^2}\left(-\frac{11}3C_A^2+\frac23 C_An_f\right)\\
b_{2,1} &=\frac{1}{\pi^2}
\left[\left(\frac{67}9-2\zeta_2\right)C_A^2 - \frac{10}{9}C_An_f\right]\\
b_{2,0} &=\frac{1}{\pi^2}\left[ 
\left(-\frac{101}{27}+\frac{11}3\zeta_2+\frac{7}2\zeta_3\right)C_A^2
             +\left(\frac{14}{27}-\frac23\zeta_2\right)C_An_f\right].
\eea
\end{subequations}
The coefficients $\bar g_{0k}$ can be straightforwardly determined as
\bea
\bar g_{0,n}=g_{0,n}-r_n
\eea
where $r_n$ can be read off Eq.~\eqref{eq:g0bardefapp}
order by order in $\as$. We obtain, setting $C_A=3$ and omitting scale 
dependence,
\begin{subequations}
\bea
\bar g_{01} =& \frac{11}{2 \pi}+ \pi\\
\bar g_{02} =& 
      \frac{133}{12} - \frac{165 \zeta_3}{4 \pi^2}  + \frac{11399}{144\pi^2}
    + \frac{3 \pi^2}{16}\nonumber \\
    & +\left( \frac{19}{8 \pi^2}+  \frac{2}{3 \pi^2} n_f \right) \log\frac{M^2}{m_t^2}\nonumber \\
    &+ n_f \left(\frac{5 \zeta_3}{6 \pi^2}-\frac{1189}{144 \pi^2} - \frac{5}{18}  \right).
\eea
\end{subequations}

\subsection{Matching}

It is necessary to compute and subtract double counting terms when matching the resummed 
expression with the fixed order results. In order to do so, one first expand 
Eqs.~(\ref{eq:gi}) in powers of $\lambda$:
\begin{subequations}
\bea
g_1(\lambda) 
=&\, \frac{A_1}{\beta_0} \left[ \lambda^2-\frac{1}{3}\lambda^3 
+ \mathcal O(\lambda^4) \right] \\
g_2(\lambda) 
=&\, \frac{A_1}{\beta_0} \left(\log\frac{M^2}{\mu_F^2}-2\gamma_E\right) \lambda
+ \left( \frac{A_2}{2\beta_0^2} + \frac{A_1}{2\beta_0} 
\left(2\gamma_E - \log\frac{M^2}{\mu_R^2}\right) \right) \lambda^2
            + \mathcal O(\lambda^3) \\
g_3(\lambda) 
=&\, \left(-\frac{A_1}{\beta_0}(\gamma_E^2+\zeta_2)
-\frac{A_2}{\beta_0^2}\gamma_E+\frac{D_2}{4\beta_0^2}\right) \lambda 
\nonumber\\
&\,+\left( \frac{A_1}{\beta_0}\gamma_E\log\frac{M^2}{\mu_R^2} 
+\frac{A_2}{2\beta_0^2}\log\frac{M^2}{\mu_F^2}
 - \frac{A_1}{4\beta_0}\left(\log^2\frac{M^2}{\mu_R^2}-\log^2\frac{\mu_F^2}{\mu_R^2}\right)
            \right) \lambda + \mathcal O(\lambda^2).
\eea
\end{subequations}
It is possible to obtain the double-counting term as the Taylor expansion in
powers of $\as$ of $g_0(\as) \exp \mathcal S(\lambda, \bar \alpha)$:
\bea
g_0(\as) \exp \mathcal S(\lambda, \bar \alpha) &= (1+ \as g_{01} + \as g_{02}^2 + \ldots)
e^{\as \mathcal  S_1 + \as^2 \mathcal S_2+ \ldots}\nonumber \\
&= 1+ (\mathcal S_1 + g_{01}) \as +
\left( \frac{\mathcal S_1^2}{2} + \mathcal S_2 + \mathcal S_1 g_{01} + g_{02} \right) \as^2 
+\mathcal O(\as^3).
\eea
In particular, one gets
\begin{align}
\as\, \mathcal S_1 
&=\left[\frac{A_1}{\beta_0} \left(\frac{\lambda}{\bar \alpha} + \log\frac{M^2}{\mu_F^2}
 - 2\gamma_E \right)\right] \lambda\\
\as^2\, \mathcal S_2 &= \Bigg[ - \frac{A_1}{3\beta_0} \frac{\lambda}{\bar \alpha}
+ \left( \frac{A_2}{2\beta_0^2} + \frac{A_1}{2\beta_0}
\left(2\gamma_E - \log\frac{M^2}{\mu_R^2}\right) \right)  
              +\bigg\{-\frac{A_1}{\beta_0}(\gamma_E^2+\zeta_2)
-\frac{A_2}{\beta_0^2}\gamma_E+\frac{D_2}{4\beta_0^2} \nonumber\\
&\qquad\qquad
            +\frac{A_1}{\beta_0}\gamma_E\log\frac{M^2}{\mu_R^2} 
+\frac{A_2}{2\beta_0^2}\log\frac{M^2}{\mu_F^2}
            - \frac{A_1}{4\beta_0}\left(\log^2\frac{M^2}{\mu_R^2}-
\log^2\frac{\mu_F^2}{\mu_R^2}\right)\bigg\} \frac{\bar \alpha}{\lambda} \Bigg] \lambda^2.
\end{align}

\section{Soft-gluon resummation formulae in SCET}

In this Section we list the perturbative expansions of the various matching 
coefficients and anomalous dimensions required to evaluate the RG-improved 
result. We use the same notation of Refs.~\cite{Becher:2007ty,Ahrens:2008nc} where the $\beta$ function is defined as
\begin{eqnarray}
   \beta(\alpha_s) &=& -2\alpha_s \left[
    \beta_0\,\frac{\alpha_s}{4\pi}
    + \beta_1 \left( \frac{\alpha_s}{4\pi} \right)^2
    + \beta_2 \left( \frac{\alpha_s}{4\pi} \right)^3 + \dots \right].
\end{eqnarray}
The expansion coefficients for the QCD $\beta$-function to four-loop order are
\begin{eqnarray}
   \beta_0 &=& \frac{11}{3}\,C_A - \frac43\,T_F n_f \,, \nonumber\\
   \beta_1 &=& \frac{34}{3}\,C_A^2 - \frac{20}{3}\,C_A T_F n_f
    - 4 C_F T_F n_f \,, \\
   \beta_2 &=& \frac{2857}{54}\,C_A^3 + \left( 2 C_F^2
    - \frac{205}{9}\,C_F C_A - \frac{1415}{27}\,C_A^2 \right) T_F n_f
    + \left( \frac{44}{9}\,C_F + \frac{158}{27}\,C_A 
    \right) T_F^2 n_f^2 \,, \nonumber\\
   \beta_3 &=& \frac{149753}{6} + 3564\zeta_3
    - \left( \frac{1078361}{162} + \frac{6508}{27}\,\zeta_3 
    \right) n_f
    + \left( \frac{50065}{162} + \frac{6472}{81}\,\zeta_3 
    \right) n_f^2
    + \frac{1093}{729}\,n_f^3 \,. \nonumber
\end{eqnarray}
The value of $\beta_3$
corresponds to $N_c=3$ and $T_F=\frac12$. 

\subsection{Drell-Yan production}\label{sect:dyapp}

\subsubsection{Two-loop matching coefficients}

The Wilson coefficient $C_V$ has the expansion \cite{Becher:2006mr}
\begin{equation}
   C_V(-M^2-i\epsilon,\mu) = 1 + \frac{C_F\alpha_s}{4\pi}
    \left( - L^2 + 3L - 8 + \frac{\pi^2}{6} \right)
    + C_F \left( \frac{\alpha_s}{4\pi} \right)^2 \left[
    C_F H_F + C_A H_A + T_F n_f H_f \right] ,
\end{equation}
where
\begin{eqnarray}
   H_F &=& \frac{L^4}{2} - 3L^3
    + \left( \frac{25}{2} - \frac{\pi^2}{6} \right) L^2
    + \left( - \frac{45}{2} - \frac{3\pi^2}{2} + 24\zeta_3 \right) L
    + \frac{255}{8} + \frac{7\pi^2}{2} - \frac{83\pi^4}{360} 
    - 30\zeta_3 \,, \nonumber\\
   H_A &=& \frac{11}{9}\,L^3
    + \left( - \frac{233}{18} + \frac{\pi^2}{3} \right) L^2
    + \left( \frac{2545}{54} + \frac{11\pi^2}{9} - 26\zeta_3 \right) L 
    \nonumber\\
   &&\mbox{}- \frac{51157}{648} - \frac{337\pi^2}{108} + \frac{11\pi^4}{45}
    + \frac{313}{9}\,\zeta_3 \,, \nonumber\\
   H_f &=& - \frac49\,L^3 + \frac{38}{9}\,L^2 
    + \left( - \frac{418}{27} - \frac{4\pi^2}{9} \right) L
    + \frac{4085}{162} + \frac{23\pi^2}{27} + \frac49\,\zeta_3 \,,
\end{eqnarray}
and $L=\ln(M^2/\mu^2)-i\pi$.

The two-loop expression for the soft function reads
\begin{equation}
   \widetilde s_{\rm DY}(L,\mu)
   = 1 + \frac{C_F\alpha_s}{4\pi} 
   \left( 2L^2 + \frac{\pi^2}{3} \right)
   + C_F \left( \frac{\alpha_s}{4\pi} \right)^2
   \left[ C_F W_F + C_A W_A + T_F n_f W_f \right] ,
\end{equation}
where
\begin{eqnarray}
   W_F &=& 2L^4 + \frac{2\pi^2}{3}\,L^2 + \frac{\pi^4}{18} 
    = \frac12 \left( 2L^2 + \frac{\pi^2}{3} \right)^2 , \nonumber\\
   W_A &=& - \frac{22}{9}\,L^3
    + \left( \frac{134}{9} - \frac{2\pi^2}{3} \right) L^2
    + \left( - \frac{808}{27} + 28\zeta_3 \right) L
    + \frac{2428}{81} + \frac{67\pi^2}{54} - \frac{\pi^4}{3}
    - \frac{22}{9}\,\zeta_3 \,, \nonumber\\
   W_f &=& \frac89\,L^3 - \frac{40}{9}\,L^2 + \frac{224}{27}\,L
    - \frac{656}{81} - \frac{10\pi^2}{27} + \frac89\,\zeta_3 \,.
\end{eqnarray}

\subsubsection{Three-loop anomalous dimensions}

Here we list expressions for the anomalous dimensions
quoting all results in the $\overline{{\rm MS}}$ 
renormalization scheme. The expansion coefficients of the anomalous 
dimensions are defined as
\begin{eqnarray}
   \Gamma_{\rm cusp}(\alpha_s) &=& \Gamma_0\,\frac{\alpha_s}{4\pi}
    + \Gamma_1 \left( \frac{\alpha_s}{4\pi} \right)^2
    + \Gamma_2 \left( \frac{\alpha_s}{4\pi} \right)^3 + \dots \,
\end{eqnarray}
and similarly for the other anomalous dimensions. 

The expansion coefficients of the cusp anomalous dimension $\Gamma_{\rm cusp}$ are \cite{Moch:2004pa}
\begin{eqnarray}\label{eq:gcuspfund}
   \Gamma_0 &=& 4 C_F \,, \nonumber\\
   \Gamma_1 &=& 4 C_F \left[ \left( \frac{67}{9} 
    - \frac{\pi^2}{3} \right) C_A - \frac{20}{9}\,T_F n_f \right] ,
    \nonumber\\
   \Gamma_2 &=& 4 C_F \Bigg[ C_A^2 \left( \frac{245}{6} 
    - \frac{134\pi^2}{27}
    + \frac{11\pi^4}{45} + \frac{22}{3}\,\zeta_3 \right) 
    + C_A T_F n_f  \left( - \frac{418}{27} + \frac{40\pi^2}{27}
    - \frac{56}{3}\,\zeta_3 \right) \nonumber\\
   &&\mbox{}+ C_F T_F n_f \left( - \frac{55}{3} + 16\zeta_3 \right) 
    - \frac{16}{27}\,T_F^2 n_f^2 \Bigg] \,.
\end{eqnarray}
The anomalous dimension $\gamma^V$ is
\begin{eqnarray}
   \gamma_0^V &=& -6 C_F \,, \nonumber\\
   \gamma_1^V &=& C_F^2 \left( -3 + 4\pi^2 - 48\zeta_3 \right)
    + C_F C_A \left( - \frac{961}{27} - \frac{11\pi^2}{3} 
    + 52\zeta_3 \right)
    + C_F T_F n_f \left( \frac{260}{27} + \frac{4\pi^2}{3} \right) ,
    \nonumber\\
   \gamma_2^V &=& C_F^3 \left( -29 - 6\pi^2 - \frac{16\pi^4}{5}
    - 136\zeta_3 + \frac{32\pi^2}{3}\,\zeta_3 + 480\zeta_5 \right) 
    \nonumber\\
   &&\mbox{}+ C_F^2 C_A \left( - \frac{151}{2} + \frac{410\pi^2}{9}
    + \frac{494\pi^4}{135} - \frac{1688}{3}\,\zeta_3
    - \frac{16\pi^2}{3}\,\zeta_3 - 240\zeta_5 \right) \nonumber\\
   &&\mbox{}+ C_F C_A^2 \left( - \frac{139345}{1458} - \frac{7163\pi^2}{243}
    - \frac{83\pi^4}{45} + \frac{7052}{9}\,\zeta_3
    - \frac{88\pi^2}{9}\,\zeta_3 - 272\zeta_5 \right) \nonumber\\
   &&\mbox{}+ C_F^2 T_F n_f \left( \frac{5906}{27} - \frac{52\pi^2}{9} 
    - \frac{56\pi^4}{27} + \frac{1024}{9}\,\zeta_3 \right) 
    \nonumber\\
   &&\mbox{}+ C_F C_A T_F n_f \left( - \frac{34636}{729}
    + \frac{5188\pi^2}{243} + \frac{44\pi^4}{45} 
    - \frac{3856}{27}\,\zeta_3 \right) \nonumber\\
   &&\mbox{}+ C_F T_F^2 n_f^2 \left( \frac{19336}{729} 
    - \frac{80\pi^2}{27} - \frac{64}{27}\,\zeta_3 \right) .
\end{eqnarray}
The anomalous dimension $\gamma^\phi$ is know to three-loop
order from the NNLO calculation of the Altarelli-Parisi splitting
functions \cite{Moch:2004pa}. The expansion coefficients are
\begin{eqnarray}
   \gamma_0^\phi &=& 3 C_F \,, \nonumber\\
   \gamma_1^\phi 
   &=& C_F^2 \left( \frac{3}{2} - 2\pi^2 + 24\zeta_3 \right) 
    + C_F C_A \left( \frac{17}{6} + \frac{22\pi^2}{9} 
    - 12\zeta_3 \right)
    - C_F T_F n_f \left( \frac{2}{3} + \frac{8\pi^2}{9} \right) ,
    \nonumber\\
   \gamma_2^\phi
   &=& C_F^3 \left( \frac{29}{2} + 3\pi^2 + \frac{8\pi^4}{5} 
    + 68\zeta_3 - \frac{16\pi^2}{3}\,\zeta_3 - 240\zeta_5 \right)
    \nonumber\\
   &&\mbox{}+ C_F^2 C_A \left( \frac{151}{4} - \frac{205\pi^2}{9}
    - \frac{247\pi^4}{135} + \frac{844}{3}\,\zeta_3
    + \frac{8\pi^2}{3}\,\zeta_3 + 120\zeta_5 \right) \nonumber\\
   &&\mbox{}+ C_F^2 T_F n_f \left( - 46 + \frac{20\pi^2}{9}
    + \frac{116\pi^4}{135} - \frac{272}{3}\,\zeta_3 \right) 
    \nonumber\\
   &&\mbox{}+ C_F C_A^2 \left( - \frac{1657}{36}
    + \frac{2248\pi^2}{81} - \frac{\pi^4}{18} 
    - \frac{1552}{9}\,\zeta_3 + 40\zeta_5 \right) \nonumber\\
   &&\mbox{}+ C_F C_A T_F n_f \left( 40 - \frac{1336\pi^2}{81}
    + \frac{2\pi^4}{45} + \frac{400}{9}\,\zeta_3 \right) \nonumber\\
   &&\mbox{}+ C_F T_F^2 n_f^2 \left( - \frac{68}{9} 
    + \frac{160\pi^2}{81} - \frac{64}{9}\,\zeta_3 \right) .
\end{eqnarray}
Using these results, one can compute the expansion coefficients
for the anomalous dimension $\gamma^W$ of the Drell-Yan soft function
from the relation $\gamma^W=2\gamma^\phi+\gamma^V$. This yields for the first
two coefficients
\begin{eqnarray}
   \gamma_0^W &=& 0 \,, \nonumber\\
   \gamma_1^W &=& C_F C_A \left( - \frac{808}{27} + \frac{11\pi^2}{9} 
    + 28\zeta_3 \right)
    + C_F T_F n_f \left( \frac{224}{27} - \frac{4\pi^2}{9} \right) .
\end{eqnarray}

\subsubsection{Renormalization-group functions}

We now give the perturbative expansions of the functions $S$ and $a_\Gamma$.
The resulting expression for $a_\Gamma$ is given by
\begin{eqnarray}\label{asol}
   a_\Gamma(\nu,\mu)
   &=& \frac{\Gamma_0}{2\beta_0}\,\Bigg\{
    \ln\frac{\alpha_s(\mu)}{\alpha_s(\nu)}
    + \left( \frac{\Gamma_1}{\Gamma_0} - \frac{\beta_1}{\beta_0} 
    \right) \frac{\alpha_s(\mu) - \alpha_s(\nu)}{4\pi} \nonumber\\ 
   &&\mbox{}+ \left[ \frac{\Gamma_2}{\Gamma_0}
    - \frac{\beta_2}{\beta_0} - \frac{\beta_1}{\beta_0}
    \left( \frac{\Gamma_1}{\Gamma_0} - \frac{\beta_1}{\beta_0} 
    \right) \right]
    \frac{\alpha_s^2(\mu) - \alpha_s^2(\nu)}{32\pi^2} + \dots \Bigg\}
    \,.
\end{eqnarray}
Similar expressions with the $\Gamma_i$ replaced by the coefficients $\gamma_i^V$ or $\gamma_i^\phi$ hold for the functions $a_{\gamma^V}$ and $a_{\gamma^\phi}$, respectively.
The expression for the Sudakov exponent $S$ are \cite{Becher:2006mr}
\begin{eqnarray}
   S(\nu,\mu) 
   &=& \frac{\Gamma_0}{4\beta_0^2}\,\Bigg\{
    \frac{4\pi}{\alpha_s(\nu)} \left( 1 - \frac{1}{r} - \ln r \right)
    + \left( \frac{\Gamma_1}{\Gamma_0} - \frac{\beta_1}{\beta_0}
    \right) (1-r+\ln r) + \frac{\beta_1}{2\beta_0} \ln^2 r \nonumber\\
   &&\mbox{}+ \frac{\alpha_s(\nu)}{4\pi} \Bigg[ 
    \left( \frac{\beta_1\Gamma_1}{\beta_0\Gamma_0} 
    - \frac{\beta_2}{\beta_0} \right) (1-r+r\ln r)
    + \left( \frac{\beta_1^2}{\beta_0^2} 
    - \frac{\beta_2}{\beta_0} \right) (1-r)\ln r \nonumber\\
   &&\hspace{1.0cm}
    \mbox{}- \left( \frac{\beta_1^2}{\beta_0^2} 
    - \frac{\beta_2}{\beta_0}
    - \frac{\beta_1\Gamma_1}{\beta_0\Gamma_0} 
    + \frac{\Gamma_2}{\Gamma_0} \right) \frac{(1-r)^2}{2} \Bigg]
    \,,
\end{eqnarray}
where $r=\alpha_s(\mu)/\alpha_s(\nu)$.

\subsection{Higgs production}

\subsubsection{Two-loop matching coefficients}

To NNLO, the short-distance coefficient $C_t$ is \cite{Kramer:1996iq,Chetyrkin:1997iv}
\begin{eqnarray}\label{Ct}
   C_t(m_t^2,\mu^2)
   &=& 1 + \frac{\alpha_s(\mu^2)}{4\pi}\,(5C_A-3C_F) \nonumber\\
   &&\mbox{}+ \left( \frac{\alpha_s(\mu^2)}{4\pi} \right)^2
    \bigg[ \frac{27}{2}\,C_F^2 
    + \left( 11\ln\frac{m_t^2}{\mu^2} - \frac{100}{3} \right) C_F C_A 
    - \left( 7\ln\frac{m_t^2}{\mu^2} - \frac{1063}{36} \right) C_A^2 
    \nonumber\\
   &&\quad\mbox{}- \frac{4}{3}\,C_F T_F - \frac{5}{6}\,C_A T_F 
   - \left( 8\ln\frac{m_t^2}{\mu^2} + 5 \right) C_F T_F n_f 
   - \frac{47}{9}\,C_A T_F n_f \bigg] \,.
\end{eqnarray} 
The two-loop expression for the Wilson coefficient $C_S$ can be extracted
from the results of \cite{Harlander:2000mg}. We write its perturbative series in the form
\begin{equation}\label{CSexp}
   C_S(-m_H^2-i\epsilon,\mu^2) = 1 + \sum_{n=1}^\infty\,c_n(L)
   \left( \frac{\alpha_s(\mu^2)}{4\pi} \right)^n \! ,
\end{equation}
where $L=\ln[(-m_H^2-i\epsilon)/\mu^2]$. The one- and two-loop coefficients are found to be
\begin{equation}\label{c1c2}
\begin{aligned}
   c_1(L) &= C_A \left( -L^2 + \frac{\pi^2}{6} \right) , \\
   c_2(L) &= C_A^2 \left[ \frac{L^4}{2} + \frac{11}{9}\,L^3
    + \left( -\frac{67}{9} + \frac{\pi^2}{6} \right) L^2 
    + \left( \frac{80}{27} - \frac{11\pi^2}{9} - 2\zeta_3 \right) L
    \right. \\
   &\qquad + \left. \frac{5105}{162} + \frac{67\pi^2}{36}
    + \frac{\pi^4}{72} - \frac{143}{9}\,\zeta_3 \right]
    + C_F T_F n_f \left( 4L - \frac{67}{3} + 16\zeta_3 \right) \\
  &\quad\mbox{}+ C_A T_F n_f \left[ -\frac{4}{9}\,L^3
   + \frac{20}{9}\,L^2 
   + \left( \frac{104}{27} + \frac{4\pi^2}{9} \right) L 
   - \frac{1832}{81} - \frac{5\pi^2}{9} - \frac{92}{9}\,\zeta_3 
   \right] .
\end{aligned}
\end{equation}
The associated soft function $\widetilde s_{\rm Higgs}$ is obtained from that 
in the Drell-Yan case by the replacement $C_F\to C_A$. This yields
\begin{equation}\label{tils}
  \widetilde s_{\rm Higgs}(L,\mu^2) 
  = 1 + \frac{\alpha_s(\mu^2)}{4\pi}\,C_A 
  \left( 2L^2 + \frac{\pi^2}{3} \right) 
  + \left( \frac{\alpha_s(\mu^2)}{4\pi} \right)^2
  \left( C_A^2\,W_A + C_A T_F n_f\,W_f \right) ,
\end{equation}
with
\begin{equation}
\begin{aligned}
   W_A &= 2L^4 - \frac{22}{9}\,L^3 + \frac{134}{9}\,L^2
    + \left( -\frac{808}{27} + 28\zeta_3 \right) L 
    + \frac{2428}{81} + \frac{67\pi^2}{54} - \frac{5\pi^4}{18} 
    - \frac{22}{9}\,\zeta_3 \,, \\
   W_f &= \frac{8}{9}\,L^3 - \frac{40}{9}\,L^2 + \frac{224}{27}\,L
    - \frac{656}{81} - \frac{10\pi^2}{27} + \frac{8}{9}\,\zeta_3 \,.
\end{aligned}
\end{equation}

\subsubsection{Three-loop anomalous dimensions}

The cusp anomalous dimension in the adjoint representation is given
(at least up to three-loop order) by $C_A/C_F$ times that in the fundamental 
representation Eq.~\eqref{eq:gcuspfund}. The explicit expressions for the evolution functions
$S$ and $a_\Gamma$ have the form shown in Sect.~\ref{sect:dyapp}.

The first three expansion coefficients
of the anomalous dimension $\gamma^S$ entering the evolution 
equation of the matching coefficient $C_S$ are
\cite{Idilbi:2005ni,Idilbi:2005er}
\begin{equation}
\begin{aligned}
  \gamma^S_0 &= 0 \,, \\
  \gamma^S_1 
  &= C_A^2 \left( -\frac{160}{27} + \frac{11\pi^2}{9}
   + 4\zeta_3 \right) 
   + C_A T_F n_f \left( -\frac{208}{27} - \frac{4\pi^2}{9} \right) 
   - 8 C_F T_F n_f \,, \\
  \gamma^S_2 
  &= C_A^3 \left[ \frac{37045}{729} + \frac{6109\pi^2}{243}
   - \frac{319\pi^4}{135} 
   + \left( \frac{244}{3} - \frac{40\pi^2}{9} \right) \zeta_3 
   - 32\zeta_5 \right] \\
  &\quad\mbox{}+ C_A^2 T_F n_f \left( -\frac{167800}{729}
   - \frac{2396\pi^2}{243} + \frac{164\pi^4}{135} 
   + \frac{1424}{27}\,\zeta_3 \right) \\
  &\quad\mbox{}+ C_A C_F T_F n_f \left( \frac{1178}{27}
   - \frac{4\pi^2}{3} - \frac{16\pi^4}{45} 
   - \frac{608}{9}\,\zeta_3 \right) + 8 C_F^2 T_F n_f \\  
  &\quad\mbox{}+ C_A T_F^2 n_f^2 \left( \frac{24520}{729}
   + \frac{80\pi^2}{81} - \frac{448}{27}\,\zeta_3 \right) 
   + \frac{176}{9} C_F T_F^2 n_f^2 \,.
\end{aligned}
\end{equation}
The first three coefficients of the anomalous
dimension $\gamma^B$, which equals one half of the 
coefficient of the $\delta(1-x)$ term in the Altarelli-Parisi splitting function
$P_{gg}(x)$, are \cite{Vogt:2004mw}
\begin{equation}
\begin{aligned}
  \gamma^B_0 
  &= \frac{11}{3}\,C_A - \frac{4}{3}\,T_F n_f
   = \beta_0 \,, \\
  \gamma^B_1 
  &= 4 C_A^2 \left( \frac{8}{3} + 3\zeta_3 \right) 
   - \frac{16}{3}\,C_A T_F n_f - 4 C_F T_F n_f \,, \\
  \gamma^B_2 
  &= C_A^3 \left[ \frac{79}{2} + \frac{4\pi^2}{9} + \frac{11\pi^4}{54} 
   + \left( \frac{536}{3} - \frac{8\pi^2}{3} \right) \zeta_3 
   - 80\zeta_5 \right] \\
  &\quad\mbox{}+ C_A^2 T_F n_f \left( \frac{233}{9} + \frac{8\pi^2}{9}
   + \frac{2\pi^4}{27} + \frac{160}{3}\,\zeta_3 \right) \\
  &\quad\mbox{}- \frac{241}{9}\,C_A C_F T_F n_f + 2 C_F^2 T_F n_f
   + \frac{58}{9}\,C_A T_F^2 n_f^2 + \frac{44}{9}\,C_F T_F^2 n_f^2 \,.
\end{aligned}
\end{equation}
From the relation
$\gamma^W = \frac{\beta(\as)}{\as} + \gamma^t 
+ \gamma^S + 2 \gamma^B$, where  $\gamma^t= \as^2 \frac{d}{d \as} \frac{\beta(\as)}{\as^2} $
we obtain
\begin{eqnarray}
   \gamma_0^W &=& 0 \,, \nonumber\\
   \gamma_1^W &=&  C_A^2 \left( - \frac{808}{27} + \frac{11\pi^2}{9} 
    + 28\zeta_3 \right)
    + C_A T_F n_f \left( \frac{224}{27} - \frac{4\pi^2}{9} \right) .
\end{eqnarray}

\cleardoublepage
\phantomsection

\renewcommand{\bibname}{References} % Al posto di Referenze puoi mettere quello che preferisci
								   % viene cambiato il nome solamente nell'indice, non della
								   % sezione, che è Riferimenti Bibliografici di default.
\fancyhf{}
\fancyhead[RE]{\nouppercase{\textbf{\textsf{\leftmark}}}}
\fancyhead[LE,RO]{\bfseries\thepage}
\renewcommand{\refname}{References} % Al posto di Referenze puoi mettere quello che preferisci
								   % viene cambiato il nome solamente nel paragrafo, ma non
								   % nell'indice. 
\addcontentsline{toc}{chapter}{\bibname}  % questo comando aggiunge all'indice la sezione 
										  % della bibliografia

\end{document}